\documentclass{ws-ijmpa}
\usepackage[super,compress]{cite}
\usepackage{graphicx}
\usepackage{here}
\def\beq{\begin{equation}}
\def\eeq{\end{equation}}
\def\bea{\begin{eqnarray}}
\def\eea{\end{eqnarray}}
\def\bq{\begin{quote}}
\def\eq{\end{quote}}

\def\nnb{\nonumber}
\def\ga{\left(}
\def\dr{\right)}

\def\lrar{\Longrightarrow}
		
\def\nnb{\nonumber}
\def\la{\langle}
\def\ra{\rangle}
\def\nin{\noindent}
\def\ba{\vspace*{-0.2cm}\begin{array}}
\def\ea{\end{array}\vspace*{-0.2cm}}

\def\b{$\bullet~$}
\def\als{\alpha_s}

\def\gg2{ \la\alpha_s G^2 \ra}
\def\gg3{g^3f_{abc}\la G^aG^bG^c \ra}
\def\ggg4{\la\als^2G^4\ra}

\def\beq{\begin{equation}}
\def\enq{\end{equation}}
\def\beqa{\begin{eqnarray}}
\def\enqa{\end{eqnarray}}
\def\nnb{\nonumber}

\def\qq{\lag\bar{q}q\rag}

\def\mix{\la \bar q G q\ra}

\def\GGG{\lag g^3G^3\rag}

\def\lb{\label}
\def\nn{\nonumber}

\newcommand{\rag}{\rangle}
\newcommand{\lag}{\langle}

\def\lv{\mathcal{L}_v}
\def\lp{\mathcal{L}_+}
\def\lid{\mbox{Li}_2}
\def\ln{\mbox{Log}}
\def\gg{\lag g^{2}_{s} G^2 \rag}
\def\ggg{\lag g^{3}_{s}G^3\rag}
\def\qGq{\lag\bar{q}Gq\rag}
\def\GG{\lag G^2 \rag}
\def\gGG{\lag g_s^2 G^2 \rag}
\def\GGG{\lag G^3 \rag}
\def\gGGG{\lag g_s^3G^3\rag}

\begin{document}
\markboth{R. Albuquerque et al.}
{XYZ-like spectra...}

\title{XYZ-like Spectra  from Laplace Sum Rule at N2LO in the Chiral Limit}
\author{R. Albuquerque}
\address{Faculty of Technology,Rio de Janeiro State University (FAT,UERJ), Brazil\\
Email address:  raphael.albuquerque@uerj.br}
\author{S. Narison}
\address{Laboratoire
Univers et Particules de Montpellier (LUPM), CNRS-IN2P3, \\
Case 070, Place Eug\`ene
Bataillon, 34095 - Montpellier, France.\\
Email address: snarison@yahoo.fr}

\author{F. Fanomezana$^{*}$, A. Rabemananjara,  D. Rabetiarivony$^{*}$ and G. Randriamanatrika$^{*}$}

\address{Institute of High-Energy Physics of Madagascar (iHEPMAD)\\
University of Ankatso,
Antananarivo 101, Madagascar}


\maketitle

\begin{history}
\end{history}

\begin{abstract}
\nin
We  present new compact integrated expressions of QCD spectral functions of heavy-light molecules and four-quark $XYZ$-like states at lowest order (LO) of perturbative (PT) QCD and up to $d=8$ condensates of the Operator Product Expansion (OPE). 
Then, by including up to next-to-next leading order (N2LO)  PT QCD corrections, which we have estimated by  assuming the factorization of the four-quark spectral functions, we improve previous  LO results from QCD spectral sum rules (QSSR), on the $XYZ$-like masses and decay constants which suffer from the ill-defined heavy quark mass. PT N3LO corrections are estimated using a geometric growth of the PT series and are included in the systematic errors. 
Our optimal results based on stability criteria are summarized in Tables\,\ref{tab:resultc} to \ref{tab:4q-resultb} and compared, in Section\,\ref{sec:confront}, with  experimental candidates and some LO QSSR results. We conclude that the masses of the $XZ$ observed states are compatible with  (almost) pure $J^{PC}=1^{+\pm}, 0^{++}$ molecule or/and four-quark states.  
The ones of the $1^{-\pm}, 0^{-\pm}$ molecule / four-quark states are about 1.5 GeV above the $Y_{c,b}$  mesons experimental candidates and hadronic thresholds. 
We also find that the couplings of
these exotics
 to the associated interpolating currents are weaker than that of ordinary $D,B$ mesons ($f_{DD}\approx 10^{-3}f_D$) and may behave numerically 
as 
$1/ \bar m_b^{3/2}$ (resp. $1/ \bar m_b$)  for the $1^{+},0^{+}$ (resp. $1^{-}, 0^{-}$)
states 
which can stimulate further theoretical studies of these decay constants. 
 
\keywords{QCD spectral sum rules, Perturbative and Non-Pertubative calculations,  Exotic mesons masses and decay constants. }
\ccode{Pac numbers: 11.55.Hx, 12.38.Lg, 13.20-v}  

\
* PhD Students.
\end{abstract}

\newpage
\section{Introduction and Experimental Facts}
\nin

 A large amount of  exotic hadrons which differ from the ``standard" $\bar cc$ charmonium and $\bar bb$ bottomium radial excitation states have been discovered in $D$ and $B$-factories through e.g. $J/\psi\pi^+\pi^-$ and $\Upsilon\pi^+\pi^-$ processes 
These states are the\,\footnote{We postpone in a future publication\,\cite{SU3} the analysis of the  states decaying to $J/\psi \phi$ such as the: $X_c(4147,4273)$  $1^{++}$ states found recently by LHCb\,\cite{LHCb1} which confirm previous CDF\,\cite{CDFX,CDFX1} results, the $X_c(4350)$ by BELLE\,\cite{BELLEX1},   the $X_c(4506,4704)$  $0^{++}$ states by LHCb\,\cite{LHCb1} and the $Y_c(4140)$ $1^{--}$ by CDF\,\cite{CDFX}.}:
\begin{description} 
\item -- $X_c(3872)$ $1^{++}$ state found by BELLE\,\cite{BELLEX}, BABAR\,\cite{BABARX1,BABARX2}, CDF\,\cite{CDF}, D0\,\cite{D0} from $B^-\to \pi^-[X\to J/\psi \pi^+\pi^-]$ and $B^-\to K^-[X\to J/\psi \pi^+\pi^-]$ decays and by LHCb \cite{LHCb} from $B^+\to K^+[X\to \psi(2S)\gamma]$ and $B^+\to K^+[X\to \psi\gamma]$ with a full width less than 1.2 MeV (90\% CL)\,\cite{PDG},
 \item -- $Y_c(4260,4360,4660)$ $1^{--}$ states by BELLE\,\cite{BELLEY}, BABAR\,\cite{BABARY}, CLEO\,\cite{CLEOY} and BESIII\,\cite{BES}\,\footnote{For a recent review on BESIII results, see e.g\,\cite{SHAN}.} discovered through the initial-state-radiation (ISR) process: $e^+e^-\to \gamma_{ISR}\pi^+\pi^- J/\psi$ and the one: $e^+e^-\to \pi^+\pi^- J/\psi$ in the charmonium region, with a respective total width of $(88.0\pm 23.5), ~(48.0\pm 15.3)$ and $(92.0^{+41.2}_{-31.9})$\,MeV,
 \item -- $Z_c(3900)$ $1^{++}$ by BELLE\,\cite{BELLEZ1,BELLEZ2} and BESIII\,\cite{BES} from  
 $e^+e^-\to \pi^\pm [Z_c\to\pi^\pm J/\psi]$. However, the $Z_c(3900)$ is now quoted in PDG\,\cite{PDG} as a $0^{++}$ state,
 \item -- $Z_c(4025)$ found by BESIII\,\cite{BESZ1,BESZ2,BESZ3}, through $e^+e^-\to D^*\bar D^*\pi,~\pi\pi h_c$. The charged [resp. neutral] one has a width of $(24.8\pm 9.5)$ [resp. $(23.0\pm 6.1)$]\,MeV,
  \item -- $Z_c(4050)$ found by BELLE\,\cite{BELLEZ2b} through $\pi^\pm J/\psi(2S)$,
  \item -- $Z_c(4430)$ from $B\to K[Z_c\to \psi'\pi^\pm]$ decays by BELLE\,\cite{BELLEZ3} and confirmed recently by LHCb \cite{LHCbZ} with a width of $(35\pm 7)$ MeV\,\cite{PDG}.
\end{description}

The observed bottomium states are the:
\begin{description} 
\item  -- $Y_b(9898,10260)$ seen by BELLE\,\cite{BELLEYb1} through $e^+e^-\to \Upsilon(5S)\to h_b(nP)\pi^+\pi^-~(n=1,2,3)$,
\item -- $Y_b(10860)$ near the $\Upsilon(5S)$ peak seen by BELLE\,\cite{BELLEYb2} where the partial widths to $\Upsilon(nS)\pi^+\pi^-~(n=1,2,3)$ of $ (1.79\pm 0.24)$ MeV and to $\Upsilon (1S)K^+K^-$ of $(0.067\pm 0.021)$ MeV are much larger than the one of a standard $\bar bb$ state,
\item  -- $Z_b(10610,10650)$ seen by BELLE\,\cite{BELLEZb} through $\Upsilon (5S)\to \Upsilon (nS)\pi^+\pi^-~ (n=1,2,3)$ and $h_b(mP)\pi^+\pi^-~(m=1,2)$ decay analyses where they have a respective total width of $(18.4\pm 2.4)$ and $(11.5\pm 2.2)$ MeV.
\end{description}

The observations of these unconventional states\,\footnote{A recent analysis of the BELLE collaboration from $\Upsilon(1S)$ inclusive decays does not confirm the existence of some of these states\,\cite{BELLE16}. },  which have some properties beyond the standard quark model (BSQM), have motivated different theoretical interpretations such as molecule and four-quark states\,\cite{MOLE1,MOLE2,MOLE3,MOLE4,MOLE5,MOLE5b,MOLE6,MOLE7,MOLE7b,MOLE8,MOLE8b,MOLE9,MOLE10,MOLE11,MOLE12,MOLE12b,MOLE13,MOLE14,MOLE15} or simply cusps/rescattering effects where no resonance is needed for explaining the data\,\cite{CUSP1,CUSP2,CUSP3}.

 The existence of molecule states has been speculated long ago for charmonium systems \cite{OKUN,ALVARO}. They can be  weakly bound states of the Van Der Vaals-type of two mesons from a long range potential  due to one meson exchange\,\cite{TORN,SUN,ZHANG}.  

 The four-quark states have been introduced earlier by \cite{JAFFE} for interpreting the complex spectra of light scalar mesons and used recently by\,\cite{MAIANI,SORBA} for explaining the $X(3872)$ meson firstly found by BELLE \cite{BELLEX}.  Recent analysis based on $1/N_c$ expansion have shown that the four-quark states should be narrow\,\cite{WEINBERG,PERIS,ROSSI} which do not then favour the four-quark interpretation of the light scalar meson $f_0(500)$, which can eventually have a large gluon component in its wave function\,\cite{VENEZIA,SNGLUE,OCHS,MENES1,MENES2,MENES3}. 

 In previous papers \cite{X1A,X1B,X2,X3A,X3B}, we have studied like various authors\, \cite{NIELSEN4,NIELSEN6,NIELSEN7,CHINESE1,CHINESE2,CHINESE3a,CHINESE3b,
WANG1,WANG2,WANG3,STEELE,BD1,BD2}, to lowest order (LO) of perturbation theory (PT) and including non-perturbative condensates of dimension $d\leq 6-8$, the masses of the $J^P=1^{\pm},$  $0^{\pm}$ molecules and four-quark states using Exponential/Borel/inverse Laplace sum rules (hereafter denoted as LSR\,\footnote{The inverse Laplace transform properties of the Exponential sum rule have been noticed in \cite{SNRAF} sum rules when NLO PT corrections
are included.
A comprehensive interpretation of the LSR using the harmonic oscillator can be found in\,\cite{BELLa,BELLb}.}) or by combining it with Finite Energy (FESR)\,\cite{FESRa,FESRb} QCD spectral sum rules (QSSR) \cite{SVZa,SVZb}\,\footnote{For reviews where complete references can be found, see e.g: \cite{SNB1,SNB2,SNB3a,SNB3b,SNB3c,RRY,CK}.}  and the double ratio of sum rules (DRSR) \cite{DRSR}\,\footnote{For some other successful applications, see\,\cite{SNFORM1,SNGh1,SNGh3,SNGh5,SNmassb,SNhl,SNmassa,SNFORM2,HBARYON1,HBARYON2,NAVARRA}. 
}. These LO results for the masses and couplings agree in many cases with the observed  $XZ$ charmonium and bottomium states and have encouraged
some authors to estimate within QSSR (but still to LO) the hadronic widths\,\cite{NIELSEN1,RAFMOLE1,NIELSEN2,NIELSEN3,RAPHAEL2} and mixing \cite{NIELSEN1,RAPHAEL1} (for reviews, see e.g. \cite{MOLE3,MOLE4,MOLE5}).

 Unfortunately, these previous results obviously suffer from the ill-defined heavy quark mass definition used at LO. The favoured numerical input values: $m_c\approx (1.23-1.26)$ GeV and $m_b\approx 4.17$ GeV used in the current literature correspond numerically to the one of the running masses though there is no reason to discard the values: $m_c\approx 1.5$ GeV and $m_b\approx 4.7$ GeV of the on-shell (pole) quark masses which are more natural because the spectral functions have been evaluated using the on-shell heavy quark propagator.

 Some of the previous LO results have been improved in conference communications\,\cite{X3C,X3D} where we have included  next-to-leading (NLO) order $\alpha_s$ and next-to-next leading (N2LO)  order $\alpha_s^2$ PT QCD corrections in the analysis.  Pursuing the analysis, we have recently improved, in\,\cite{SNX}, the
existing LO analysis\,\cite{NIELSEN,STEELEX,TURC1,TURC2,WANG} interpreting, as a molecule or four-quark state, the recent experimental candidate $X(5568)$ seen by D0\,\cite{D0X} through the sequential decay to $B^0_s\pi^\pm:~B^0_s\to J/\psi\,\phi,~J/\psi\to\mu^+\mu-,~\phi\to K^+K^-$,  where a $J^{P}=0^{+}$ is favoured,  but this observation was not confirmed by LHCb\,\cite{LHCbX}. A conclusion which is consistent with our findings in\,\cite{SNX}. 

In this paper, we pursue and complete the previous program by reconsidering the existing estimates of the masses of the XYZ-like states obtained at LO from QSSR, namely the spin one and spin zero $\bar DD$, $\bar BB$-like 
molecules (see Table\,\ref{tab:current}) and four-quark $[Qq\bar Q\bar q]$ (see Table \,\ref{tab:4qcurrent}) states. In so doing,   we include the NLO and N2LO PT contributions to the QCD two-point correlator by assuming the factorization of the four-quark correlator into a convolution of two quark correlators built from bilinear quark currents. We add to it the contribution of the order $\alpha_s^3$ N3LO contribution estimated from a geometric growth of the PT series\,\cite{SNZ}. To these new higher order (HO) PT contributions, we add the contributions of  condensates having a dimension ($d\leq 6$) already available in the literature but rederived in this paper. Due to the uncertainties on the size (violation of the factorization assumption\,\cite{SNTAU,LNT,JAMI2a,JAMI2b,JAMI2c} and mixing of operators\,\cite{SNT})  and incomplete contributions (only one class of contributions are only computed in the literature) of higher dimension $(d\geq 8)$, we do not include them into the analysis but only consider their effects as a source of errors in the truncation of the Operator Product Expansion (OPE). 

 Our results are summarized in Tables\,\ref{tab:resultc} to \ref{tab:4q-resultb}  and in the last section : Summary and Conclusions.

 A confrontation with different experimental candidates is given in Section\,\ref{sec:confront}.  

\section{QCD expressions of the Spectral Functions}
\nin

 Compared to previous LO QCD expressions of the spectral functions given in the literature, we provide integrated compact expressions which are more easier to handle for the numerical analysis. These expressions are tabulated in the Appendices. 

 The PT expression of the spectral function obtained using on-shell renormalization has been transformed into the  $\overline{MS}$-scheme by using the relation between the  $\overline{MS}$ running mass $\overline{m}_Q(\mu)$ and the on-shell mass (pole) ~$M_Q$ , to order $\alpha_s^2$ \cite{TAR,COQUEa,COQUEb,SNPOLEa,SNPOLEb,BROAD2a,AVDEEV,BROAD2b,CHET2a,CHET2b}:
\bea
M_Q &=& \overline{m}_Q(\mu)\Big{[}
1+\frac{4}{3} a_s+ (16.2163 -1.0414 n_l)a_s^2\nnb\\
&&+\ln{\ga\frac{\mu}{ M_Q}\dr^2} \ga a_s+(8.8472 -0.3611 n_l) a_s^2\dr\nnb\\
&&+\ln^2{\ga\frac{\mu}{ M_Q}\dr^2} \ga 1.7917 -0.0833 n_l\dr a_s^2...\Big{]},
\label{eq:pole}
\eea
for $n_l$ light flavours where $\mu$ is the arbitrary subtraction point and $a_s\equiv \alpha_s/\pi$.

 Higher order PT corrections are obtained using the factorization assumption of the four-quark correlators into a convolution of bilinear current correlators
as we shall discuss later on.
\section{QSSR analysis of the  Heavy-Light Molecules}
\subsection{Molecule  currents and the QCD two-point function}
\nin

For describing these molecule states, we shall consider the usual lowest dimension local interpolating currents where each bilinear current has the quantum number of the corresponding open $D(0^-)$, $D^*_0(0^+)$, $D^*(1^-)~,D_1(1^+)$  states and the analogous states in the $b$-quark channel\,\footnote{For convenience, we shall not consider colored and more general combinations of interpolating operators discussed e.g in \cite{NIELSEN3,STEELE} as well as higher dimension ones involving derivatives. }. 
The previous  assignment is consistent with the definition of a molecule to be a weakly bound state of two mesons within a Van der Vaals force other than a gluon exchange. This feature can justify the approximate  use (up to order $1/N_c$) of the factorization of the four-quark currents as a convolution of two bilinear quark-antiquark currents when estimating the HO PT corrections.
These states and the corresponding interpolating currents are given in Table \ref{tab:current}.  

{\scriptsize
\begin{center}
\begin{table}[hbt]
\setlength{\tabcolsep}{1.8pc}
\newlength{\digitwidth} \settowidth{\digitwidth}{\rm 0}
\catcode`?=\active \def?{\kern\digitwidth}

 \tbl{
Interpolating currents with a definite $C$-parity describing the molecule-like 
states. $Q\equiv$ $c$ (resp. $b$) for the $\bar DD$ (resp. $\bar BB$)-like molecules.  $q\equiv u,d$.  
}
    {\footnotesize
\begin{tabular}{lcl}

&\\
\hline
\hline
States& $J^{PC}$&Molecule Currents  $\equiv{\cal O}_{mol}(x)$  \\
\hline
\\
&$\bf 0^{++}$& \\
$\bar DD,~\bar BB$  &
&$( \bar{q} \gamma_5 Q ) (\bar{Q} \gamma_5 q)$ \\
%
$\bar D^*D^*,\bar B^*B^*$ & 
&$( \bar{q} \gamma_\mu Q ) (\bar{Q} \gamma^\mu q)$ \\ 
$\bar D^*_0D^*_0,~\bar B_0^*B^*_0$&  & $( \bar{q} Q ) (\bar{Q} q)$\\ 
\\
&$\bf 1^{++}$& \\
$\bar D^*D,~\bar B^*B$& 
&$ \frac{i}{\sqrt{2}} \Big[ (\bar{Q} \gamma_\mu q) ( \bar{q} \gamma_5 Q ) 
     - (\bar{q} \gamma_\mu Q) ( \bar{Q} \gamma_5 q ) \Big]$\\ 
 $\bar D^*_0D_1,~\bar B^*_0B_1$  & 
&$ \frac{1}{\sqrt{2}} \Big[ ( \bar{q} Q ) (\bar{Q} \gamma_\mu \gamma_5 q)
     + ( \bar{Q} q ) (\bar{q} \gamma_\mu \gamma_5 Q) \Big]$\\ 
\\
&$\bf 0^{-\pm}$& \\
     $\bar D^*_0D,~\bar B_0^*B$ & & $ \frac{1}{\sqrt{2}} \Big[ ( \bar{q} Q ) (\bar{Q} \gamma_5 q) 
     \pm ( \bar{Q} q ) (\bar{q} \gamma_5 Q) \Big]$\\
$\bar D^*D_1,~\bar B^*B_1$ & 
&$ \frac{1}{\sqrt{2}} \Big[ ( \bar{Q} \gamma_\mu q ) (\bar{q} \gamma^\mu \gamma_5 Q)
     \mp ( \bar{Q} \gamma_\mu \gamma_5 q ) (\bar{q} \gamma^\mu Q) \Big]$\\
     \\
&$\bf 1^{-\pm}$& \\
$\bar D^*_0D^*,~\bar B^*_0B^*$  & 
&$ \frac{1}{\sqrt{2}} \Big[ ( \bar{q} Q ) (\bar{Q} \gamma_\mu q) 
     \mp ( \bar{Q} q ) (\bar{q} \gamma_\mu Q) \Big]$\\ 
     $\bar DD_1,~\bar BB_1$  &
 &    $ \frac{i}{\sqrt{2}} \Big[ ( \bar{Q} \gamma_\mu \gamma_5 q ) (\bar{q} \gamma_5 Q)
     \pm ( \bar{q} \gamma_\mu \gamma_5 Q ) (\bar{Q} \gamma_5 q) \Big]$\\
     \\
\hline
\hline
\end{tabular}
}
\label{tab:current}
\end{table}
\end{center}
}
\nin

 The  two-point correlators associated to the (axial)-vector interpolating operators  are:

\bea
\Pi^{\mu\nu}_{mol}(q)&\equiv&i\int d^4x ~e^{iq.x}\lag 0
|T[{\cal O}^\mu_{mol}(x){\cal O}_{mol}^{\nu\dagger}(0)]
|0\rag\nnb\\
&=&-\Pi^{(1)}_{mol}(q^2)(g^{\mu\nu}-\frac{q^\mu q^\nu}{q^2})+\Pi^{(0)}_{mol}(q^2)\frac{q^\mu
q^\nu}{ q^2}~,
\lb{2po}
\eea
The two invariants, $\Pi^{(1)}_{mol}$ and $\Pi^{(0)}_{mol}$, appearing in 
Eq.~(\ref{2po}) are independent and have respectively the quantum numbers 
of the spin 1 and 0 mesons. 

$\Pi^{(0)}_{mol}$ is related via  Ward identities\,\cite{SNB1,SNB2} to the (pseudo)scalar two-point functions $\psi^{(s,p)}(q^2)$ 
built directly from the (pseudo)scalar currents given in Table\,\ref{tab:current}:
\beq 
\psi^{(s,p)}_{mol}(q^2)=i\int d^4x ~e^{iq.x}\lag 0
|T[{\cal O}^{(s,p)}_{mol}(x){\cal O}^{(s,p)}_{mol}(0)]
|0\rag~,
\label{2po5}
\eeq
with which we shall work in the following.

Thanks  to their analyticity properties, the invariant functions $\Pi^{(1,0)}_{mol}(q^2)$ in Eq.~(\ref{2po}) and the two-point correlator $ \psi^{(s,p)}_{mol}(q^2)$ in Eq.\,\ref{2po5} obey the dispersion relation:
\beq
\Pi^{(1,0)}_{mol}(q^2), ~\psi^{(s,p)}_{mol}(q^2)=\frac{1}{\pi}\int_{4m_c^2}^\infty dt \frac{{\rm Im}\:\Pi^{(1,0)}_{mol}(t),~ {\rm Im}\:\psi^{(s,p)}_{mol}(t)}{ t-q^2-i\epsilon}+\cdots \;,
\lb{ope}
\enq
where $\mbox{Im}\:\Pi^{(1,0)}_{mol}(t),~ {\rm Im}\:\psi^{(s,p)}_{mol}(t)$ are the spectral functions and $\cdots$ indicate subtraction points which are polynomial in $q^2$. 
\subsection{LO PT and NP corrections to the molecule spectral functions}
The new different LO integrated expressions including non-perturbative (NP) corrections up to dimension $d$=6-8 used in the analysis are tabulated in Appendix~A. 

Compared to the ones in the literature, the expressions of the spectral functions are
in integrated and compact forms which are more easier to handle for the numerical phenomenological analysis. However, one should note that most of the expressions given in the literature do not agree each others. Due to the 
few informations given by the authors on their derivation, it is difficult to trace back the origin of such discrepancies. Hopefully, within  the accuracy of the approach, such discrepancies affect only slightly the final results if the errors are taken properly. 

In the chiral limit $m_q=0$ and $\la\bar uu\ra=\la\bar dd\ra$, we have checked that the orthogonal combinations of $\bar D^*D, \bar B^*B (1^{++}), \bar D_0^* D_1,\bar B_0^* B_1 (0^{--})$ and $\bar D^*D_1, \bar B^*B_1 (0^{--} )$ molecules give the same results up to the $d=6$ contributions. This is due to the presence of one $\gamma_5$ matrix in the current which neutralizes the different traces appearing in each pair. This is not the case of the $\bar D^*_0D^*, \bar B^*_0B^*$ (without $\gamma_5$) and $\bar DD_1, \bar BB_1$ (with two  $\gamma_5$). 

\subsection{$1/q^2$ tachyonic gluon mass and large order PT corrections}
\nin
The $1/q^2$ corrections due to a tachyonic gluon  mass  discussed in \cite{CNZ1,CNZ2} (for reviews see: \,\cite{ZAK1,ZAK2}) will not be included here. Instead, we shall consider the fact that they are dual to the sum of the large order PT series\,\cite{SNZ} such that, with the inclusion of the N3LO term estimated from the geometric growth of the QCD PT series \cite{SNZ} as a source of the PT errors,  we expect to give a good approximation of these uncalculated higher order terms. The estimate of these errors is given in Tables\,\ref{tab:errorc} to \,\ref{tab:4q-errorb}.
\subsection{NLO and N2LO PT corrections using factorization}
\nin
Assuming a factorization of the four-quark interpolating current as a natural consequence of the molecule
definition of the state, we can write the corresponding spectral function as a convolution of the
spectral functions associated to quark bilinear current\,\footnote{It is called properly sesquilinear instead of bilinear  current as it is a formed by a quark field and its anti-particle. We thank Professor Raoelina Andriambololona for this remark.} as illustrated by the Feynman diagrams in Fig.\,\ref{fig:factor}. In this way, we obtain \cite{PICH}\,\footnote{For some applications to the $\bar BB$ mixing, see e.g. \cite{BBAR1,BBAR2,BBAR3}.} for the  $\bar DD^*$ and $\bar D^*_0D^*$ spin 1 states:
\bea
\frac{1}{ \pi}{\rm Im} \Pi^{(1)}_{mol}(t)&=& \theta (t-4M_Q^2)\ga \frac{1}{ 4\pi}\dr^2 t^2 \int_{M_Q^2}^{(\sqrt{t}-M_Q)^2}\hspace*{-0.5cm}dt_1\int_{M_Q^2}^{(\sqrt{t}-\sqrt{t_1})^2} \hspace*{-1cm}dt_2\nnb\\
&&\times~\lambda^{3/2}\frac{1}{ \pi}{\rm Im} \Pi^{(1)}(t_1) \frac{1}{ \pi}{\rm Im} \psi^{(s,p)}(t_2)~.
\label{eq:convolution}
\eea
For the  $\bar DD$ spin 0 state, one has:
\bea
\frac{1}{ \pi}{\rm Im} \psi^{(s)}_{mol}(t)&=& \theta (t-4M_Q^2)\ga \frac{1}{4\pi}\dr^2 t^2 \int_{m_Q^2}^{(\sqrt{t}-M_Q)^2}\hspace*{-0.5cm}dt_1\int_{m_Q^2}^{(\sqrt{t}-\sqrt{t_1})^2}  \hspace*{-1cm}dt_2~\nnb\\
&&\times~\lambda^{1/2}\ga \frac{t_1}{ t}+ \frac{t_2}{ t}-1\dr^2\nnb\\
&&\times ~\frac{1}{ \pi}{\rm Im}\psi^{(p)}(t_1) \frac{1}{ \pi} {\rm Im} \psi^{(p)}(t_2),
\eea
and for the $\bar D^*D^*$ spin 0 state:
\bea
\frac{1}{ \pi}{\rm Im} \psi_{mol}(t)&=& \theta (t-4M_Q^2)\ga \frac{1}{4\pi}\dr^2 t^2 \int_{m_Q^2}^{(\sqrt{t}-M_Q)^2}\hspace*{-0.5cm}dt_1\int_{m_Q^2}^{(\sqrt{t}-\sqrt{t_1})^2}  \hspace*{-1cm}dt_2~\nnb\\
&&\times~\lambda^{1/2}\Big{[}\ga \frac{t_1}{ t}+ \frac{t_2}{ t}-1\dr^2
+\frac{8t_1t_2}{ t^2}\Big{]}\nnb\\
&&\times ~\frac{1}{ \pi}{\rm Im} \Pi^{(1)}(t_1) \frac{1}{ \pi} {\rm Im} \Pi^{(1)}(t_2),
\eea
where:
\beq
\lambda=\ga 1-\frac{\ga \sqrt{t_1}- \sqrt{t_2}\dr^2}{ t}\dr \ga 1-\frac{\ga \sqrt{t_1}+ \sqrt{t_2}\dr^2}{ t}\dr~,
\eeq
is the phase space factor and $M_Q$ is the on-shell heavy quark mass. 
Im $ \Pi^{(1)}(t)$ is the spectral function associated to the bilinear $\bar c\gamma_\mu (\gamma_5)q$  vector or axial-vector current, while Im $\psi^{(5)}(t)$ is associated to the 
$\bar c(\gamma_5)q$  scalar or pseudoscalar current\,\footnote{In the limit where the light quark mass $m_q=0$, the PT expressions of the vector (resp. scalar) and axial-vector (resp. pseudoscalar) spectral functions are the same.}. 
This representation simplifies the evaluation of the PT $\alpha_s^n$-corrections as we can use the PT expression of the spectral functions for heavy-light bilinear currents known to order $\alpha_s$ (NLO) from \cite{BROAD} and to order $\alpha_s^2$ (N2LO) from \cite{CHETa,CHETb} which are available as a Mathematica Program named  Rvs.
From this above representation, the anomalous dimension of the correlator comes from  the (pseudo)scalar current and the corresponding renormalization group invariant interpolating current reads to NLO\,\footnote{The spin 0 current built from two (axial)-vector currents has no anomalous dimension.}:
\beq
 \bar {\cal O}^{(s,p)}_{mol}(\mu)= a_s(\mu)^{4/\beta_1} {\cal O}^{(s,p)}_{mol}~,~~~~~~~~  \bar {\cal O}^{(1)}_{mol}(\mu)= a_s(\mu)^{2/\beta_1} {\cal O}^{(1)}_{mol}~.
 \eeq
  Within the above procedure, we have checked that we reproduce the factorized PT LO contributions obtained using for example the PT expressions of $\bar D^*_0D^*_0$ and $\bar D^*_0D^*$ given in Appendix A.
\subsection{Parametrization of the Spectral Function within MDA}
\nin

 We shall  use the Minimal Duality Ansatz (MDA) given in Eq. \ref{eq:duality} for parametrizing the spectral function (generic notation):
\beq
\frac{1}{\pi}\mbox{ Im}\Pi_{mol}(t)\simeq f^2_{mol}M^8_{mol}\delta(t-M_{mol}^2)
  \ + \
  ``\mbox{QCD continuum}" \theta (t-t_c),
\label{eq:duality}
\eeq
where $f_{mol}$ is the decay constant defined as:
\beq
\la 0| {\cal O}^{(s,p)}_{mol}|mol\ra=f^{(s,p)}_{mol}M^4_{mol}~,~~~~~~~~~~~~\la 0| {\cal O}^\mu_{mol}|mol\ra=f^{(1)}_{mol}M^5_{mol}\epsilon_\mu~,
\label{eq:coupling}
\eeq
respectively for spin 0 and 1 molecule states with  $\epsilon_\mu$ the vector polarization.
The higher states contributions are smeared by the ``QCD continuum" coming from the discontinuity of the QCD diagrams and starting from a constant threshold $t_c$.  

 Noting that in the previous definition in Table \ref{tab:current},  the bilinear (pseudo)scalar current acquires an anomalous dimension due to its normalization, thus the decay constants run  to order $\alpha_s^2$ as\,\footnote{The coupling of the (pseudo)scalar molecule built from two (axial)-vector currents has no anomalous dimension and does not run.}:
\beq
f^{(s,p)}_{mol}(\mu)=\hat f^{(s,p)}_{mol} \ga -\beta_1a_s\dr^{4/\beta_1}/r_m^2~,~~~~f^{(1)}_{mol}(\mu)=\hat f^{(1)}_{mol} \ga  -\beta_1a_s\dr^{2/\beta_1}/r_m~,
\label{eq:fhat}
\eeq
where we have introduced the renormalization group invariant coupling $\hat f_{mol}$; 
$-\beta_1=(1/2)(11-2n_f/3)$ is the first coefficient of the QCD $\beta$-function for $n_f$ flavours and  $a_s\equiv (\alpha_s/\pi)$.  The QCD corrections numerically read;
\bea
r_m(n_f=4)=1+1.014 a_s +1.389a_s^2,~~~~
r_m(n_f=5)=1+1.176a_s +1.501a_s^2.  
\eea
\subsection{The inverse Laplace  transform sum rule (LSR)}
\nin
The exponential sum rules firstly derived by SVZ\,\cite{SVZa,SVZb} have been called Borel sum rules due to the factorial suppression factor of the condensate contributions in the OPE. Their quantum mechanics version have been studied by Bell-Bertlmann in \cite{BELLa,BELLb,BELLc} through the harmonic oscillator where $\tau$ has the property of an imaginary time, while the derivation of their radiative corrections has been firstly shown by Narison-de Rafael\, \cite{SNRAF} to have the properties of the inverse Laplace sum rule (LSR). The LSR and its ratio read\,\footnote{The last equality in Eq.\,\ref{eq:ratioLSR} is obtained when one uses MDA in Eq.\,\ref{eq:duality} for parametrizing the spectral function.}:
\beq
{\cal L}_{mol}(\tau,t_c,\mu)=\int_{4M_Q^2}^{t_c}dt~e^{-t\tau}\frac{1}{\pi} \mbox{Im}\Pi^{(1,0)}_{mol}(t,\mu)~,
\label{eq:LSR}
\eeq
\beq\label{eq:ratioLSR}
{\cal R}_{mol}(\tau,t_c,\mu) = \frac{\int_{4M_Q^2}^{t_c} dt~t~ e^{-t\tau}\frac{1}{\pi}\mbox{Im}\Pi^{(1,0)}_{mol}(t,\mu)}
{\int_{4M_Q^2}^{t_c} dt~ e^{-t\tau} \frac{1}{\pi} \mbox{Im}\Pi^{(1,0)}_{mol}(t,\mu)}\simeq M_R^2~,
\eeq
where $\mu$ is the subtraction point which appears in the approximate QCD series when radiative corrections are included and $\tau$ is the sum rule variable replacing $q^2$. 
Similar sum rules are obtained for the (pseudo)scalar two-point function $\psi^{(s,p)}(q^2)$. 
The variables $\tau,\mu$ and $t_c$ are, in principle, free parameters. 
We shall use stability criteria (if any), with respect to these free 3 parameters,  for extracting the optimal results. 
\subsection{Tests of MDA and Stability Criteria}
\nin

 In the standard Minimal Duality Ansatz (MDA) given in Eq. \ref{eq:duality} for parametrizing the spectral function,
 the ``QCD continuum" threshold $t_c$ is constant and is independent on the subtraction point $\mu$ \,\footnote{Some model with a $\mu$-dependence of $t_c$ has been discussed e.g in\,\cite{LUCHA}.}. One should notice that this standard MDA with constant $t_c$ describes quite well the properties of the lowest ground state as explicitly demonstrated in \cite{SNFB12a,SNFB12b} and in various examples\,\cite{SNB1,SNB2}, while it has been also successfully tested in the large $N_c$ limit of QCD in \cite{PERISa,PERISb}. 

 Ref. \cite{SNFB12a,SNFB12b}  has explicitly tested  this simple model by confronting the predictions of the integrated spectral function within this simple parametrization with the full data measurements. One can notice  
in Figs. 1 and 2 of Ref.  \cite{SNFB12a,SNFB12b} the remarkable agreement of the model predictions and of the measured data of the $J/\psi$ charmonium and $\Upsilon$ bottomium systems for a large range of the inverse sum rule variable $\tau$. Though it is difficult to estimate with a good precision the systematic error related to this simple model, this feature indicates the ability of the model for reproducing accurately the data. We expect that the same feature is reproduced for the case of the XYZ discussed here where complete data are still lacking.

In order to extract an optimal information for the lowest resonance parameters from this rather crude  description of the spectral function and from the approximate QCD expression, one often applies the stability criteria at which an optimal result can be extracted. This stability is signaled by the existence of a stability plateau, an extremum or an inflexion point (so-called ``sum rule window") versus the changes of the external sum rule variables $\tau$ and $t_c$ where the simultaneous  requirement on the dominance over the continuum contribution and on the convergence of the OPE is automatically satisfied. This optimization criterion demonstrated in series of papers by Bell-Bertlmann \cite{BELLa,BELLb,BELLc}, in the case of the $\tau$-variable, by taking the examples of harmonic oscillator and charmonium sum rules and extended to the case of the $t_c$-parameter in \cite{SNB1,SNB2} gives a more precise meaning of  the so-called ``sum rule window" originally discussed by SVZ \cite{SVZa,SVZb} and used in the sum rules literature. Similar applications of the optimization method to the pseudoscalar $D$ and $B$ open meson states have been successful when compared with results from some other determinations as discussed in Ref.\,\cite{SNFB12a,SNFB12b} and reviewed in\,\cite{SNB1,SNB2,SNFB01,SNREV14,SNREV15} and in some other recent reviews\,\cite{ROSNERb,LATT13}. 

 In this paper, we shall add to the previous well-known $\tau$- and $t_c$-stability criteria, the one associated  to the requirement of stability  versus the arbitrary subtraction constant $\mu$ often put by hand  in the current literature  and which is often the source of large errors from the PT series in the sum rule analysis.  The $\mu$-stability procedure has been applied recently in\,\cite{SNFB12a,SNFB12b,SNFB13,SNLIGHT,SNREV14,SNREV15,SNFB14}\,\footnote{Some other alternative approaches for optimizing the PT series can be found in \cite{STEVENSONa,STEVENSONb,STEVENSONc,STEVENSONd,STEVENSONe}.} which gives a much better meaning on the choice of $\mu$-value at which the observable is extracted, while the errors  in the determinations of the results have been reduced due to a better control of the $\mu$ region of variation which is not the case in the existing literature.

{\scriptsize
\begin{table}[hbt]
\setlength{\tabcolsep}{2pc}
 \tbl{QCD input parameters:
the original errors for 
$\la\alpha_s G^2\ra$, $\la g^3  G^3\ra$ and $\rho \la \bar qq\ra^2$ have been multiplied by about a factor 3 for a conservative estimate of the errors (see also the text). }  
    {\small
 {\begin{tabular}{@{}lll@{}} \toprule
&\\
\hline
\hline
Parameters&Values& Ref.    \\
\hline
$\alpha_s(M_\tau)$& $0.325(8)$&\cite{SNTAU,BNPa,BNPb}\\
$\overline{m}_c(m_c)$&$1261(12)$ MeV &average \cite{SNmass02,SNH10a,SNH10b,SNH10c,PDG,IOFFEa,IOFFEb}\\
$\overline{m}_b(m_b)$&$4177(11)$ MeV&average \cite{SNmass02,SNH10a,SNH10b,SNH10c,PDG}\\
$\hat \mu_q$&$(253\pm 6)$ MeV&\cite{SNB1,SNmassa,SNmassb,SNmass98a,SNmass98b,SNLIGHT}\\
$M_0^2$&$(0.8 \pm 0.2)$ GeV$^2$&\cite{JAMI2a,JAMI2b,JAMI2c,HEIDa,HEIDb,HEIDc,SNhl}\\
$\la\alpha_s G^2\ra$& $(7\pm 3)\times 10^{-2}$ GeV$^4$&
\cite{SNTAU,LNT,SNIa,SNIb,YNDU,BELLa,BELLb,BELLc,SNH10a,SNH10b,SNH10c,SNG1,SNG2,SNGH}\\
$\la g^3  G^3\ra$& $(8.2\pm 2.0)$ GeV$^2\times\la\alpha_s G^2\ra$&
\cite{SNH10a,SNH10b,SNH10c}\\
$\rho \alpha_s\la \bar qq\ra^2$&$(5.8\pm 1.8)\times 10^{-4}$ GeV$^6$&\cite{SNTAU,LNT,JAMI2a,JAMI2b,JAMI2c}\\
\hline\hline
\end{tabular}}
}
\label{tab:param}
\end{table}
} 
\subsection{QCD Input Parameters}
\hspace*{0.5cm} 
The QCD parameters which shall appear in the following analysis will be the charm and bottom quark masses $m_{c,b}$ (we shall neglect  the light quark masses $q\equiv u,d$),
the light quark condensate $\qq$,  the gluon condensates $ \lag
\alpha_sG^2\rag
\equiv \la \alpha_s G^a_{\mu\nu}G_a^{\mu\nu}\ra$ 
and $ \la g^3G^3\ra
\equiv \la g^3f_{abc}G^a_{\mu\nu}G^{b,\nu}_{\rho}G^{c,\rho\mu}\ra$, 
the mixed condensate $\la\bar qGq\ra
\equiv {\la\bar qg\sigma^{\mu\nu} (\lambda_a/2) G^a_{\mu\nu}q\ra}=M_0^2\la \bar qq\ra$ 
and the four-quark 
 condensate $\rho\alpha_s\la\bar qq\ra^2$, where
 $\rho\simeq 3-4$ indicates the deviation from the four-quark vacuum 
saturation. Their values are given in Table \ref{tab:param}. 

We shall work with the running
light quark condensates and masses, which read to leading order in $\alpha_s$: 
\beq
{\la\bar qq\ra}(\tau)=-{\hat \mu_q^3  \ga-\beta_1a_s\dr^{2/{
\beta_1}}},~~~~~~~~~
{\la\bar q Gq\ra}(\tau)=-{M_0^2{\hat \mu_q^3} \ga-\beta_1a_s\dr^{1/{3\beta_1}}}~,
\label{d4g}
\eeq
where $\beta_1=-(1/2)(11-2n_f/3)$ is the first coefficient of the $\beta$ function 
for $n_f$ flavours; $a_s\equiv \alpha_s(\tau)/\pi$; 
$\hat\mu_q$ is the spontaneous RGI light quark condensate \cite{FNR}. 
We shall use:
\beq   
\alpha_s(M_\tau)=0.325(8) \lrar  \alpha_s(M_Z)=0.1192(10)
\label{eq:alphas}
\eeq
from $\tau$-decays \cite{SNTAU,BNPa,BNPb}\,\footnote{A recent update is done in\,\cite{PICHTAU} where the same central value is obtained and more complete references are given.}
 which agree with the 2016 world average\,\cite{BETHKE}: 
\beq
\alpha_s(M_Z)=0.1181(11)~. 
\eeq
The value of the running $\la \bar qq\ra$ condensate is deduced from  the well-known GMOR relation: 
\beq
(m_u+m_d)\la \bar uu+\bar dd\ra=-m_\pi^2f_\pi^2~,
\eeq
where $f_\pi=130.4(2)$ MeV \cite{ROSNERb}. The value of $(\overline{m}_u+\overline{m}_d)(2)=(7.9\pm 0.6)$ MeV obtained in  \cite{SNmassa,SNmassb} agrees with the PDG  in \cite{PDG}  and lattice averages in \cite{LATT13}. Then, we deduce the RGI light quark spontaneous mass $\hat\mu_q$ given  in Table~\ref{tab:param}. 

 For the heavy quarks, we shall use the running mass and the corresponding value of $\alpha_s$ evaluated at the scale $\mu$. These sets of correlated parameters are given in Table \ref{tab:alfa} for different values of $\mu$ and for a given number of flavours $n_f$.

 For the $\la \alpha_s G^2\ra$ condensate, we have enlarged the original error by a factor about 3 in order to have
a conservative result for recovering the original SVZ estimate and the alternative extraction in \cite{IOFFEa,IOFFEb} from charmonium sum rules.
However, a direct comparison of this  range of values obtained within short QCD series (few terms) with the one from lattice calculations \cite{BALIa} obtained within a long QCD series\,\cite{BALIb} can be misleading. 

Some other estimates of the gluon and four-quark condensates using $\tau$-decay and $e^+e^-\to I=1$ hadrons data can be found in\,\cite{DAVIER,BOITOa,FESRa,FESRb}. Due to the large uncertainties induced by the different resummations of the QCD series and by the less-controlled effects of some eventual duality violation, we do not consider explicitly these values in the following analysis. However, we shall see later on that the effects of the gluon and four-quark condensates on the values of the decay constants and masses are almost negligible though they play an important r\^ole in the stability analysis. 

{\scriptsize
\begin{table}[hbt]
\setlength{\tabcolsep}{1.1pc}
 \tbl{
$\alpha_s(\mu)$ and correlated values of $\overline{m}_Q(\mu)$ used in the analysis for different values of the subtraction scale $\mu$. The error in $\overline{m}_Q(\mu)$ has been induced by the one of $\alpha_s(\mu)$ to which one has added the error on their determination given in Table\,\ref{tab:param}. }
    {\small
{\begin{tabular}{@{}llll@{}} \toprule
&\\
\hline
\hline
Input for $\bar DD,...,~[cq\bar c\bar q],$ : $n_f=4$\\
\hline
$\mu$[GeV]&$\alpha_s(\mu)$&& $\overline{m}_c(\mu)$[GeV]\\
Input: $\overline{m}_c(m_c)$&0.4084(144)&&1.26\\
1.5&0.3649(110)&&1.176(5)\\
2&0.3120(77)&&1.069(9)\\
2.5&0.2812(61)&&1.005(10)\\
3.0&0.2606(51)&&0.961(10)\\
3.5&0.2455(45)&&0.929(11)\\
4.0&0.2339(41)&&0.903(11) \\
4.5&0.2246(37)&&0.882(11)\\
5.0&0.2169(35)&&0.865(11)\\
5.5&0.2104(33)&&0.851(12)\\
6.0&0.2049(30)&&0.838(12)\\
\hline
Input for $\bar BB,...,~[bq\bar b\bar q]$ : $n_f=5$\\
\hline
$\mu$[GeV]&$\alpha_s(\mu)$&& $\overline{m}_b(\mu)$[GeV]\\
3&0.2590(26)&&4.474(4)\\
3.5&0.2460(20)&&4.328(2)\\
Input: $\overline{m}_b(m_b)$&0.2320(20)&&4.177\\
4.5&0.2267(20)&&4.119(1)\\
5.0&0.2197(18)&&4.040(1)\\
5.5&0.2137(17)&&3.973(2)\\
6.0&0.2085(16)&&3.914(2)\\
6.5&0.2040(15)&&3.862(2)\\
7.0&0.2000(15)&&3.816(3)\\
\hline
\hline
\end{tabular}}
}
\label{tab:alfa}
\end{table}
} 
\nin

\section{Accuracy of the Factorization Assumption}\label{sec:factor}
\subsection{PT Lowest order tests}
\begin{figure}[hbt] 
\begin{center}
{\includegraphics[width=6.29cm  ]{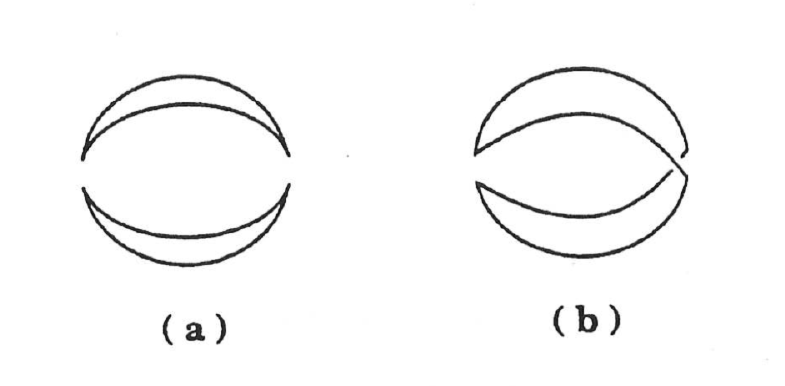}}
\caption{
\scriptsize 
{\bf (a)} Factorized contribution to the four-quark correlator at lowest order of PT; {\bf (b)} Non-factorized contribution at lowest order of PT (the figure comes from\,\cite{PICH}).
}
\label{fig:factor} 
\end{center}
\end{figure} 
\nin
To lowest order of PT QCD, the four-quark correlator can be subdivided  into its factorized (Fig.\,\ref{fig:factor}a) and its non-factorized (Fig.\,\ref{fig:factor}b) parts.  In the following, we shall further test the factorization assumption if one does it at lowest order (LO) of PT by taking the example of the  $\bar M^*_0M^*(1^-)$  molecule states
where $M\equiv D$ (resp. $B$) meson in the charm (resp. bottom) quark channels. The factorized expression corresponds to the value $\epsilon=0$ and the full one to $\epsilon=1$ in the QCD expressions of the spectral functions given in Appendix A. 

\subsubsection*{\b  Decay Constants and Masses of the $ \bar M^*_0M^*(1^-)$ Molecules}
\begin{figure}[hbt] 
\begin{center}
{\includegraphics[width=6.29cm  ]{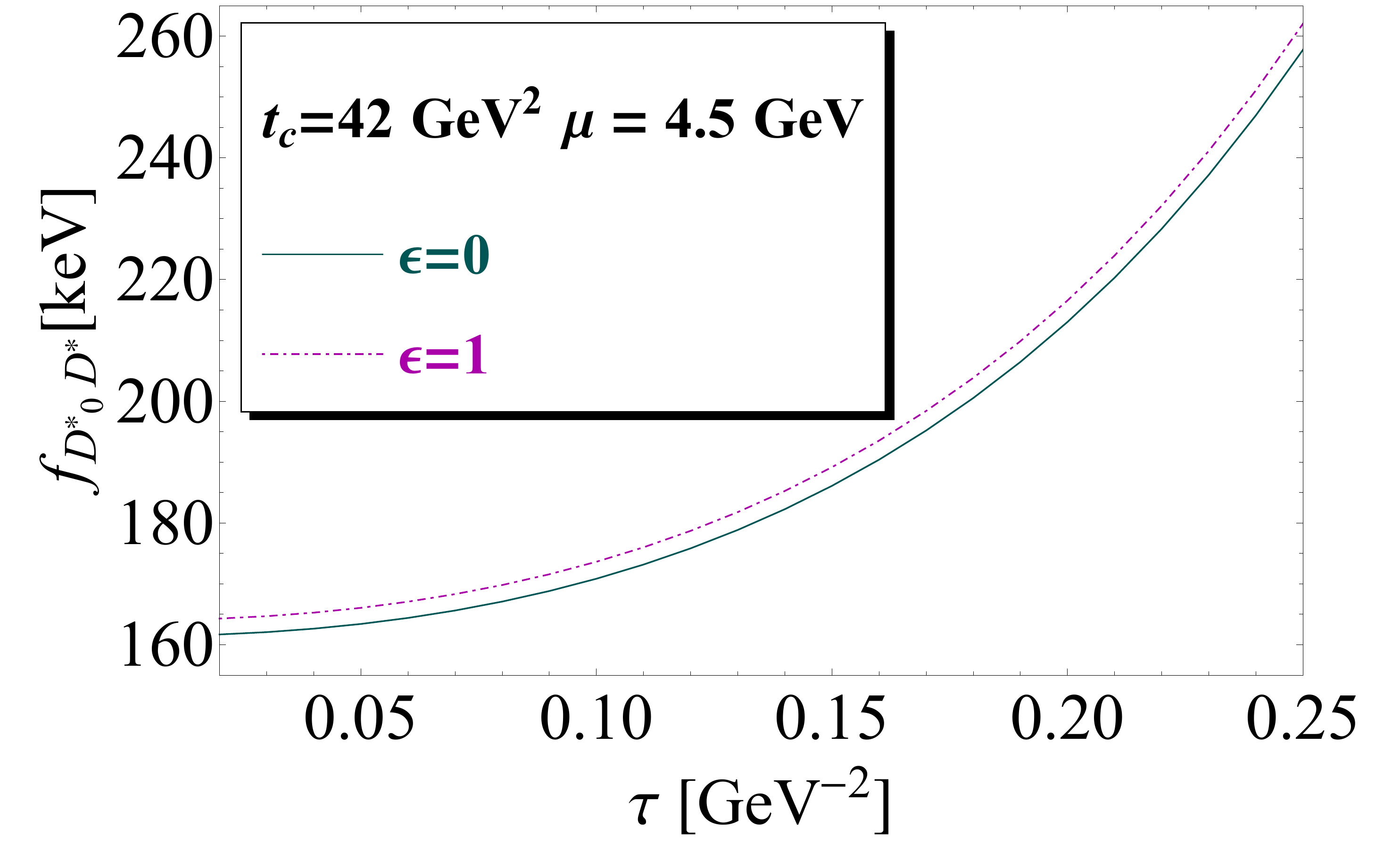}}
{\includegraphics[width=6.29cm  ]{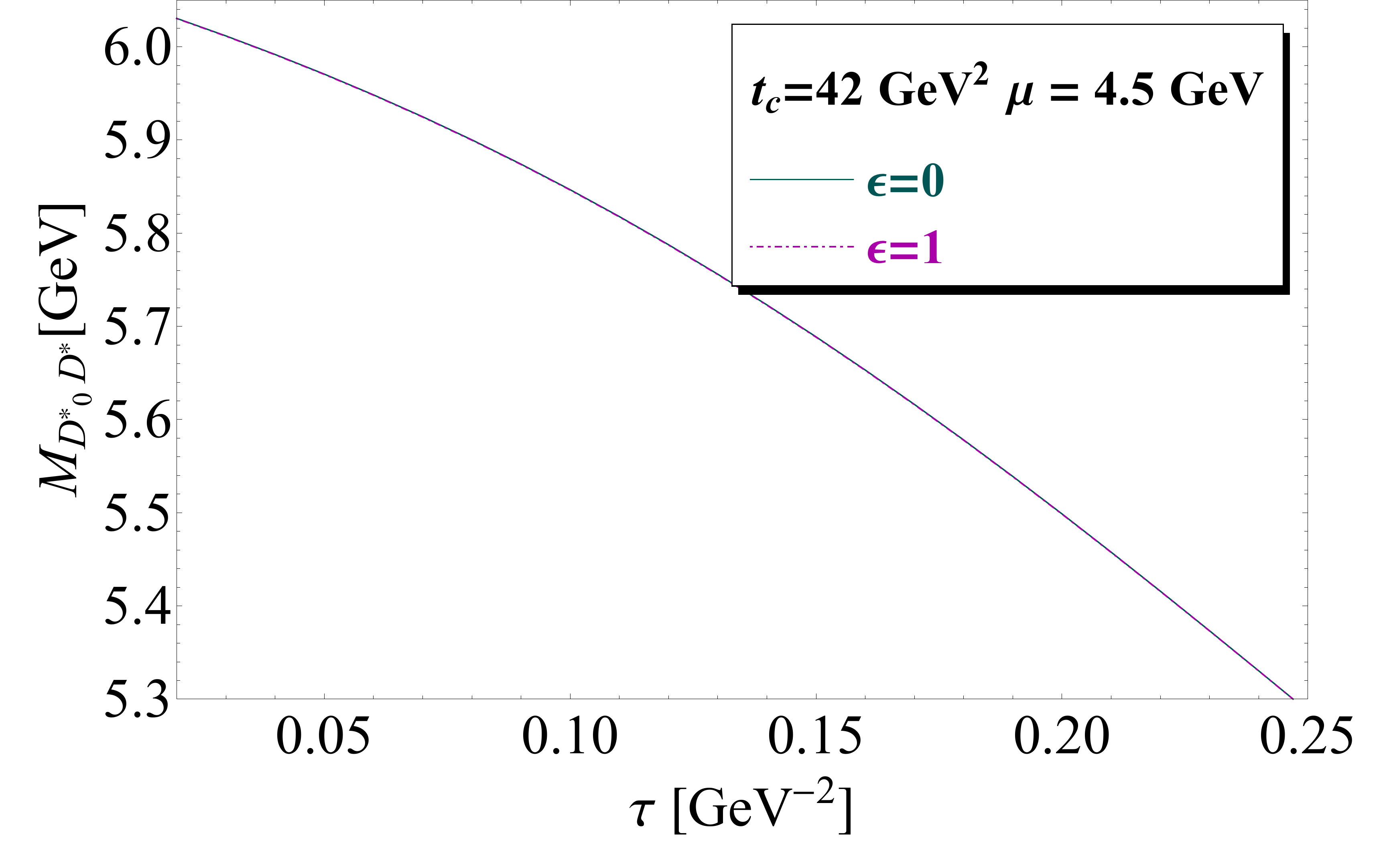}}
\scriptsize\centerline {\hspace*{-3cm} b)\hspace*{6cm} b) 
}
\caption{
\scriptsize 
{\bf a)} Factorized ($\epsilon=0$) and full ($\epsilon=1$) lowest order PT contributions to $f_{D^*_0D^*}$  as function of $\tau$ for a given value of $t_c=42$ GeV$^2$, $\mu=4.5$ GeV, $\overline{m}_c(\overline{m}_c)=1.26$ GeV and using the QCD parameters in Tables\,\ref{tab:param} and \ref{tab:alfa}; {\bf b)} The same as a) but for the mass $M_{D^*_0D^*}$.
}
\label{fig:dstar0dstar-b01} 
\end{center}
\end{figure}

\nin
We study in Fig. {\ref{fig:dstar0dstar-b01}} the effect of factorization for a given value of $t_c=42$ GeV$^2$ and $\mu=4.5$ GeV at lowest order of PT for the $ \bar D^*_0D^*(1^-)$ molecule.  An analogous analysis is done for the $B^*_0B^*$ molecule which presents the same qualitative behaviour
as the one in Fig. {\ref{fig:dstar0dstar-b01}}.
\nin
\subsubsection*{\b Conclusions from the PT lowest order analysis} 
\nin
We conclude from the previous two examples that assuming a factorization of the PT contributions
at LO induces an almost negligible effect on the decay constant ($\simeq 1.5\%$) and mass ($\simeq 7\times 10^{-4}$) determinations for the $\bar D^*_0D^*$ and $\bar B^*_0B^*$ vector molecules. 
\subsection{Factorization tests for PT$\oplus$NP contributions at LO}
One can notice, from the QCD expression including NP contributions, that the factorization assumption modifies the structure of the
OPE due to the vanishing of some contributions at LO. 
\subsubsection*{\b  Decay Constants and Masses of the $ \bar M^*_0M^*(1^{--})$ Molecules}
\begin{figure}[hbt] 
\begin{center}
{\includegraphics[width=6.29cm  ]{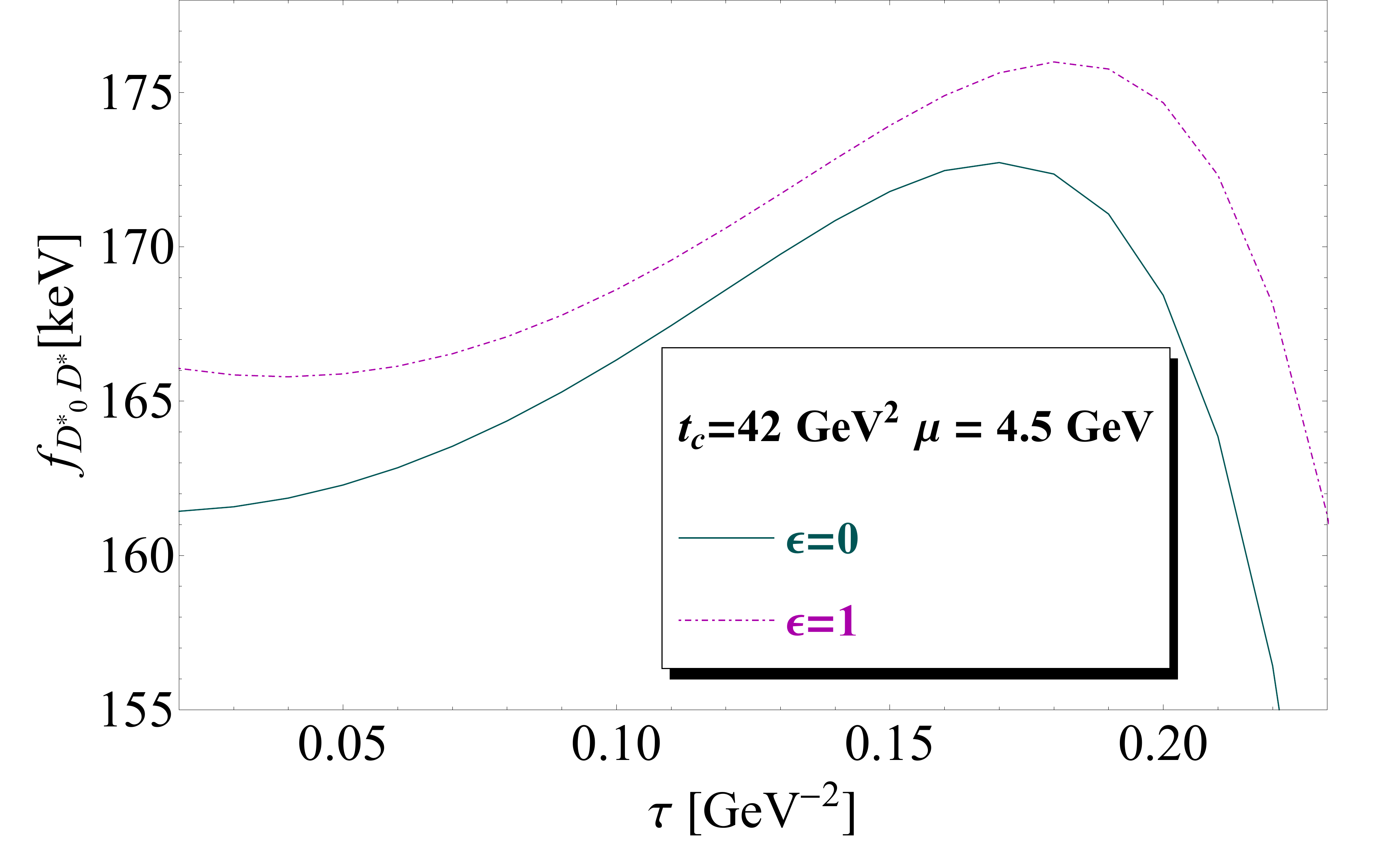}}
{\includegraphics[width=6.29cm  ]{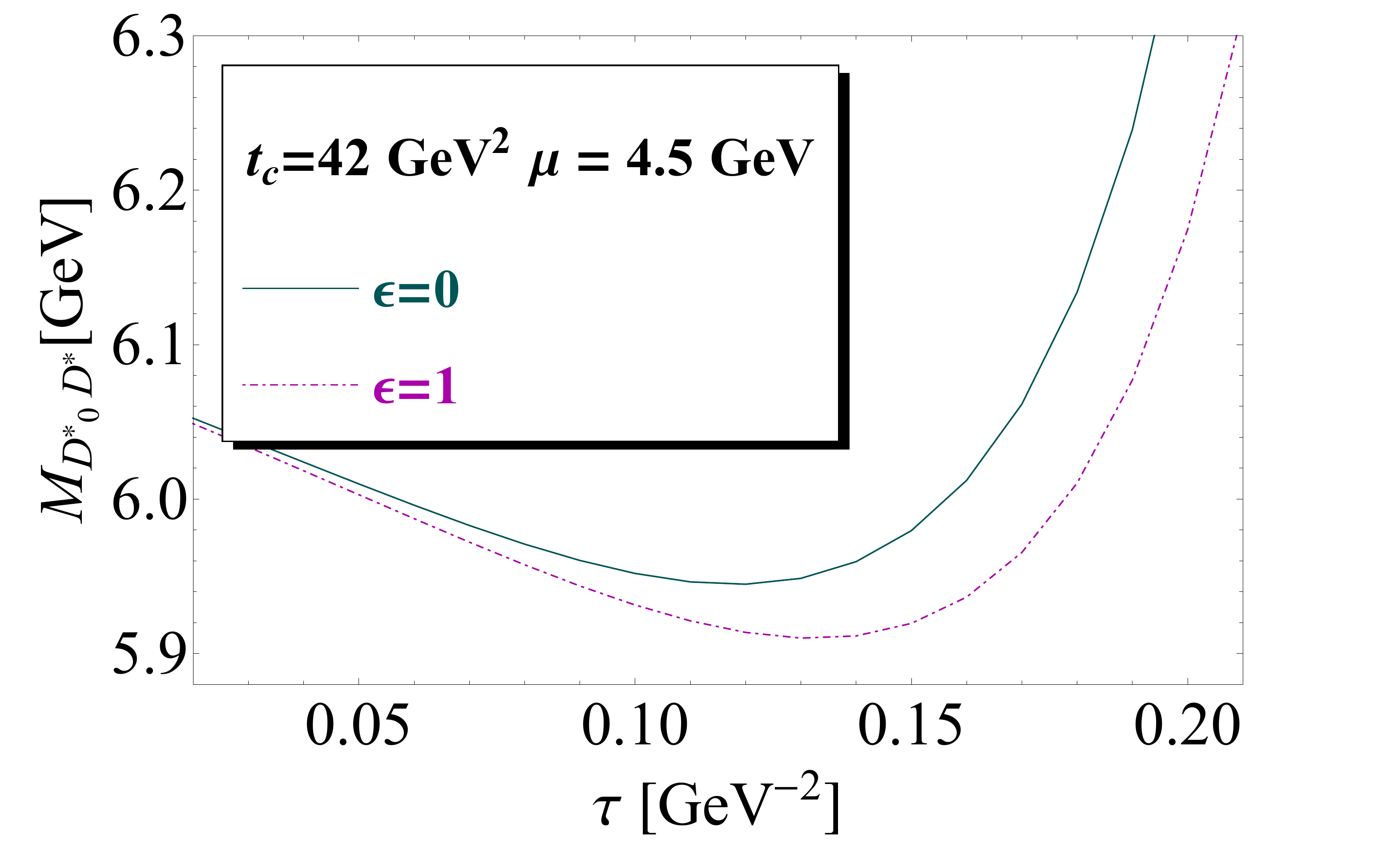}}
\scriptsize\centerline {\hspace*{-3cm} b)\hspace*{6cm} b) 
}
\caption{
\scriptsize 
{\bf a)} Factorized ($\epsilon=0$) and full ($\epsilon=1$) lowest order PT contributions to $f_{D^*_0D^*}$  as function of $\tau$ for a given value of $t_c=42$ GeV$^2$, $\mu=4.5$ GeV, $\overline{m}_c(\overline{m}_c)=1.26$ GeV and using the QCD parameters in Tables\,\ref{tab:param} and \ref{tab:alfa}; {\bf b)} The same as a) but for the mass $M_{D^*_0D^*}$.
}
\label{fig:dstar0dstar-b01a} 
\end{center}
\end{figure}

\nin
We study in Fig. {\ref{fig:dstar0dstar-b01a}} the effect of factorization for PT$\oplus$NP at LO for a given value of $t_c=42$ GeV$^2$ and $\mu=4.5$ GeV at lowest order of PT for the $ \bar D^*_0D^*(1^-)$ molecule.  An analogous analysis is done for the $\bar B^*_0B^*$ molecule which presents the same qualitative behaviour
as the one in Fig. {\ref{fig:dstar0dstar-b01a}}.
\nin
\subsubsection*{\b Conclusions for the PT$\oplus$NP analysis}
One can notice from Fig.\,\ref{fig:dstar0dstar-b01a} that the effect of factorization of the PT$\oplus$NP at LO is about 2.2\% for the 
decay constant and 0.5\% for the mass which is quite tiny. However, to avoid this (small) effect, we shall work in the following with the full non-factorized PT$\oplus$NP of the  LO expressions. 
\subsection{Test at NLO of PT}
\subsubsection*{\b Example of the $B^0\bar B^0$ four-quark correlator}
\begin{figure}[hbt] 
\begin{center}
{\includegraphics[width=6.29cm  ]{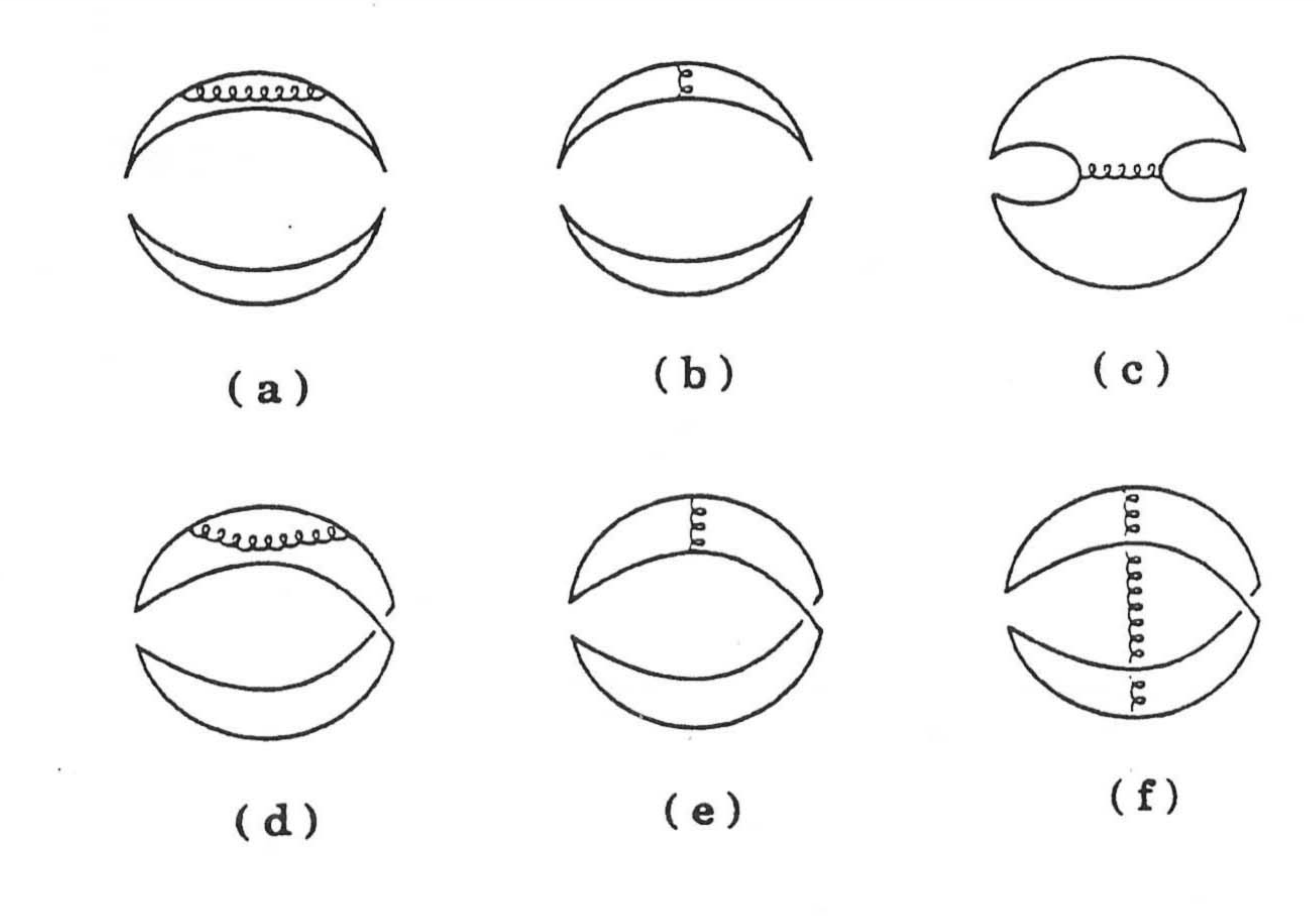}}
\caption{
\scriptsize 
{\bf (a,b)} Factorized contributions to the four-quark correlator at NLO of PT; {\bf (c to f)} Non-factorized contributions  at NLO of PT (the figure comes from\,\cite{PICH}).
}
\label{fig:factoras} 
\end{center}
\end{figure}

For extracting the PT $\alpha_s^n$ corrections to the correlator and due to the technical complexity of the calculations, we shall assume that these radiative corrections are dominated by the ones from the factorized diagrams (Fig.\,\ref{fig:factoras}a,b) while we neglect the ones from non-factorized diagrams (Fig.\,\ref{fig:factoras}c to f). This fact has been proven explicitly by \,\cite{BBAR2,BBAR3}  in the case of the $\bar B^0B^0$ systems (very similar correlator as the ones discussed in the following) where the non-factorized $\alpha_s$ corrections do not exceed 10\% of the total 
$\alpha_s$ contributions. 
\subsubsection*{\b Conclusions of the PT NLO analysis} 
\nin
We expect from the previous LO examples that the masses of the molecules are known with a good accuracy while, for the coupling, we shall have in mind the systematics induced by the radiative corrections estimated by keeping only the factorized diagrams. The contributions of the factorized diagrams will be extracted from the convolution integrals given in Eq.\,\ref{eq:convolution}. Here, the suppression of the NLO corrections will be more pronounced for the extraction of the meson masses from the ratio of sum rules compared to the case of the $\bar B^0B^0$ systems. 
\nin
\section{The  $(0^{++})$ Heavy-Light Scalar Molecule States}
We shall study the $\bar DD,~\bar D^*D^*$ and $\bar D^*_0D^*_0$ and their beauty analogue using the same
approaches and strategies. The qualitative behaviours of the curves in these different channels are very similar such that we shall only illustrate explicitly the analysis for the $\bar DD$ and $\bar BB$ molecules and will only quote the results for the others. 
\subsection{Decay constant and mass of the $ \bar DD$ molecule}

\subsubsection*{$\bullet$ $\tau$ and $t_c$ stabilities}
\begin{figure}[hbt] 
\begin{center}
{\includegraphics[width=6.29cm  ]{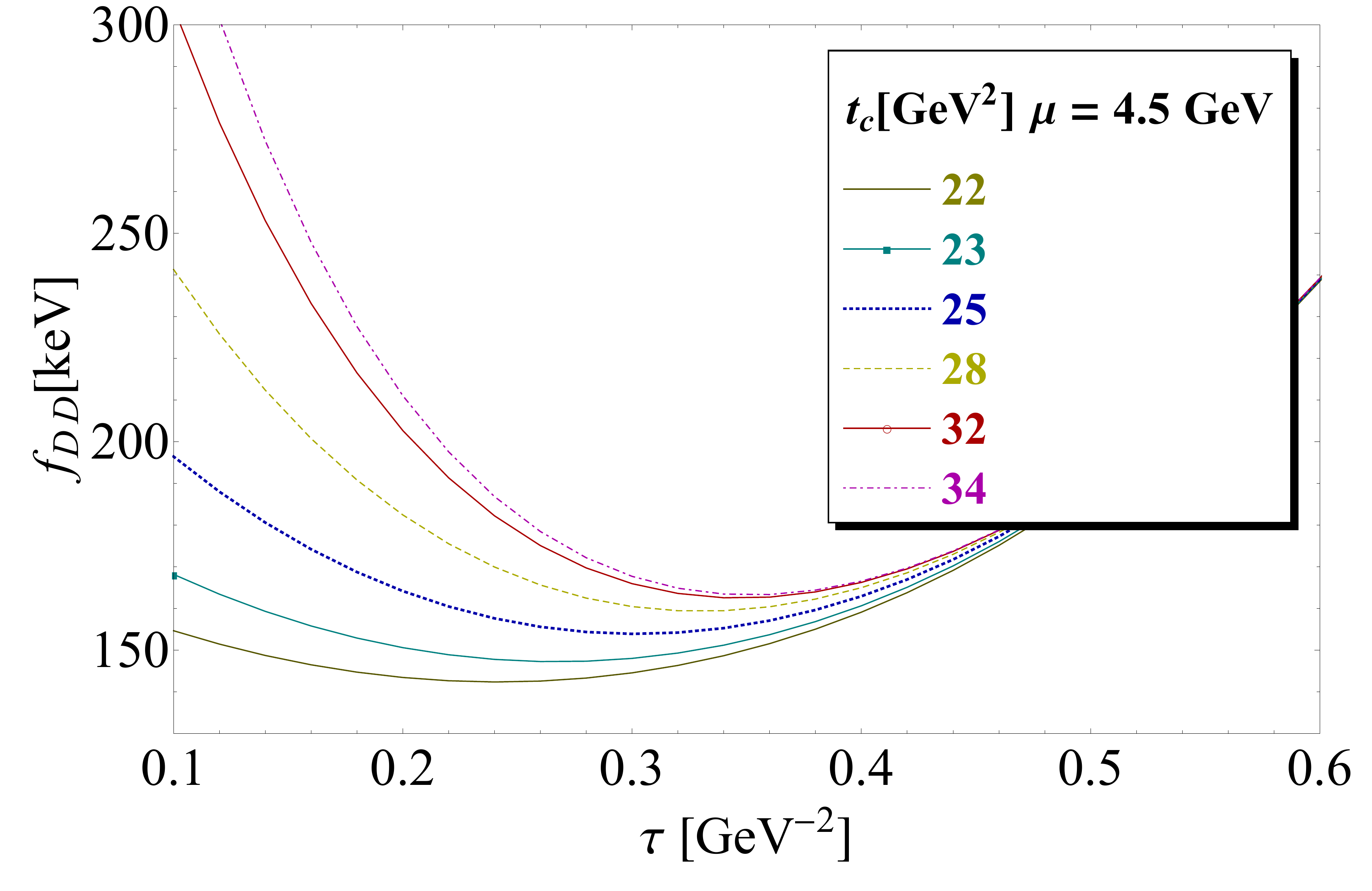}}
{\includegraphics[width=6.29cm  ]{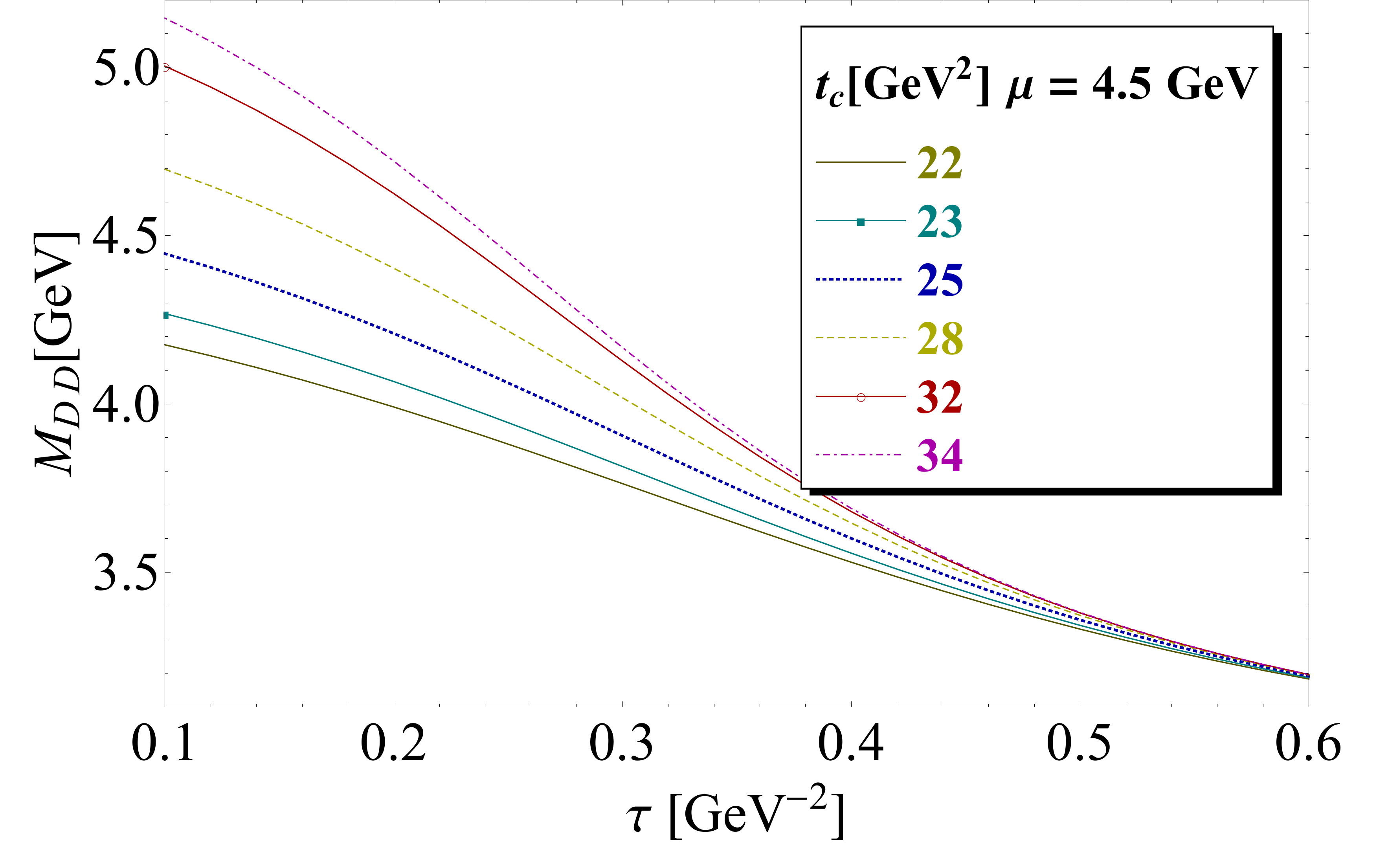}}
\centerline {\hspace*{-3cm} a)\hspace*{6cm} b) }
\caption{
\scriptsize 
{\bf a)} $f_{DD}$  at LO as function of $\tau$ for different values of $t_c$, for $\mu=4.5$ GeV  and for the QCD parameters in Tables\,\ref{tab:param} and \ref{tab:alfa}; {\bf b)} The same as a) but for the mass $M_{DD}$.
}
\label{fig:d-lo} 
\end{center}
\end{figure}

\begin{figure}[hbt] 
\begin{center}
{\includegraphics[width=6.29cm  ]{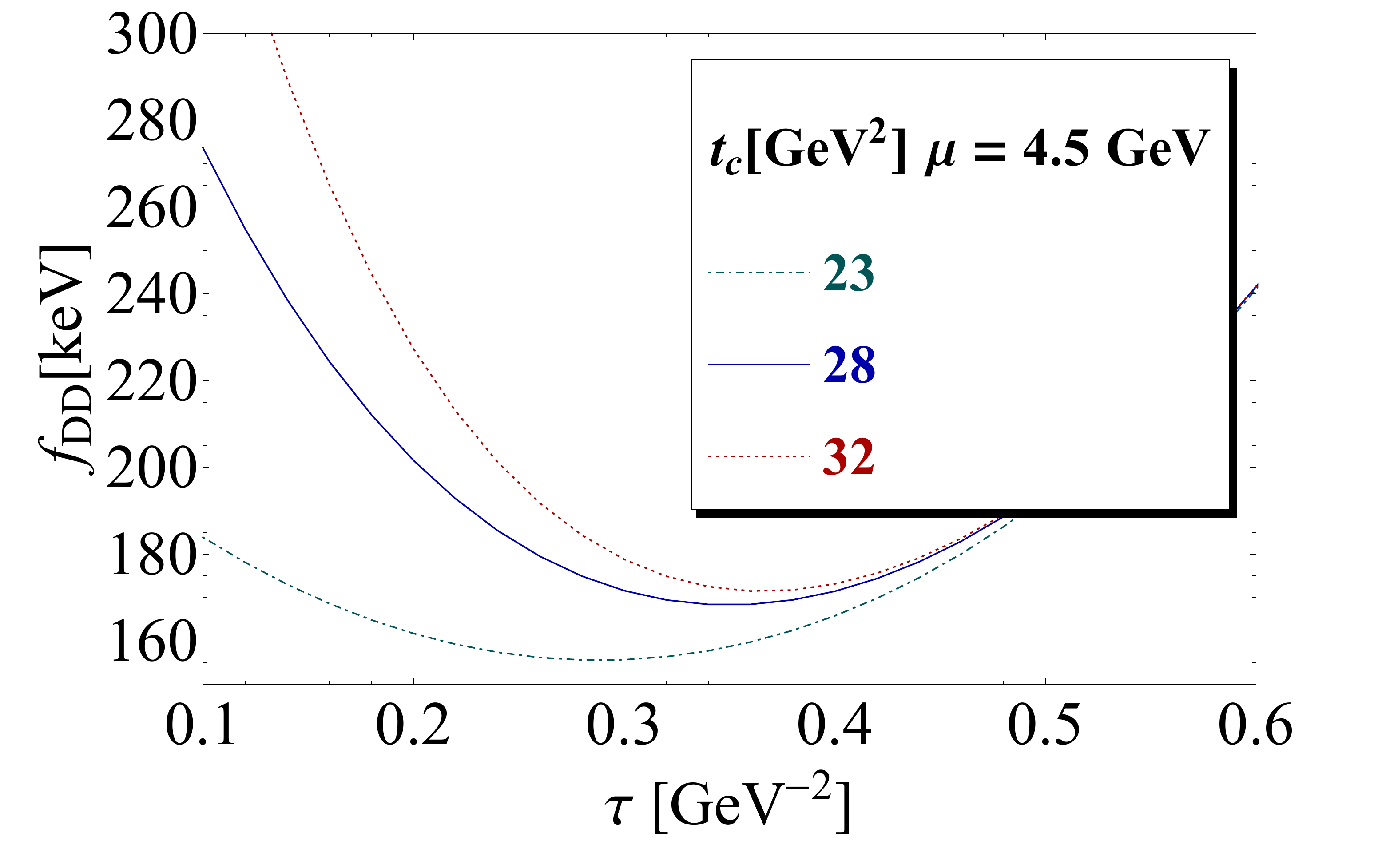}}
{\includegraphics[width=6.29cm  ]{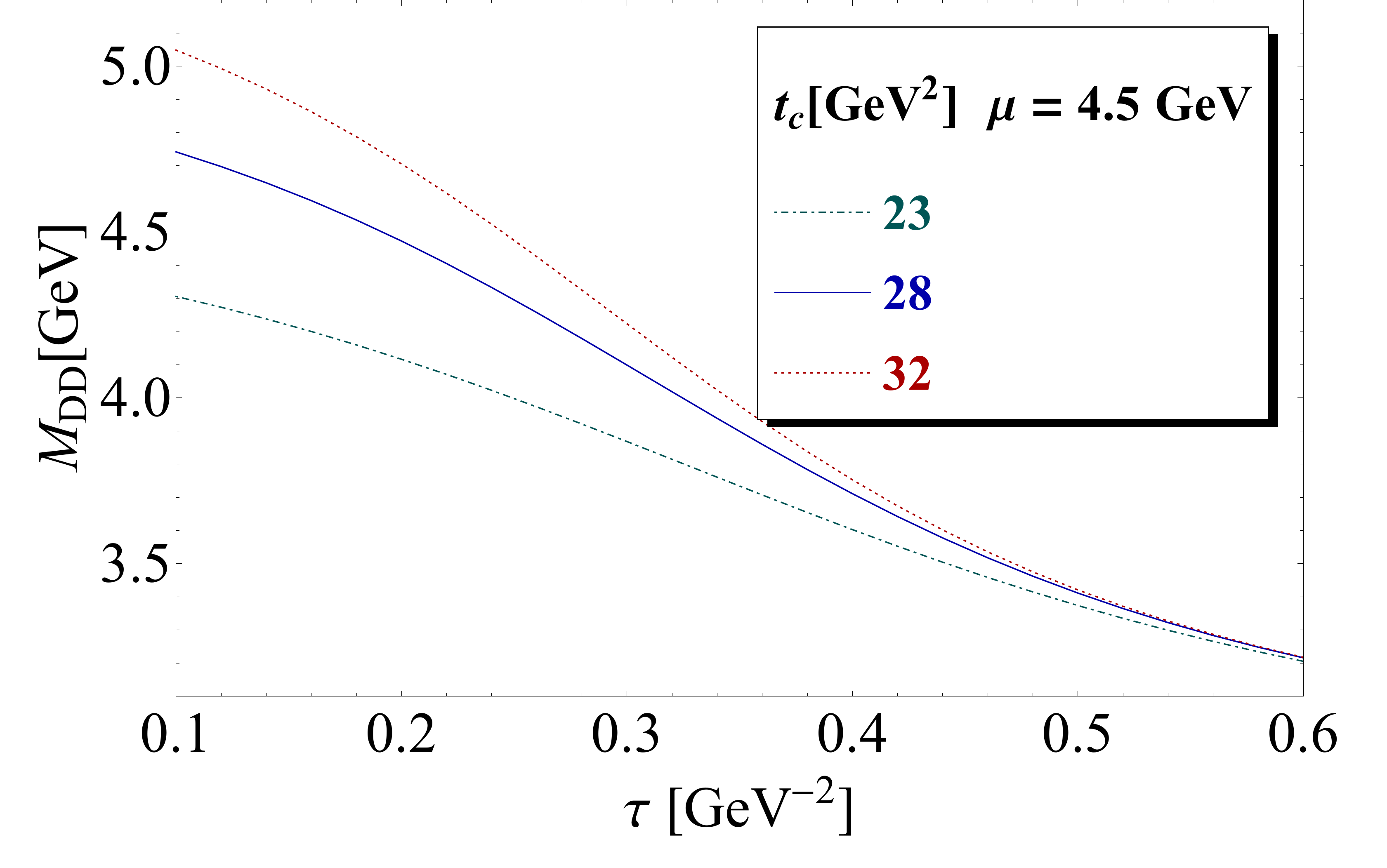}}
\centerline {\hspace*{-3cm} a)\hspace*{6cm} b) }
\caption{
\scriptsize 
{\bf a)} $f_{DD}$ at NLO  as function of $\tau$ for different values of $t_c$, for $\mu=4.5$ GeV  and for the QCD parameters in Tables\,\ref{tab:param} and \ref{tab:alfa}; {\bf b)} The same as a) but for the mass $M_{DD}$.
}
\label{fig:d-nlo} 
\end{center}
\end{figure}

\begin{figure}[hbt] 
\begin{center}
{\includegraphics[width=6.29cm  ]{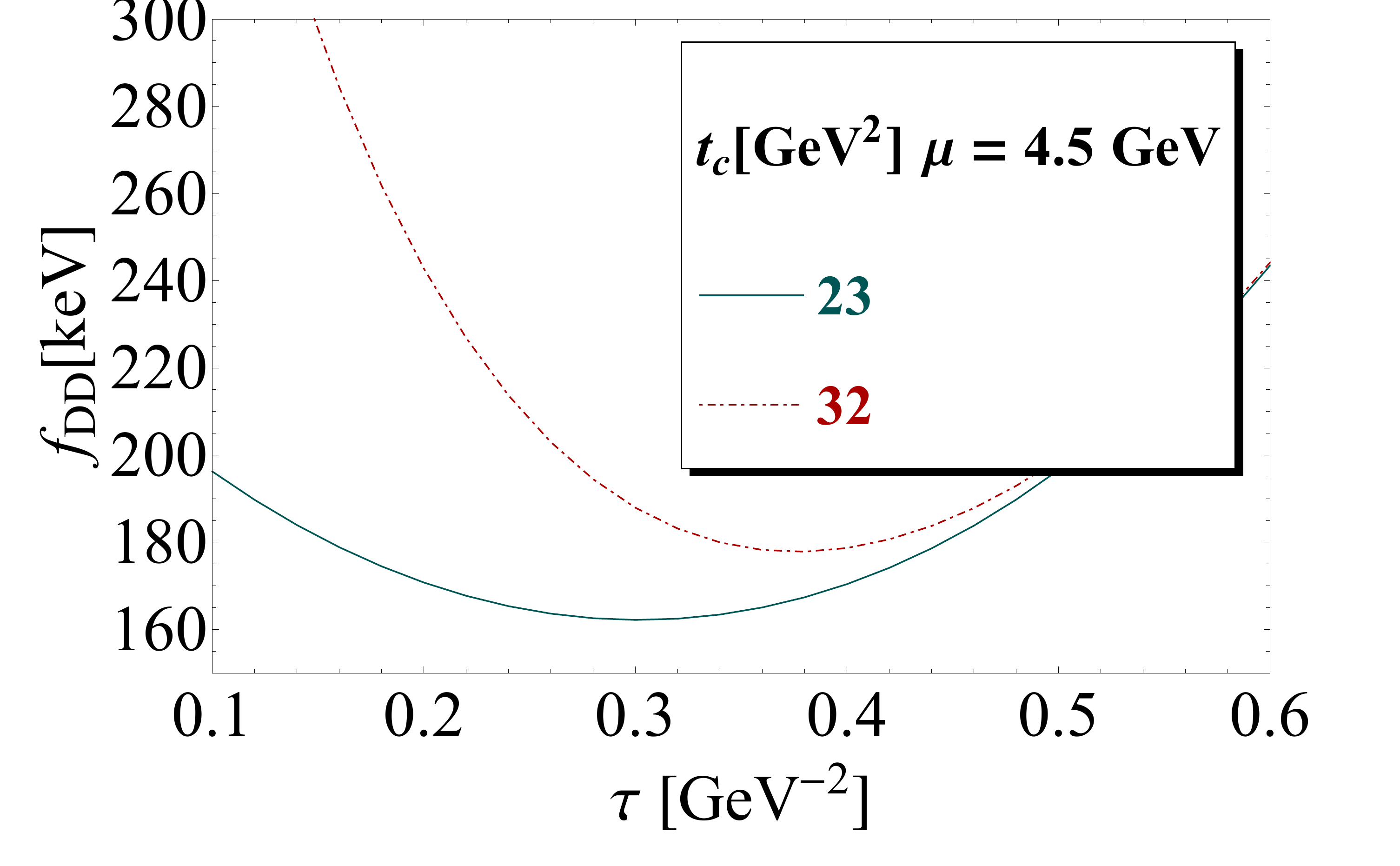}}
{\includegraphics[width=6.29cm  ]{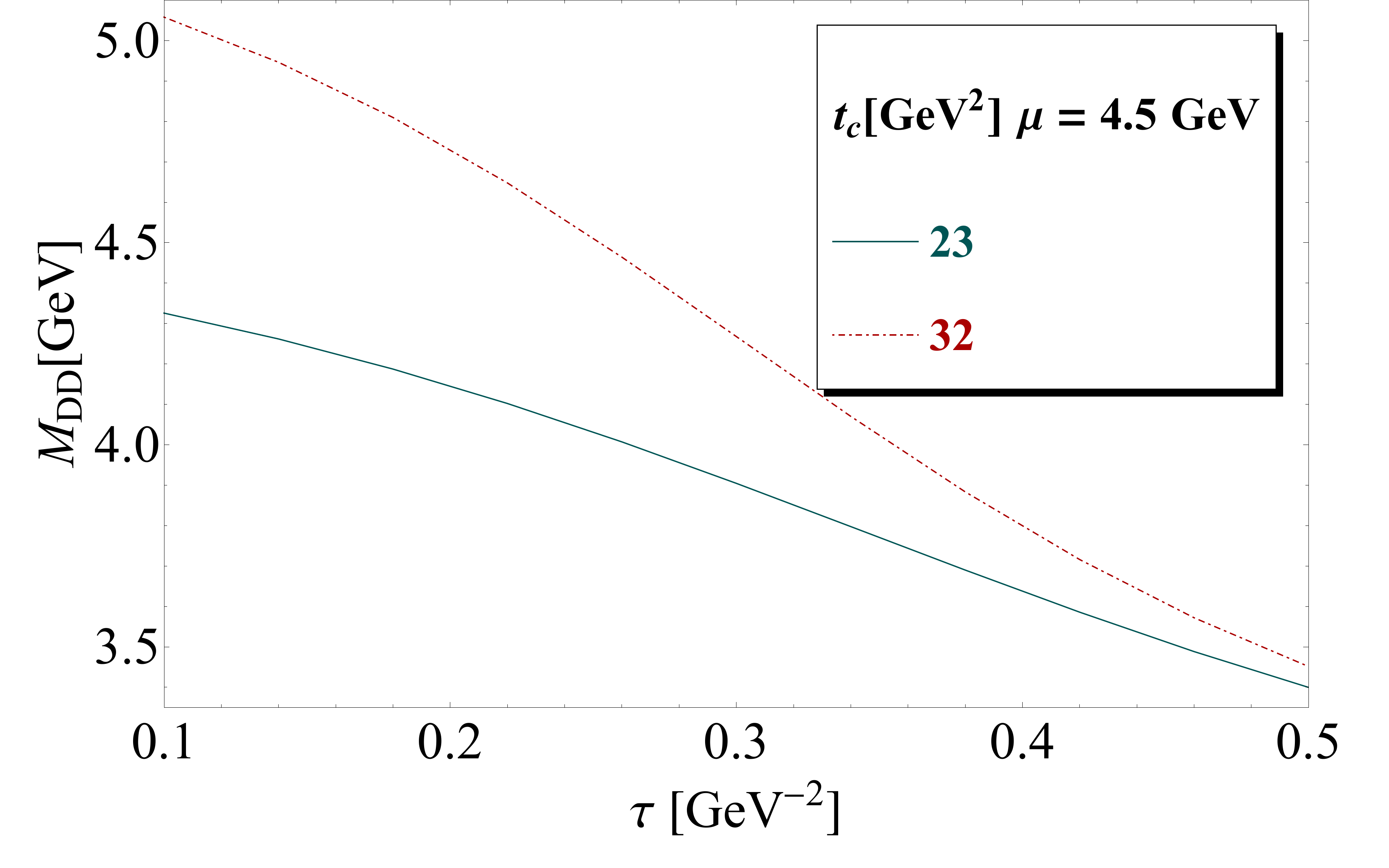}}
\centerline {\hspace*{-3cm} a)\hspace*{6cm} b) }
\caption{
\scriptsize 
{\bf a)} $f_{DD}$ at N2LO  as function of $\tau$ for different values of $t_c$, for $\mu=4.5$ GeV  and for the QCD parameters in Tables\,\ref{tab:param} and \ref{tab:alfa}; {\bf b)} The same as a) but for the mass $M_{DD}$.
}
\label{fig:d-n2lo} 
\end{center}
\end{figure}

\nin
 We study the behaviour of the coupling\,\footnote{Here and in the following : decay constant is the same as : coupling.} $f_{DD}$  and mass $M_{DD}$ in terms of the LSR variable $\tau$ at different values of $t_c$ as shown in Fig.\ref{fig:d-lo} at LO, in Fig.\,\ref{fig:d-nlo} at NLO  and in Fig.\,\ref{fig:d-n2lo} at N2LO.
\nin
 We consider, as a final and conservative result, the one corresponding to the beginning of the $\tau$-stability ($\tau\simeq$ 0.25 GeV$^{-2}$) for $t_c$=22 GeV$^2$ until the one where $t_c$-stability starts to be reached for $t_c\simeq$ 32 GeV$^2$ and for $\tau\simeq$ 0.35 GeV$^{-2}$. In these stability regions, the requirement that the pole contribution is larger than the one of the continuum  is automatically satisfied (see e.g.\,\cite{MOLE5}). 
\nin
\nin
   \subsubsection*{\b Running versus the pole quark mass definitions}
\begin{figure}[hbt] 
\begin{center}
{\includegraphics[width=6.29cm  ]{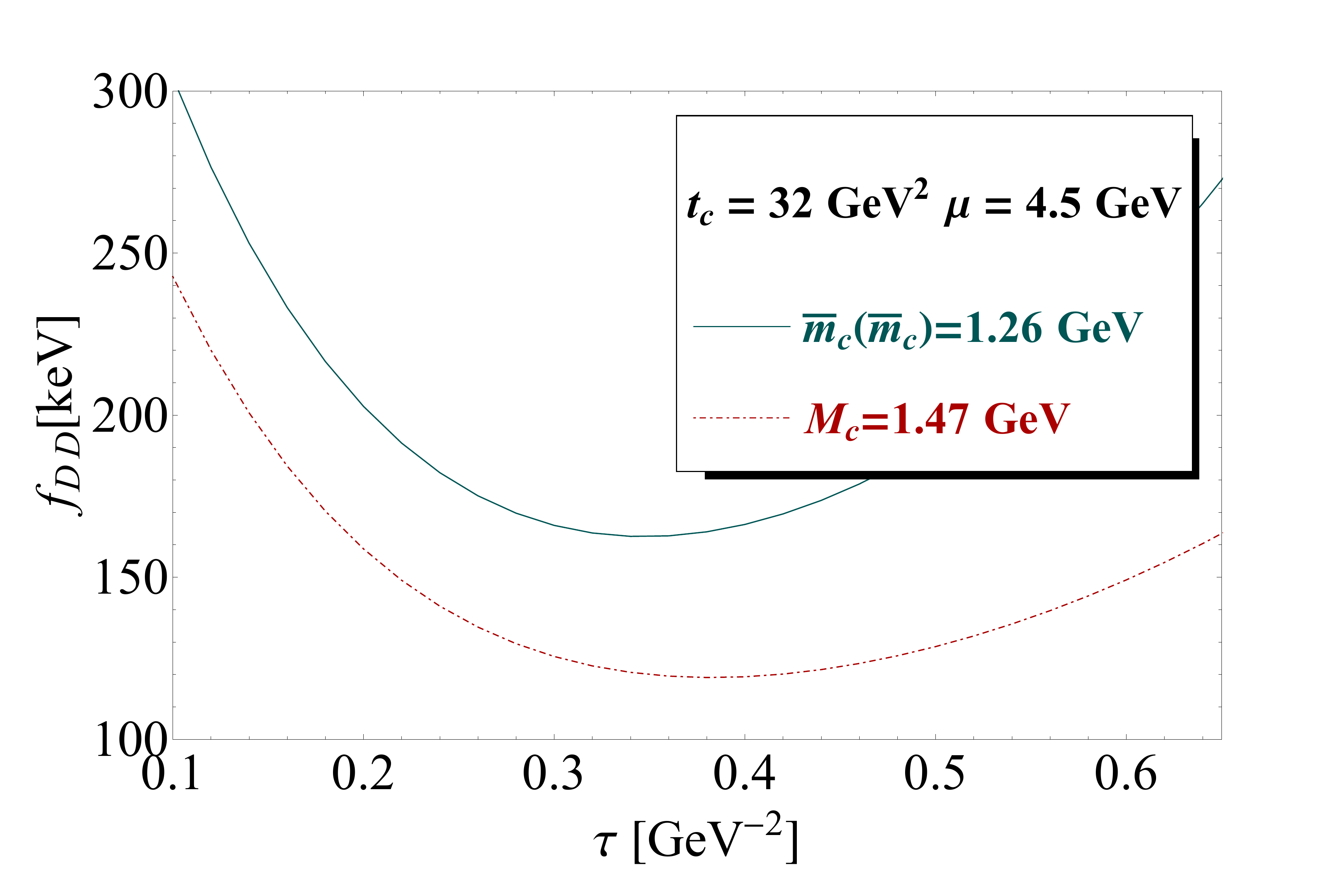}}
{\includegraphics[width=6.29cm  ]{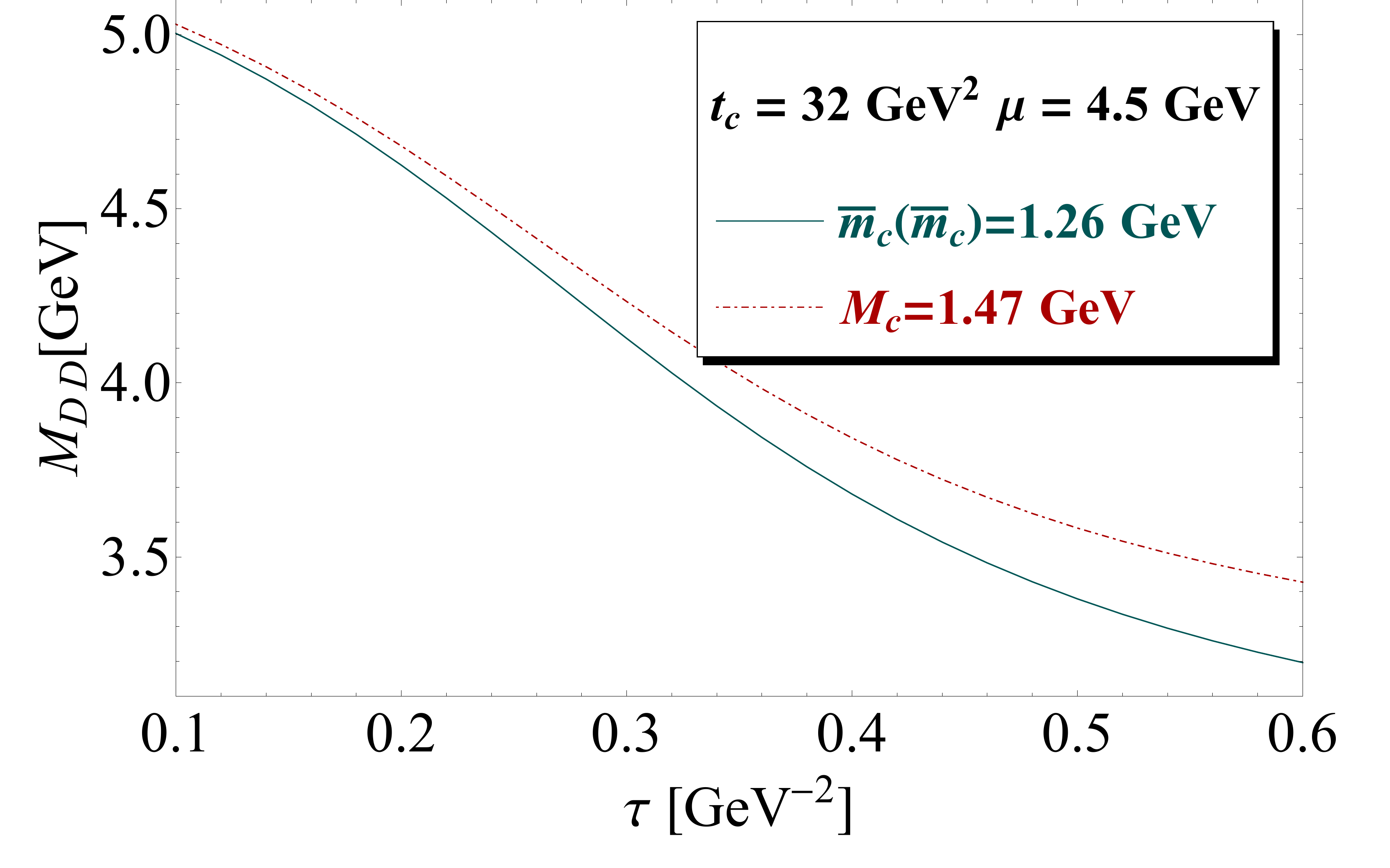}}
\centerline {\hspace*{-3cm} a)\hspace*{6cm} b) }
\caption{
\scriptsize 
{\bf a)} $f_{DD}$  at LO as function of $\tau$ for  $t_c=32$ GeV$^2$, for $\mu=4.5$ GeV, for  values of the running $\overline{m}_c(\overline{m}_c)=1.26$ GeV and pole mass $M_c=1.47$ GeV. We use     
the QCD parameters in Tables\,\ref{tab:param} and \ref{tab:alfa}; {\bf b)} The same as a) but for the mass $M_{DD}$.
}
\label{fig:dd-const} 
\end{center}
\end{figure}

   \nin
   We show in Fig.\,\ref{fig:dd-const} the effect of the definitions (running and pole) of the heavy quark mass used in the analysis at LO which is relatively important. The difference should be added as errors in the LO analysis.  This source of errors is never considered in the current literature.
\nin
\subsubsection*{$\bullet$ Convergence of the PT series} 
\begin{figure}[hbt] 
\begin{center}
{\includegraphics[width=6.29cm  ]{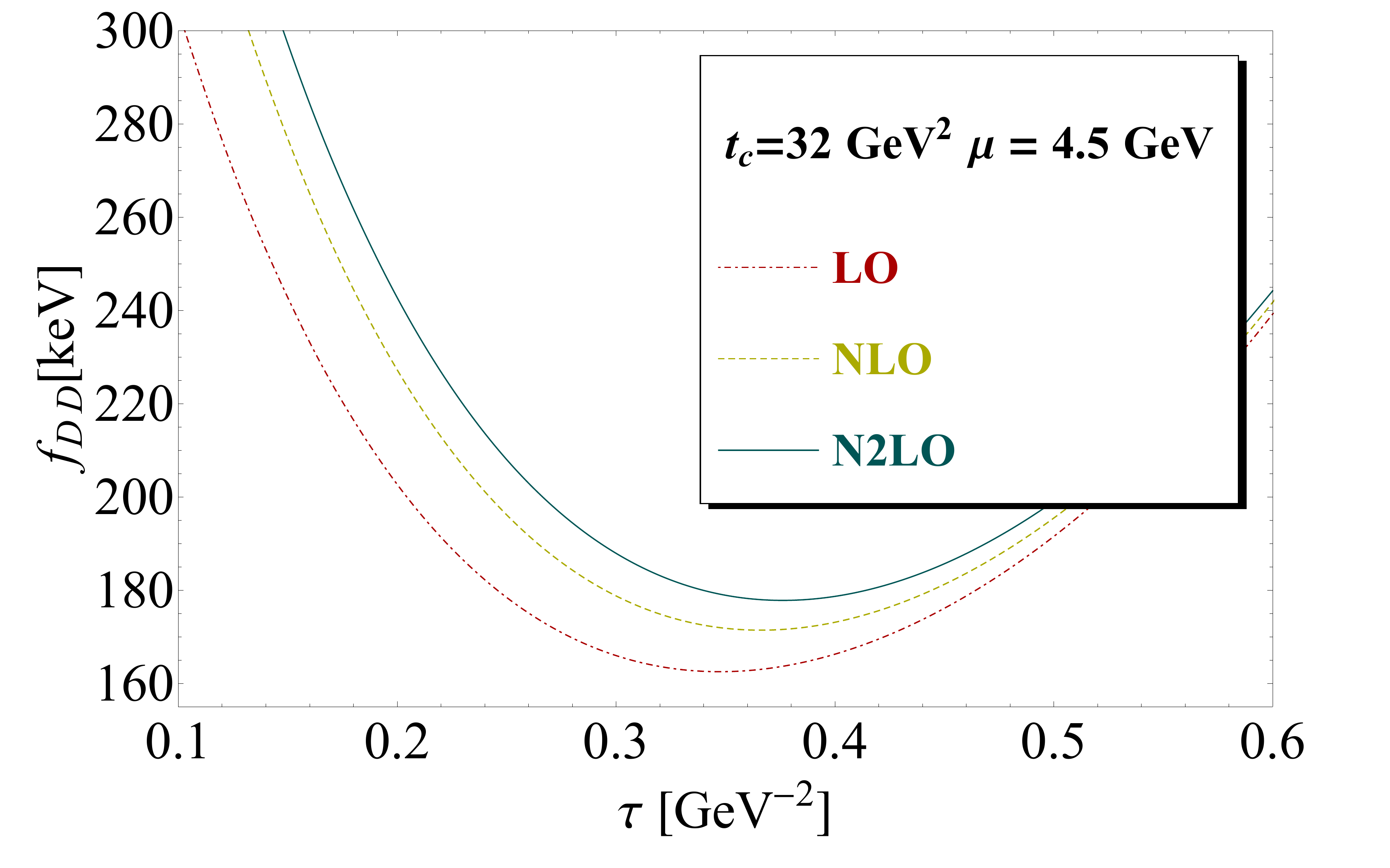}}
{\includegraphics[width=6.29cm  ]{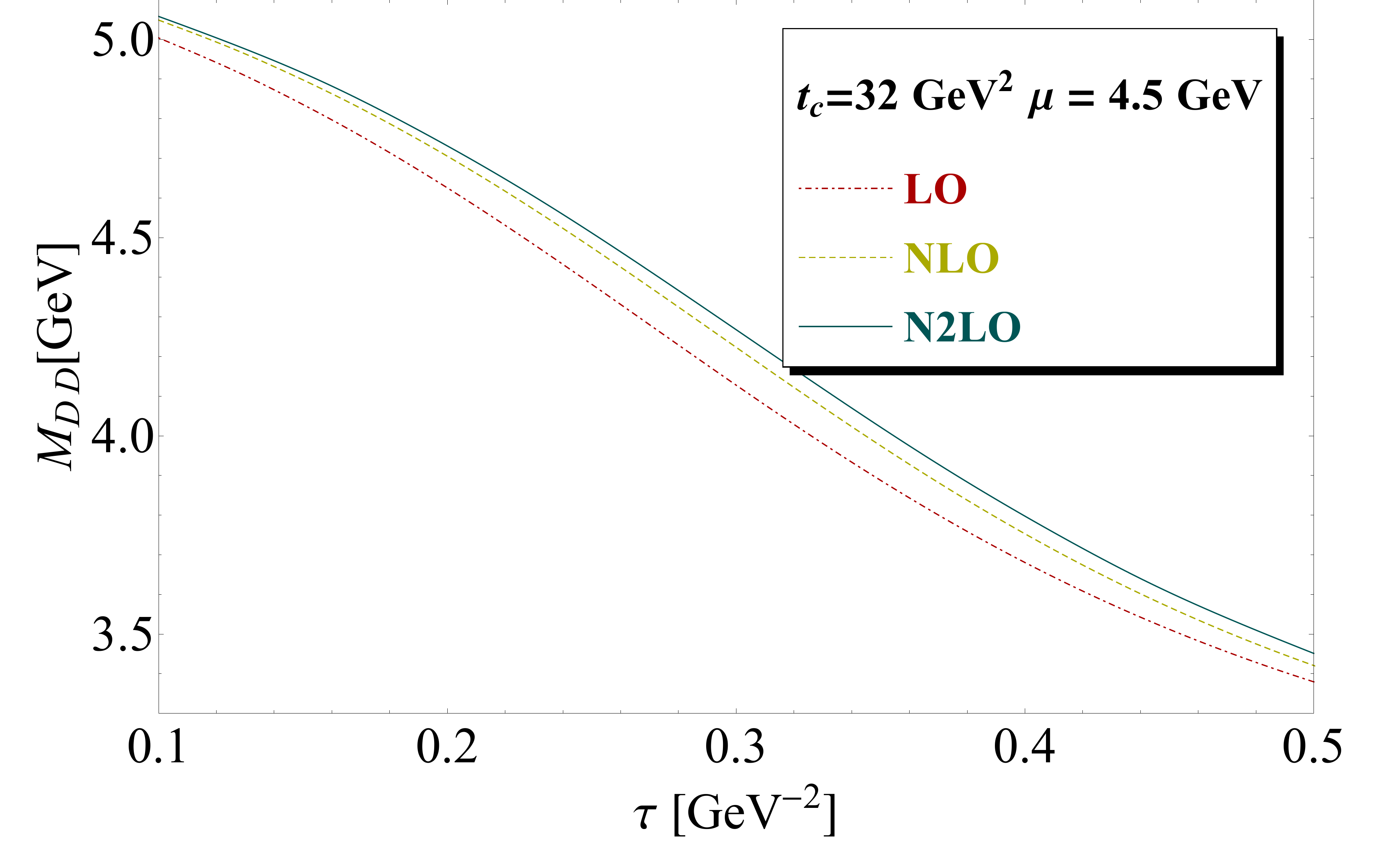}}
\centerline {\hspace*{-3cm} a)\hspace*{6cm} b) }
\caption{
\scriptsize 
{\bf a)} $f_{DD}$  as function of $\tau$ for a given value of $t_c=32$ GeV$^2$, for $\mu=4.5$ GeV, for different truncation of the PT series  and for the QCD parameters in Tables\,\ref{tab:param} and \ref{tab:alfa}; {\bf b)} The same as a) but for the mass $M_{DD}$.
}
\label{fig:d-pt} 
\end{center}
\end{figure}

\nin
Using  $t_c=32$ GeV$^2$, we study in Fig. {\ref{fig:d-pt}} the convergence of the PT series for a given value of $\mu=4.5$ GeV.  We observe (see Table\,\ref{tab:resultc}) that from NLO to N2LO the mass decreases by about only 1 per mil indicating the good convergence of the PT series.
\nin
\begin{figure}[hbt] 
\begin{center}
{\includegraphics[width=6.2cm  ]{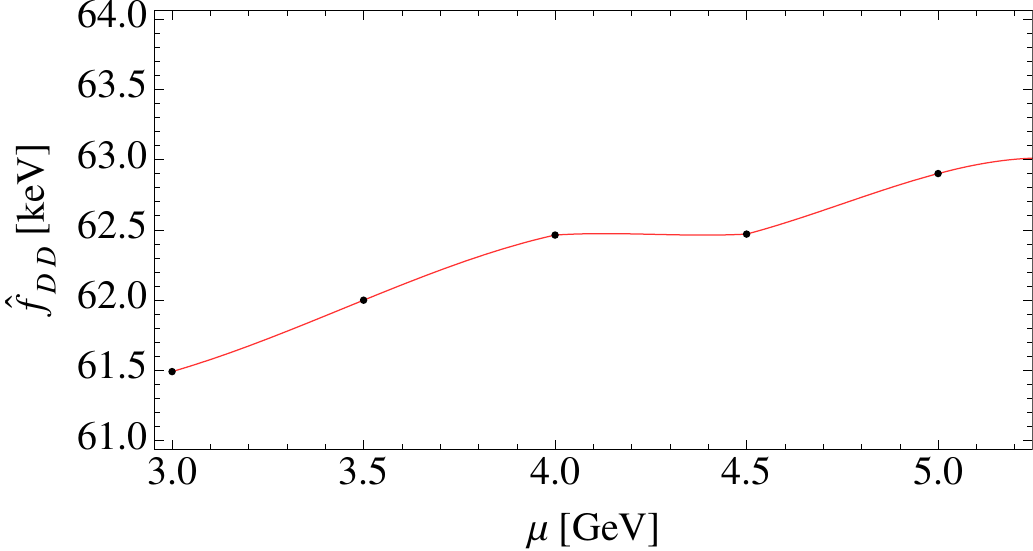}}
{\includegraphics[width=6.2cm  ]{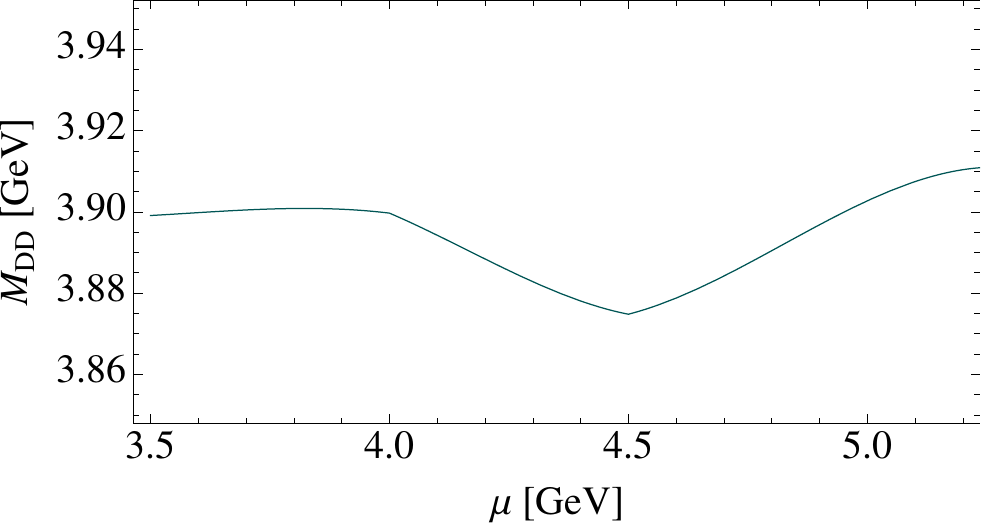}}
\centerline {\hspace*{-3cm} a)\hspace*{6cm} b) }
\caption{
\scriptsize 
{\bf a)} Renormalization group invariant coupling $\hat f_{DD}$ at NLO as function of $\mu$, for the corresponding $\tau$-stability region, for $t_c\simeq 32$ GeV$^2$ and for the QCD parameters in Tables\,\ref{tab:param}  and \ref{tab:alfa};  {\bf b)} The same as a) but for the  mass $ M_{DD}$.
}
\label{fig:d-mu} 
\end{center}
\end{figure} 
\nin
\subsubsection*{$\bullet$ $\mu$-stability}
\nin
We improve our previous results by using different values of $\mu$ (Fig. {\ref{fig:d-mu}}). Using the fact that 
the final result must be independent of the arbitrary parameter $\mu$ (plateau / inflexion point for the coupling and minimum for the mass), we consider as an optimal result the one at  $\mu\simeq 4.5$ GeV where we deduce the result given in Table\,\ref{tab:resultc}.
\subsection{Coupling and mass of the $ \bar BB$ molecule}
\begin{figure}[hbt] 
\begin{center}
{\includegraphics[width=6.34cm  ]{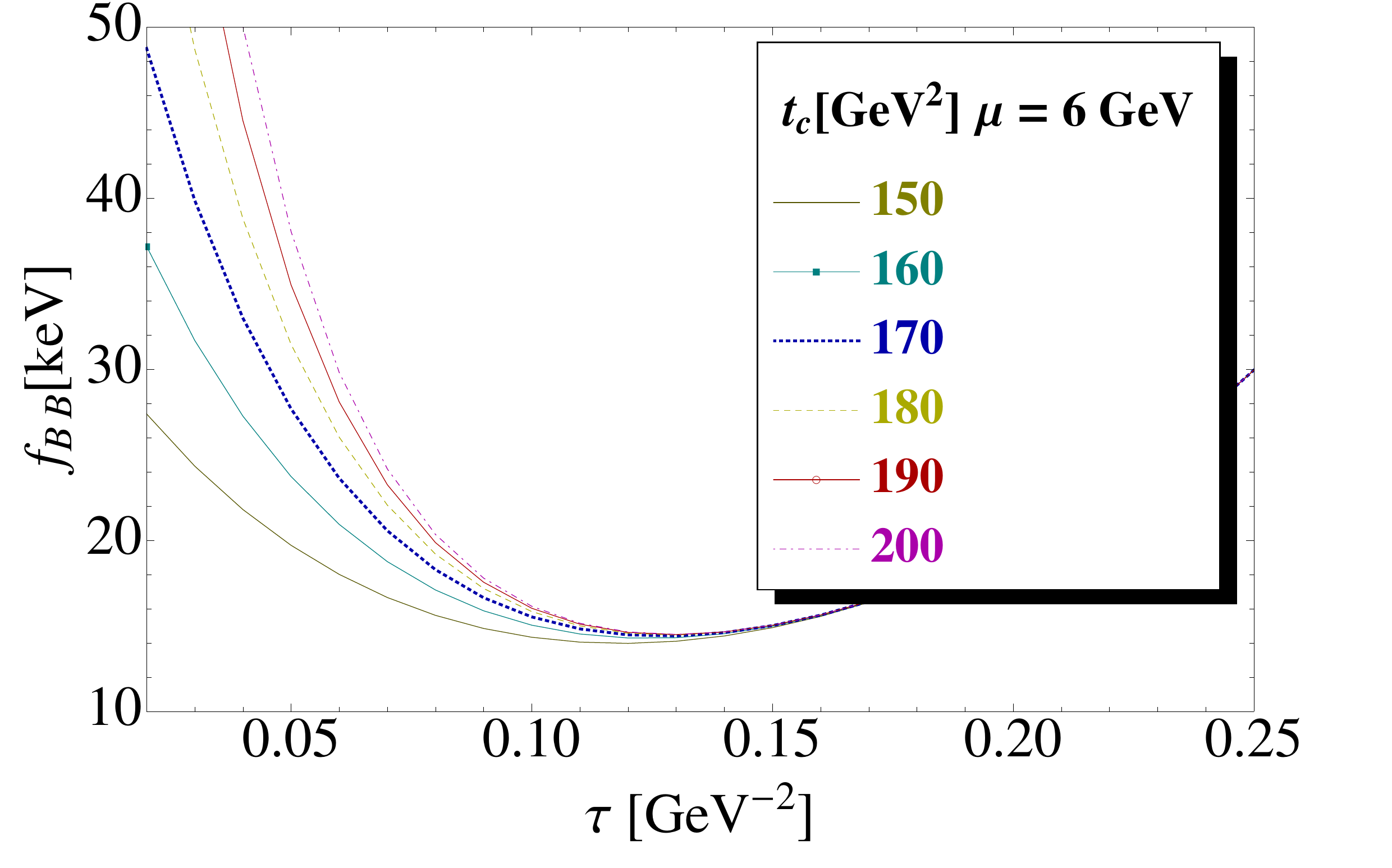}}
{\includegraphics[width=6.24cm  ]{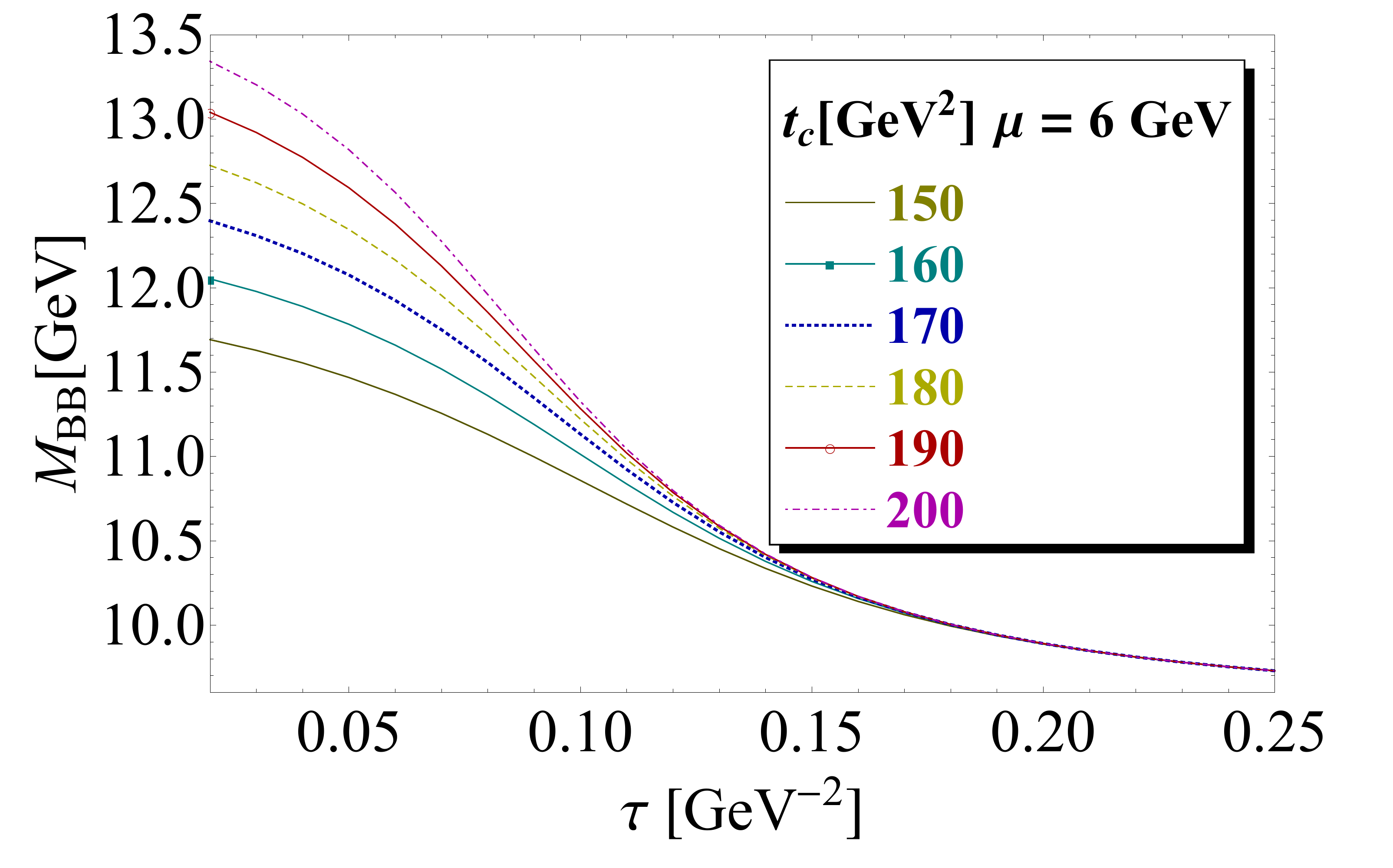}}
\centerline {\hspace*{-3cm} a)\hspace*{6cm} b) }
\caption{
\scriptsize 
{\bf a)} $f_{BB}$  at LO as function of $\tau$ for different values of $t_c$, for $\mu=6$ GeV  and for the QCD parameters in Tables\,\ref{tab:param} and \ref{tab:alfa}; {\bf b)} The same as a) but for the mass $M_{BB}$.
}
\label{fig:b-lo} 
\end{center}
\end{figure}

\begin{figure}[hbt] 
\begin{center}
{\includegraphics[width=6.29cm  ]{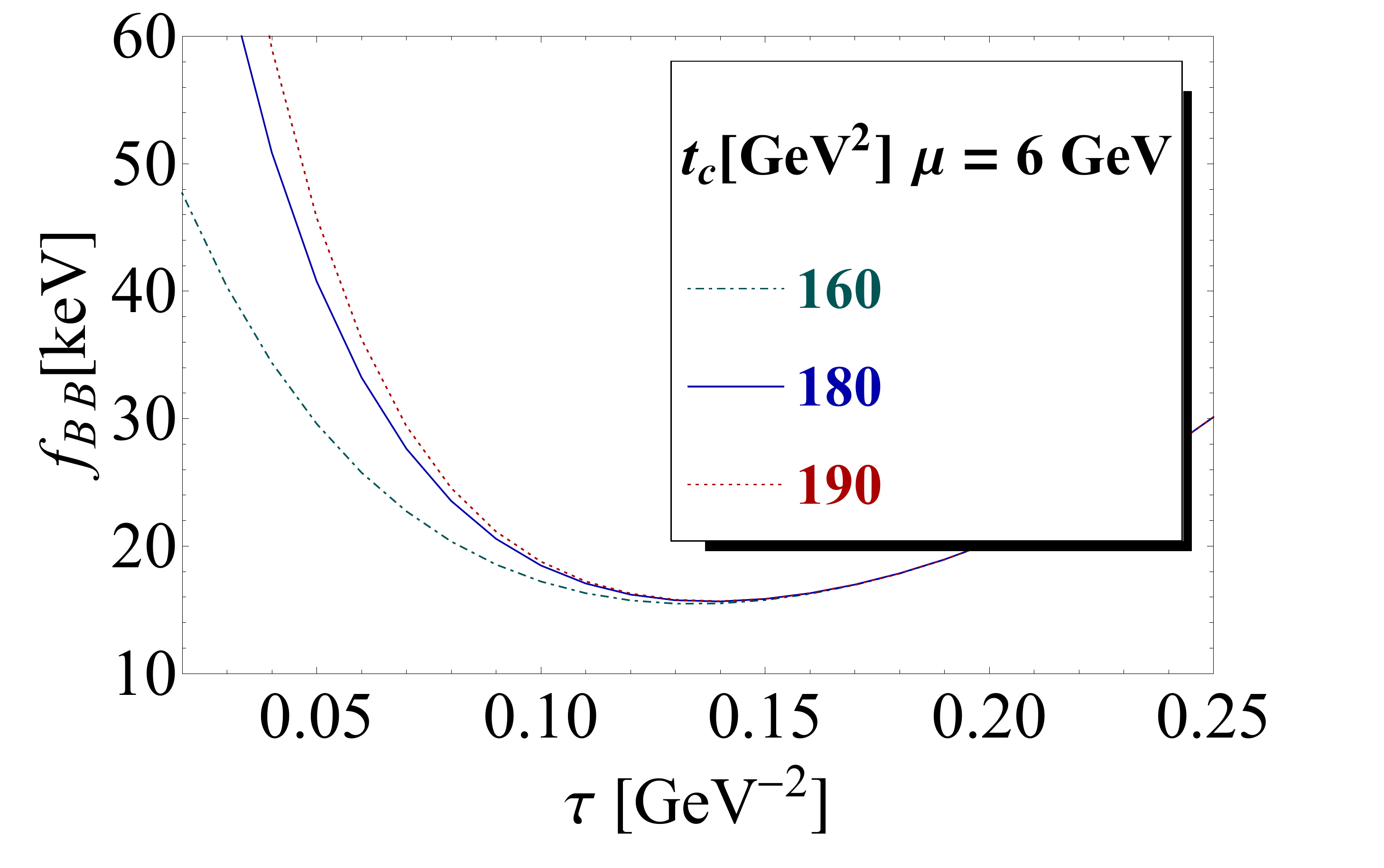}}
{\includegraphics[width=6.29cm  ]{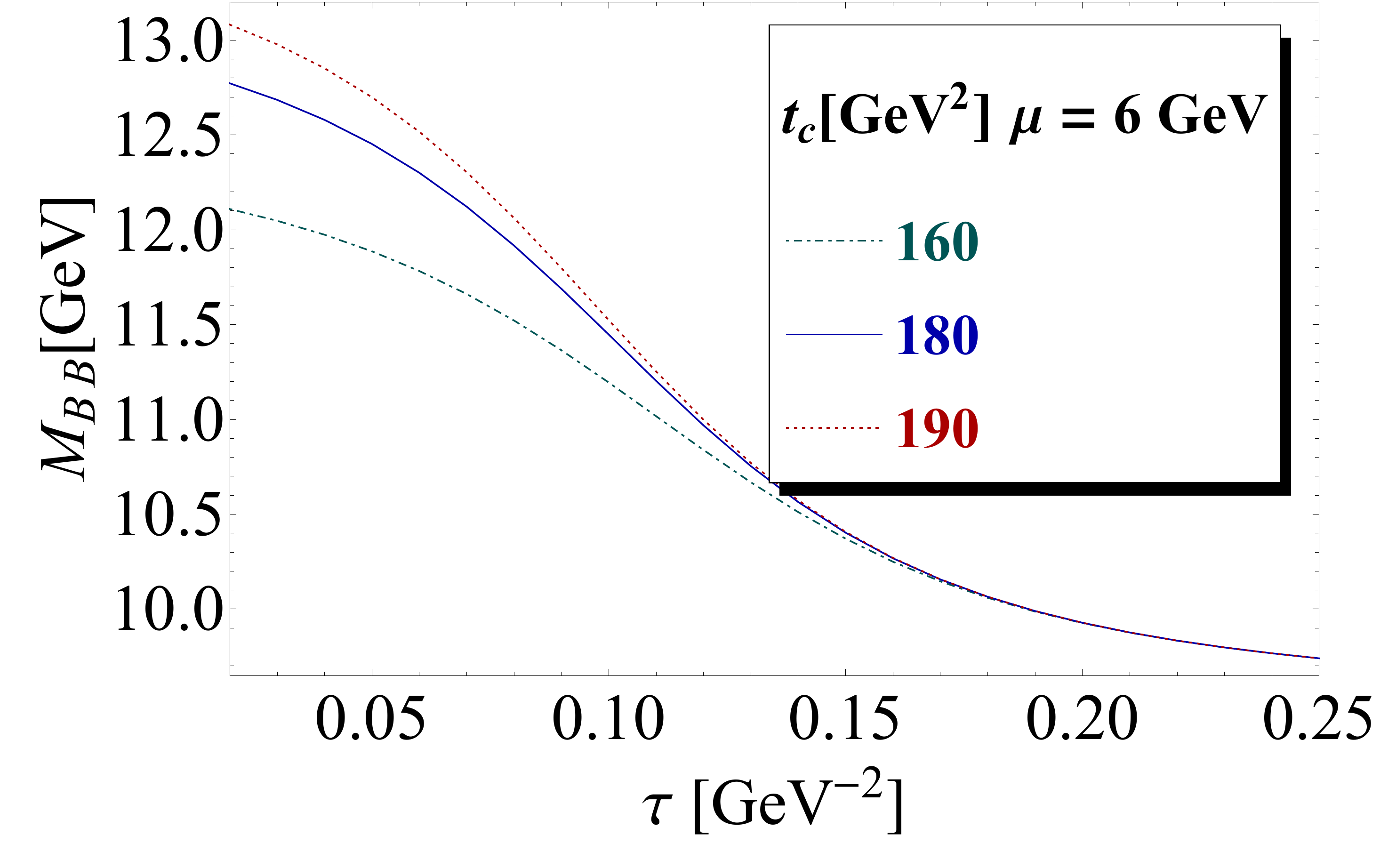}}
\centerline {\hspace*{-3cm} a)\hspace*{6cm} b) }
\caption{
\scriptsize 
{\bf a)} $f_{BB}$ at NLO  as function of $\tau$ for different values of $t_c$, for $\mu=6$ GeV  and for the QCD parameters in Tables\,\ref{tab:param} and \ref{tab:alfa}; {\bf b)} The same as a) but for the mass $M_{BB}$.
}
\label{fig:b-nlo} 
\end{center}
\end{figure}

\begin{figure}[hbt] 
\begin{center}
{\includegraphics[width=6.29cm  ]{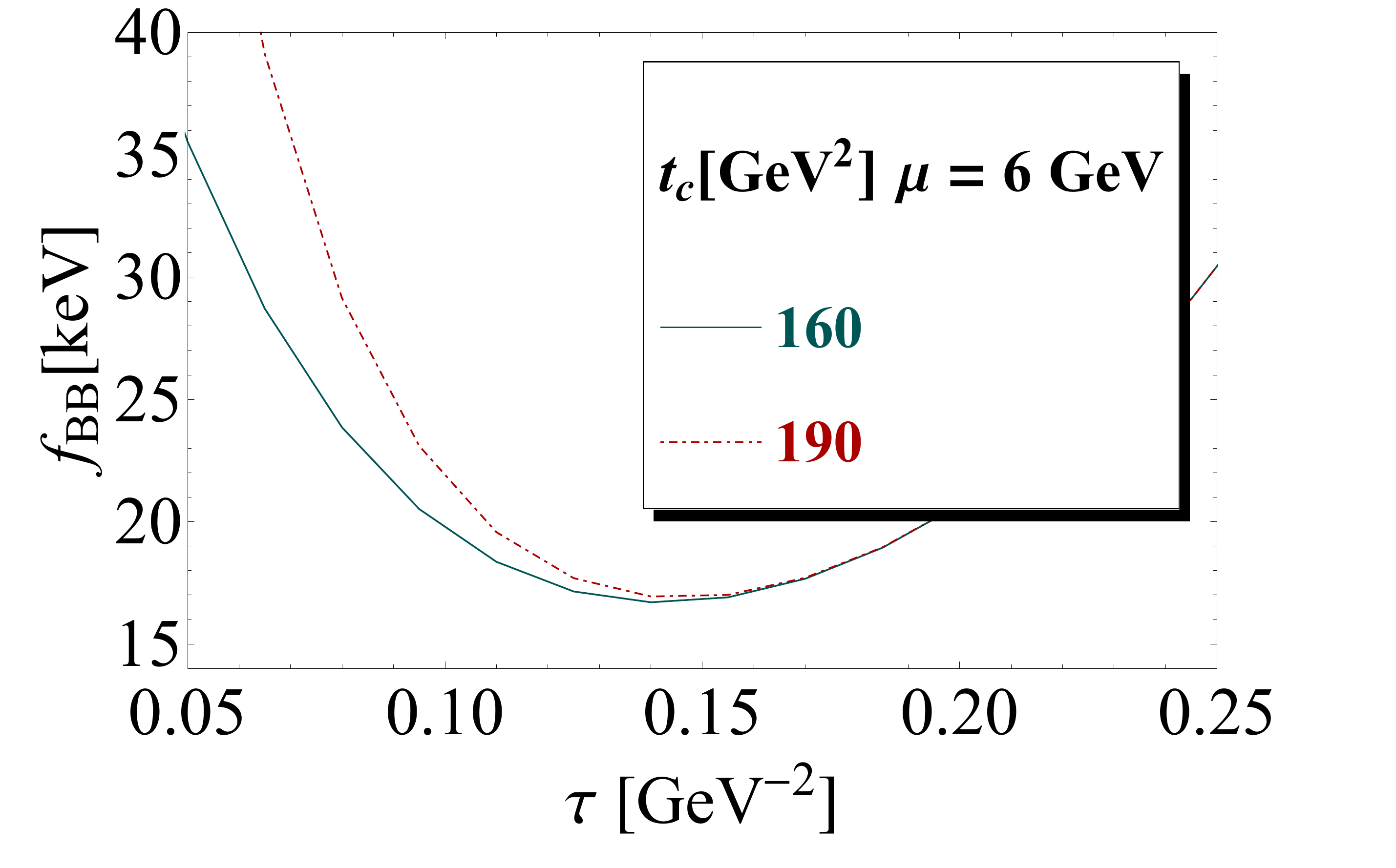}}
{\includegraphics[width=6.29cm  ]{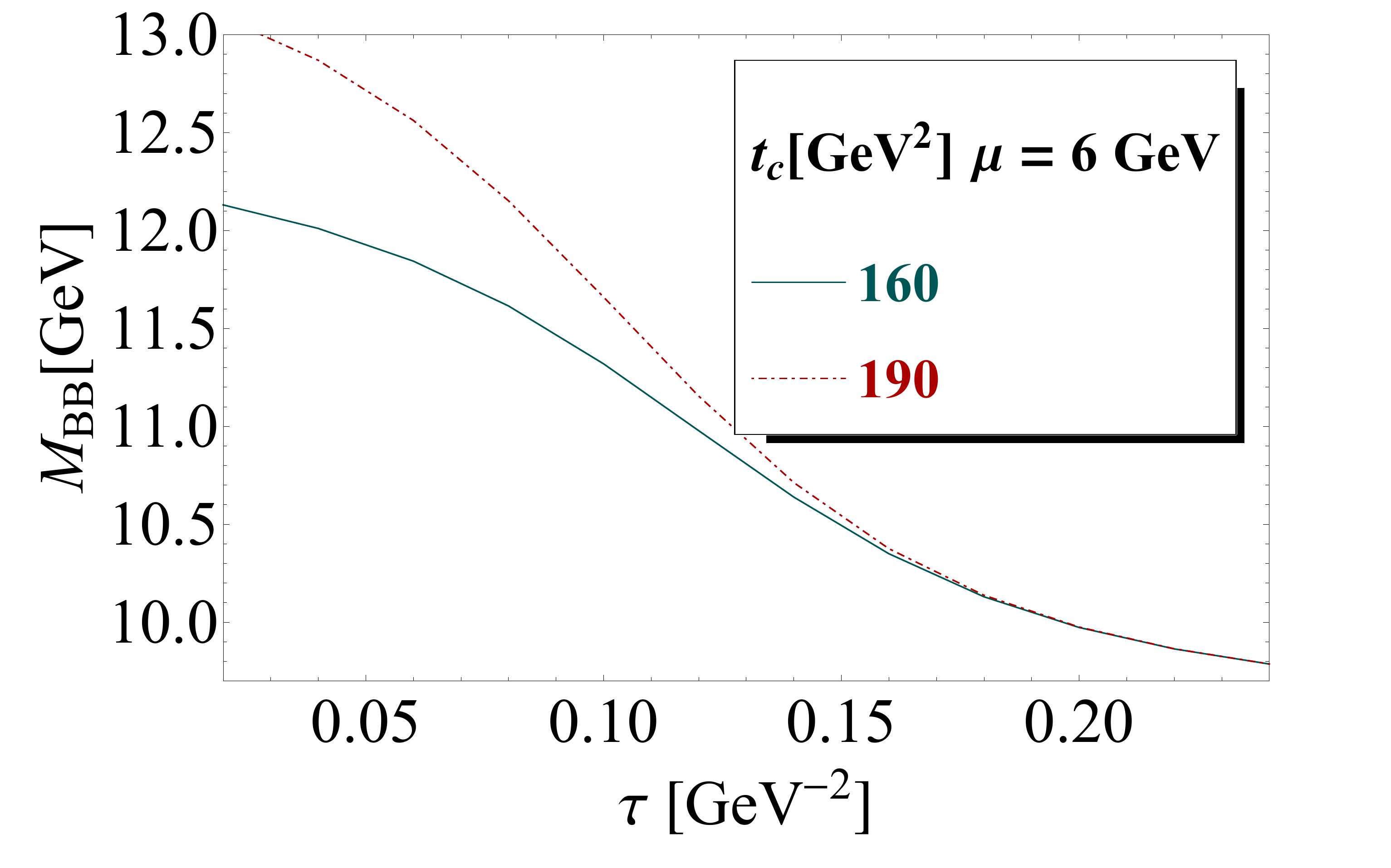}}
\centerline {\hspace*{-3cm} a)\hspace*{6cm} b) }
\caption{
\scriptsize 
{\bf a)} $f_{BB}$ at N2LO  as function of $\tau$ for different values of $t_c$, for $\mu=6$ GeV  and for the QCD parameters in Tables\,\ref{tab:param} and \ref{tab:alfa}; {\bf b)} The same as a) but for the mass $M_{BB}$.
}
\label{fig:b-n2lo} 
\end{center}
\end{figure}

\begin{figure}[hbt] 
\begin{center}
{\includegraphics[width=6.29cm  ]{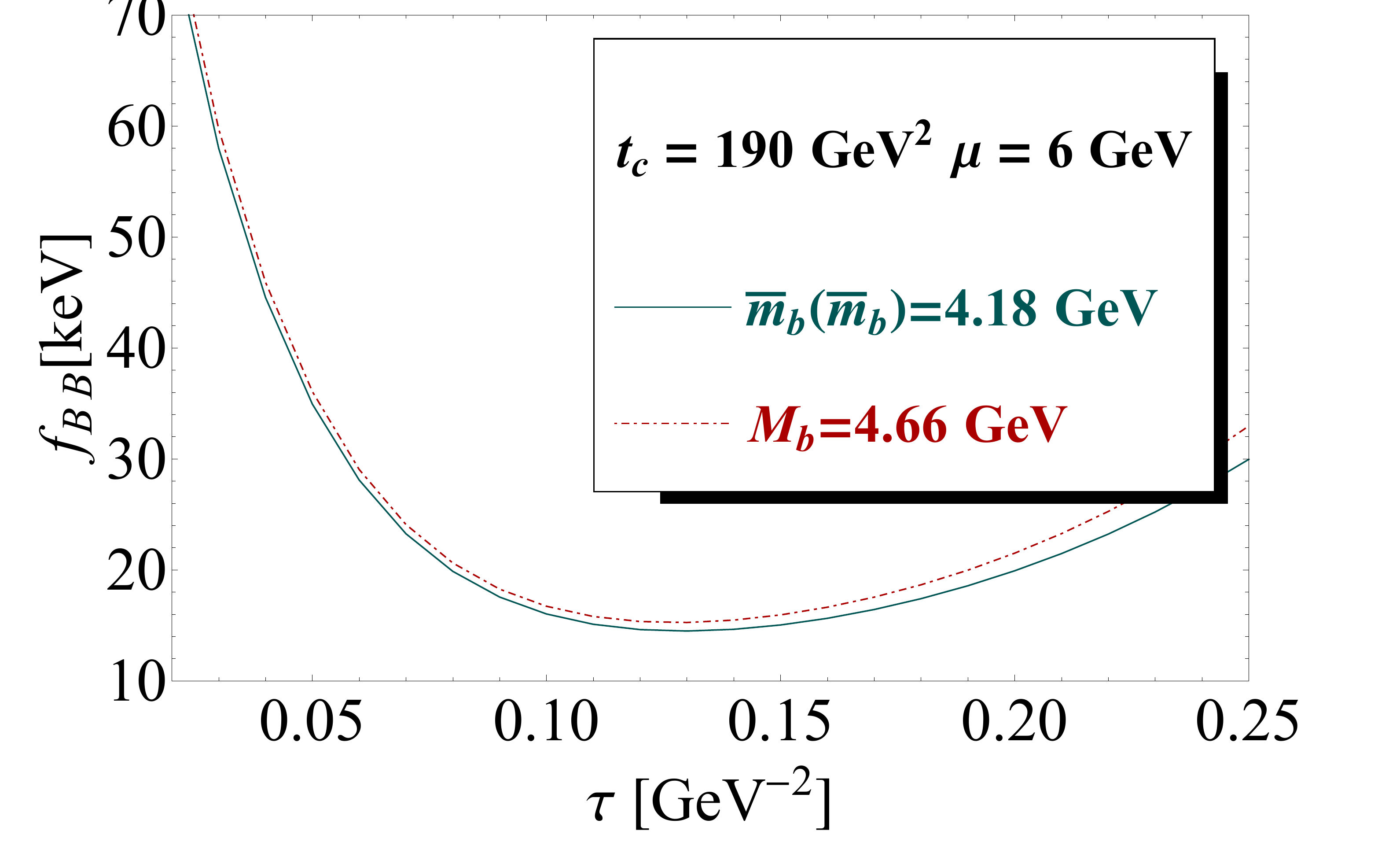}}
{\includegraphics[width=6.29cm  ]{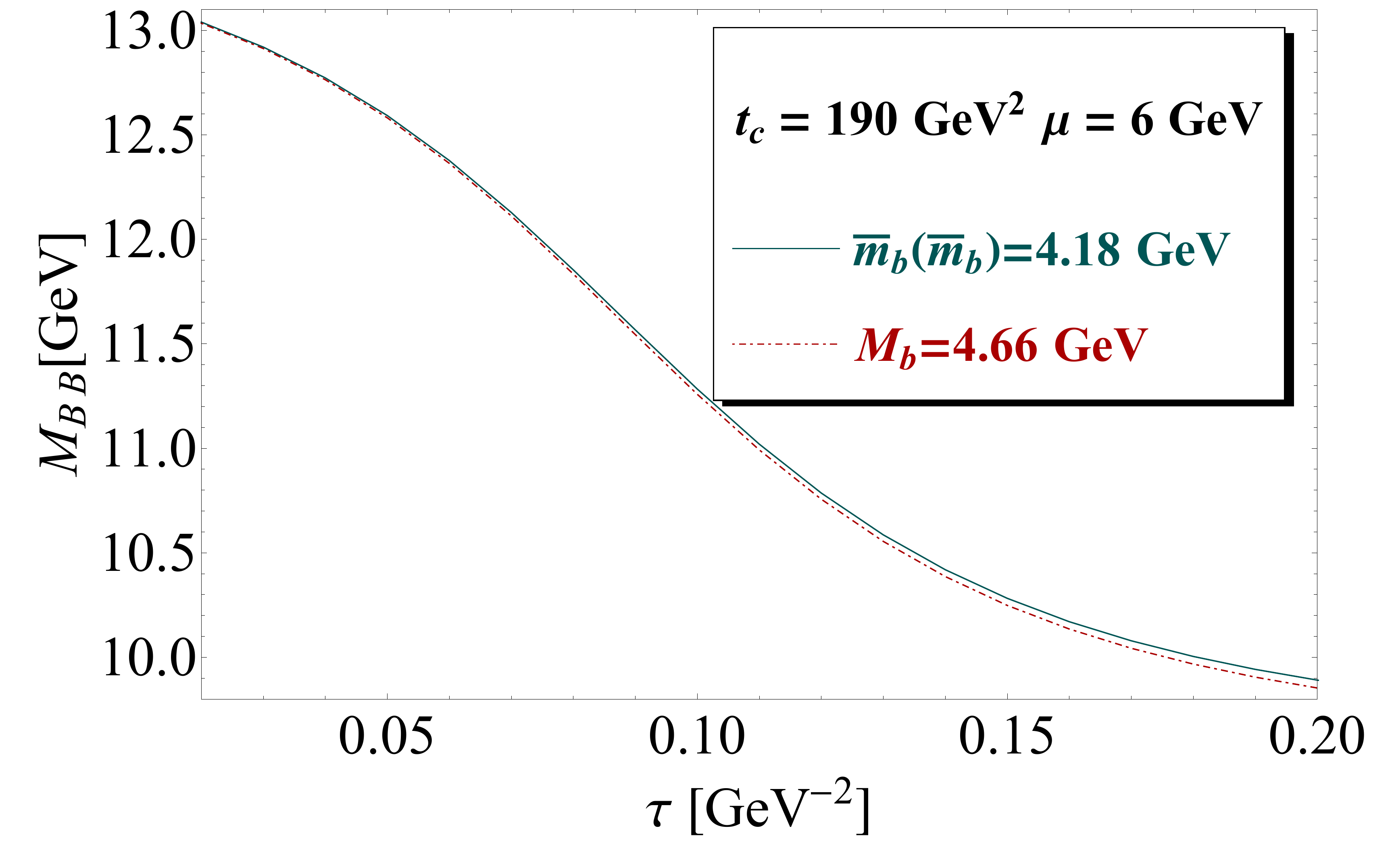}}
\centerline {\hspace*{-3cm} a)\hspace*{6cm} b) }
\caption{
\scriptsize 
{\bf a)} $f_{BB}$  at LO as function of $\tau$ for  $t_c=190$ GeV$^2$, for $\mu=6$ GeV, for  values of the running $\overline{m}_b(\overline{m}_b)=4.18$ GeV and pole mass $M_b=4.66$ GeV. We use     
the QCD parameters in Tables\,\ref{tab:param} and \ref{tab:alfa}; {\bf b)} The same as a) but for the mass $M_{BB}$.
}
\label{fig:bb-const} 
\end{center}
\end{figure}

\begin{figure}[hbt] 
\begin{center}
{\includegraphics[width=6.29cm  ]{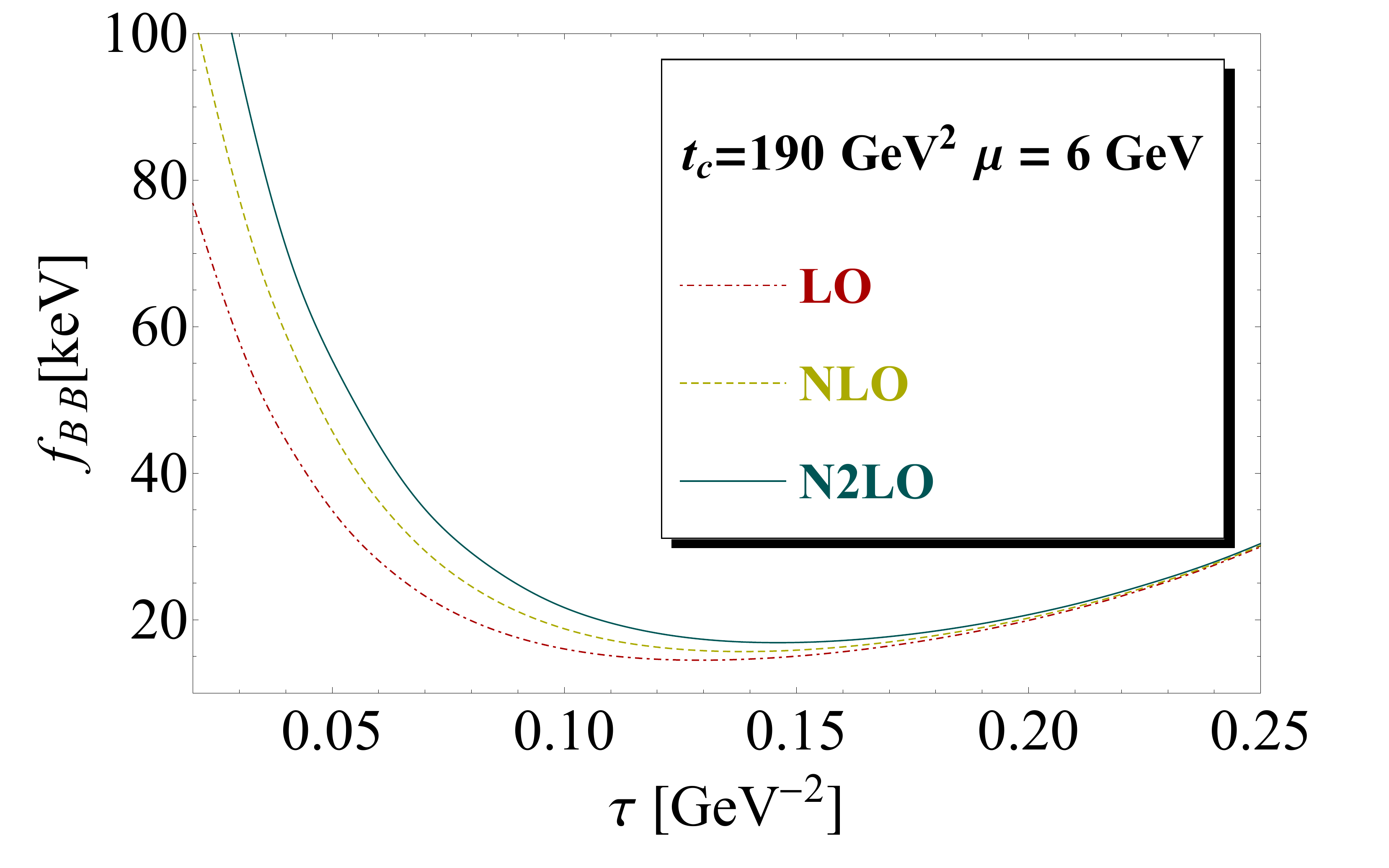}}
{\includegraphics[width=6.29cm  ]{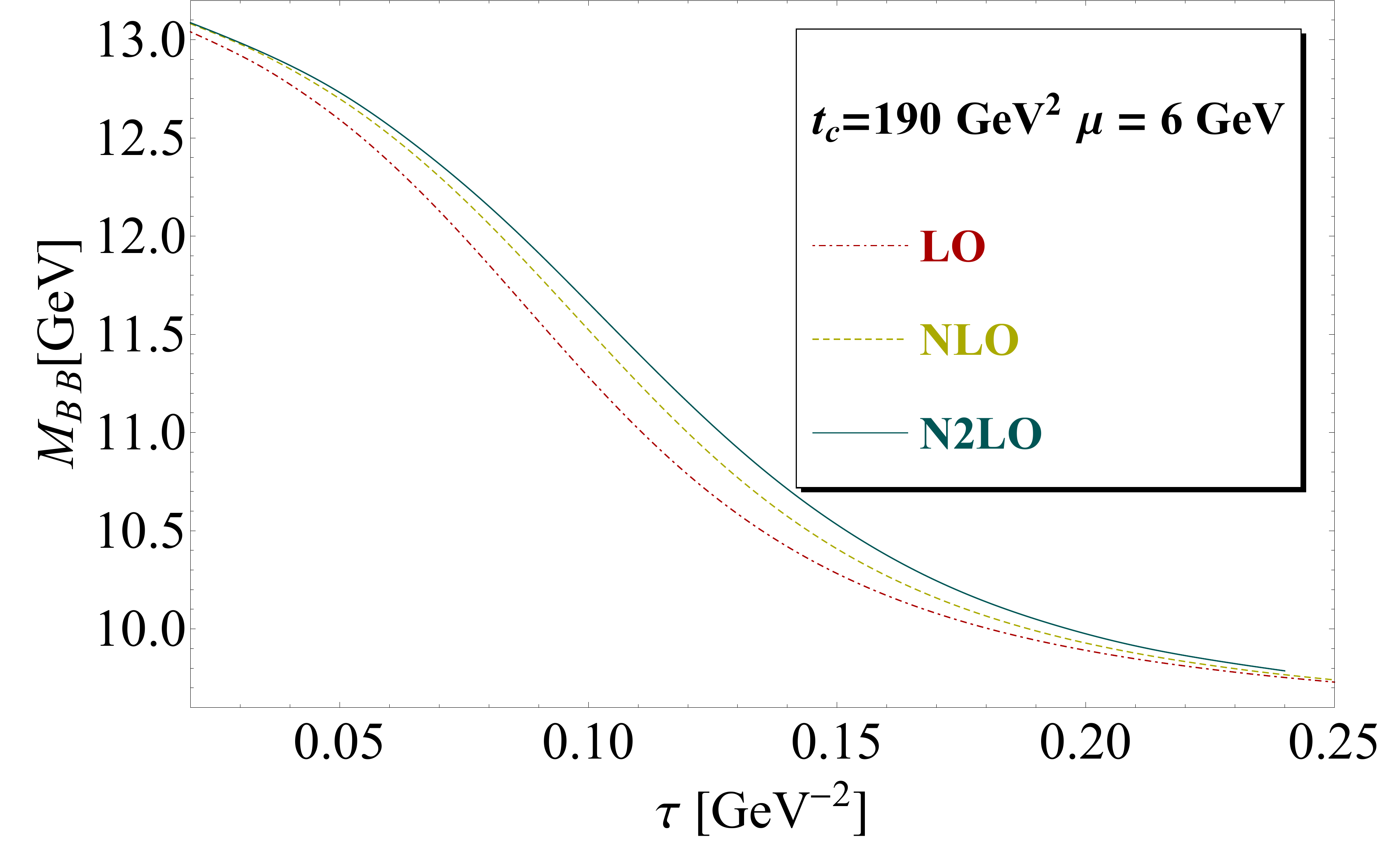}}
\centerline {\hspace*{-3cm} a)\hspace*{6cm} b) }
\caption{
\scriptsize 
{\bf a)} $f_{BB}$  as function of $\tau$ for a given value of $t_c=190$ GeV$^2$, for $\mu=6$ GeV, for different truncation of the PT series  and for the QCD parameters in Tables\,\ref{tab:param} and \ref{tab:alfa}; {\bf b)} The same as a) but for the mass $M_{BB}$.
}
\label{fig:b-pt} 
\end{center}
\end{figure}

\begin{figure}[hbt] 
\begin{center}
{\includegraphics[width=6.2cm  ]{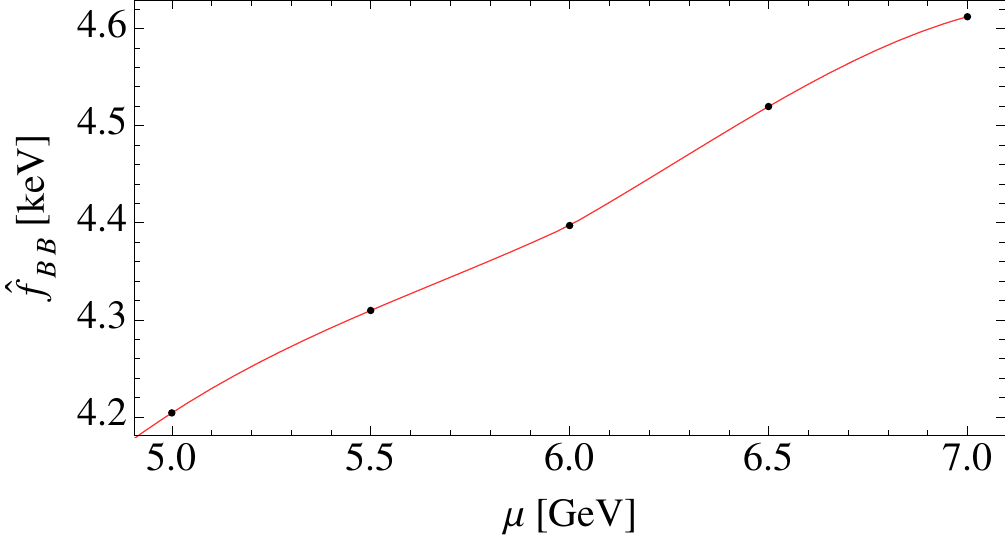}}
{\includegraphics[width=6.2cm  ]{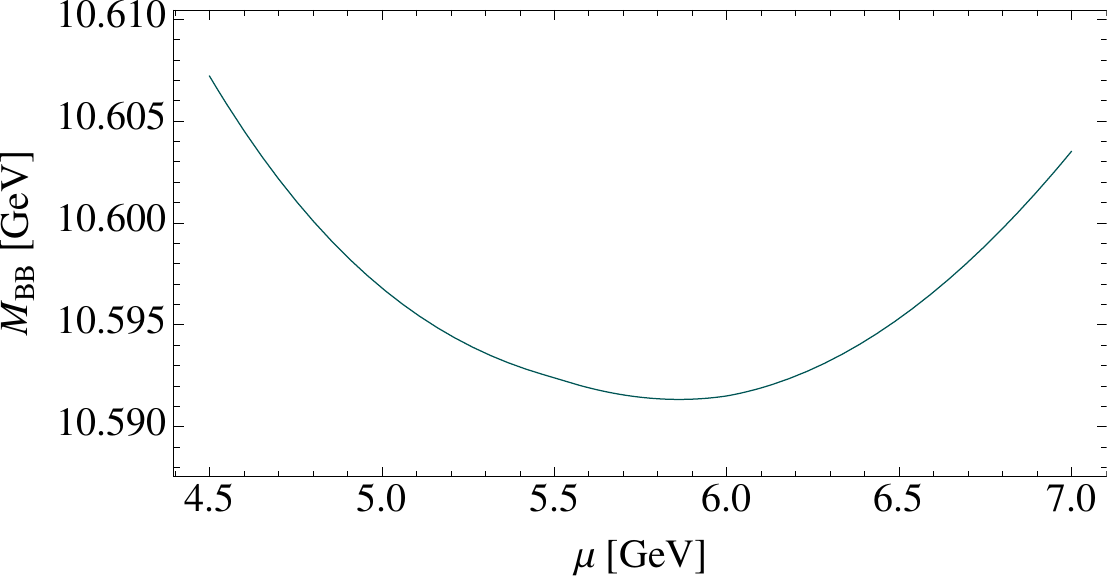}}
\centerline {\hspace*{-3cm} a)\hspace*{6cm} b) }
\caption{
\scriptsize 
{\bf a)} Renormalization group invariant coupling $\hat f_{BB}$ at NLO as function of $\mu$, for the corresponding $\tau$-stability region, for $t_c\simeq 190$ GeV$^2$ and for the QCD parameters in Tables\,\ref{tab:param}  and \ref{tab:alfa};  {\bf b)} The same as a) but for the  mass $ M_{BB}$.
}
\label{fig:b-mu} 
\end{center}
\end{figure}

We extend the analysis to the $b$-quark sector which we show in Figs.\,\ref{fig:b-lo} to \ref{fig:b-mu}. The optimal results of the analysis  given  in Tables\,\ref{tab:errorb} and \ref{tab:resultb}, are obtained at N2LO for the set: 
\beq
\tau\simeq 0.15~{\rm GeV}^{-2}, ~~~~t_c\simeq  (160-190)~{\rm GeV}^2~~~~ {\rm and}~~~ ~\mu\simeq  5.5~\rm{ GeV}. 
\eeq
One can notice from Figs.\,\ref{fig:b-lo} to \ref{fig:b-n2lo} that the value of $\tau$ at which the optimal results are obtained shifts at LO from
0.08 to 0.1~ GeV$^{-2}$. Comparing the $c$ and $b$ channels, one finds that at N2LO, the values of $\tau $ is about $(0.3 \sim 0.4)$ GeV$^{-2}$ and $\mu$ about 4.5 GeV for the charm channel. 
\nin
\nin
\nin
\nin
\nin
\nin
\section{The  $(1^{+\pm})$ Heavy-Light Axial-Vector Molecule  States}
We shall study here the masses and couplings of the $J^{PC}=1^{+\pm}$ axial-vector molecule states :
$\bar D^*D,~\bar B^*B$ and $\bar D^*_0D_1,~\bar B^*_0B_1$. States with opposite $C$-parity are degenerated in this channel. 

The analysis (shapes of different curves) is very similar to the one of the $\bar DD$ and $\bar BB$ channels and will not be repeated here. We shall only quote the results in Tables\,\ref{tab:errorc} to \ref{tab:resultb} obtained at N2LO for the set of parameters: 
\beq
\tau\simeq (0.30-0.37)~{\rm GeV}^{-2}, ~~~~t_c\simeq  (23-32)~{\rm GeV}^2~~~~ {\rm and}~~~ ~\mu\simeq  4.7~\rm{ GeV},
\eeq
for the $c$ channel and:
\beq
\tau\simeq (0.12-0.14)~{\rm GeV}^{-2}, ~~~~t_c\simeq  (140-170)~{\rm GeV}^2~~~~ {\rm and}~~~ ~\mu\simeq  6~\rm{ GeV}, 
\eeq
for the $b$ channel.  

We observe in Tables\,\ref{tab:resultc} and \ref{tab:resultb} a good convergence of the results from NLO to N2LO where the corresponding variations are smaller than the errors of the masses and couplings determinations.
\section{The  $(0^{-\pm})$ Heavy-Light Pseudoscalar Molecule States}
Here, we shall analyze the masses and couplings of the pseudoscalar $\bar D^*_0D,~\bar D^*D_1$
and their beauty analogue, which will be illustrated by the case of $\bar D^*_0D$ and $\bar B^*_0B$.
States with opposite $C$-parities are degenerated in this channel. 
\subsection{Coupling and mass of the $ \bar D^*_0D$ molecule}
\subsubsection*{$\bullet$ $\tau$ and $t_c$ stabilities}
\begin{figure}[hbt] 
\begin{center}
{\includegraphics[width=6.32cm  ]{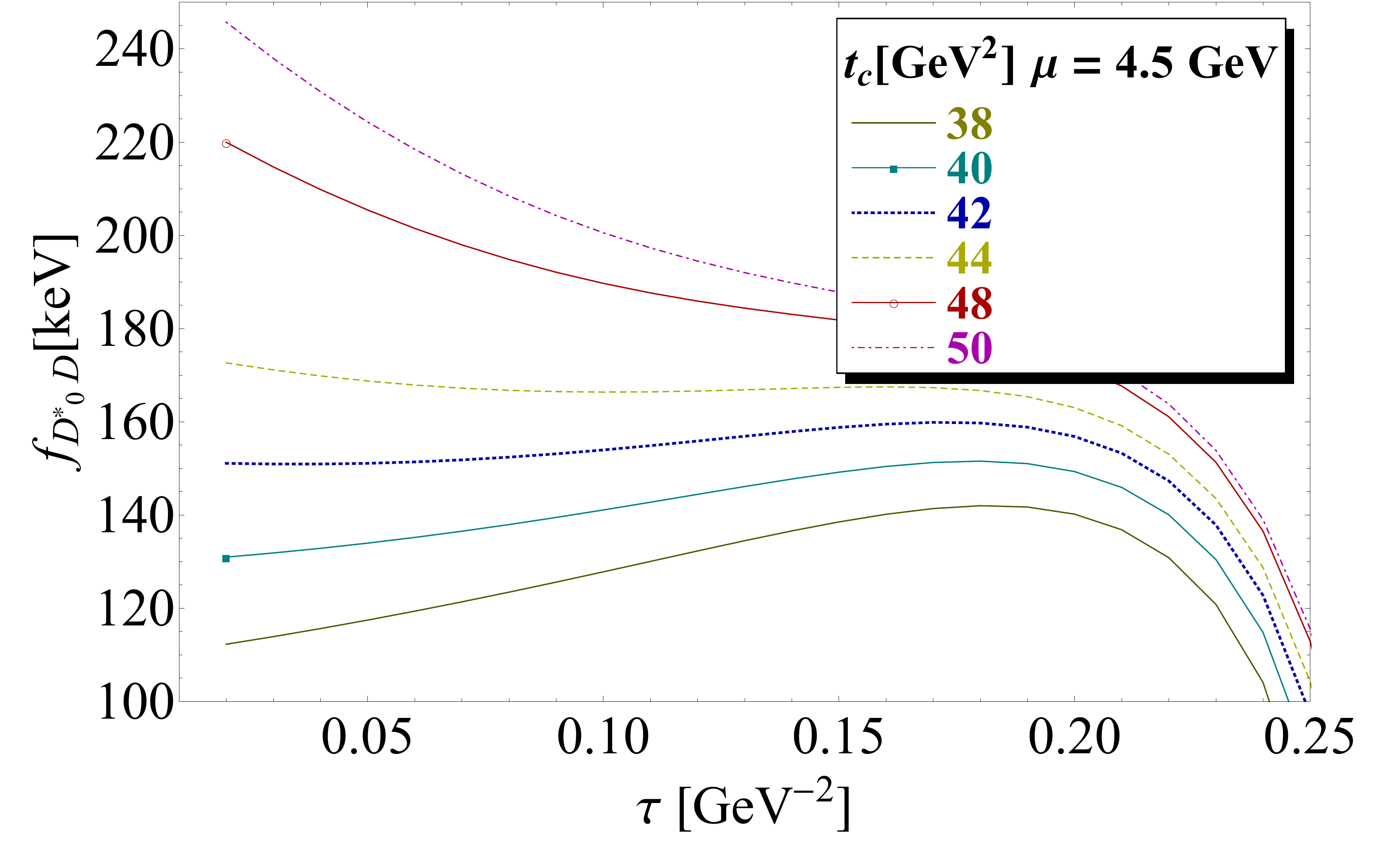}}
{\includegraphics[width=6.26cm  ]{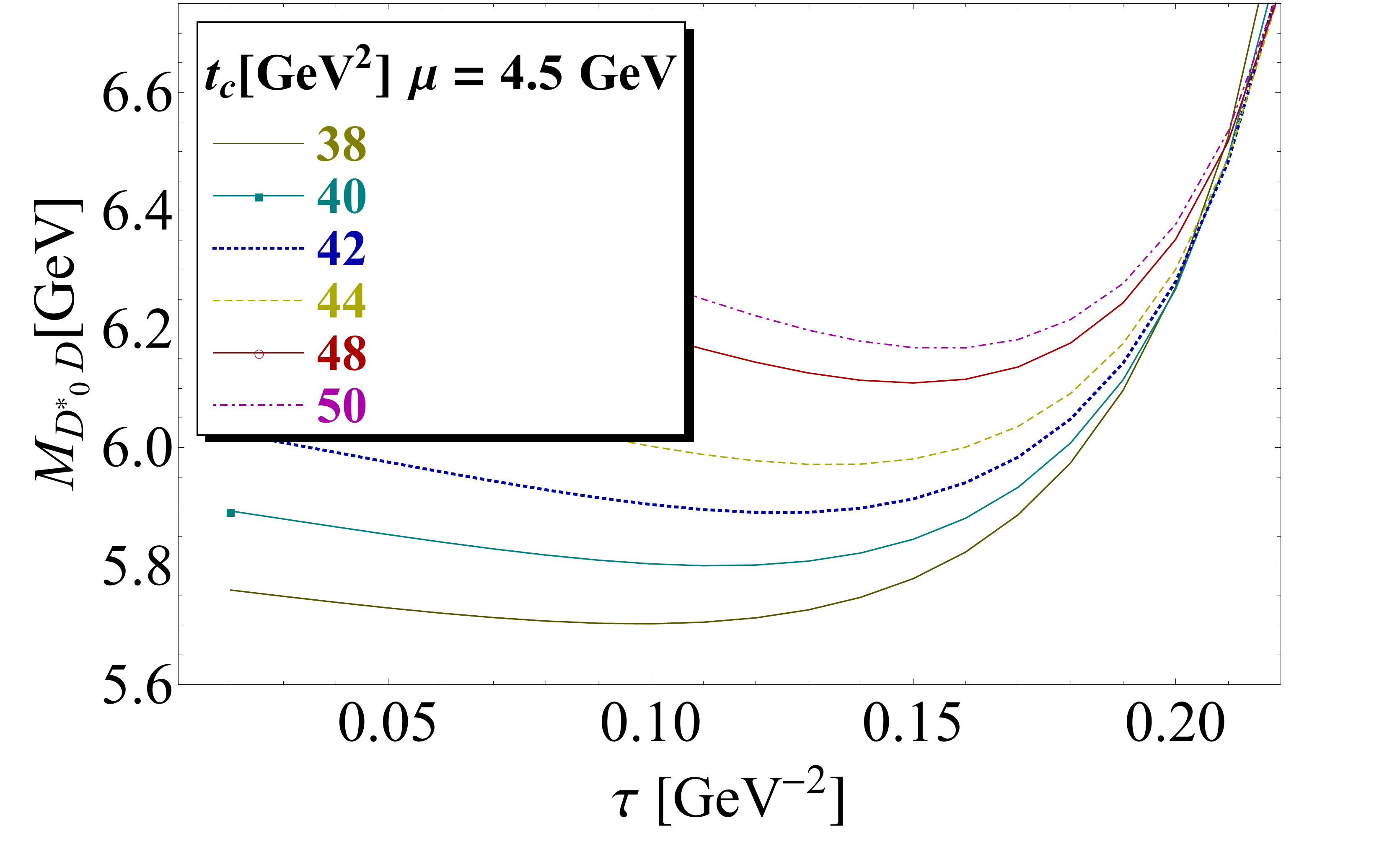}}
\centerline {\hspace*{-3cm} a)\hspace*{6cm} b) }
\caption{
\scriptsize 
{\bf a)} $f_{D^*_0D}$  at LO as function of $\tau$ for different values of $t_c$, for $\mu=4.5$ GeV  and for the QCD parameters in Tables\,\ref{tab:param} and \ref{tab:alfa}; {\bf b)} The same as a) but for the mass $M_{D^*_0D}$.
}
\label{fig:dstar0d-lo} 
\end{center}
\end{figure}

\begin{figure}[hbt] 
\begin{center}
{\includegraphics[width=6.29cm  ]{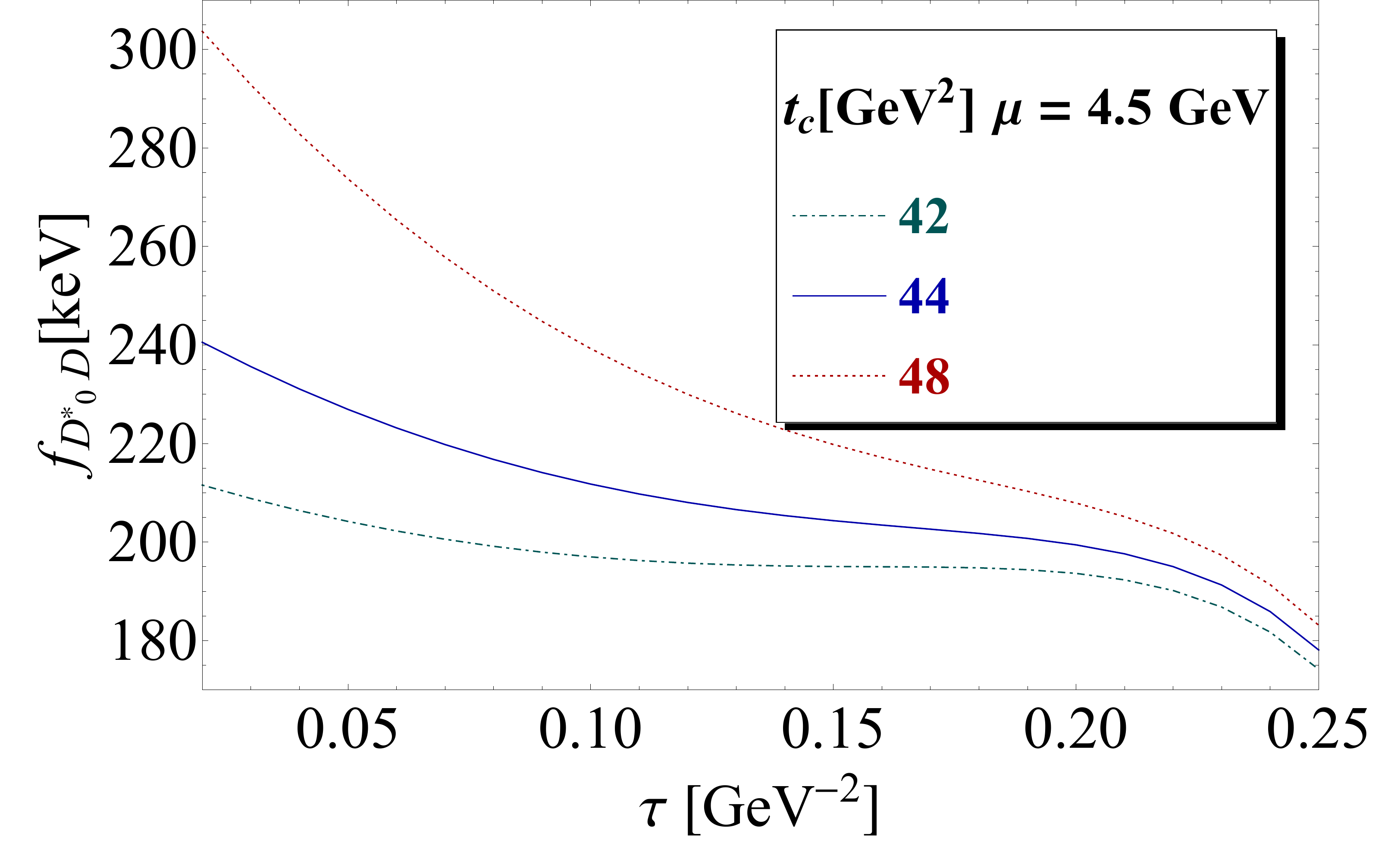}}
{\includegraphics[width=6.29cm  ]{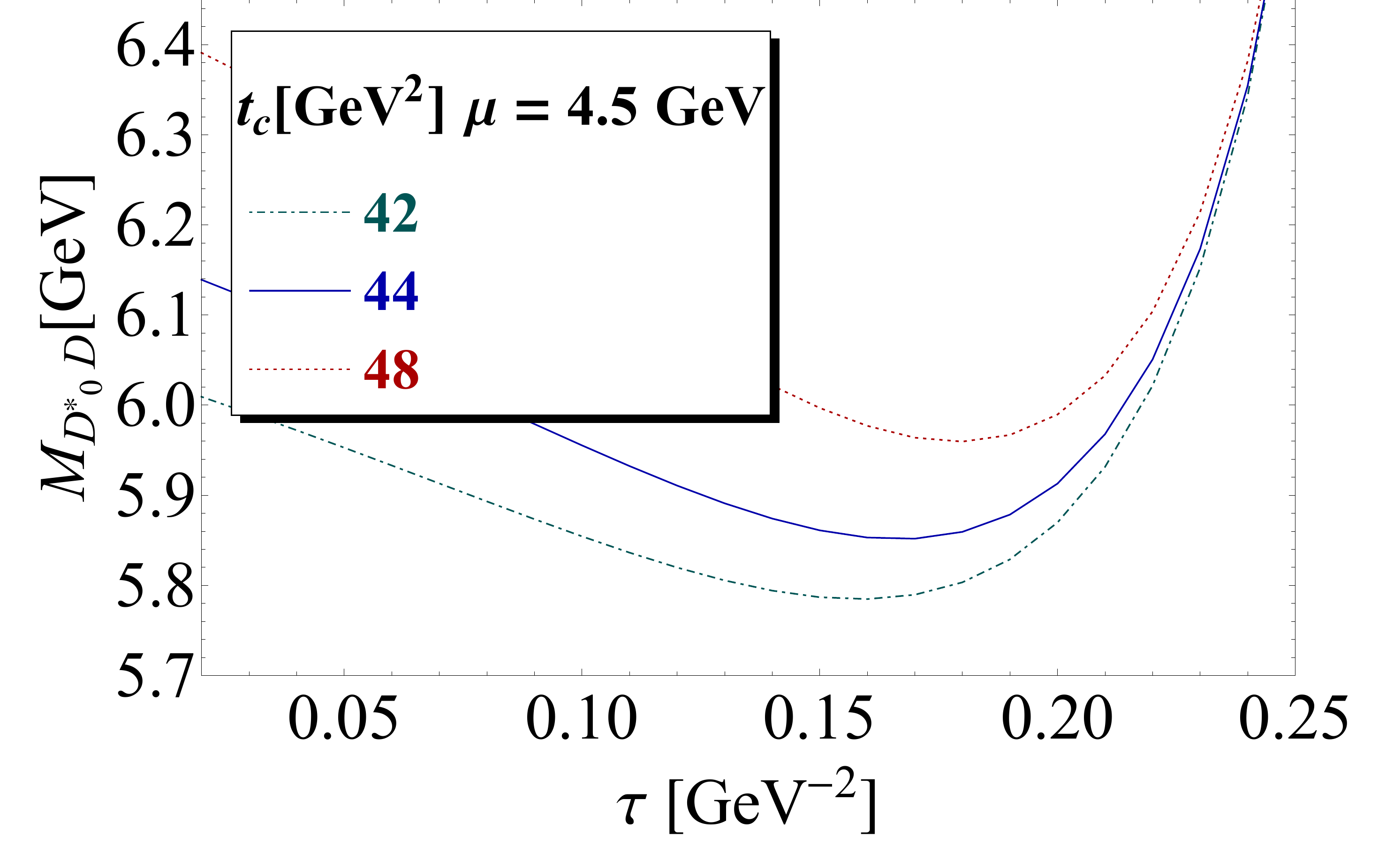}}
\centerline {\hspace*{-3cm} a)\hspace*{6cm} b) }
\caption{
\scriptsize 
{\bf a)} $f_{D^*_0D}$ at NLO  as function of $\tau$ for different values of $t_c$, for $\mu=4.5$ GeV  and for the QCD parameters in Tables\,\ref{tab:param} and \ref{tab:alfa}; {\bf b)} The same as a) but for the mass $M_{D^*_0D}$.
}
\label{fig:dstar0d-nlo} 
\end{center}
\end{figure}

\begin{figure}[hbt] 
\begin{center}
{\includegraphics[width=6.29cm  ]{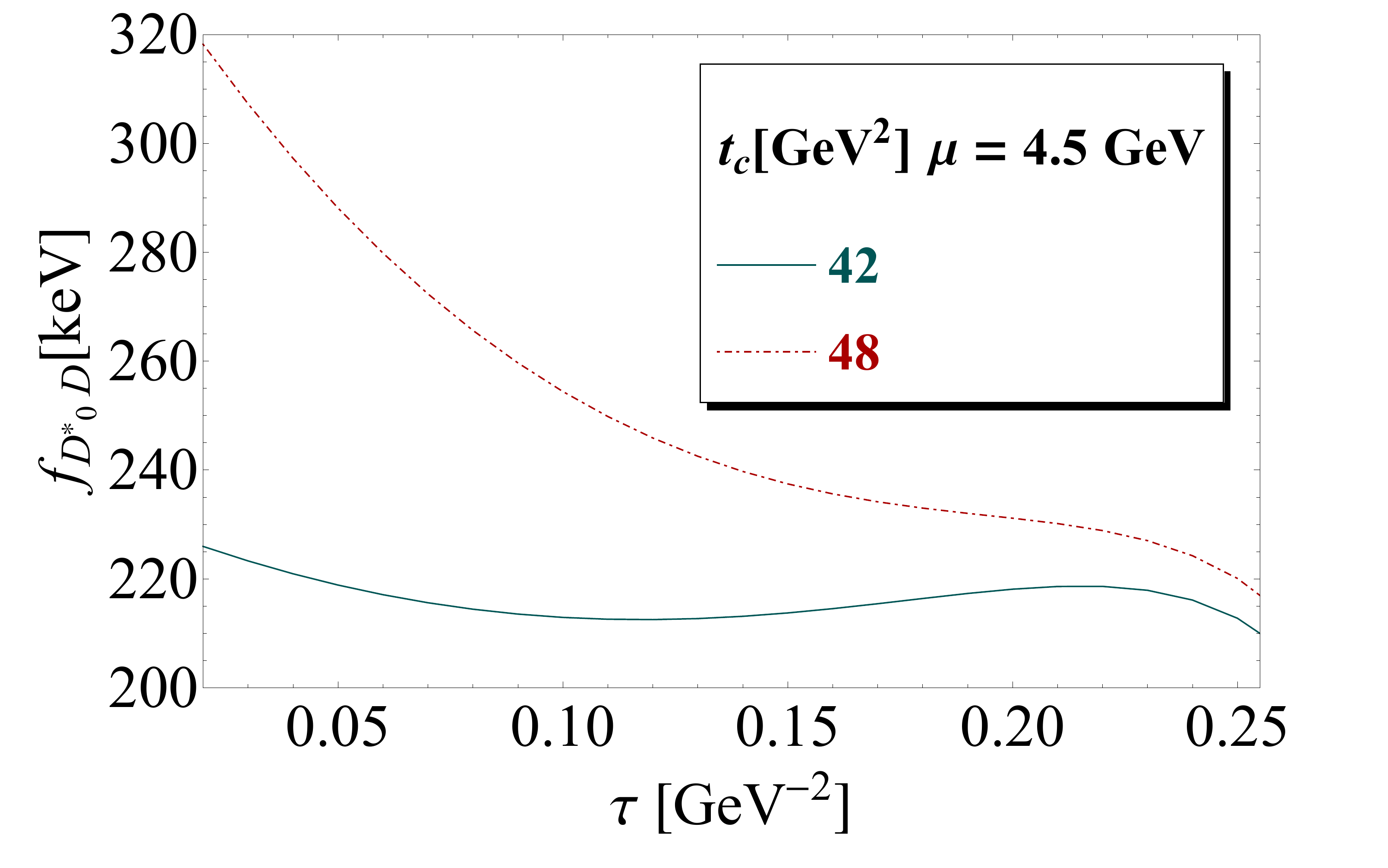}}
{\includegraphics[width=6.29cm  ]{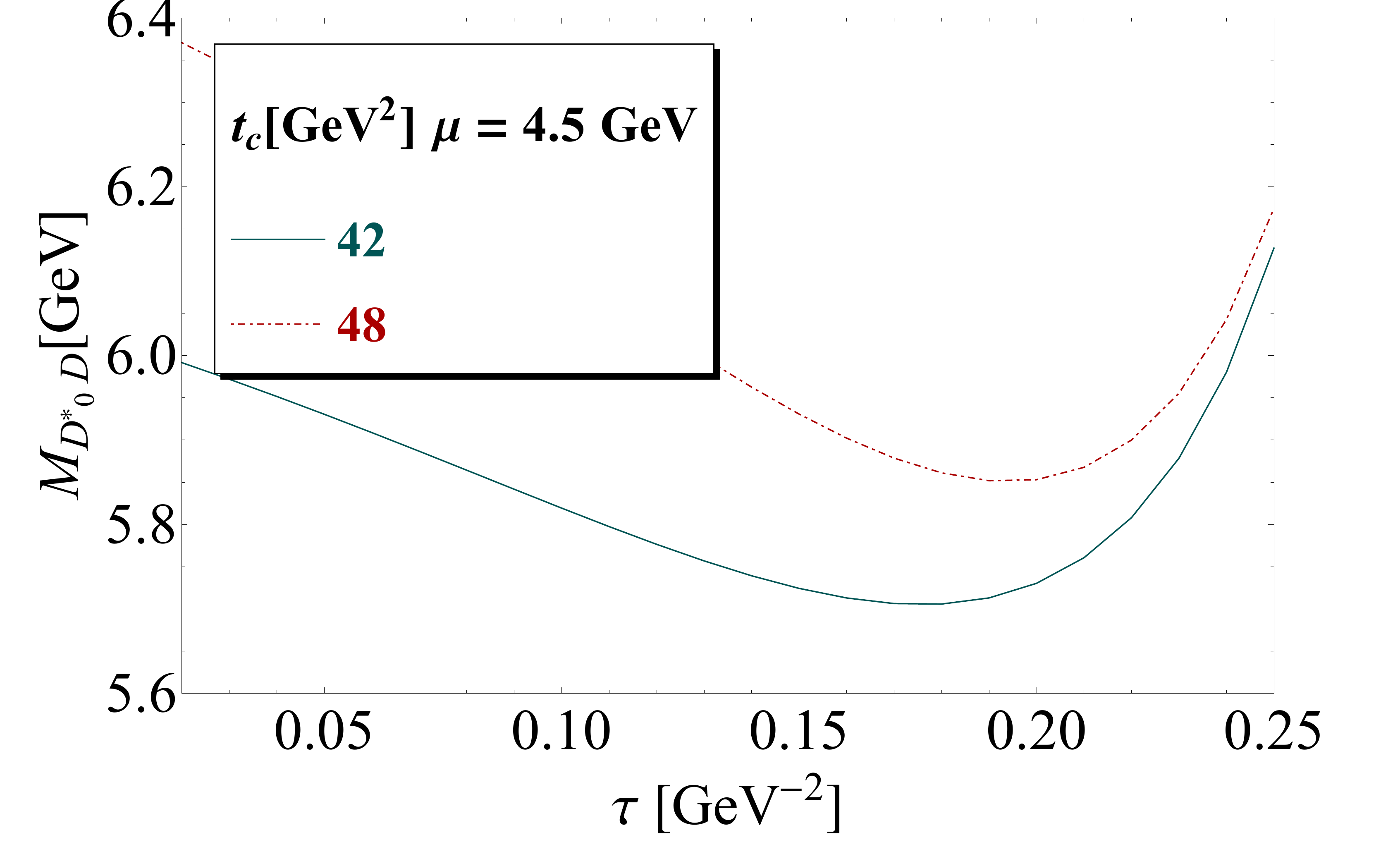}}
\centerline {\hspace*{-3cm} a)\hspace*{6cm} b) }
\caption{
\scriptsize 
{\bf a)} $f_{D^*_0D}$ at N2LO  as function of $\tau$ for different values of $t_c$, for $\mu=4.5$ GeV  and for the QCD parameters in Tables\,\ref{tab:param} and \ref{tab:alfa}; {\bf b)} The same as a) but for the mass $M_{D^*_0D}$.
}
\label{fig:dstar0d-n2lo} 
\end{center}
\end{figure}

\nin
 We study the behaviour of the coupling $f_{D^*_0D}$  and mass $M_{D^*_0D}$ in terms of the LSR variable $\tau$ at different values of $t_c$ as shown in Fig.\,\ref{fig:dstar0d-lo} at LO, in Fig.\,\ref{fig:dstar0d-nlo} at NLO and in Fig.\,\ref{fig:dstar0d-n2lo} at N2LO. 
\nin
 We consider as a final and conservative result the one corresponding to the beginning of the $\tau$-stability for $t_c$=42 GeV$^2$ until the one where $t_c$-stability starts to be reached for $t_c\simeq$ 48 GeV$^2$. In these stability regions, the requirement that the pole contribution is larger than the one of the continuum (see e.g.\,\cite{MOLE5}) is automatically satisfied. 
\nin
\nin

   \subsubsection*{ \b Running versus the pole quark mass definitions}
\begin{figure}[hbt] 
\begin{center}
{\includegraphics[width=6.29cm  ]{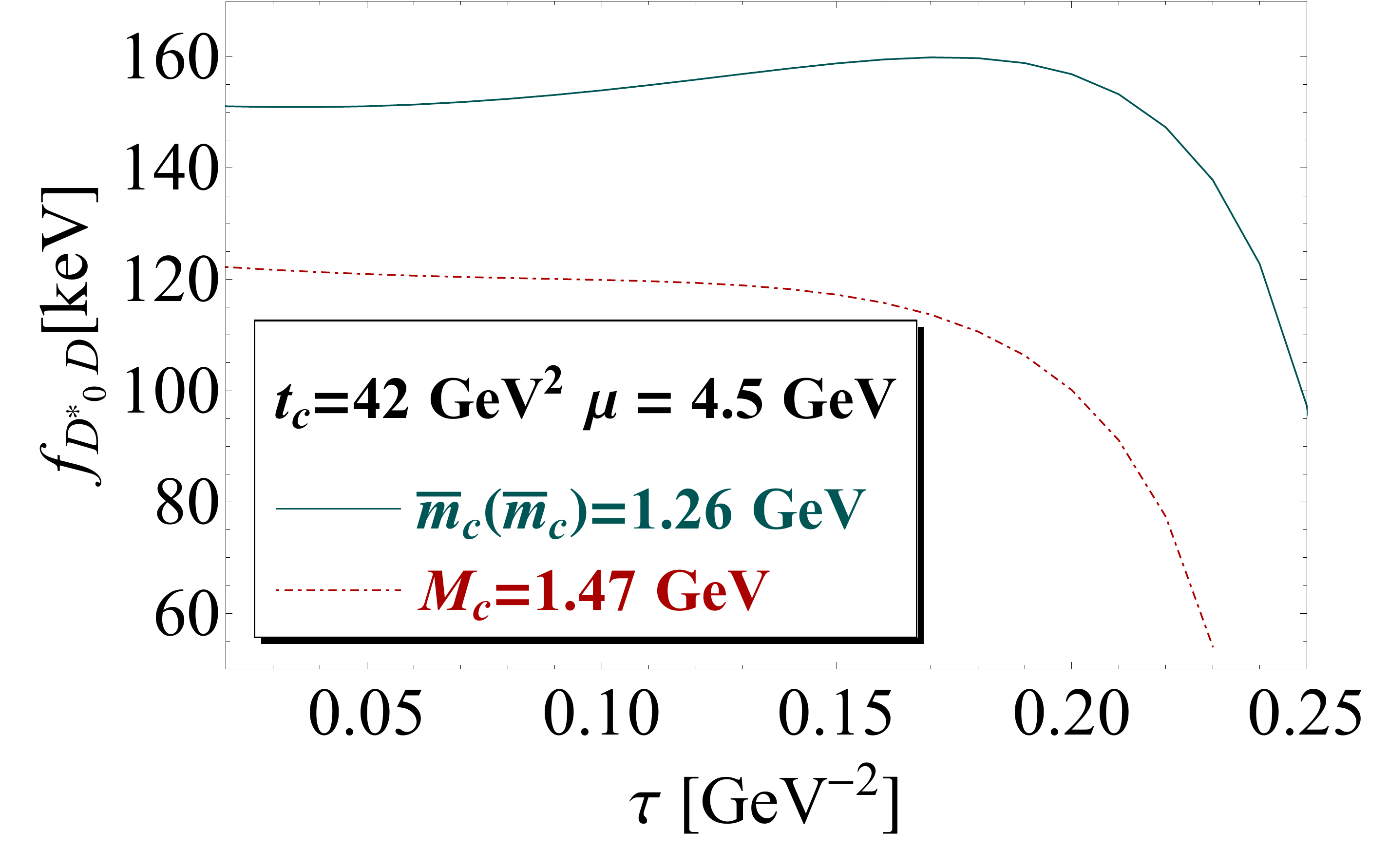}}
{\includegraphics[width=6.30cm  ]{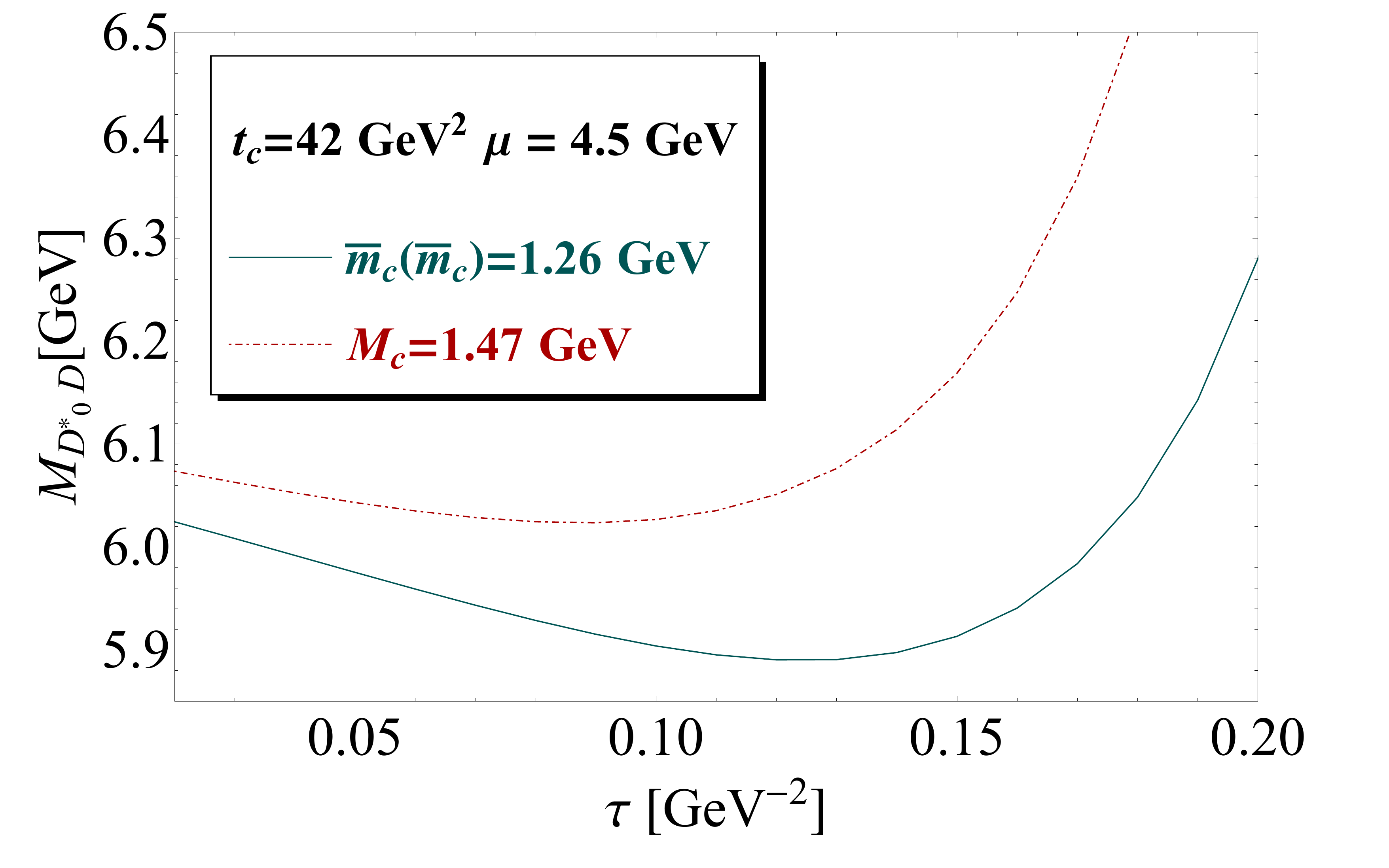}}
\centerline {\hspace*{-3cm} a)\hspace*{6cm} b) }
\caption{
\scriptsize 
{\bf a)} $f_{D^*_0D}$  at LO as function of $\tau$ for  $t_c=42$ GeV$^2$, for $\mu=4.5$ GeV, for  values of the running $\overline{m}_c(\overline{m}_c)=1.26$ GeV and pole mass $M_c=1.47$ GeV. We use     
the QCD parameters in Tables\,\ref{tab:param} and \ref{tab:alfa}; {\bf b)} The same as a) but for the mass $M_{D^*_0D}$.
}
\label{fig:dstar0d-const} 
\end{center}
\end{figure}

   \nin
   We show in Fig.\,\ref{fig:dstar0d-const} the effect of the definitions (running or pole) of the heavy quark mass used in the analysis at LO which is important for the coupling and the mass. 
\nin
\subsubsection*{$\bullet$ Convergence of the PT series} 
\begin{figure}[hbt] 
\begin{center}
{\includegraphics[width=6.29cm  ]{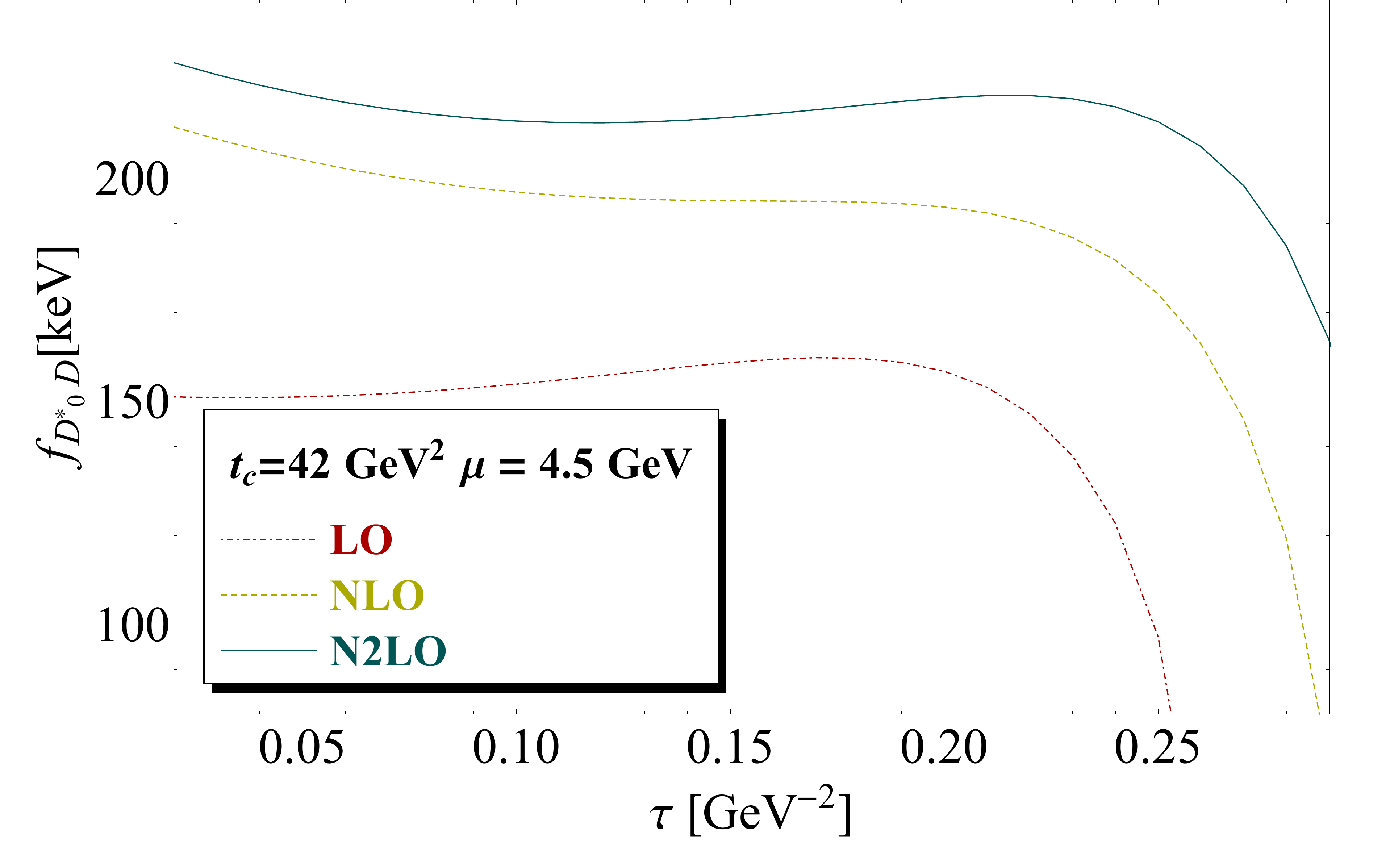}}
{\includegraphics[width=6.29cm  ]{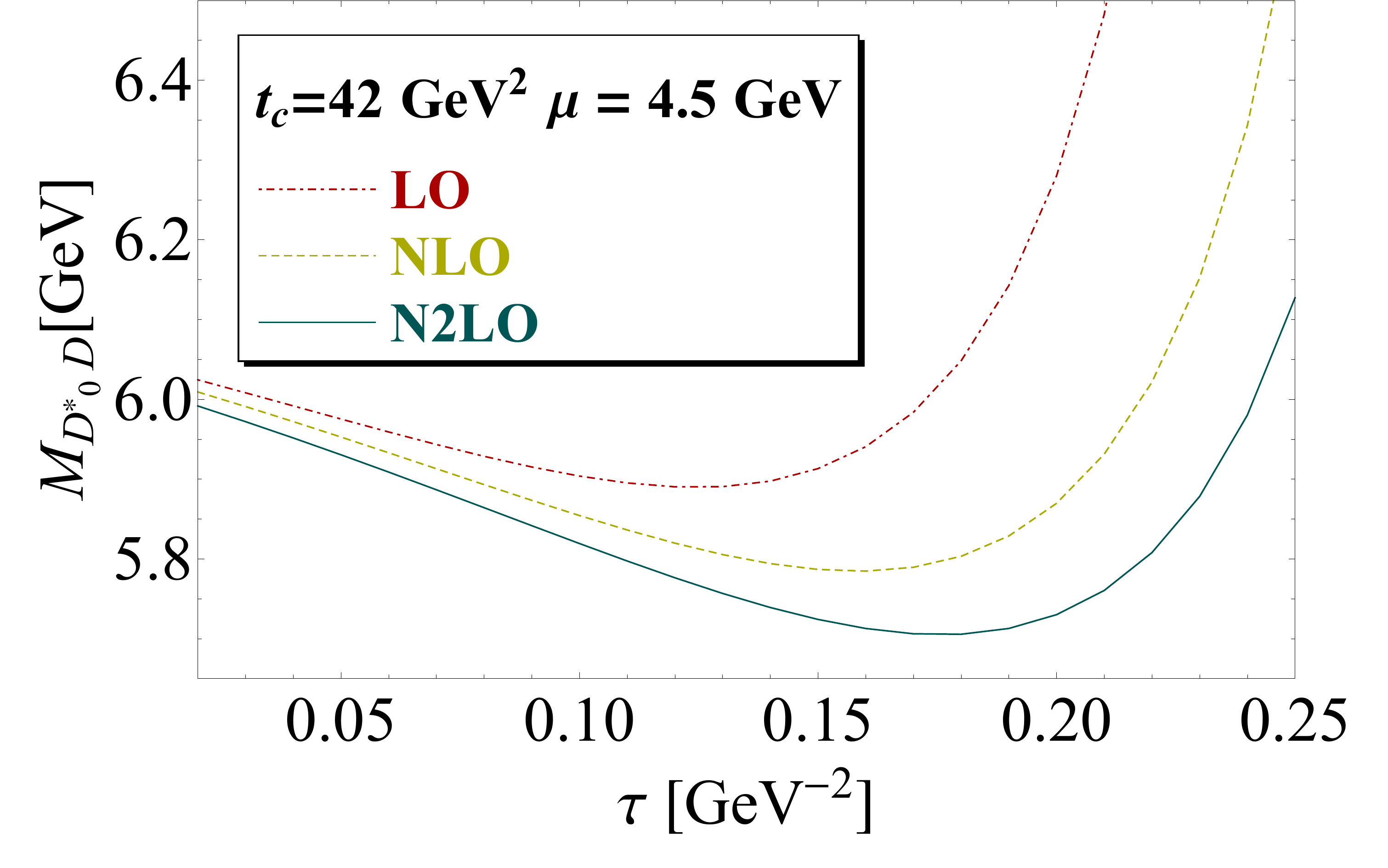}}
\centerline {\hspace*{-3cm} a)\hspace*{6cm} b) }
\caption{
\scriptsize 
{\bf a)} $f_{D^*_0D}$  as function of $\tau$ for a given value of $t_c=42$ GeV$^2$, for $\mu=4.5$ GeV, for different truncation of the PT series  and for the QCD parameters in Tables\,\ref{tab:param} and \ref{tab:alfa}; {\bf b)} The same as a) but for the mass $M_{D^*_0D}$.
}
\label{fig:dstar0d-pt} 
\end{center}
\end{figure}

\nin
Using  $t_c\simeq  42$ GeV$^2$, we study in Fig. {\ref{fig:dstar0d-pt}} the convergence of the PT series for a given value of $\mu=4.5$ GeV.  We observe that from NLO to N2LO the mass decreases by about only 1.5$\%$ indicating the good convergence of the PT series.
\nin
\begin{figure}[hbt] 
\begin{center}
{\includegraphics[width=6.2cm  ]{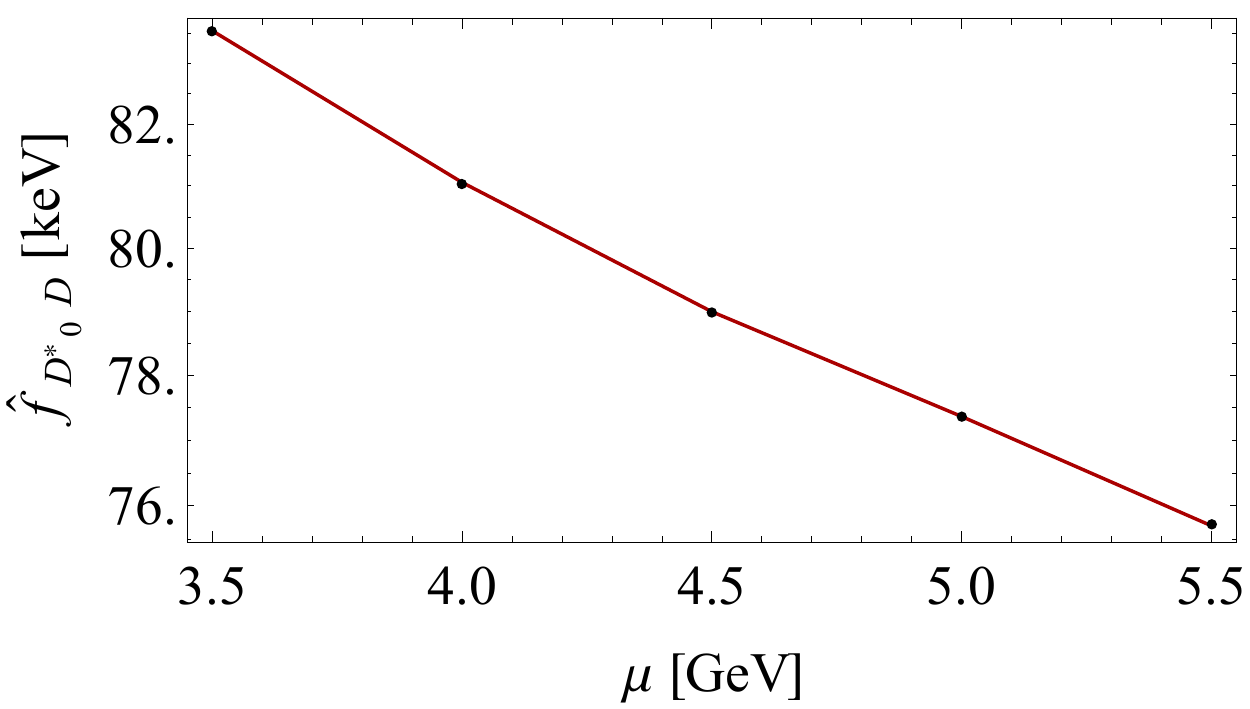}}
{\includegraphics[width=6.2cm  ]{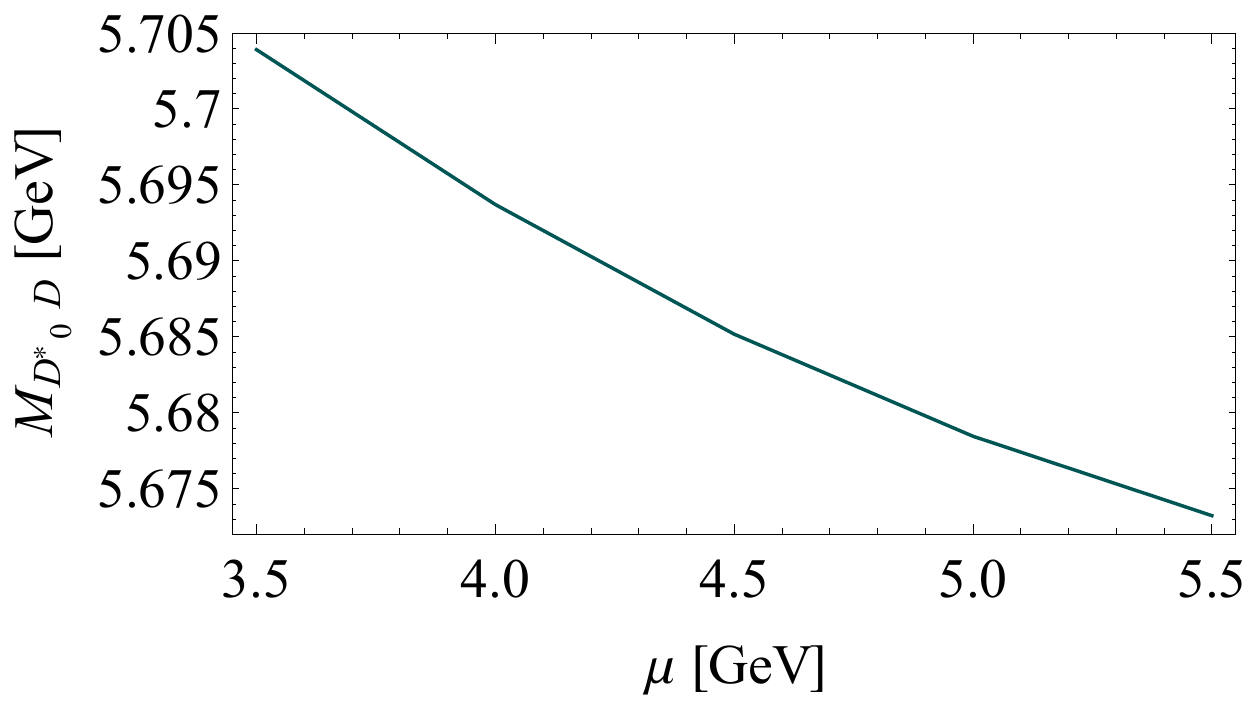}}
\centerline {\hspace*{-3cm} a)\hspace*{6cm} b) }
\caption{
\scriptsize 
{\bf a)} Renormalization group invariant coupling $\hat f_{D^*_0D}$ at NLO as function of $\mu$, for the corresponding $\tau$-stability region, for $t_c\simeq 42$ GeV$^2$ and for the QCD parameters in Tables\,\ref{tab:param}  and \ref{tab:alfa};  {\bf b)} The same as a) but for the  mass $ M_{D^*_0D}$.
}
\label{fig:dstar0d-mu} 
\end{center}
\end{figure} 
\nin
\subsubsection*{$\bullet$ $\mu$-stability}
\nin
We improve our previous results by using different values of $\mu$ (Fig. {\ref{fig:dstar0d-mu}}). Using the fact that 
the final result must be independent of the arbitrary parameter $\mu$, we consider as an optimal result the one at the inflexion point
for $\mu\simeq 4.5$ GeV at which we deduce the result in Table\,\ref{tab:resultc}.
\subsection{Coupling and mass of the $ \bar B^*_0B$ molecule}
\begin{figure}[hbt] 
\begin{center}
{\includegraphics[width=6.29cm  ]{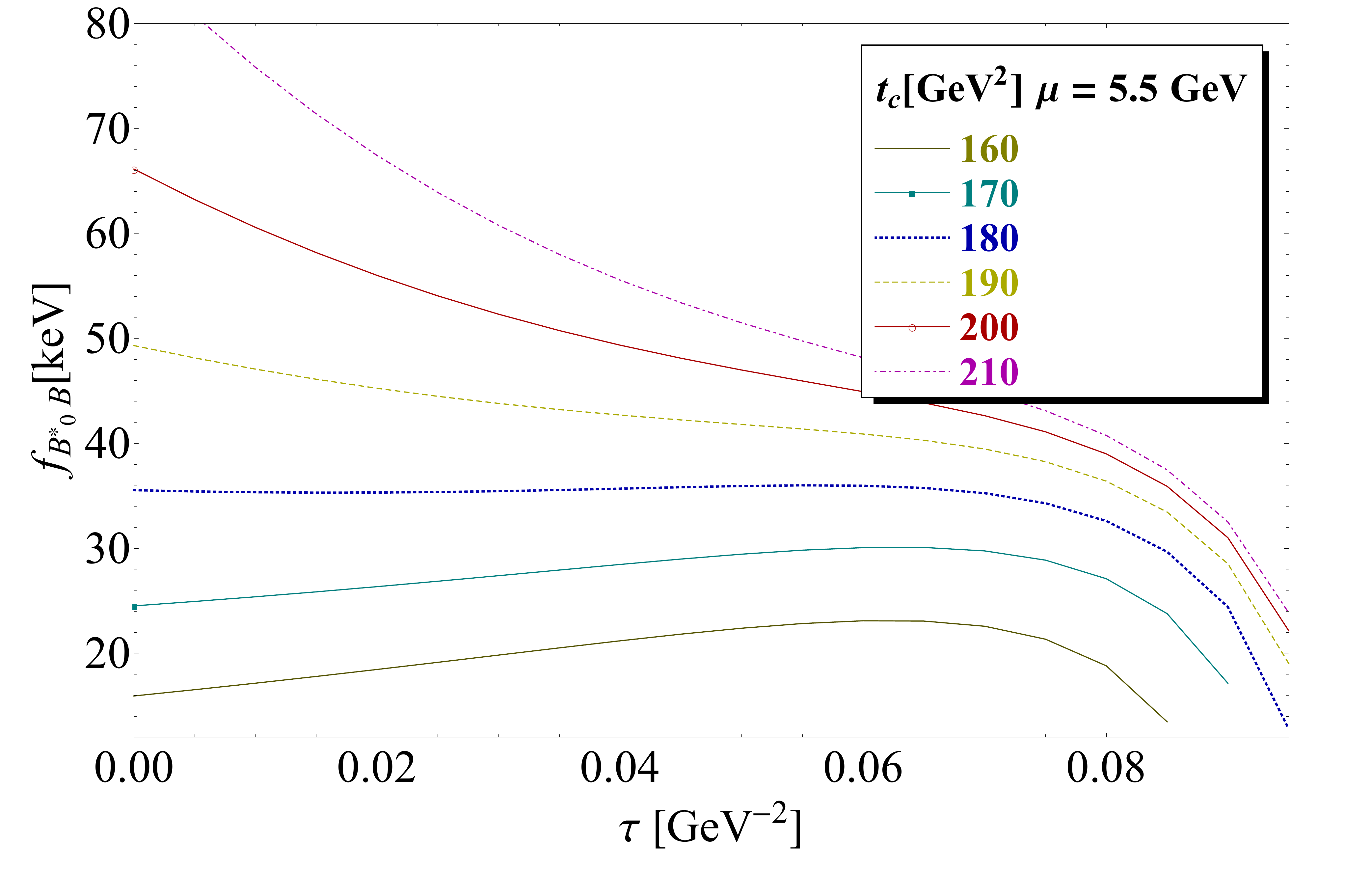}}
{\includegraphics[width=6.29cm  ]{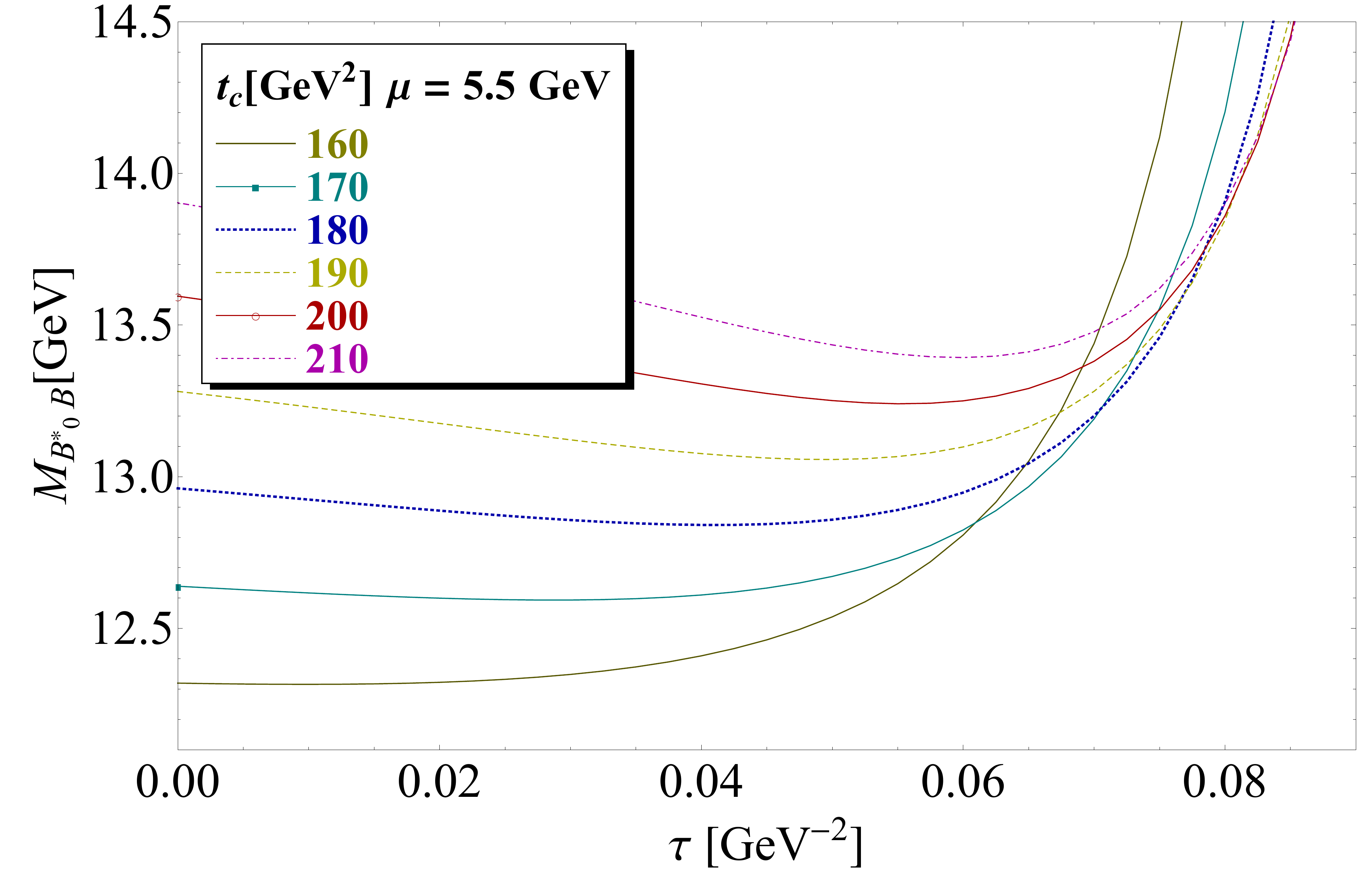}}
\centerline {\hspace*{-3cm} a)\hspace*{6cm} b) }
\caption{
\scriptsize 
{\bf a)} $f_{B^*_0B}$  at LO as function of $\tau$ for different values of $t_c$, for $\mu=5.5$ GeV  and for the QCD parameters in Tables\,\ref{tab:param} and \ref{tab:alfa}; {\bf b)} The same as a) but for the mass $M_{B^*_0B}$.
}
\label{fig:bstar0b-lo} 
\end{center}
\end{figure}

\begin{figure}[hbt] 
\begin{center}
{\includegraphics[width=6.29cm  ]{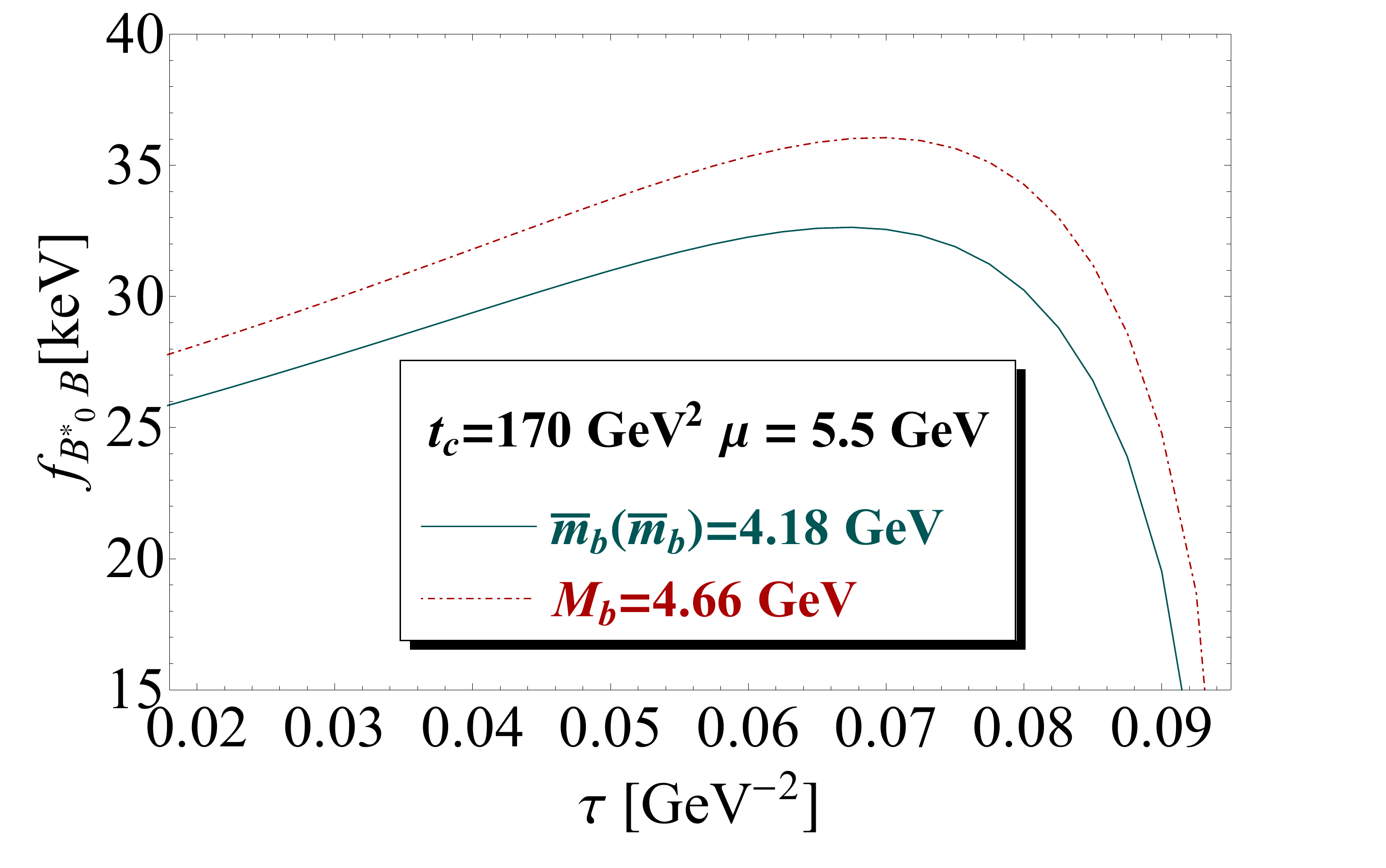}}
{\includegraphics[width=6.29cm  ]{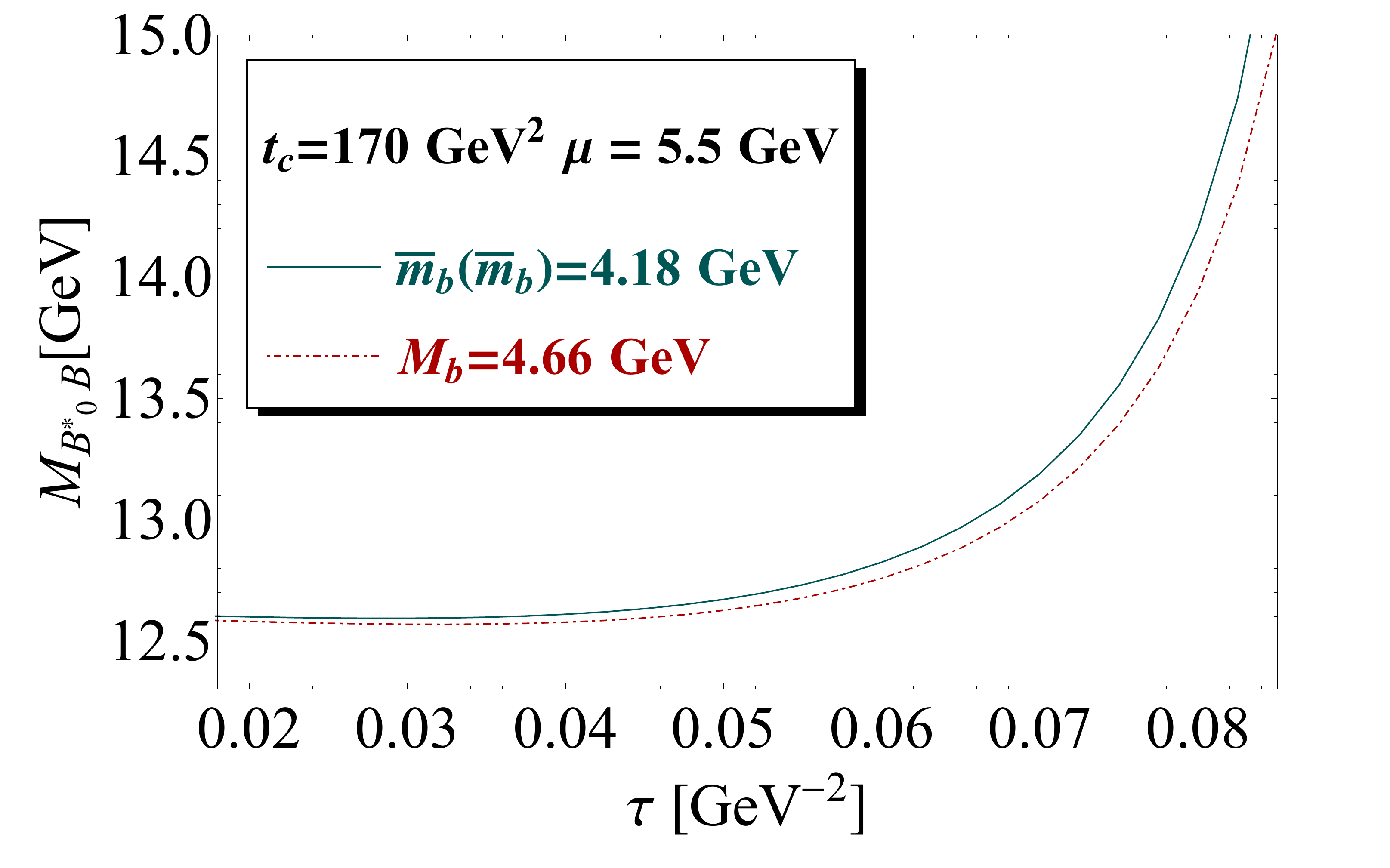}}
\centerline {\hspace*{-3cm} a)\hspace*{6cm} b) }
\caption{
\scriptsize 
{\bf a)} $f_{B^*_0B}$  at LO as function of $\tau$ for  $t_c=170$ GeV$^2$, for $\mu=5.5$ GeV, for  values of the running $\overline{m}_b(\overline{m}_b)=4.18$ GeV and pole mass $M_b=4.66$ GeV. We use     
the QCD parameters in Tables\,\ref{tab:param} and \ref{tab:alfa}; {\bf b)} The same as a) but for the mass $M_{B^*_0B}$.
}
\label{fig:bstar0b-const} 
\end{center}
\end{figure}

\begin{figure}[hbt] 
\begin{center}
{\includegraphics[width=6.29cm  ]{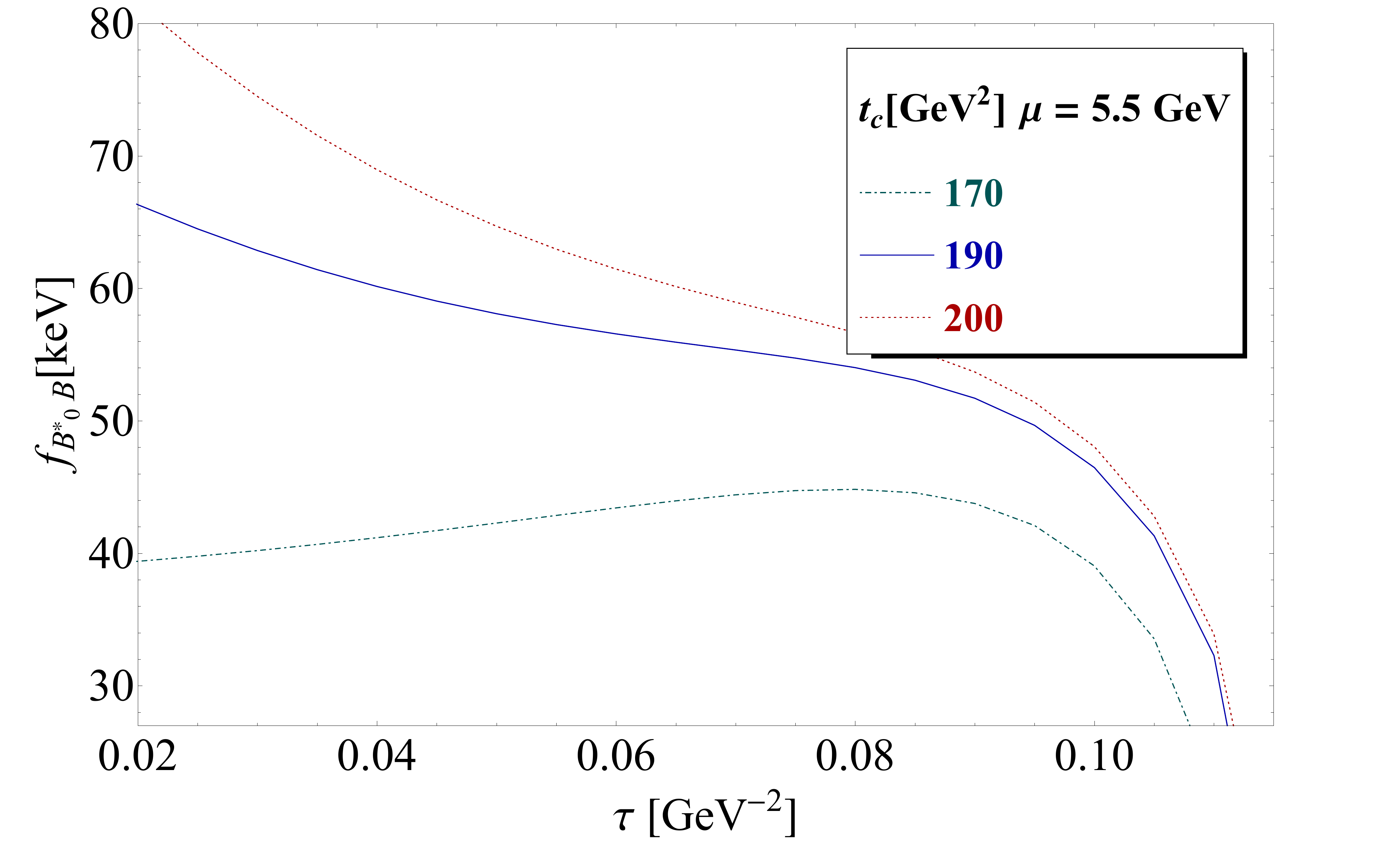}}
{\includegraphics[width=6.29cm  ]{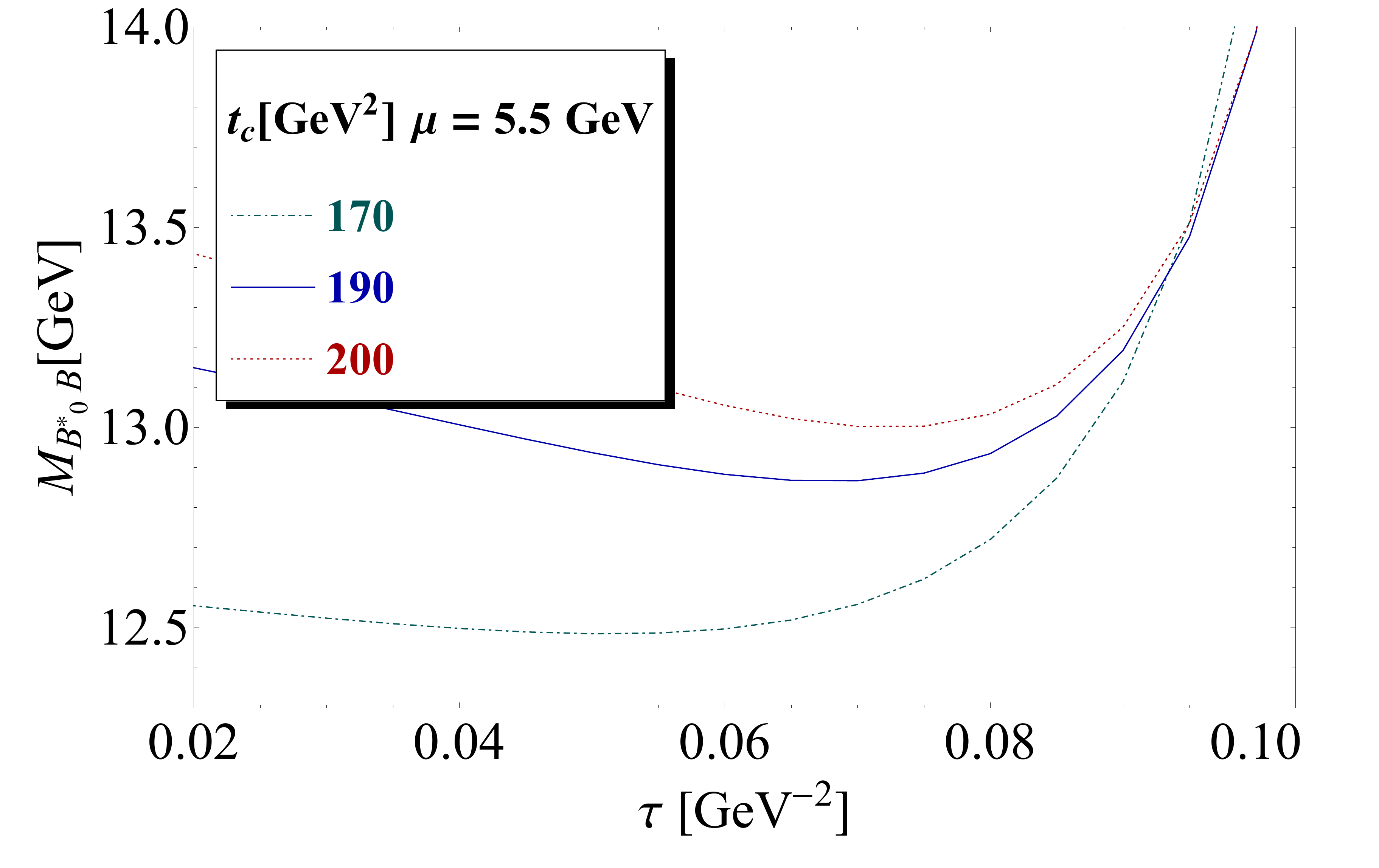}}
\centerline {\hspace*{-3cm} a)\hspace*{6cm} b) }
\caption{
\scriptsize 
{\bf a)} $f_{B^*_0B}$ at NLO  as function of $\tau$ for different values of $t_c$, for $\mu=5.5$ GeV  and for the QCD parameters in Tables\,\ref{tab:param} and \ref{tab:alfa}; {\bf b)} The same as a) but for the mass $M_{B^*_0B}$.
}
\label{fig:bstar0b-nlo} 
\end{center}
\end{figure}

\begin{figure}[hbt] 
\begin{center}
{\includegraphics[width=6.29cm  ]{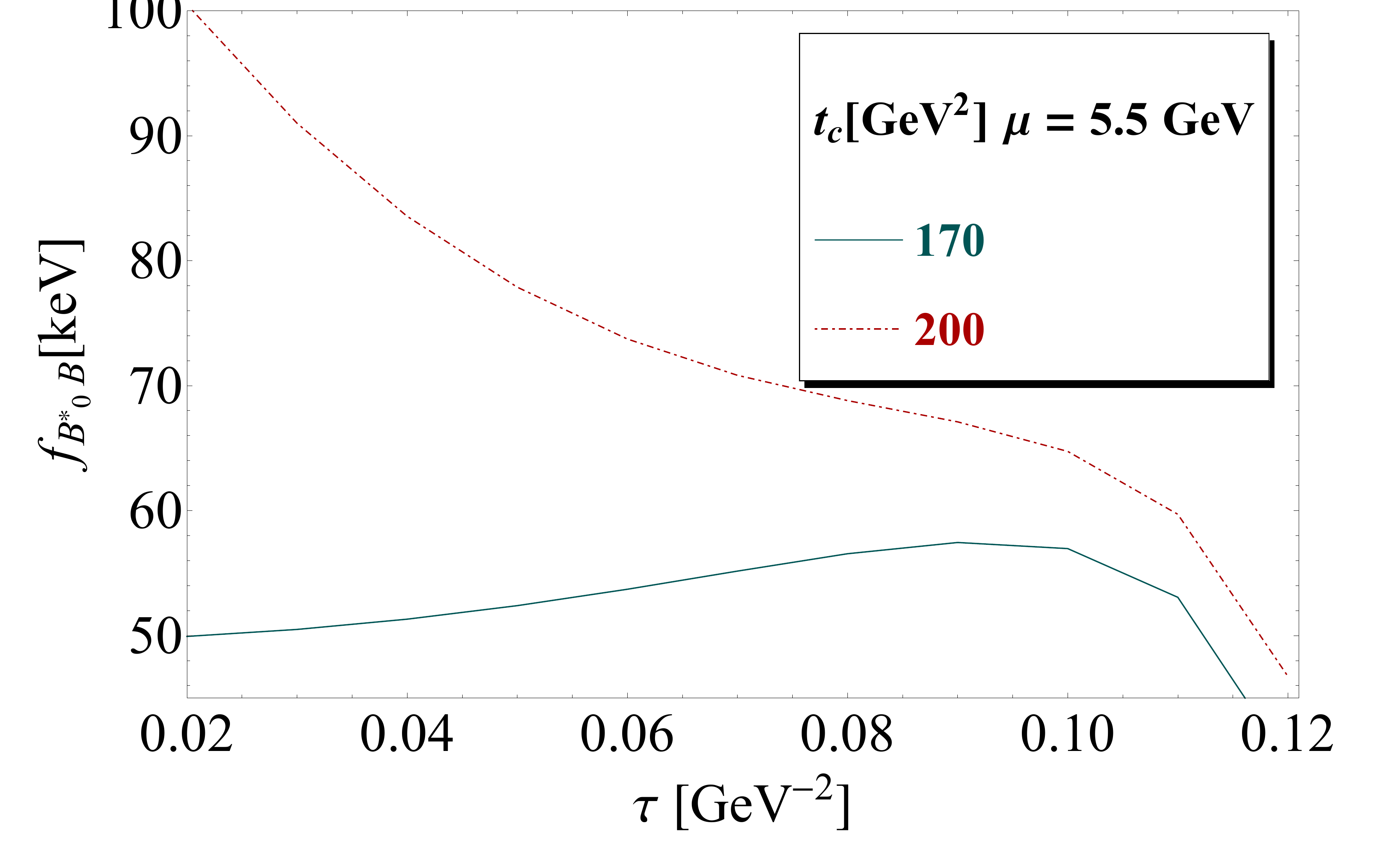}}
{\includegraphics[width=6.29cm  ]{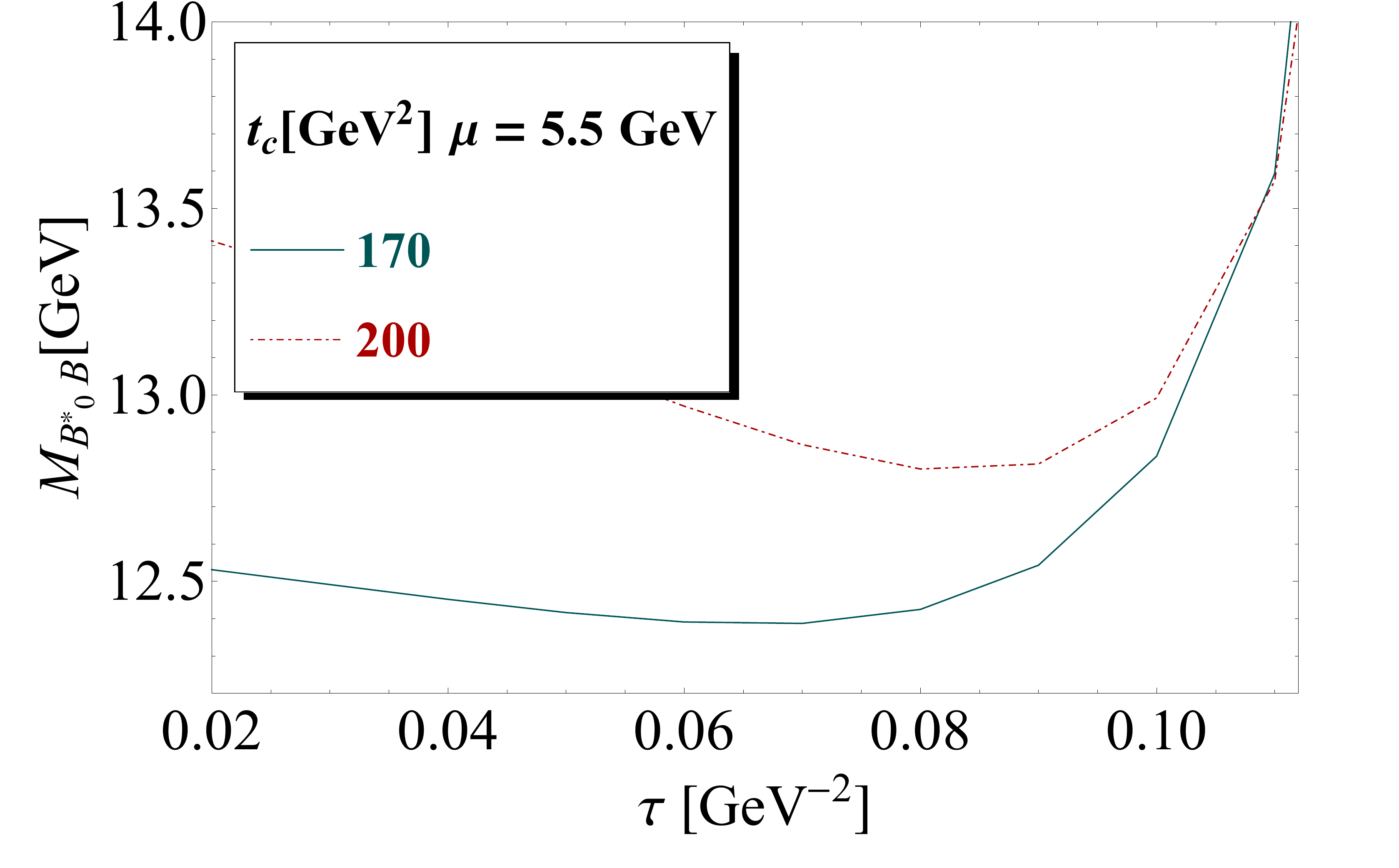}}
\centerline {\hspace*{-3cm} a)\hspace*{6cm} b) }
\caption{
\scriptsize 
{\bf a)} $f_{B^*_0B}$ at N2LO  as function of $\tau$ for different values of $t_c$, for $\mu=5.5$ GeV  and for the QCD parameters in Tables\,\ref{tab:param} and \ref{tab:alfa}; {\bf b)} The same as a) but for the mass $M_{B^*_0B}$.
}
\label{fig:bstar0b-n2lo} 
\end{center}
\end{figure}

\begin{figure}[hbt] 
\begin{center}
{\includegraphics[width=6.29cm  ]{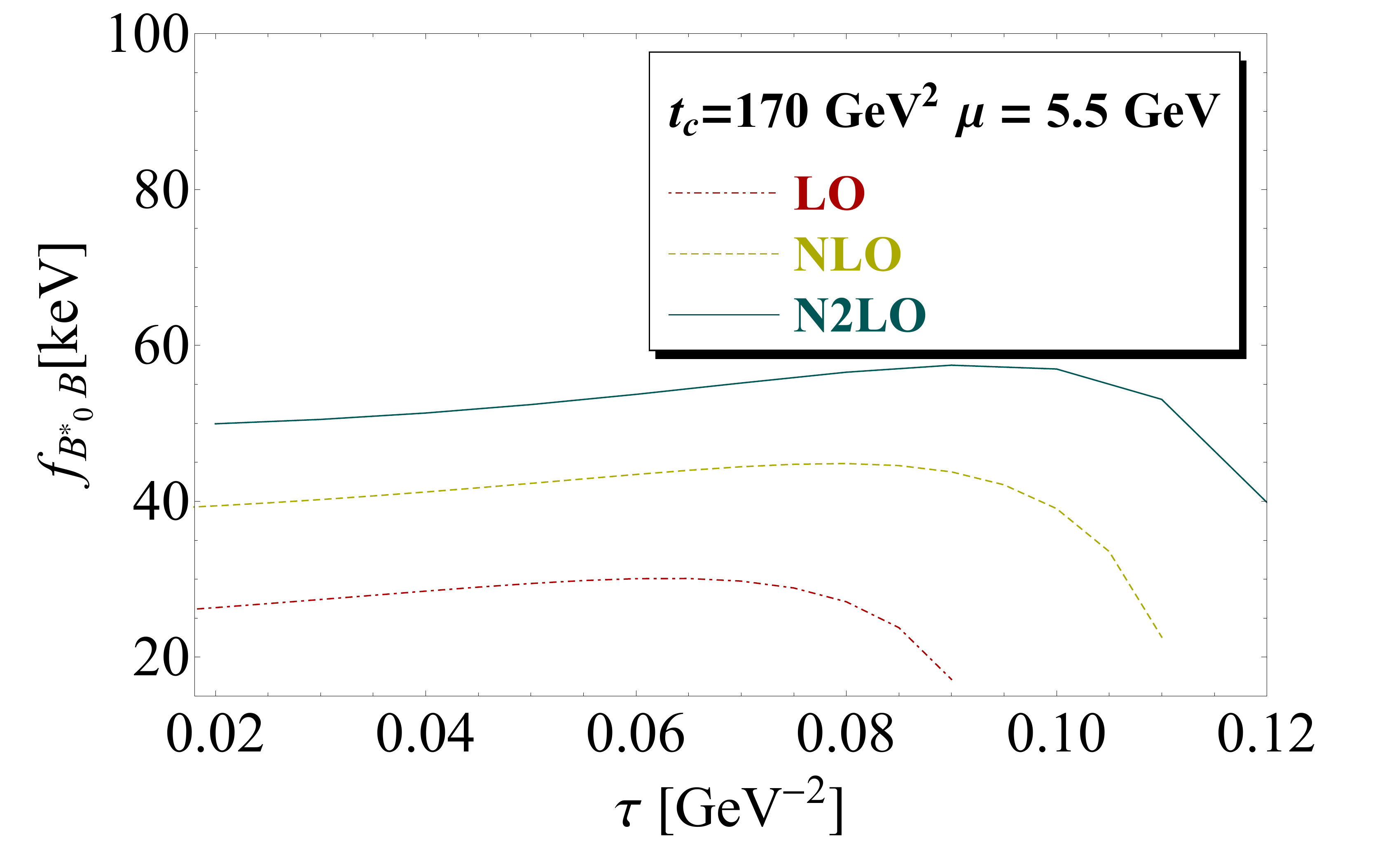}}
{\includegraphics[width=6.29cm  ]{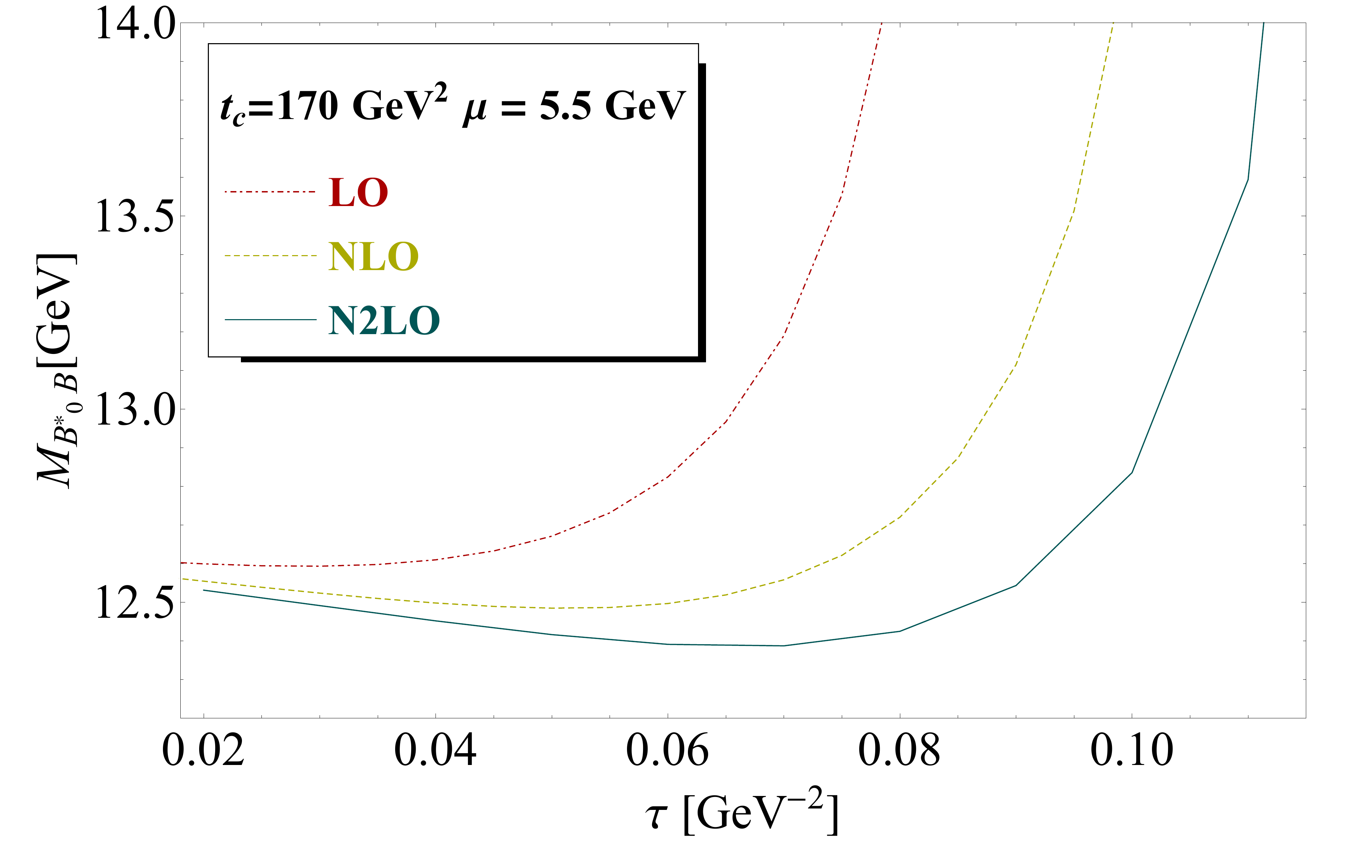}}
\centerline {\hspace*{-3cm} a)\hspace*{6cm} b) }
\caption{
\scriptsize 
{\bf a)} $f_{B^*_0B}$  as function of $\tau$ for a given value of $t_c=170$ GeV$^2$, for $\mu=5.5$ GeV, for different truncation of the PT series  and for the QCD parameters in Tables\,\ref{tab:param} and \ref{tab:alfa}; {\bf b)} The same as a) but for the mass $M_{B^*_0B}$.
}
\label{fig:bstar0b-pt} 
\end{center}
\end{figure}

\begin{figure}[hbt] 
\begin{center}
{\includegraphics[width=6.2cm  ]{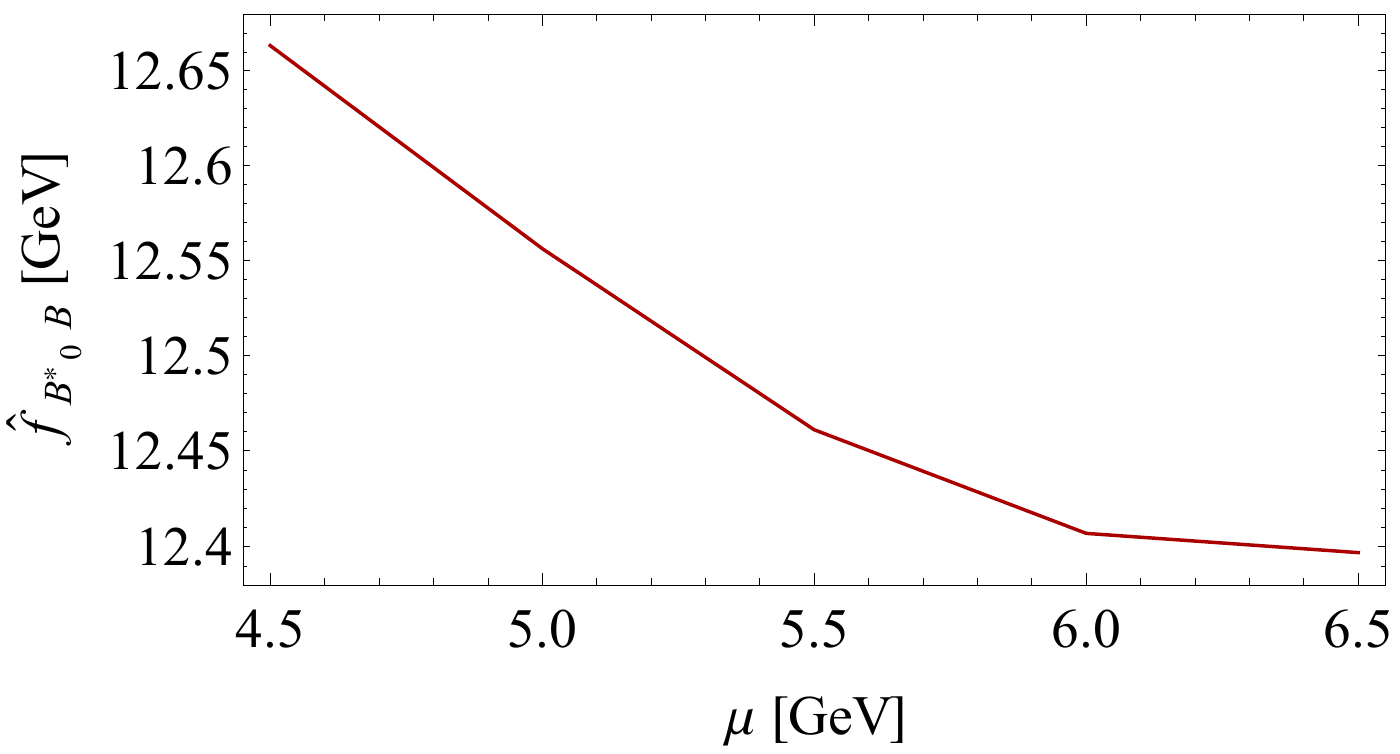}}
{\includegraphics[width=6.2cm  ]{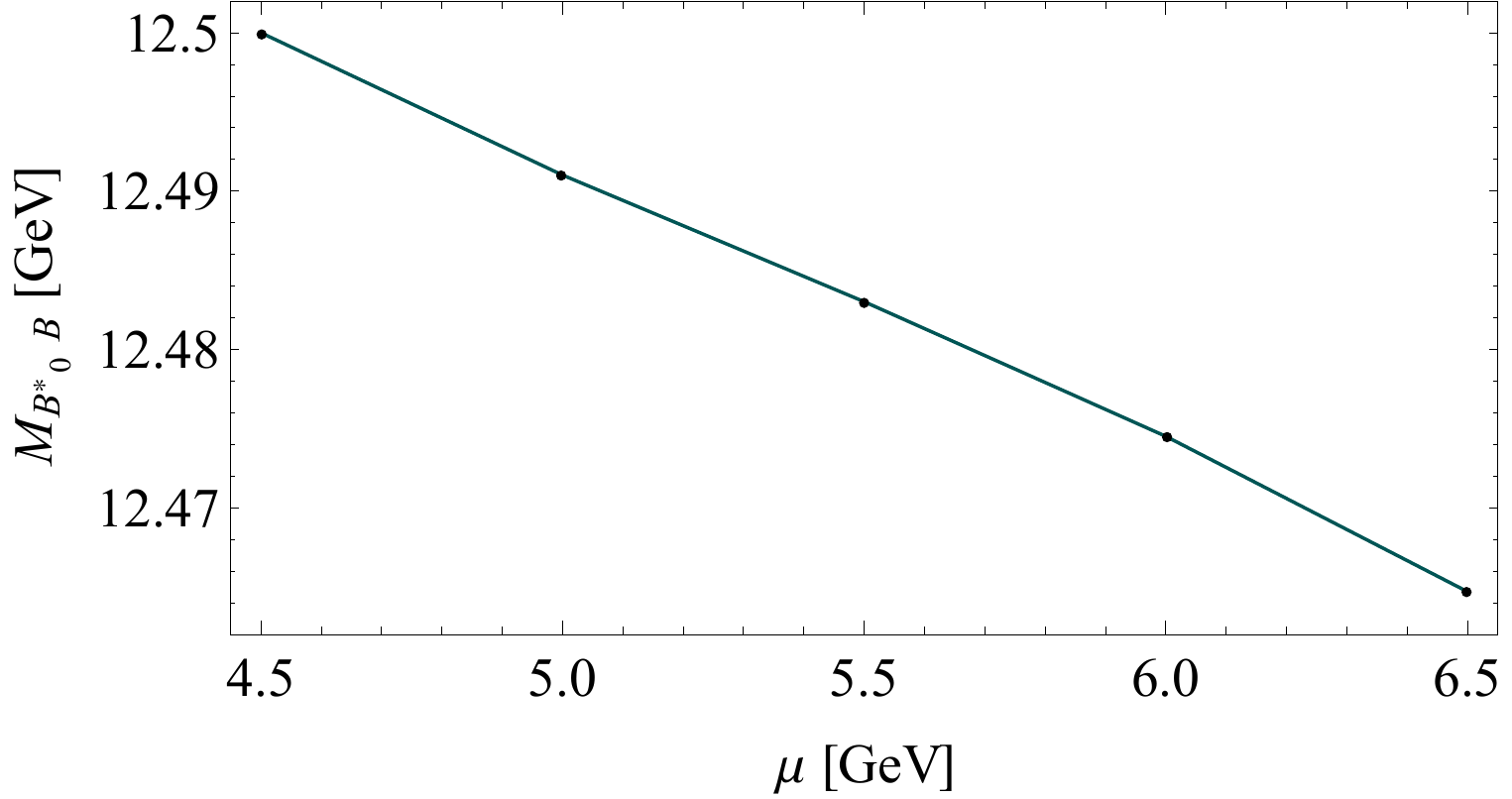}}
\centerline {\hspace*{-3cm} a)\hspace*{6cm} b) }
\caption{
\scriptsize 
{\bf a)} Renormalization group invariant coupling $\hat f_{B^*_0B}$ at NLO as function of $\mu$, for the corresponding $\tau$-stability region, for $t_c\simeq 170$ GeV$^2$ and for the QCD parameters in Tables\,\ref{tab:param}  and \ref{tab:alfa};  {\bf b)} The same as a) but for the  mass $ M_{B^*_0B}$.
}
\label{fig:bstar0b-mu} 
\end{center}
\end{figure}

We extend the analysis to the $b$-quark sector which we show in Figs.\,\ref{fig:bstar0b-lo} to \ref{fig:bstar0b-mu}. 
The result is shown in Table\,\ref{tab:resultb}. At N2LO, it corresponds to the set of parameters:
\beq
\tau\simeq (0.07-0.09)~{\rm GeV}^{-2}, ~~~~t_c\simeq  (170-200)~{\rm GeV}^2~~~~ {\rm and}~~~ ~\mu\simeq  5.5~\rm{ GeV}. 
\eeq

\nin
\nin

\nin
\nin
\nin
\nin

\section{The  $(1^{-\pm})$  Heavy-Light Vector Molecule states}
We shall study the $ \bar D^*_0D^*,~ \bar DD_1$ $(1^{--})$, their beauty analogue 
and their orthogonal combinations $(1^{-+})$ having positive $C$-parity using the currents in Table\,\ref{tab:current}.
The analysis (shapes of the curves) are very similar to the 
one of the $D^*_0D$ and $B^*_0B$ and will not be reported here. The results of the analysis are summarized in Tables\,\ref{tab:errorc}
to \ref{tab:resultb}. At N2LO, they correspond to the set of parameters:
\beq
\tau\simeq (0.15-0.21)~{\rm GeV}^{-2}, ~~~~t_c\simeq  (42-48)~{\rm GeV}^2~~~~ {\rm and}~~~ ~\mu\simeq  4.5~\rm{ GeV}, 
\eeq
in the $c$-channel and :
\beq
\tau\simeq (0.07-0.09)~{\rm GeV}^{-2}, ~~~~t_c\simeq  (170-200)~{\rm GeV}^2~~~~ {\rm and}~~~ ~\mu\simeq  5.5~\rm{ GeV}, 
\eeq
in the $b$-channel.
\section{The heavy-light four-quark states}

\subsection{The QCD interpolating currents}
The four-quark states $\big[ \,Q q \bar{Q} \bar{q} \,\big]$ will be described by the interpolating currents 
given in Table\,\ref{tab:4qcurrent}:
{\scriptsize
\begin{center}
\begin{table}[hbt]
\setlength{\tabcolsep}{2.2pc}

 \tbl{
Interpolating currents with a definite $P$-parity describing the four-quark 
states. $Q\equiv$ $c$ (resp. $b$) in the charm (resp. bottom) channel.  $q\equiv u,d$.  
}
    {\footnotesize
\begin{tabular}{ll}

&\\
\hline
\hline
$J^{P}$&Four-Quark Currents  $\equiv{\cal O}_{4q}(x)$  \\
\hline
\\
$\bf 0^{+}$&$\epsilon_{abc}\epsilon_{dec}  \bigg[
		\big( q^T_a \: C\gamma_5 \:Q_b \big) \big( \bar{q}_d \: \gamma_5 C \: \bar{Q}^T_e \big) 
		+ k \big( q^T_a \: C \:Q_b \big) \big( \bar{q}_d \: C \: \bar{Q}^T_e \big) \bigg] $\\		
$\bf 1^{+}$& $\epsilon_{abc}\epsilon_{dec}  \bigg[
		\big( q^T_a \: C\gamma_5 \:Q_b \big) \big( \bar{q}_d \: \gamma_\mu C \: \bar{Q}^T_e \big) 
		+ k \big( q^T_a \: C \:Q_b \big) \big( \bar{q}_d \: \gamma_\mu\gamma_5 C \: \bar{Q}^T_e \big) \bigg]  $ \\
	
$\bf 0^{-}$&	$\epsilon_{abc}\epsilon_{dec}  \bigg[
		\big( q^T_a \: C\gamma_5 \:Q_b \big) \big( \bar{q}_d \: C \: \bar{Q}^T_e \big) 
		+ k \big( q^T_a \: C \:Q_b \big) \big( \bar{q}_d \:\gamma_5 C \: \bar{Q}^T_e \big) \bigg] $\\	
$\bf 1^{-}$&$  \epsilon_{abc}\epsilon_{dec}  \big[
		\big( q^T_a \: C\gamma_5 \:Q_b \big) \big( \bar{q}_d \: \gamma_\mu\gamma_5 C \: \bar{Q}^T_e \big) 
		+ k \big( q^T_a \: C \:Q_b \big) \big( \bar{q}_d \: \gamma_\mu C \: \bar{Q}^T_e \big) \big]$\\
		\\
\hline
\hline
\end{tabular}
}
\label{tab:4qcurrent}
\end{table}
\end{center}
}
\nin
The corresponding spectral functions are defined analogously to  Eq.\,\ref{2po} as: $\frac{1}{\pi}{\rm Im}\Pi_{4q}^{(1)}(t)$ for spin 1  and $\frac{1}{\pi}{\rm Im}\psi_{4q}^{(s,p)}(t)$ from Eq.\,\ref{2po5} for spin 0 mesons. $k$ is the mixing of the two operators. We shall take the optimal choice $k=0$ as  demonstrated in\,\cite{X3A,X3B}. The expressions of the spectral functions to LO of PT and including the contributions of condensates of dimension $d\leq 8$ are given  in \ref{app:4q}. 

\subsection{Coupling and mass of the $ S_c(0^{+})$ four-quark state}
\begin{figure}[hbt] 
\begin{center}
{\includegraphics[width=6.29cm  ]{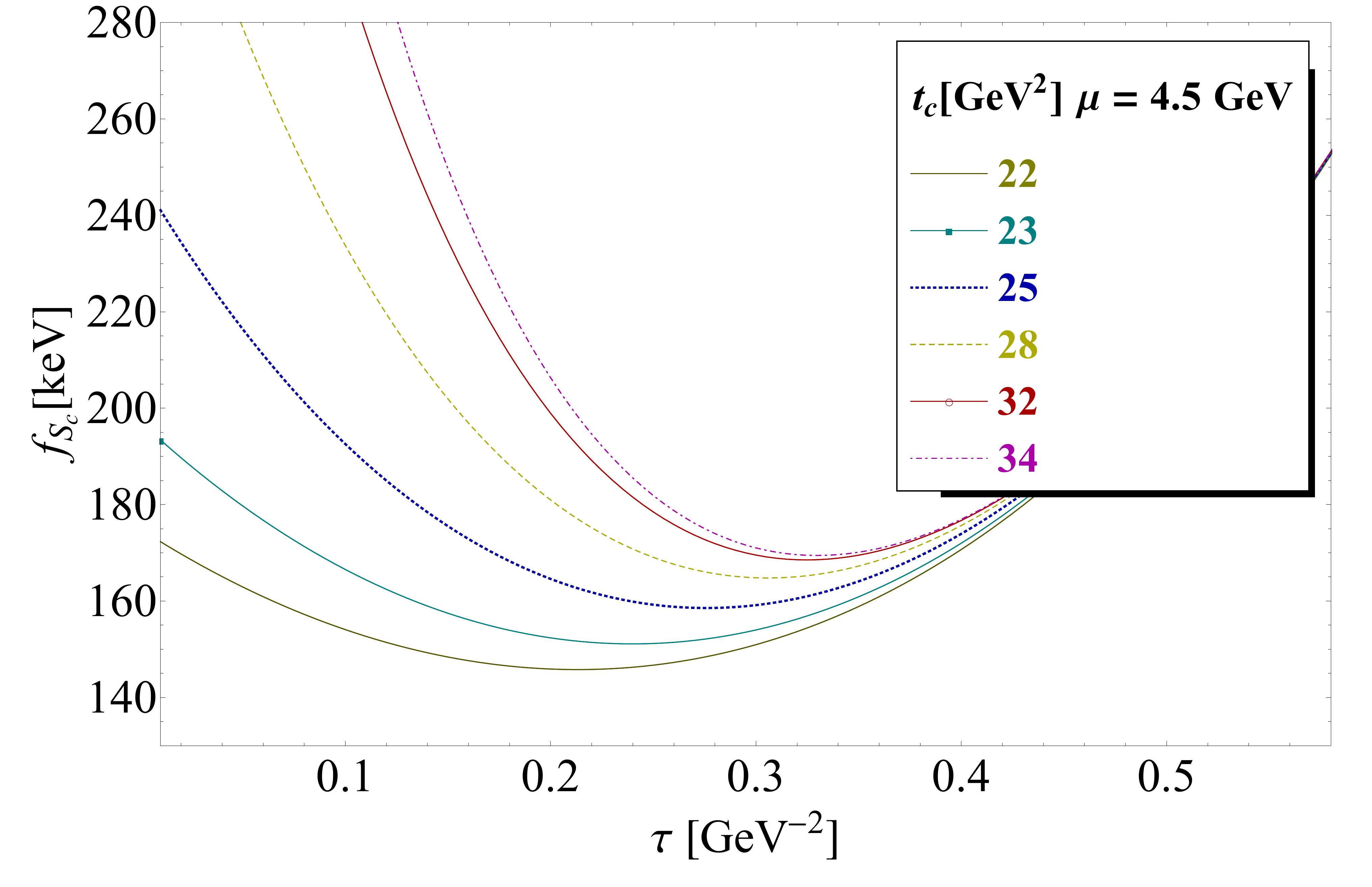}}
{\includegraphics[width=6.29cm  ]{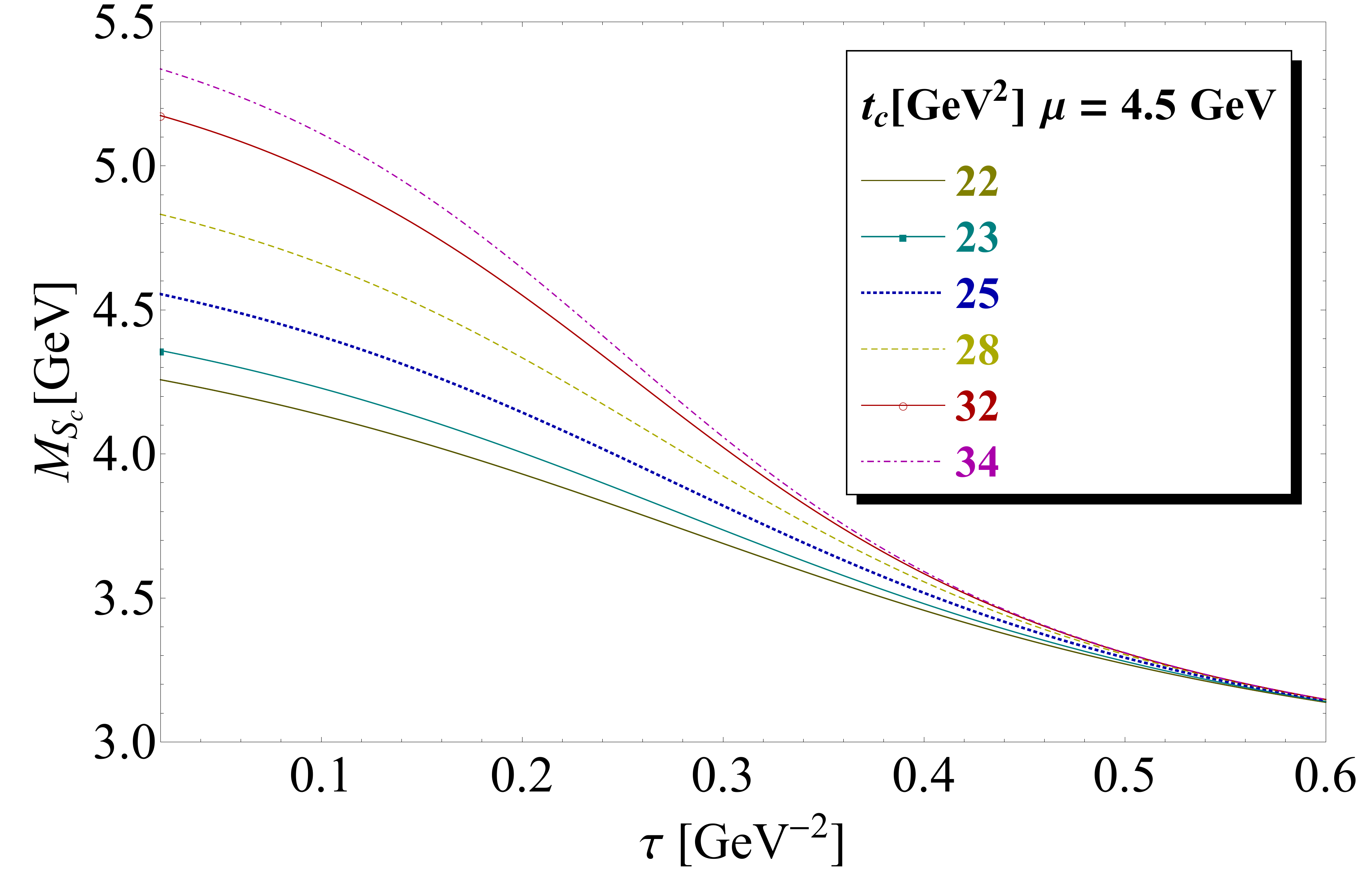}}
\centerline {\hspace*{-3cm} a)\hspace*{6cm} b) }
\caption{
\scriptsize 
{\bf a)} $f_{S_c}$  at LO as function of $\tau$ for different values of $t_c$, for $\mu=4.5$ GeV  and for the QCD parameters in Tables\,\ref{tab:param} and \ref{tab:alfa}; {\bf b)} The same as a) but for the mass $M_{S_c}$.
}
\label{fig:sc-lo} 
\end{center}
\end{figure}

\begin{figure}[hbt] 
\begin{center}
{\includegraphics[width=6.29cm  ]{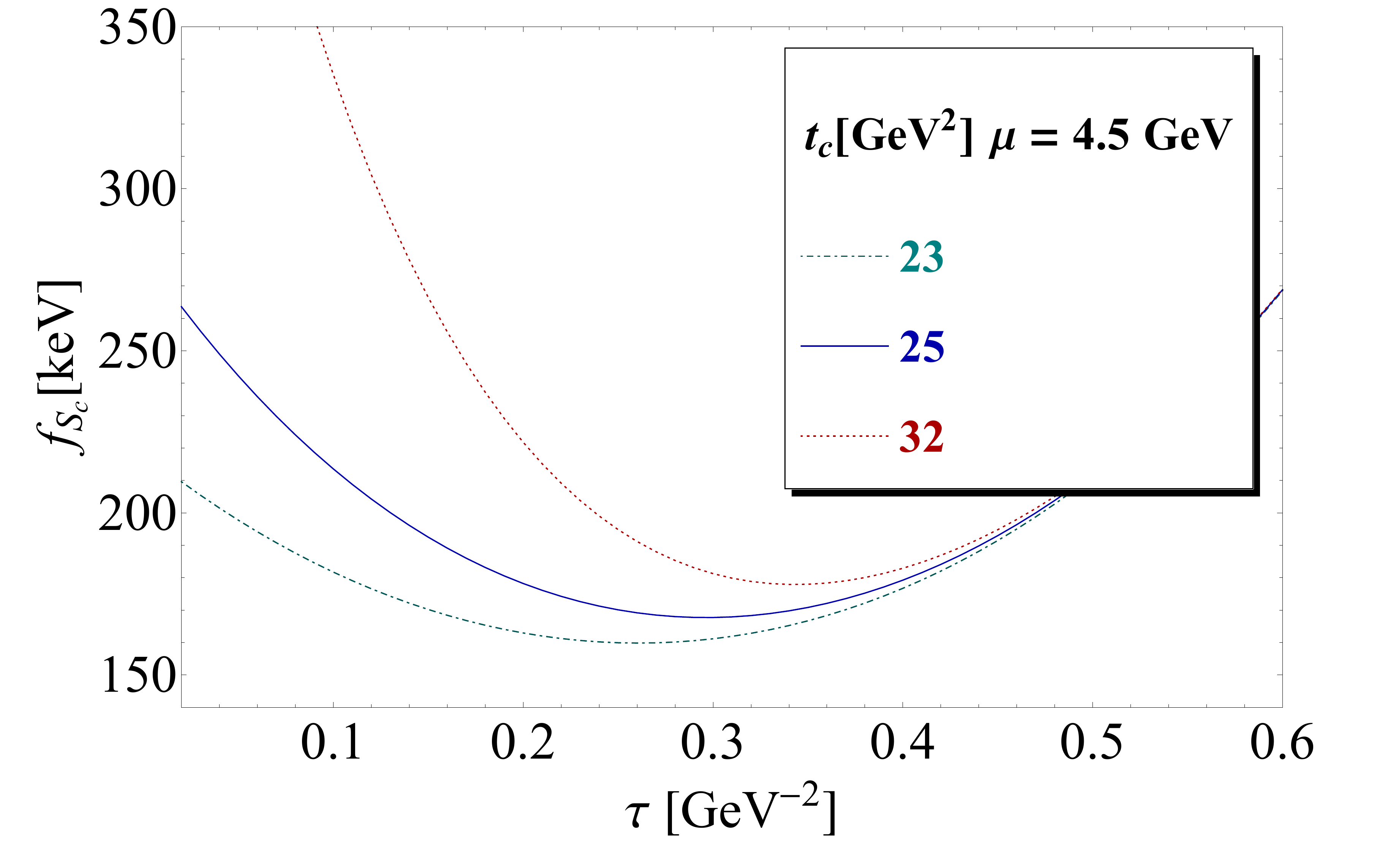}}
{\includegraphics[width=6.29cm  ]{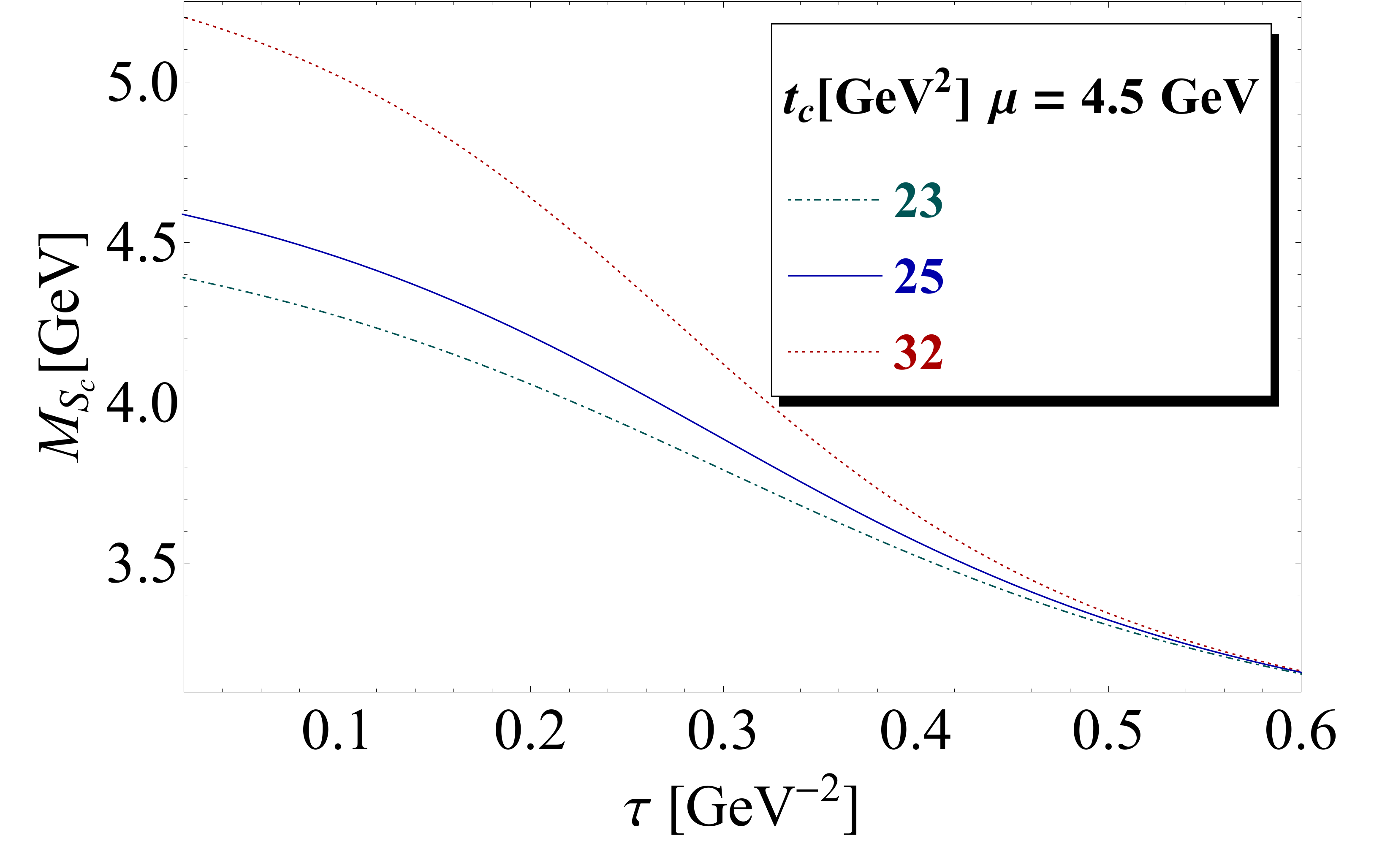}}
\centerline {\hspace*{-3cm} a)\hspace*{6cm} b) }
\caption{
\scriptsize 
{\bf a)} $f_{S_c}$ at NLO  as function of $\tau$ for different values of $t_c$, for $\mu=4.5$ GeV  and for the QCD parameters in Tables\,\ref{tab:param} and \ref{tab:alfa}; {\bf b)} The same as a) but for the mass $M_{S_c}$.
}
\label{fig:sc-nlo} 
\end{center}
\end{figure}

\begin{figure}[hbt] 
\begin{center}
{\includegraphics[width=6.29cm  ]{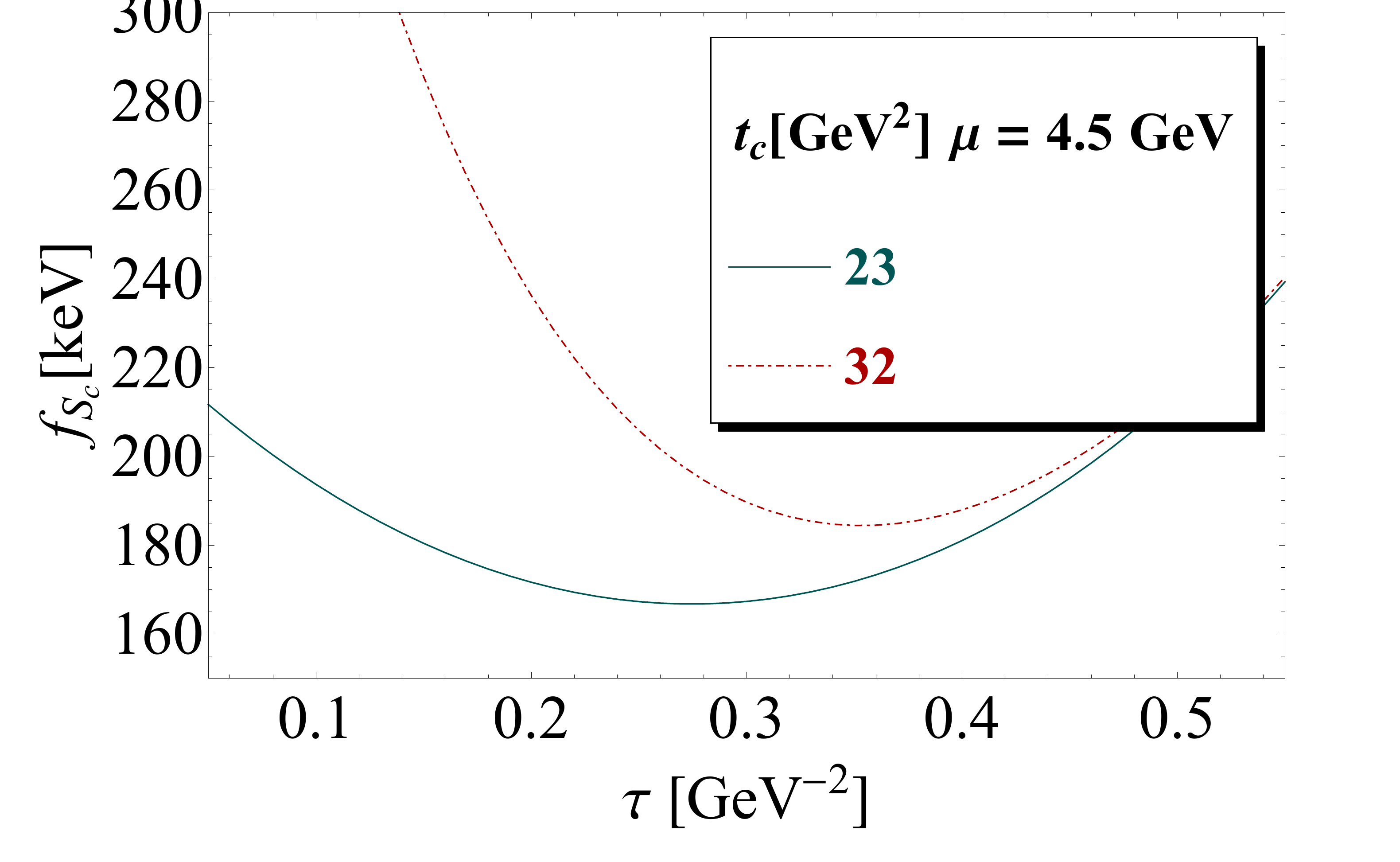}}
{\includegraphics[width=6.29cm  ]{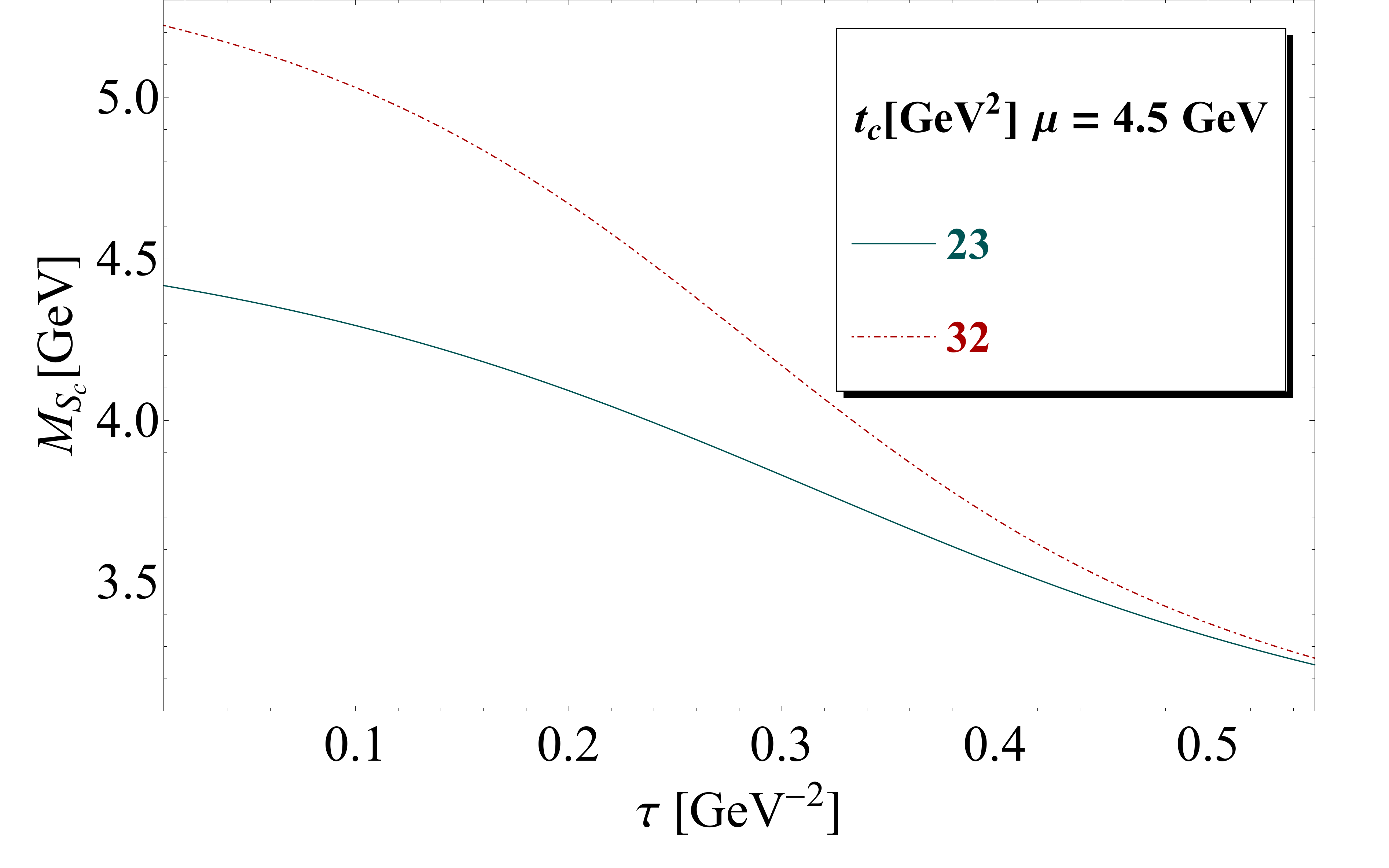}}
\centerline {\hspace*{-3cm} a)\hspace*{6cm} b) }
\caption{
\scriptsize 
{\bf a)} $f_{S_c}$ at N2LO  as function of $\tau$ for different values of $t_c$, for $\mu=4.5$ GeV  and for the QCD parameters in Tables\,\ref{tab:param} and \ref{tab:alfa}; {\bf b)} The same as a) but for the mass $M_{S_c}$.
}
\label{fig:sc-n2lo} 
\end{center}
\end{figure}

\begin{figure}[hbt] 
\begin{center}
{\includegraphics[width=6.29cm  ]{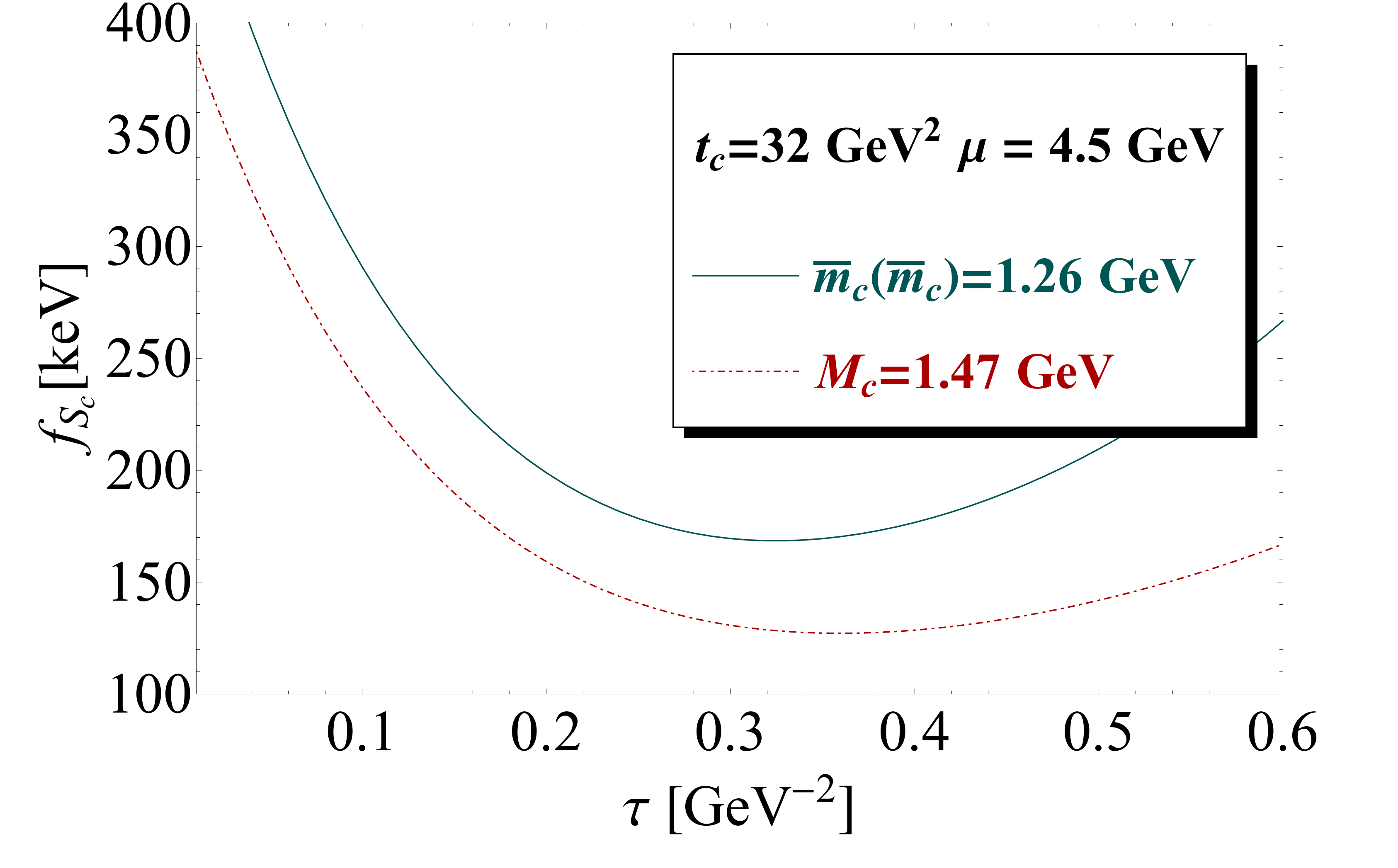}}
{\includegraphics[width=6.29cm  ]{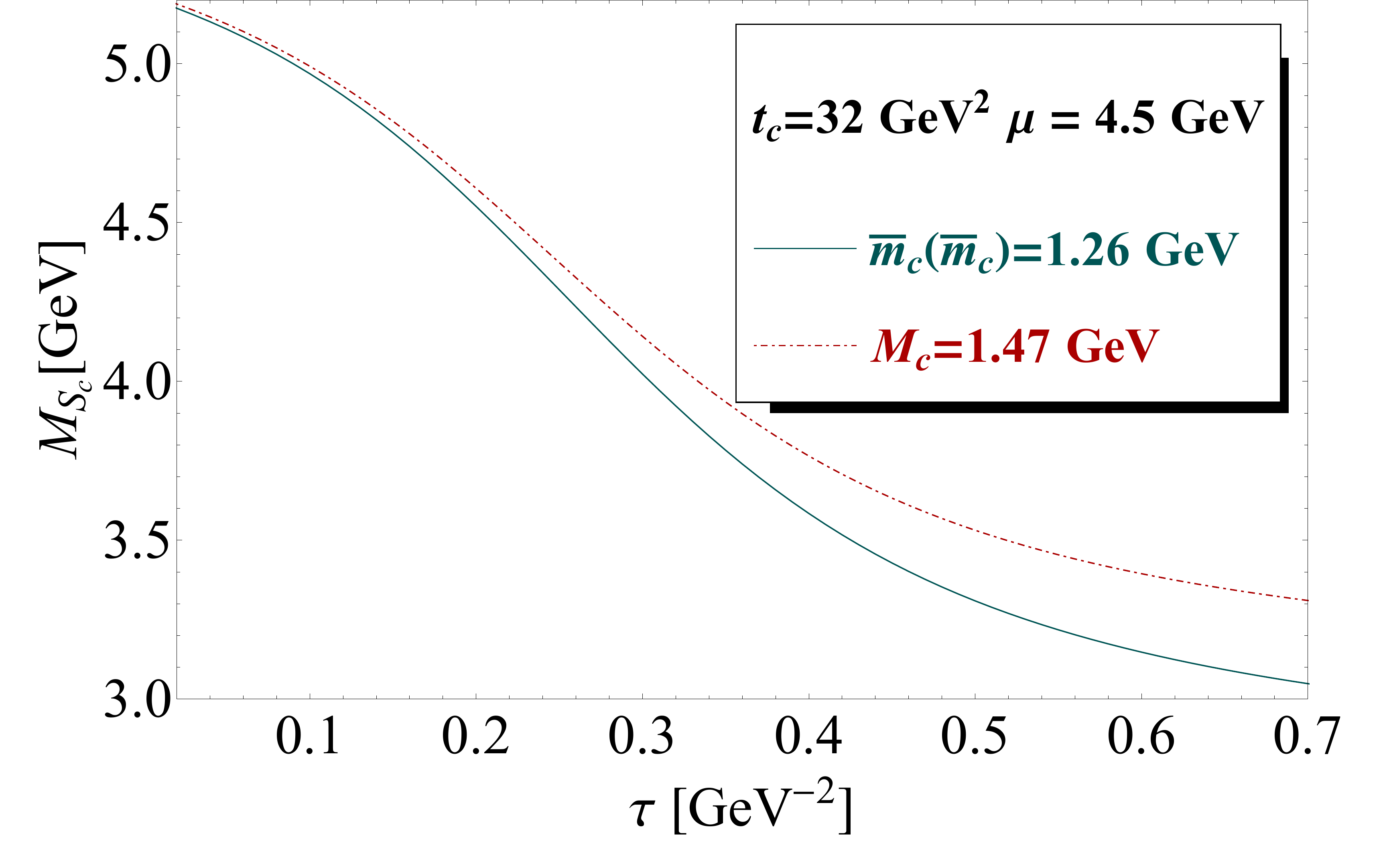}}
\centerline {\hspace*{-3cm} a)\hspace*{6cm} b) }
\caption{
\scriptsize 
{\bf a)} $f_{S_c}$  at LO as function of $\tau$ for  $t_c=32$ GeV$^2$, for $\mu=4.5$ GeV, for  values of the running $\overline{m}_c(\overline{m}_c)=1.26$ GeV and pole mass $M_c=1.47$ GeV. We use     
the QCD parameters in Tables\,\ref{tab:param} and \ref{tab:alfa}; {\bf b)} The same as a) but for the mass $M_{S_c}$.
}
\label{fig:sc-const} 
\end{center}
\end{figure}

\begin{figure}[hbt] 
\begin{center}
{\includegraphics[width=6.29cm  ]{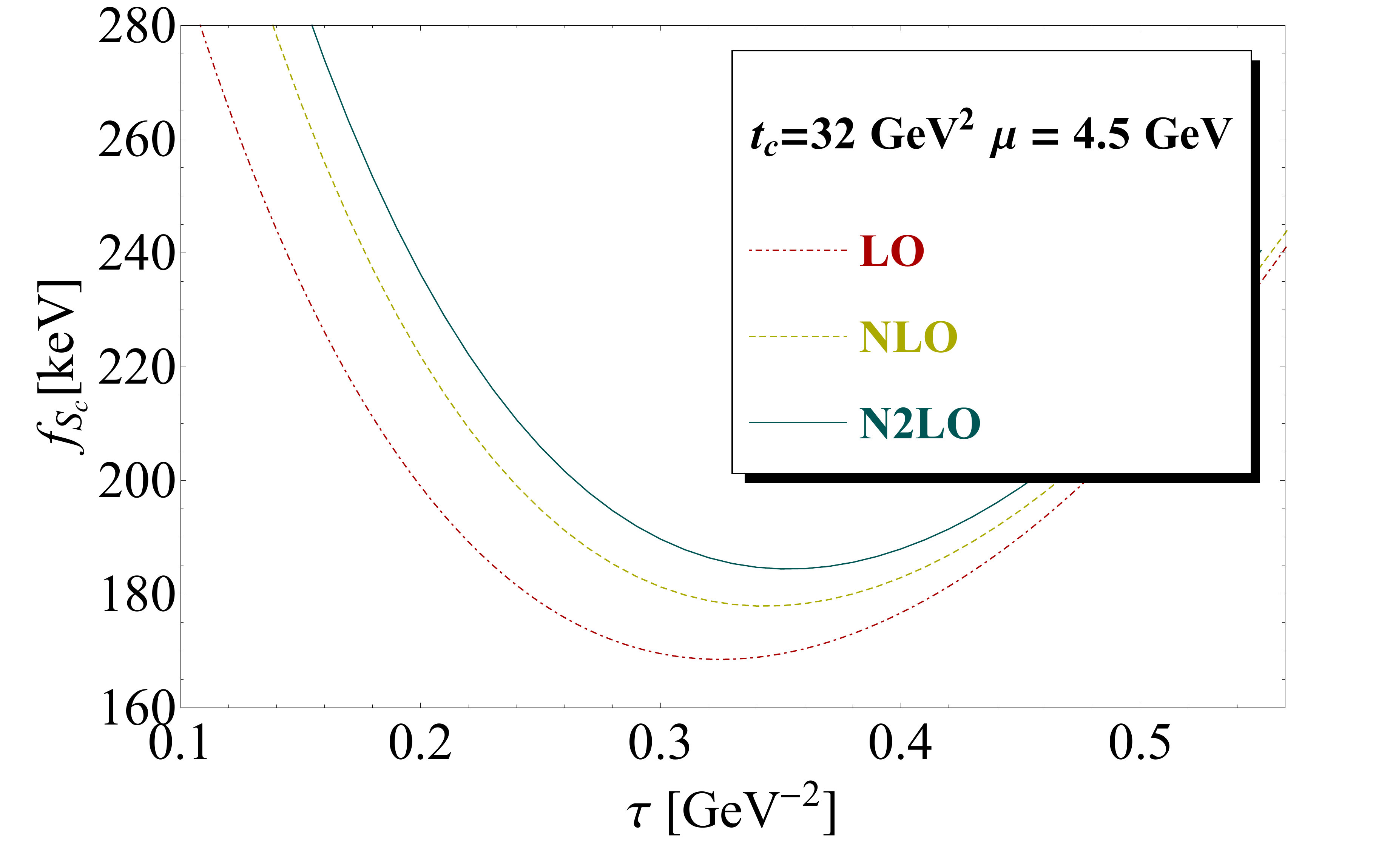}}
{\includegraphics[width=6.29cm  ]{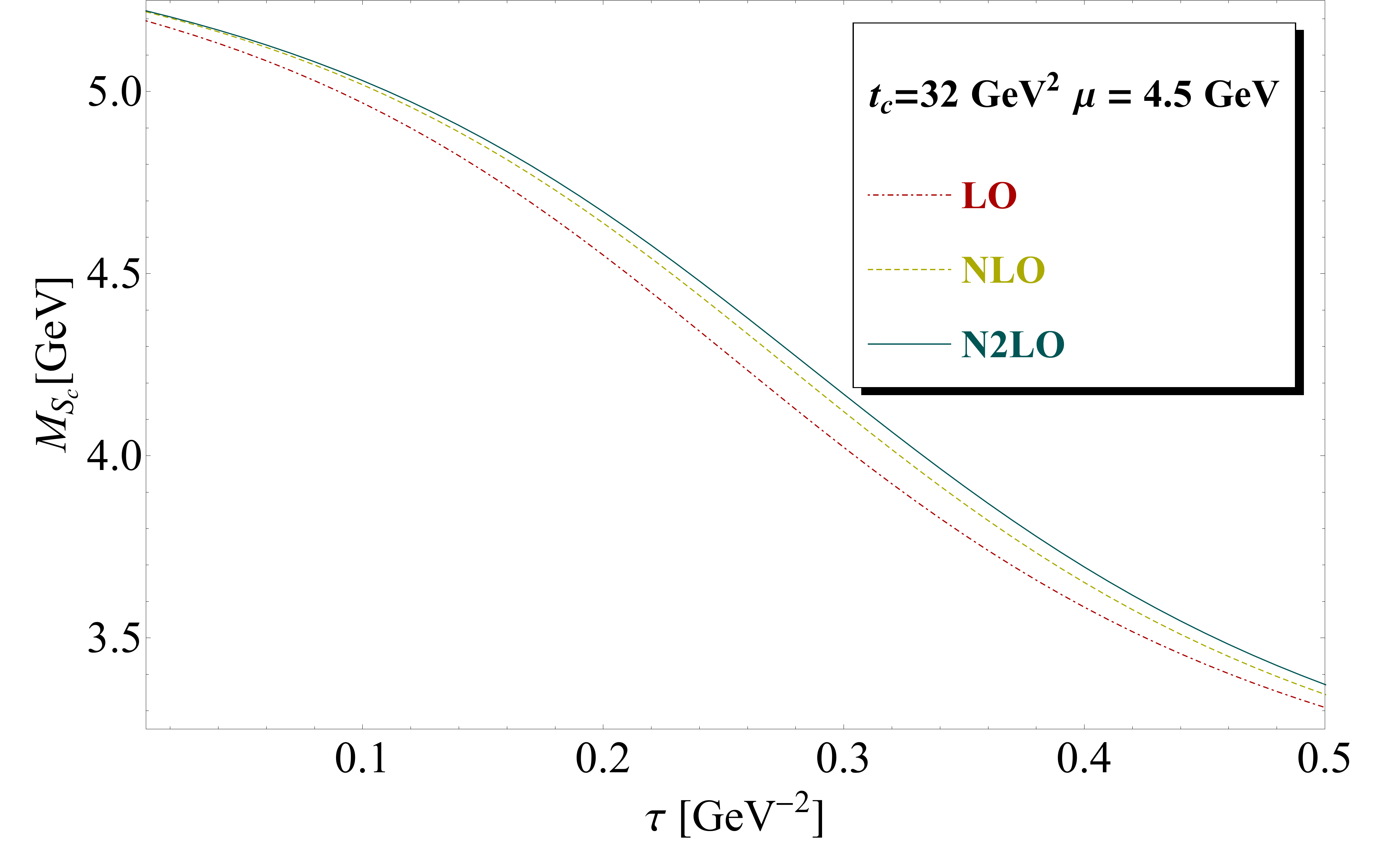}}
\centerline {\hspace*{-3cm} a)\hspace*{6cm} b) }
\caption{
\scriptsize 
{\bf a)} $f_{S_c}$  as function of $\tau$ for a given value of $t_c=32$ GeV$^2$, for $\mu=4.5$ GeV, for different truncation of the PT series  and for the QCD parameters in Tables\,\ref{tab:param} and \ref{tab:alfa}; {\bf b)} The same as a) but for the mass $M_{S_c}$.
}
\label{fig:sc-pt} 
\end{center}
\end{figure}

\begin{figure}[hbt] 
\begin{center}
{\includegraphics[width=6.2cm  ]{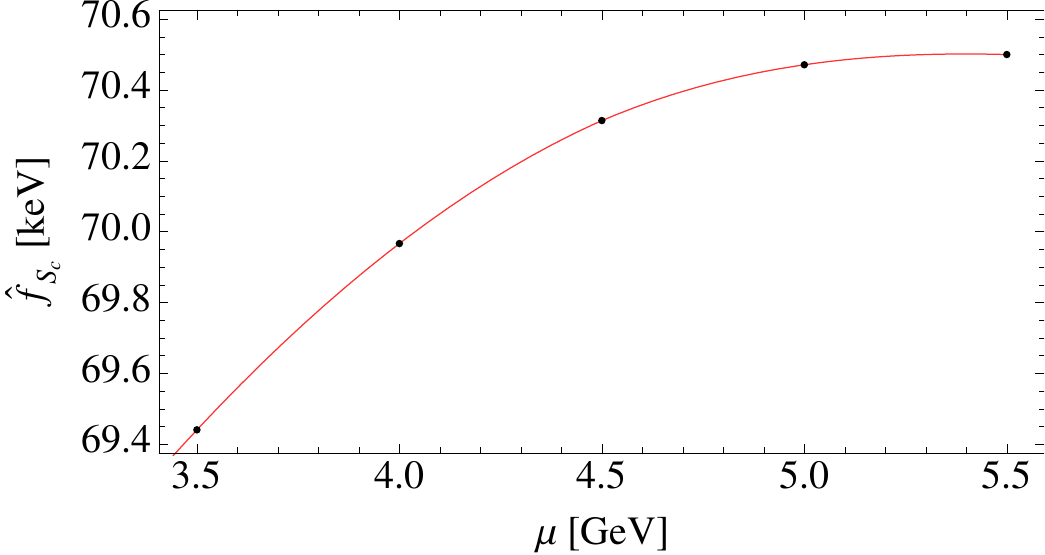}}
{\includegraphics[width=6.2cm  ]{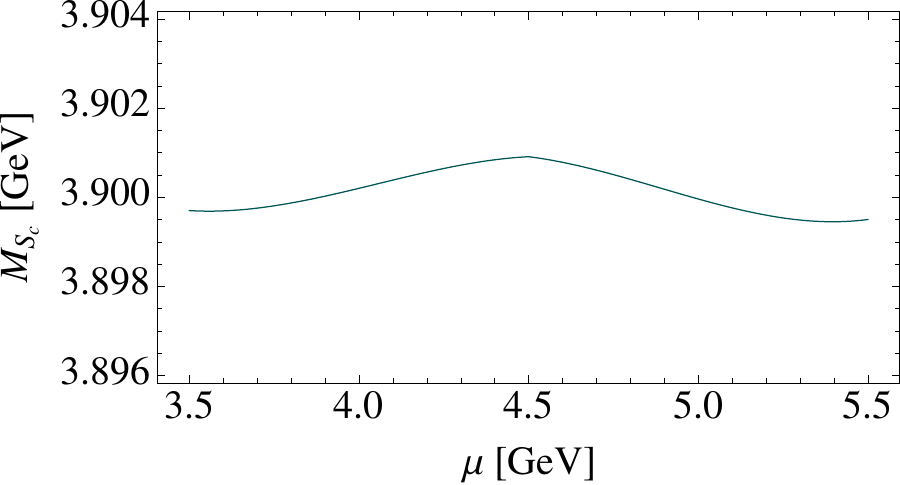}}
\centerline {\hspace*{-3cm} a)\hspace*{6cm} b) }
\caption{
\scriptsize 
{\bf a)} Renormalization group invariant coupling $\hat f_{S_c}$ at NLO as function of $\mu$, for the corresponding $\tau$-stability region, for $t_c\simeq 32$ GeV$^2$ and for the QCD parameters in Tables\,\ref{tab:param}  and \ref{tab:alfa};  {\bf b)} The same as a) but for the  mass $ M_{S_c}$.
}
\label{fig:sc-mu} 
\end{center}
\end{figure}

Like in the previous case of the molecule states, we study the coupling and mass of the scalar $S_c(0^+)$ four-quark state which we show in Figs.\,\ref{fig:sc-lo} to \ref{fig:sc-mu}. We shall see that the analysis of the four-quark states is very similar to the one of the molecules and present analogous features (presence of minimas or/and inflexion points, good convergence of the PT series and the OPE). The results are summarized in Tables\,\ref{tab:4q-errorc} and \,\ref{tab:4q-resultc}. At N2LO, the corresponding set of parameters are:
\beq
\tau\simeq (0.3-0.4)~{\rm GeV}^{-2}, ~~~~t_c\simeq  (23-32)~{\rm GeV}^2~~~~ {\rm and}~~~ ~\mu\simeq  4.5~\rm{ GeV}, 
\eeq
\nin
\nin
\nin

\nin
\nin
\nin
\subsection{Coupling and mass of the $ S_b(0^{+})$ four-quark state}
We extend the analysis to the $b$-quark sector. The related curves are very similar to the ones of the $S_c$ and $\bar BB$ molecules and will not be reported here.
The results are summarized in Tables\,\ref{tab:4q-errorb} and \,\ref{tab:4q-resultb}. At N2LO, the corresponding set of parameters are:
\beq
\tau\simeq (0.13-0.14)~{\rm GeV}^{-2}, ~~~~t_c\simeq  (160-190)~{\rm GeV}^2~~~~ {\rm and}~~~ ~\mu\simeq  5.5~\rm {GeV}, 
\eeq

\subsection{Couplings and masses of the $ A_{c,b}(1^{+})$ four-quark states}
The study of the couplings and masses of the axial-vector $A_{c,b}(1^+)$  four-quark states presents analogous features as the ones of the  $ S_{c,b}(0^{+})$ four-quark states. The results are summarized in Tables\,\ref{tab:4q-errorc},\,\ref{tab:4q-errorb},\,\ref{tab:4q-resultc} and\,\ref{tab:4q-resultb}.
At N2LO, the corresponding set of parameters are:
\beq
\tau\simeq (0.3-0.4)~{\rm GeV}^{-2}, ~~~~t_c\simeq  (23-32)~{\rm GeV}^2~~~~ {\rm and}~~~ ~\mu\simeq  4.5~\rm{ GeV}, 
\eeq
for the $c$-quark channel and:
\beq
\tau\simeq (0.11-0.14)~{\rm GeV}^{-2}, ~~~~t_c\simeq  (140-170)~{\rm GeV}^2~~~~ {\rm and}~~~ ~\mu\simeq  5.5~\rm{ GeV}, 
\eeq
for the $b$-quark channel.
\subsection{Coupling and mass of the $ \pi_c(0^{-})$  four-quark state}
\begin{figure}[hbt] 
\begin{center}
{\includegraphics[width=6.29cm  ]{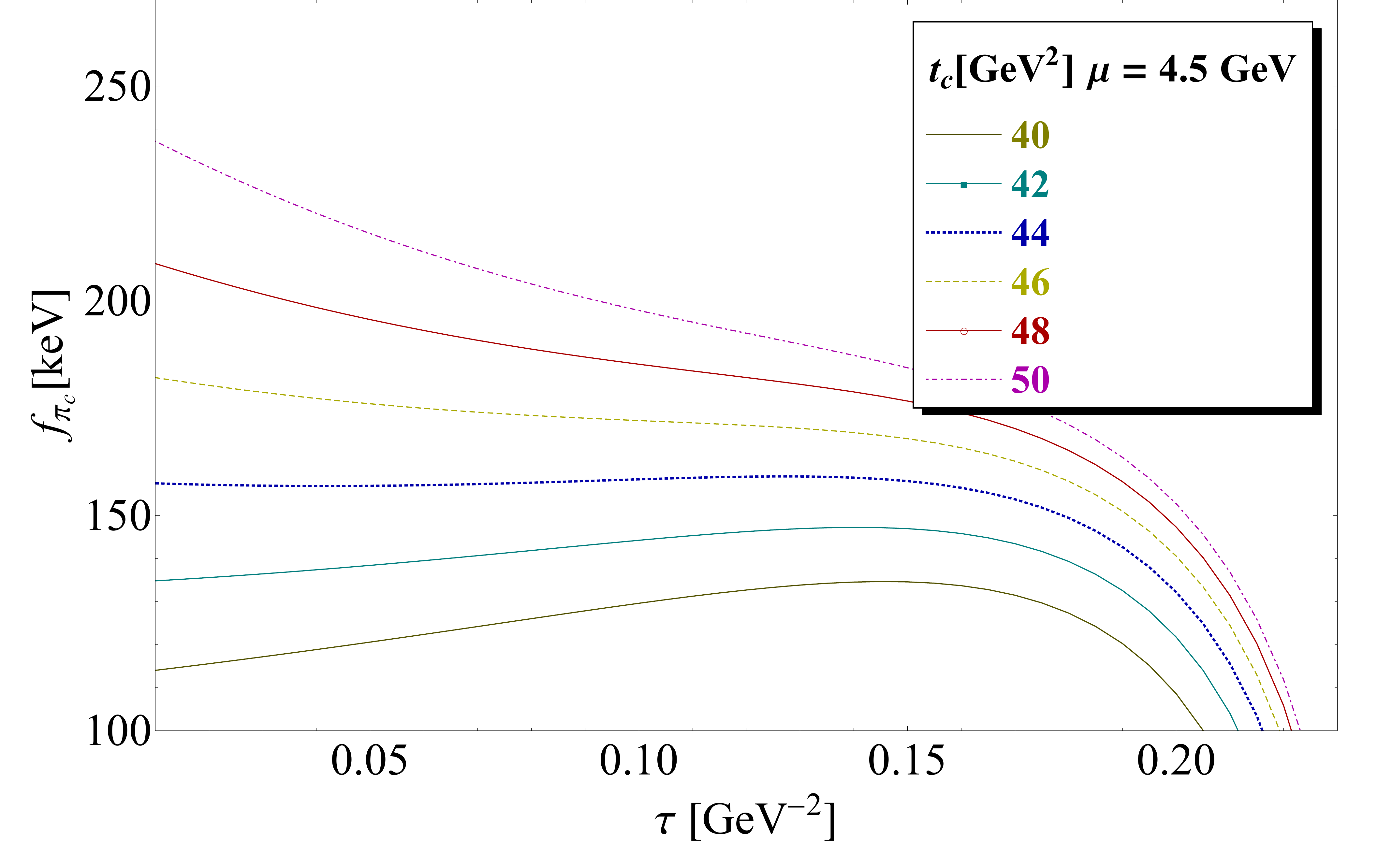}}
{\includegraphics[width=6.29cm  ]{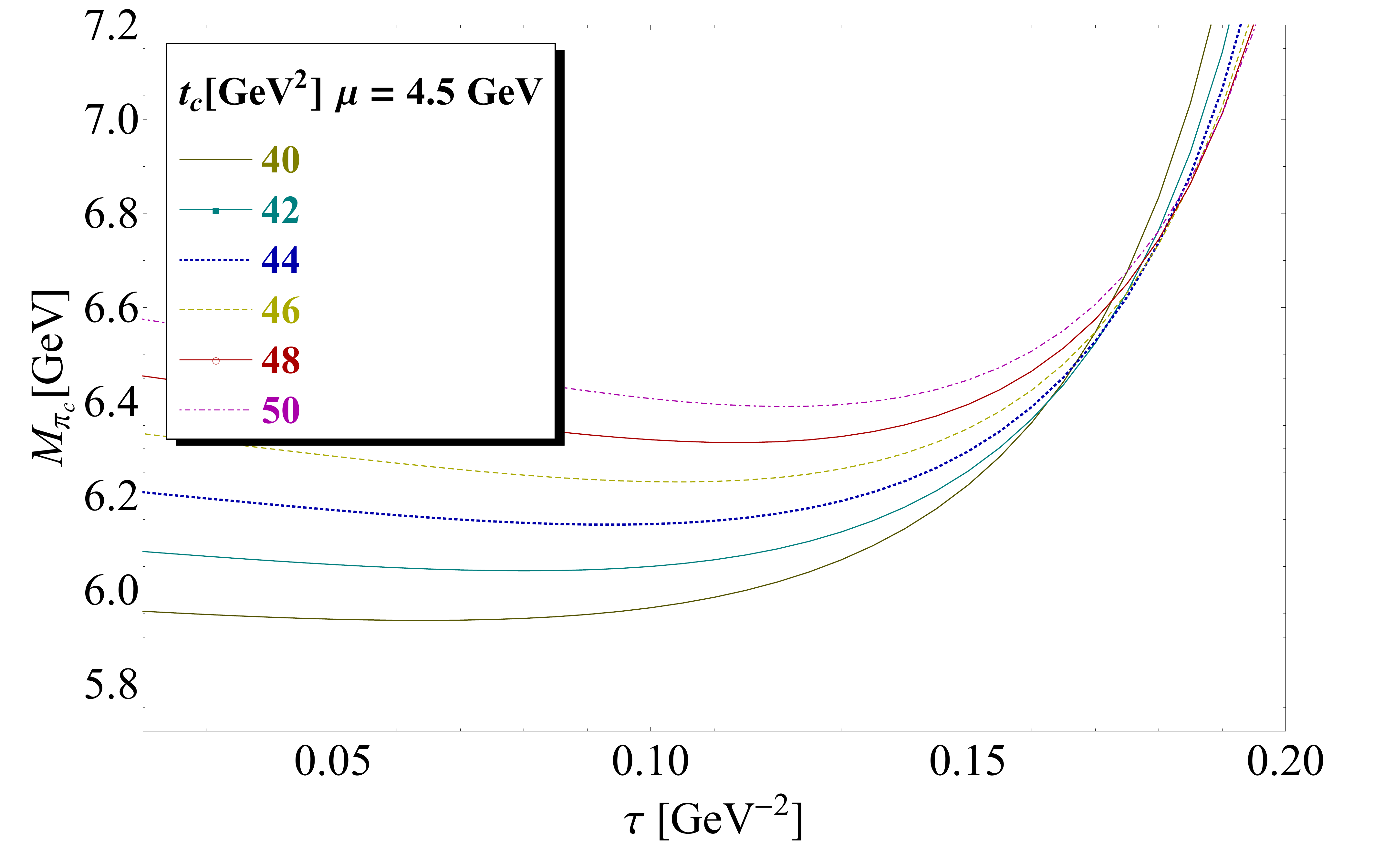}}
\centerline {\hspace*{-3cm} a)\hspace*{6cm} b) }
\caption{
\scriptsize 
{\bf a)} $f_{\pi_c}$  at LO as function of $\tau$ for different values of $t_c$, for $\mu=4.5$ GeV  and for the QCD parameters in Tables\,\ref{tab:param} and \ref{tab:alfa}; {\bf b)} The same as a) but for the mass $M_{\pi_c}$.
}
\label{fig:pc-lo} 
\end{center}
\end{figure}

\begin{figure}[hbt] 
\begin{center}
{\includegraphics[width=6.29cm  ]{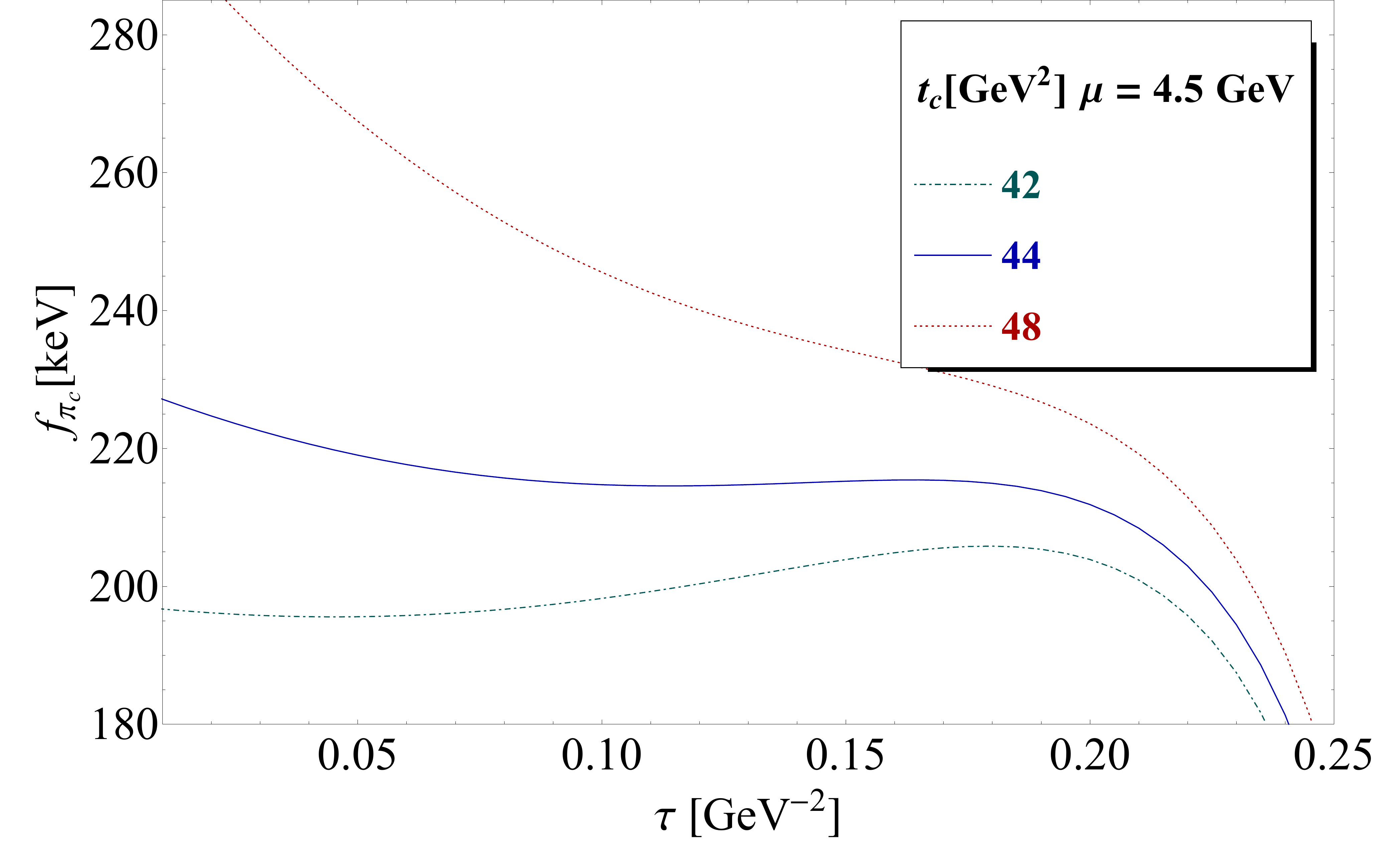}}
{\includegraphics[width=6.29cm  ]{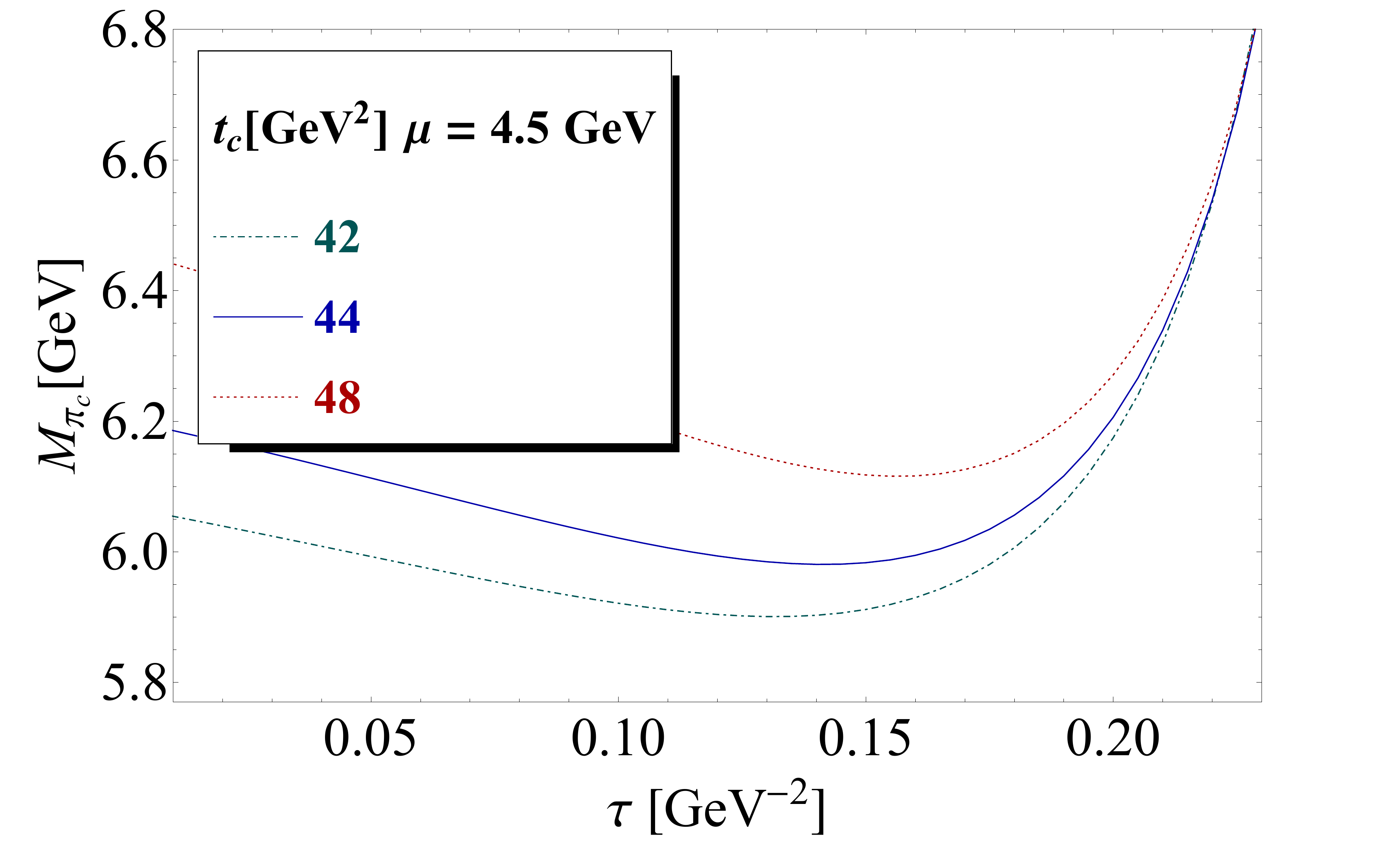}}
\centerline {\hspace*{-3cm} a)\hspace*{6cm} b) }
\caption{
\scriptsize 
{\bf a)} $f_{\pi_c}$ at NLO  as function of $\tau$ for different values of $t_c$, for $\mu=4.5$ GeV  and for the QCD parameters in Tables\,\ref{tab:param} and \ref{tab:alfa}; {\bf b)} The same as a) but for the mass $M_{\pi_c}$.
}
\label{fig:pc-nlo} 
\end{center}
\end{figure}

\begin{figure}[hbt] 
\begin{center}
{\includegraphics[width=6.29cm  ]{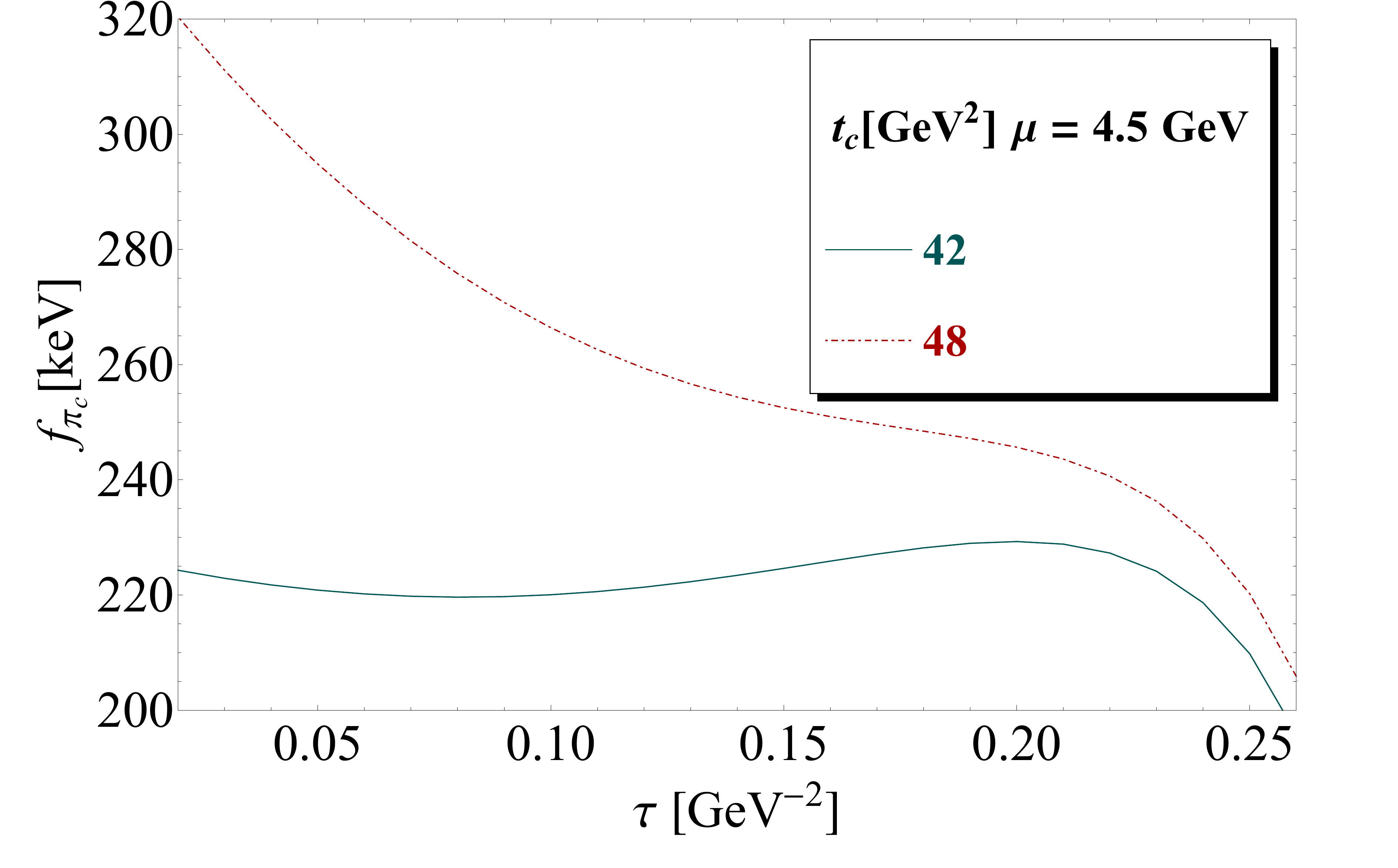}}
{\includegraphics[width=6.29cm  ]{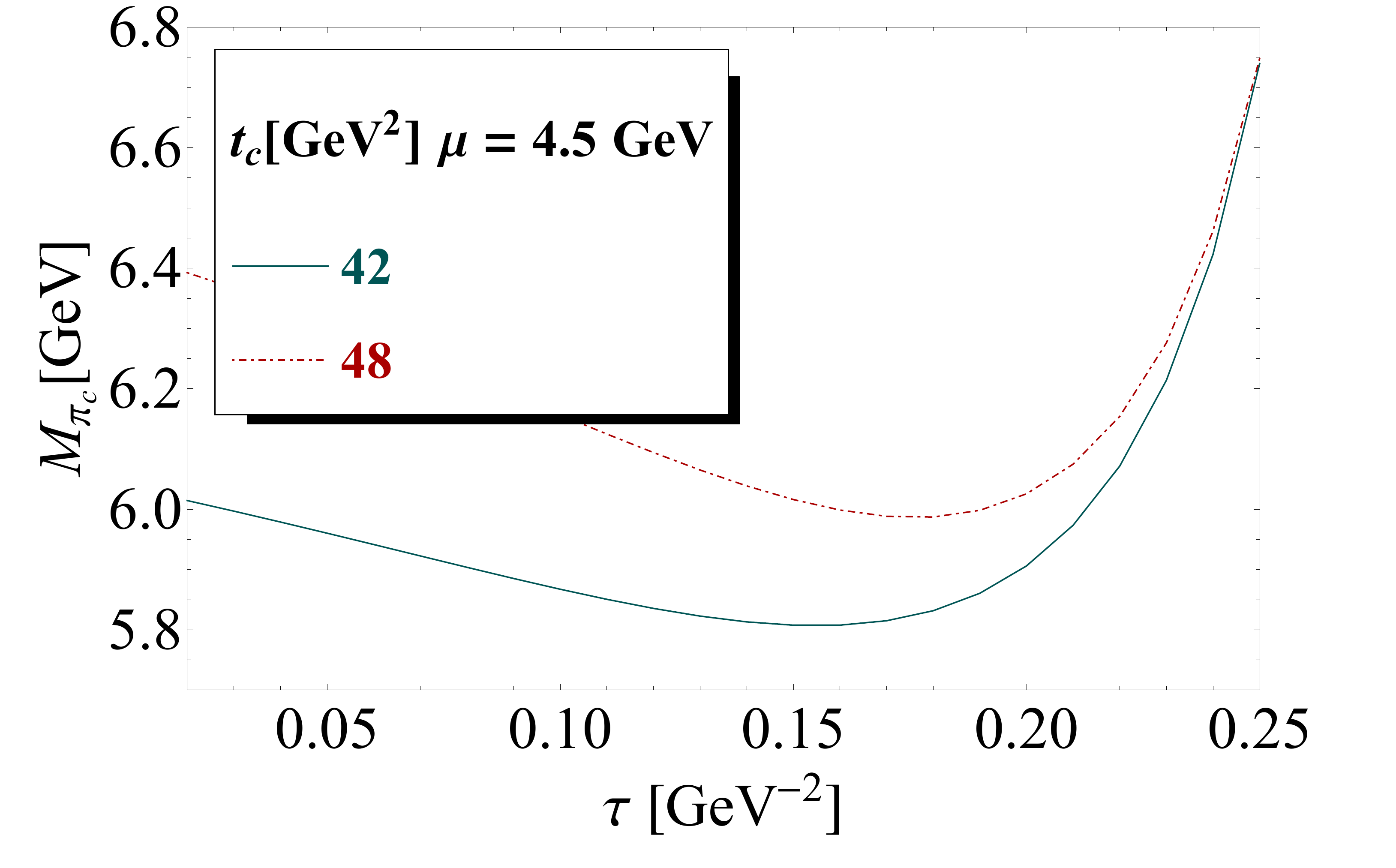}}
\centerline {\hspace*{-3cm} a)\hspace*{6cm} b) }
\caption{
\scriptsize 
{\bf a)} $f_{\pi_c}$ at N2LO  as function of $\tau$ for different values of $t_c$, for $\mu=4.5$ GeV  and for the QCD parameters in Tables\,\ref{tab:param} and \ref{tab:alfa}; {\bf b)} The same as a) but for the mass $M_{\pi_c}$.
}
\label{fig:pc-n2lo} 
\end{center}
\end{figure}

\begin{figure}[hbt] 
\begin{center}
{\includegraphics[width=6.29cm  ]{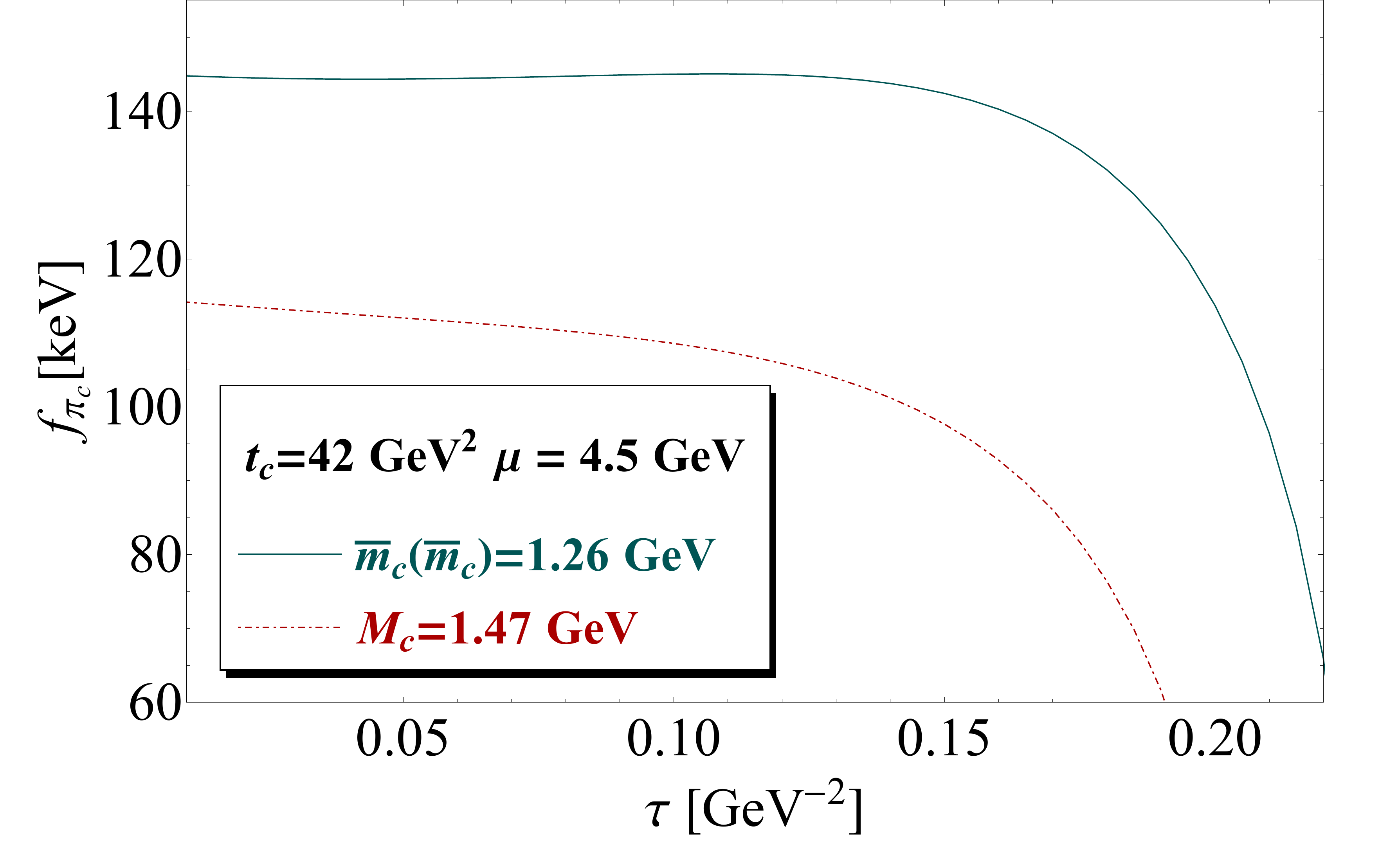}}
{\includegraphics[width=6.29cm  ]{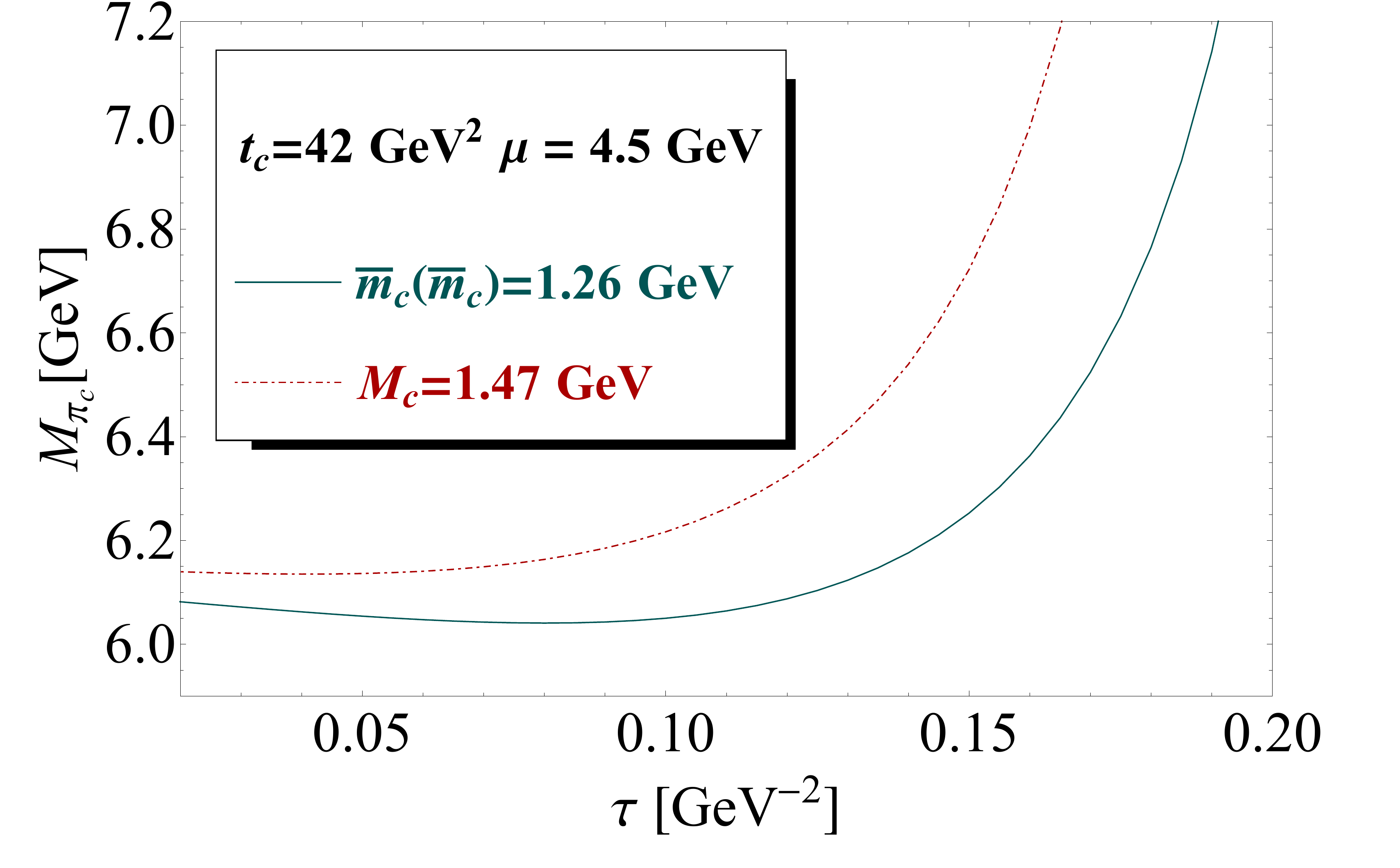}}
\centerline {\hspace*{-3cm} a)\hspace*{6cm} b) }
\caption{
\scriptsize 
{\bf a)} $f_{\pi_c}$  at LO as function of $\tau$ for  $t_c=42$ GeV$^2$, for $\mu=4.5$ GeV, for  values of the running $\overline{m}_c(\overline{m}_c)=1.26$ GeV and pole mass $M_c=1.47$ GeV. We use     
the QCD parameters in Tables\,\ref{tab:param} and \ref{tab:alfa}; {\bf b)} The same as a) but for the mass $M_{\pi_c}$.
}
\label{fig:pc-const} 
\end{center}
\end{figure}

\begin{figure}[hbt] 
\begin{center}
{\includegraphics[width=6.29cm  ]{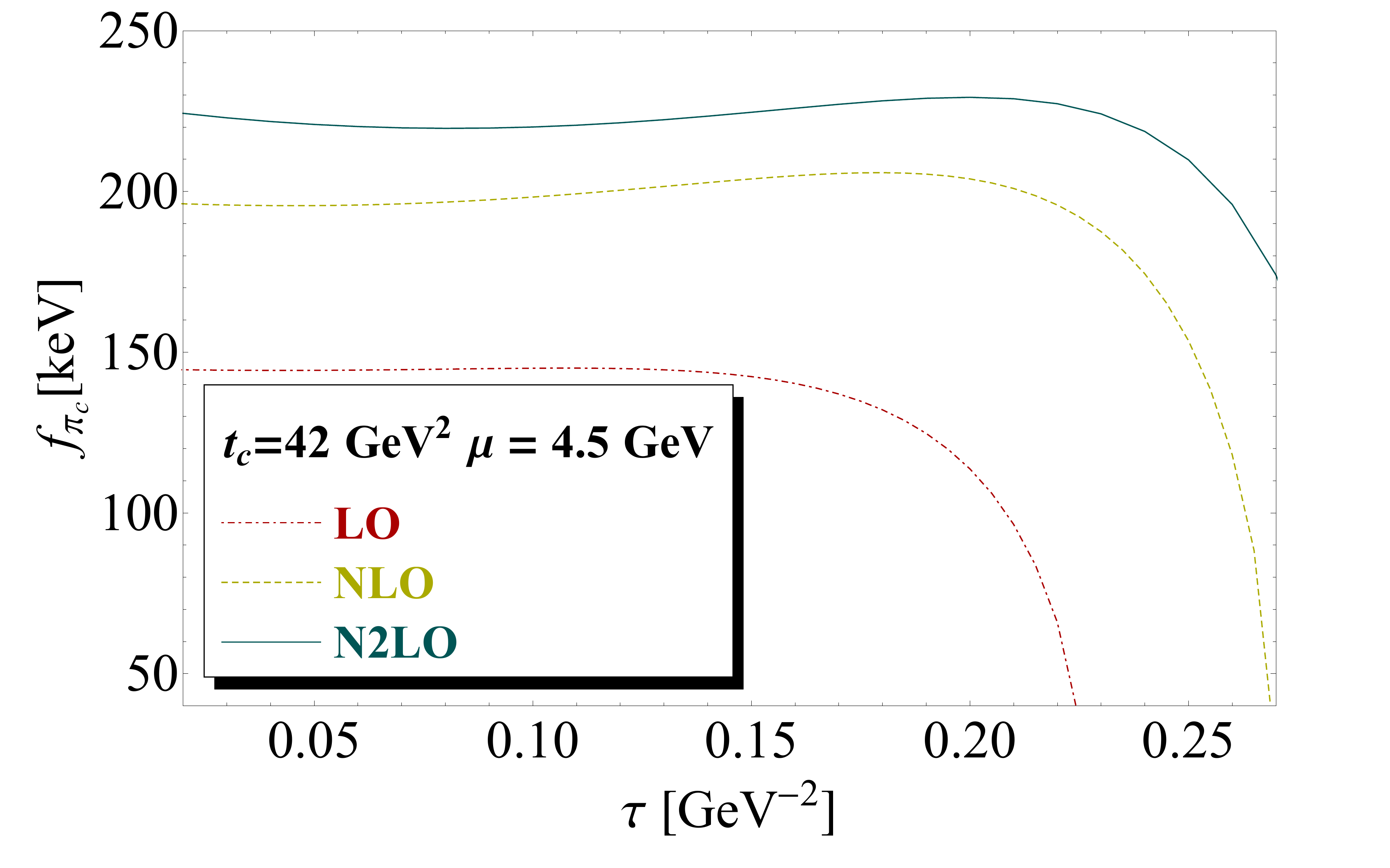}}
{\includegraphics[width=6.29cm  ]{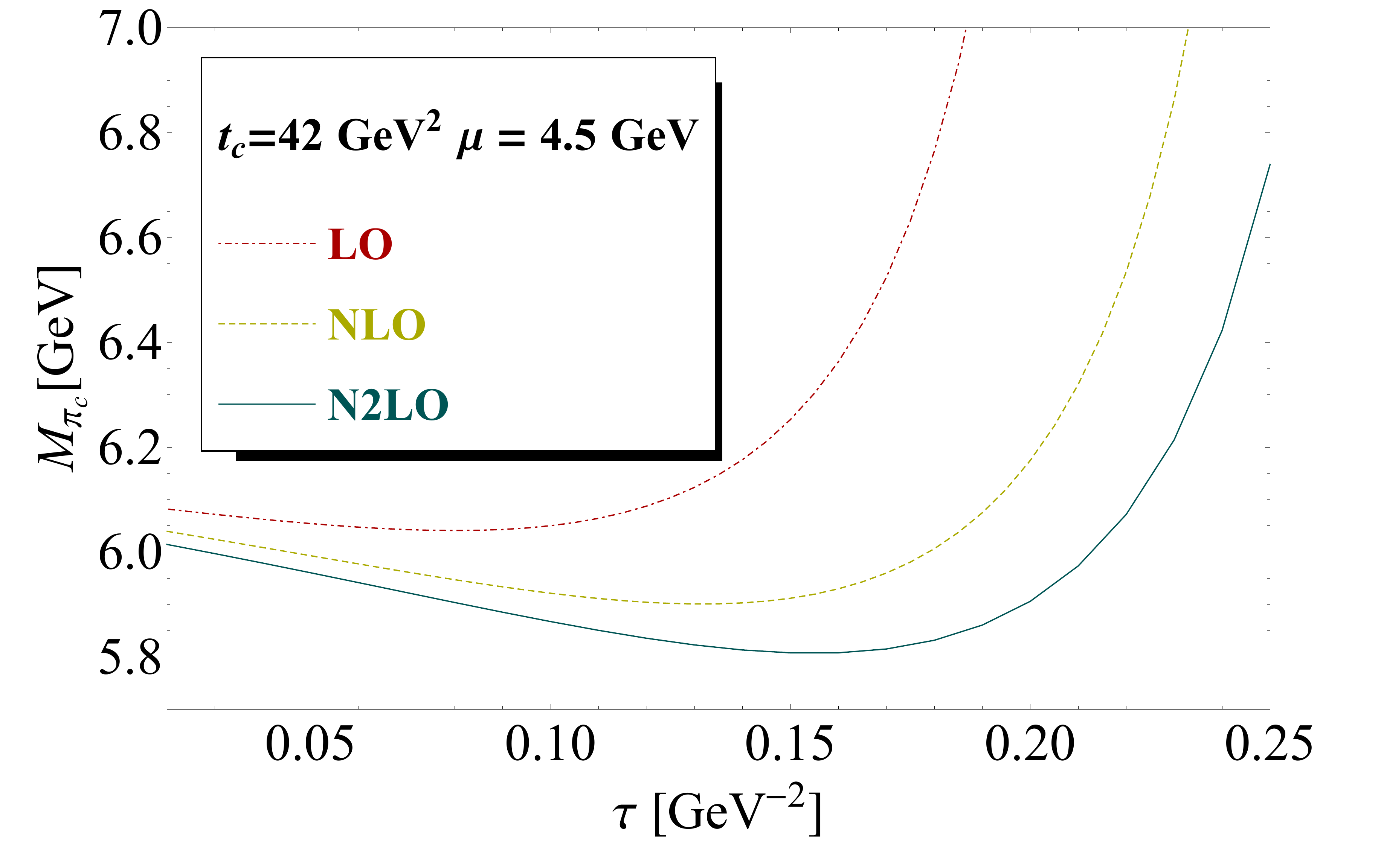}}
\centerline {\hspace*{-3cm} a)\hspace*{6cm} b) }
\caption{
\scriptsize 
{\bf a)} $f_{\pi_c}$  as function of $\tau$ for a given value of $t_c=42$ GeV$^2$, for $\mu=4.5$ GeV, for different truncation of the PT series  and for the QCD parameters in Tables\,\ref{tab:param} and \ref{tab:alfa}; {\bf b)} The same as a) but for the mass $M_{\pi_c}$.
}
\label{fig:pc-pt} 
\end{center}
\end{figure}

\begin{figure}[hbt] 
\begin{center}
{\includegraphics[width=6.2cm  ]{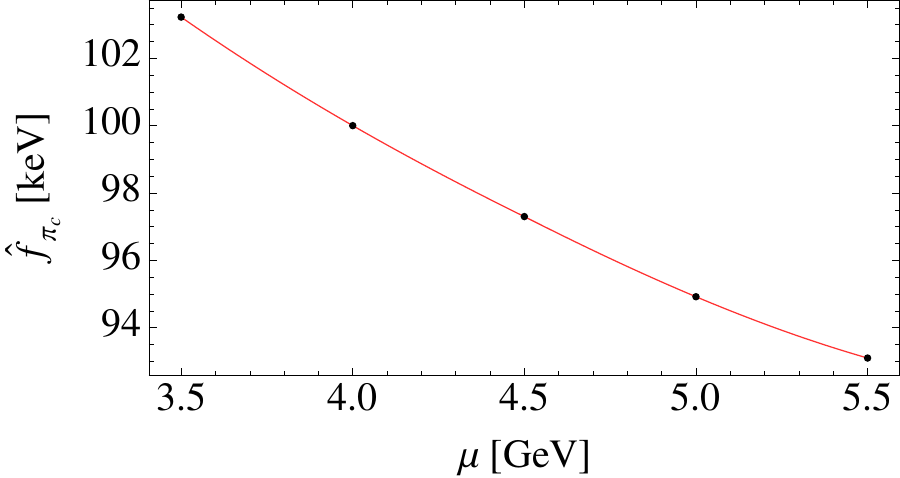}}
{\includegraphics[width=6.2cm  ]{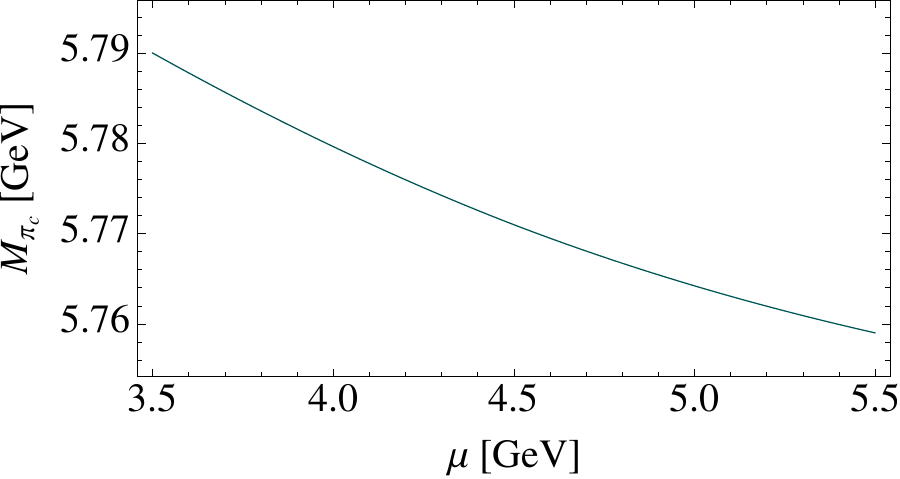}}
\centerline {\hspace*{-3cm} a)\hspace*{6cm} b) }
\caption{
\scriptsize 
{\bf a)} Renormalization group invariant coupling $\hat f_{\pi_c}$ at NLO as function of $\mu$, for the corresponding $\tau$-stability region, for $t_c\simeq 42$ GeV$^2$ and for the QCD parameters in Tables\,\ref{tab:param}  and \ref{tab:alfa};  {\bf b)} The same as a) but for the  mass $ M_{\pi_c}$.
}
\label{fig:pc-mu} 
\end{center}
\end{figure}

Like in the previous cases, we study the coupling and mass of the pseudoscalar $\pi_c(0^-)$ four-quark state which we show in Figs.\,\ref{fig:pc-lo} to \ref{fig:pc-mu}. The results are summarized in Tables\,\ref{tab:4q-errorc} and \,\ref{tab:4q-resultc}.  At N2LO, the corresponding set of parameters are:
\beq
\tau\simeq (0.15-0.22)~{\rm GeV}^{-2}, ~~~~t_c\simeq  (42-48)~{\rm GeV}^2~~~~ {\rm and}~~~ ~\mu\simeq  4.5~\rm{ GeV}, 
\eeq
for the $c$-quark channel.

\nin
\nin
\nin

\nin
\nin
\nin
\subsection{Coupling and mass of the $ \pi_b(0^{-})$ four-quark state}
We extend the analysis to the $b$-quark sector. The results are summarized in Tables\,\ref{tab:4q-errorb} and \,\ref{tab:4q-resultb}.
At N2LO, it corresponds to the set of parameters:
\beq
\tau\simeq (0.05-0.09)~{\rm GeV}^{-2}, ~~~~t_c\simeq  (180-220)~{\rm GeV}^2~~~~ {\rm and}~~~ ~\mu\simeq  6~\rm{ GeV}, 
\eeq
for the $b$-quark channel.

\subsection{Couplings and masses of the $ V_{c,b}(1^{-})$  four-quark state}
Like in the previous cases, we study the coupling and mass of the vector $V_c(1^-)$ four-quark state. The results are summarized in Tables\,\ref{tab:4q-errorc},\,\ref{tab:4q-errorb},\,\ref{tab:4q-resultc} and \,\ref{tab:4q-resultb}. At N2LO, it corresponds to the set of parameters:
\beq
\tau\simeq (0.15-0.20)~{\rm GeV}^{-2}, ~~~~t_c\simeq  (42-48)~{\rm GeV}^2~~~~ {\rm and}~~~ ~\mu\simeq  4.5~\rm{ GeV}, 
\eeq
for the $c$-quark channel and:
\beq
\tau\simeq (0.06-0.09)~{\rm GeV}^{-2}, ~~~~t_c\simeq  (170-200)~{\rm GeV}^2~~~~ {\rm and}~~~ ~\mu\simeq  5.5~\rm{ GeV}, 
\eeq
for the $b$-quark channel.

{\scriptsize
\begin{table}[hbt]
\setlength{\tabcolsep}{0.02pc}
\tbl{Different sources of errors for the estimate of the $0^{++}$ and $1^{++}$ $\bar DD$-like molecule masses (in units of MeV) and couplings $f_{MM}(\mu)$ (in units of keV). }
   {\scriptsize
 {\begin{tabular}{@{}llllllllllll@{}} 
&\\
\hline
\hline
\bf Inputs $[GeV]^d$&$\Delta M_{DD}$&$\Delta f_{DD}$&$\Delta M_{D^*D^*}$&$\Delta f_{D^*D^*}$
&$\Delta M_{D^*D}$&$\Delta f_{D^* D}$&$\Delta M_{D^*_0D_1}$&$\Delta f_{D^*_0D_1}$&$\Delta M_{D^*_0D^*_0}$&$\Delta f_{D^*_0D^*_0}$\\
\hline 
{\it LSR parameters}&\\
$(t_c,\tau)$&6 &7.9&17.4& 9.4& 3.4  &7.6& 12.2& 2.6&0.67 &3.5\\
$\mu=(4.5\pm 0.5)$&23&0.32&22.5&3.1&  26 &2.5&20.5&0.2&2.0&2.0\\
{\it QCD inputs}&\\
$\bar m_c$&10.48&4.43&7.44&7.12&10.28 &4.05&4.78&2.18& 7.78&2.30\\
$\alpha_s$&11.66&3.56&11.46&5.83&  11.74&3.36&16.32&1.36&14.10&1.24\\
$N3LO$&0.0&0.35&0.0&0.99&0.0&0.07&0.0&8.67&0.00&11.16&\\
$\la\bar qq\ra$&6.83&1.94&5.05&1.73& 7.89&1.63&12.5&3.8&6.63&4.69\\
$\la\alpha_s G^2\ra$&1.65&0.63&1.23&0.76&  0.21&0.04&0.63&0.1&0.05&1.74\\
$M_0^2$&5.64&0.16&4.08&2.36& 5.84&1.18&16.09&1.69&13.68&0.23\\
$\la\bar qq\ra^2$&3.0&10.4&2.0 &25& 9.5&10.5&23.5&11.4&6.3&16.5\\
$\la g^3G^3\ra$&0.07&0.1&0.11&0.22&0.03&0.07&0.04&0.13&0.51&0.16\\
$d\geq8$&19.0&2.1&176&37&52&9.0&158.5&5.6&47&7.5\\
{\it Total errors}&35.5&14.55&178.95&46.77& 61.82 &16.94&164.16&16.35&53.70&22.39\\
\hline\hline
\end{tabular}}
\label{tab:errorcplus}
}
\end{table}
} 
{\scriptsize
\begin{table}[hbt]
\setlength{\tabcolsep}{0.03pc}
\tbl{Different sources of errors for the estimate of the $0^{++}$ and $1^{++}$ $\bar BB$-like molecule masses (in units of MeV) and couplings $f_{MM}(\mu)$ (in units of keV). }
     {\scriptsize
 {\begin{tabular}{@{}llllllllllll@{}} 
&\\
\hline
\hline
\bf Inputs $[GeV]^d$&$\Delta M_{\bar BB}$&$\Delta f_{\bar BB}$&$\Delta M_{B^*B^*}$&$\Delta f_{B^*B^*}$
&$\Delta M_{B^*B}$&$\Delta f_{B^* B}$&$\Delta M_{B^*_0B_1}$&$\Delta f_{B^*_0B_1}$&$\Delta M_{B^*_0B^*_0}$&$\Delta f_{B^*_0B^*_0}$\\
\hline 
{\it LSR parameters}&\\
$(t_c,\tau)$&54 &1.6&102&5.3&44 &5.3&122&0.2&0.50& 0.16&\\
$\mu=(6.0\pm 0.5)$&5&2.01&7&0.01&  71 &0.3&43&0.04&3.0&0.5&\\
{\it QCD inputs}&\\
$\bar m_b$&2.08&0.08&2.10&0.15&2.85 &0.08&1.66&0.08&2.07&0.05&\\
$\alpha_s$&10.51&0.28&11.01&0.49& 12.68&0.31&15.16&0.25&16.30&0.19&\\
$N3LO$&0.0&0.12&0.0&0.22&0.02&0.14&0.01&1.13&0.00&1.11\\
$\la\bar qq\ra$&4.24&0.20&2.85&0.21& 8.43&0.12&4.05&0.41&3.55&0.23&\\
$\la\alpha_s G^2\ra$&0.50&0.02&0.37&0.02&  0.03&0.0&0.03&0.0&1.43&0.02&\\
$M_0^2$&1.0&1.1&37.0&0.14& 58&0.13&20.0&0.20&11.07&0.15&\\
$\la\bar qq\ra^2$&16.0&1.64&8.0&0.92& 22.9&1.15&9.5&1.88&1.95&1.58&\\
$\la g^3G^3\ra$&0.02&0.0&0.03&0.0&0.0&0.0&0.04&0.0&0.01&0.0&\\
$d\geq8$&1.0&1.17&37.0&0.07&107&0.07&0.05&0.01&111&3&\\
{\it Total errors}&57.72&3.46&115.71&5.41&  150.18&5.44&132.18&2.26&112.88&3.62\\
\hline\hline
\end{tabular}}
\label{tab:errorbplus}
}
\end{table}
} 
{\scriptsize
\begin{table}[hbt]
\setlength{\tabcolsep}{0.33pc}
\tbl{Different sources of errors for the estimate of the $0^{-\pm}$ and $1^{--}$ $\bar DD$-like
 molecule masses (in units of MeV) and couplings $f_{MM}(\mu)$ (in units of keV). The errors for the $1^{-+}$ $\bar D^*_0D^*$ and $\bar DD_1$ states are similar to the $1^{--}$ case except for the $\qq$ and $\la\alpha_s G^2\ra$ condensates where they are equal to zero in the latter.}  
    {\scriptsize
 {\begin{tabular}{@{}llllllllll@{}} 
&\\
\hline
\hline
\bf Inputs $[GeV]^d$&$\Delta M_{D^*_0 D}$&$\Delta f_{D^*_0 D}$&$\Delta M_{D^* D_1}$&$\Delta f_{D^* D_1}$
&$\Delta M_{D^*_0 D^*}$&$\Delta f_{D^*_0 D^*}$&$\Delta M_{DD_1}$&$\Delta f_{DD_1}$
\\
\hline 
{\it LSR parameters}&\\
$(t_c,\tau)$&115&15.86&88.6&25.3&83.41&11.44&150&18.70
\\
$\mu=(4.5\pm0.5)$&7.00&3.96&7.00&9.54&3.83&3.34&8.25&3.50
\\
{\it QCD inputs}&\\
$\bar m_c$& 14.62&5.19&15.15&10.98& 14.79&5.08&13.71&4.62
\\
$\alpha_s$&3.92&2.20&4.02&4.78& 4.77&2.19&4.58&2.09
\\
$N3LO$&2.88 &4.47 & 2.36&20.92 &2.40  &10.87 &2.14 &6.04 
\\
$\la\bar qq\ra$&0.00&0.00&0.00&0.00& 0.00&0.00&0.00&0.00
\\
$\la\alpha_s G^2\ra$&9.33&1.56&2.75&1.00& 1.42&0.18&0.00&0.00
\\
$M_0^2$&0.00&0.00&0.00&0.00& 0.00&0.00&9.54&1.14
\\
$\la\bar qq\ra^2$&68.93&4.85&72.09&9.99& 53.78&3.22&56.43&3.16
\\
$\la g^3G^3\ra$&0.85&0.11&0.95&0.24&0.81&0.10&0.75&0.10
\\
$d\geq8$&36.0&4.00&80.70&6.70&6.92&0.31&8.0&0.8
\\
{\it Total errors}&140.17&19.00&140.94&37.70&100.81&17.36&161.62&11.63 \\
\hline\hline
\end{tabular}}
\label{tab:errorc}
}
\end{table}
} 
{\scriptsize
\begin{table}[hbt]
\setlength{\tabcolsep}{0.33pc}
 \tbl{Different sources of errors for the estimate of the $0^{-\pm}$ and $1^{--}$ $\bar BB$-like  molecule masses (in units of MeV) and couplings $f_{MM}(\mu)$ (in units of keV). The errors for the $1^{-+}$ $\bar B^*_0B^*$ and $\bar BB_1$ are very similar to the ones of the $1^{--}$ states except for $\qq$ and $\la\alpha_s G^2\ra$ which are zero here. }  
    {\scriptsize
 {\begin{tabular}{@{}llllllllll@{}} 
&\\
\hline
\hline
\bf Inputs $[GeV]^d$&$\Delta M_{B^*_0 B}$&$\Delta f_{B^*_0 B}$&$\Delta M_{B^* B_1}$&$\Delta f_{B^* B_1}$
&$\Delta M_{B^*_0 B^*}$&$\Delta f_{B^*_0 B^*}$&$\Delta M_{B B_1}$&$\Delta f_{BB_1}$
\\
\hline 
{\it LSR parameters}&\\
$(t_c,\tau)$&254.25 &8.66&213.75&14.97&260.85 &9.17&249&10.57
\\
$\mu=(5.5\pm0.5)$&8.50&1.11&8.00&2.17&8.25&1.11&8.5&1.0
\\
{\it QCD inputs}&\\
$\bar m_b$&1.59&0.19&1.49&0.36&1.45 &0.18&1.60&0.19
\\
$\alpha_s$&3.35&0.52&3.22&1.00& 3.24&0.49&3.40&0.56
\\
$N3LO$&2.60&4.79&1.97 &6.98 & 1.96 &4.74 & 2.23&4.23
\\
$\la\bar qq\ra$&0.00&0.00&0.00&0.00&  0.00&0.00&0.00&0.00
\\
$\la\alpha_s G^2\ra$&1.47&0.055&0.63&0.051&  0.055&0.00&0.00&0.00
\\
$M_0^2$&0.00&0.00&0.00& 0.00&0.00&0.00&1.95&0.050
\\
$\la\bar qq\ra^2$&49.32&0.65&44.05&1.15& 39.12&0.50&49.32&0.63
\\
$\la g^3G^3\ra$&0.035&0.00&0.04&0.002&0.031&0.00&0.04&0.00
\\
$d\geq8$&22.00&0.79&53.40&8.60&32.0&1.77&39.0&2.0
\\
{\it Total errors}&260.11&10.04&224.86&18.81&265.86&10.56&257.0&11.64
\\
\hline\hline
\end{tabular}}
\label{tab:errorb}
}
\end{table}
} 
\vspace*{-0cm}
{\scriptsize
\begin{table}[hbt]
\setlength{\tabcolsep}{0.65pc}
 \tbl{Different sources of errors for the estimate of the four-quark $[cq\bar c\bar q]$ (pseudo)scalar $S_c(\pi_c)$ and (axial) vector $V_c(A_c)$ states,  masses (in units of MeV) and couplings $f(\mu)$ (in units of keV); $q\equiv u,d$.  }  
    {\scriptsize
 {\begin{tabular}{@{}llllllllll@{}} 
&\\
\hline
\hline
\bf Inputs $[GeV]^d$&$\Delta M_{S_c}$&$\Delta f_{S_c}$&$\Delta M_{A_c}$&$\Delta f_{A_c}$&$\Delta M_{\pi_c}$&$\Delta f_{\pi_c}$&$\Delta M_{V_c}$&$\Delta f_{V_c}$
\\
\hline 
{\it LSR parameters}&\\
$(t_c,\tau)$&0.2&9.3&0.23&9.31&101.3&5.66&90.01&14.00
\\
$\mu=(4.5\pm 0.5)$&0.94&8.12& 26.86&8.33&7.7&4.35&6.31&4.24
\\
{\it QCD inputs}&\\
$\bar m_c$&10.22&4.97&10.04&4.61& 15.46&6.25&14.59&5.69
\\
$\alpha_s$&11.75&4.05&11.73&3.85&4.01&2.60&3.34&2.28
\\
$N3LO$&0.00&0.41&0.38&0.72 &3.55&4.56&2.89&8.61
\\
$\la\bar qq\ra$&7.58&1.96&8.10&1.77&0.0&0.0&0.13&0.19
\\
$\la\alpha_s G^2\ra$&0.52&0.35&0.50&0.048&3.73&0.66&0.84&0.21
\\
$M_0^2$&6.27&1.39&6.12&1.91&0.0&0.0&6.66&0.84
\\
$\la\bar qq\ra^2$&1.4&14.62&93.94&14.87&73.37&6.09&71.97&5.50
\\
$\la g^3G^3\ra$&0.03&0.09&0.03&0.09&0.92&0.14&0.82&0.12
\\
$d\geq8$&5.60&0.07&83.0&22.0&10.0&1.0&37.0&3.0
\\
{\it Total errors}&54.25&20.34&129.53&30.08&126.83&12.50&122.35&19.13
\\
\hline\hline
\end{tabular}}
\label{tab:4q-errorc}
}
\end{table}
} 
{\scriptsize
\begin{table}[hbt]
\setlength{\tabcolsep}{0.65pc}
 \tbl{Different sources of errors for the estimate of the four-quark $[bq\bar b \bar q]$  (pseudo)scalar $S_b(\pi_b)$ and (axial) vector $V_b(A_b)$ states,  masses (in units of MeV) and couplings $f(\mu)$ (in units of keV); $q\equiv u,d$. }  
    {\scriptsize
 {\begin{tabular}{@{}llllllllll@{}} 
&\\
\hline
\hline
\bf Inputs $[GeV]^d$&$\Delta M_{S_b}$&$\Delta f_{S_b}$&$\Delta M_{A_b}$&$\Delta f_{A_b}$&$\Delta M_{\pi_b}$&$\Delta f_{\pi_b}$&$\Delta M_{V_b}$&$\Delta f_{V_b}$
\\
\hline
{\it LSR parameters}&\\
$(t_c,\tau)$&0.10&0.14&8.9&0.87&235&9&213.6&7.60
\\
$\mu=(5.5\pm 0.5)$&2.9&1.0&43.65 &0.99 &9.0&1.65&7.75&1.28
\\
{\it QCD inputs}&\\
$\bar m_b$&2.85&0.09& 2.59&0.091 & 2.01&0.29&1.47&0.21
\\
$\alpha_s$&13.20&0.35&12.30&0.35 &4.06&0.79&3.12&0.56
\\
$N3LO$&0.00&0.13&0.30&0.00 &4.54&3.88&1.10&3.94
\\
$\la\bar qq\ra$&7.39&0.15&7.33&0.16 &0.0&0.0&0.18&0.022
\\
$\la\alpha_s G^2\ra$&0.10&0.005&0.12& 0.0035&0.54&0.02&0.29&0.014
\\
$M_0^2$&7.02&0.16&7.46&0.17&0.0&0.0&1.23&0.036
\\
$\la\bar qq\ra^2$&1.2&1.47&112.82 &1.42 &69.01&1.24&43.77&0.66
\\
$\la g^3G^3\ra$&0.002&0.0& 0.0035&0.001&0.05&0.002&0.032&0.001
\\
$d\geq8$&108&3&120.4&2.3&68.0&3.35&105.0&1.5
\\
{\it Total errors}& 109.37&3.52 &171.68 &3.03 &254.43&10.60&242.16&8.83
\\
\hline\hline
\end{tabular}}
\label{tab:4q-errorb}
}
\end{table}
} 
{\scriptsize
\begin{table}[hbt]
\setlength{\tabcolsep}{0.22pc}
 \tbl{$\bar DD$-like molecules masses, invariant and running couplings  from LSR within stability criteria at LO to N2LO of PT. The errors are the quadratic sum of the ones in Tables\,\ref{tab:errorcplus} and\,\ref{tab:errorc}.}  
{\scriptsize{
\begin{tabular}{@{}lll   lll  lll  l cc@{}}
\hline
\hline
                \bf Nature& \multicolumn{3}{c}{$\bf{\hat f_X}$ \bf [keV]} 
                 & \multicolumn{3}{c }{$\bf{ f_X(4.5)}$ \bf [keV]} 
                 &  \multicolumn{3}{c}{\bf Mass  [MeV]} 
                  &\bf Threshold
                 & \bf Exp.
                 \\
\cline{2-4} \cline{5-7}\cline{8-10}
                 & \multicolumn{1}{l}{LO} 
                 & \multicolumn{1}{l}{NLO} 
                 & \multicolumn{1}{l }{\bf N2LO} 
                      & \multicolumn{1}{l}{LO} 
                 & \multicolumn{1}{l}{NLO} 
                 & \multicolumn{1}{l }{\bf N2LO} 
                   & \multicolumn{1}{l}{LO} 
                 & \multicolumn{1}{l}{NLO} 
                 & \multicolumn{1}{l}{\bf N2LO} 
                  \\
\hline
 $\bf {J^{PC}=0^{++}}$ &&&&&&&&&&&--\\
$\bar DD$&56&60&\bf 62(6)&155&164&\bf 170(15)&3901&3901&\bf 3898(36)&3739
\\
$\bar D^*D^*$&--&--&--&269&288&\bf 302(47)&3901&3903&\bf 3903(179)&4020&\\
$ D^*_0D^*_0$&27&42&\bf 50(8)&74&116&\bf 136(22)&4405&4402&\bf 4398(54)&4636\\
\\
$\bf {J^{PC}=1^{+\pm}} $ &&&&&&&&&&&$X_c,Z_c$\\
$\bar D^*D$&87&93&\bf 97(10)&146&154&\bf 161(17)&3901&3901&\bf 3903(62)&3880\\
$\bar D^*_0D_1$&48&71&\bf 83(10)&81&118&\bf 137(16)&4394&4395&\bf 4401(164)&4739\\
\\
$\bf  {J^{PC}=0^{-\pm}} $ &&&&&&&&&&&--\\
$\bar D^*_0D$&68&88&\bf 94(7)&190&240&\bf 257(19)&5956&5800&\bf 5690(140)&4188\\
$\bar D^*D_1$&--&--&--&382&490&\bf 564( 38)&6039&5898&\bf 5797(141)&4432&\\
\\
$\bf { J^{PC}=1^{--}} $ &&&&&&&&&&&$Y_c$\\
$\bar D^*_0D^*$&112&143&$\bf 157(10)$&186&238&\bf 261(17) &6020&5861&\bf 5748(101)&4328\\
$\bar DD_1$&98&126&\bf 139(13)&164&209&\bf 231(21)&5769&5639&\bf 5544(162)&4291\\
$\bf { J^{PC}=1^{-+}} $ &&&&&&&&&&&$Y_c$\\
$\bar D^*_0D^*$&105&135&\bf 150(13)& 174 & 224 &\bf  249(22) &6047 & 5920 & \bf 5828(132)&4328 \\
$\bar DD_1$&97&128&\bf 145(15)&162&213&\bf 241(25)&5973&5840&\bf 5748 (179) \\
\\
\hline
\hline
\end{tabular}
}}
\label{tab:resultc}
\end{table}
}
\vspace*{-0cm}
{\scriptsize
\begin{table}[hbt]
\setlength{\tabcolsep}{0.22pc}
 \tbl{$\bar BB$-like molecules masses, invariant and running couplings  from LSR within stability criteria from LO to N2LO of PT. The errors are the quadratic sum of the ones in Tables\,\ref{tab:errorbplus} and\,\ref{tab:errorb}. }
{\scriptsize{
\begin{tabular}{@{}lll   lll  lll  l cc@{}}
\hline
\hline
                 \bf Nature& \multicolumn{3}{c}{$\bf{\hat f_X}$ \bf [keV]} 
                 & \multicolumn{3}{c }{$\bf{ f_X(5.5)}$ \bf [keV]} 
                 &  \multicolumn{3}{c}{\bf Mass  [MeV]} 
                  &\bf Threshold
                 & \bf Exp.
                 \\
\cline{2-4} \cline{5-7}\cline{8-10}
                 & \multicolumn{1}{l}{LO} 
                 & \multicolumn{1}{l}{NLO} 
                 & \multicolumn{1}{l }{\bf N2LO} 
                      & \multicolumn{1}{l}{LO} 
                 & \multicolumn{1}{l}{NLO} 
                 & \multicolumn{1}{l }{\bf N2LO} 
                   & \multicolumn{1}{l}{LO} 
                 & \multicolumn{1}{l}{NLO} 
                 & \multicolumn{1}{l}{\bf N2LO} 
                  \\
\hline
 $\bf{ J^{PC}=0^{++}}$ &&&&&&&&&&&--\\
$\bar BB$&4.0&4.4&\bf 5(1)&14.4&15.6&\bf 17(4)&10605&10598&\bf 10595(58)&10559\\
$\bar B^*B^*$&--&--&--&27&30&\bf 32(5)&10626&10646&\bf 10647(184)&10650\\
$ B^*_0B^*_0$&2.1&3.2&\bf 4(1)&7.7&11.3&\bf 14(4)&10653&10649&\bf 10648(113)&--\\
\\
$\bf { J^{PC}=1^{+\pm} }$ &&&&&&&&&&&$X_b,Z_b$\\
$\bar B^*B$&7&8&\bf 9(3)&14&16&\bf 17(5)&10680&10673&\bf 10646(150)&10605\\
$\bar B^*_0B_1$&4&6&\bf 7(1)&8&11&\bf 14(2)&10670&10679&\bf 10692(132)&--\\
\\
$\bf { J^{PC}=0^{-\pm}} $ &&&&&&&&&&&--\\
$\bar B^*_0B$&11&16&\bf 20(3)&39&55&\bf 67(10)&12930&12737&\bf 12562(260) &--
\\
$\bar B^*B_1$&--&--&-- &71&105&\bf 136(19) &12967&12794&\bf 12627(225)&11046\\
\\
$\bf  {J^{PC}=1^{--}} $ &&&&&&&&&&& $Y_b$\\
$\bar B^*_0B^*$&21&29&\bf 35(6) &39&54&\bf 66(11) &12936&12756&\bf 12592(266) &--\\
$\bar BB_1$&21&29&\bf 35(7)&39&54&\bf 65(12)&12913&12734&\bf 12573(257)&11000\\
$\bf  {J^{PC}=1^{-+}} $ &&&&&&&&&&& $Y_b$\\
$\bar B^*_0B^*$&20&29&\bf 34(4)&38&54&\bf 64(8)&12942&12774&\bf 12617(220) &-- \\
$\bar BB_1$& 20&29&\bf 35(5)&	37&53&\bf 65(9) &12974& 12790&\bf 12630(236)					&11000\\

\hline
\hline
\end{tabular}
}}
\label{tab:resultb}
\end{table}
}
{\scriptsize
\begin{table}[hbt]
\setlength{\tabcolsep}{0.50pc}
\tbl{Four-quark masses, invariant and running couplings  from LSR within stability criteria from LO to N2LO of PT. The errors are the quadratic sum of the ones in Tables\,\ref{tab:errorcplus} and\,\ref{tab:errorc}.}  
{\scriptsize{
\begin{tabular}{@{}lll   lll  lll  l c@{}}
\hline
\hline
                \bf Nature& \multicolumn{3}{c}{\bf{$\hat f_X$} \bf [keV]} 
                 & \multicolumn{3}{c }{\bf{ $f_X(4.5)$} \bf [keV]} 
                 &  \multicolumn{3}{c}{\bf Mass  [MeV]} 
                 & Exp.
                 \\
\cline{2-4} \cline{5-7}\cline{8-10}
                 & \multicolumn{1}{l}{LO} 
                 & \multicolumn{1}{l}{NLO} 
                 & \multicolumn{1}{l }{\bf N2LO} 
                      & \multicolumn{1}{l}{LO} 
                 & \multicolumn{1}{l}{NLO} 
                 & \multicolumn{1}{l }{\bf N2LO} 
                   & \multicolumn{1}{l}{LO} 
                 & \multicolumn{1}{l}{NLO} 
                 & \multicolumn{1}{l}{\bf N2LO} 
                  \\
\hline
 \bf{$c$}\bf-quark \\
  $S_c(0^{+})$&62&67&\bf 70(7)&173&184&\bf 191(20)&3902&3901&\bf 3898(54)&--  \\
    $A_c(1^{+})$&100&106&\bf 112(18)&166&176&\bf 184(30)&3903&3890&\bf 3888(130)&$X_c,Z_c$\\
    $\pi_c(0^{-})$&84&106&\bf 113(5)&233&292&\bf 310(13)&6048&5872&\bf 5750(127)&--  \\
         $V_c(1^{-})$&123&162&\bf 178(11)&205&268&\bf 296(19)&6062&5904&\bf 5793(122)&$Y_c$\\
 \hline
\hline
\end{tabular}
}}
\label{tab:4q-resultc}
\end{table}
}

{\scriptsize
\begin{table}[hbt]
\setlength{\tabcolsep}{0.52pc}
\tbl{Four-quark masses, invariant and running couplings from LSR within stability criteria from LO to N2LO of PT. The errors are the quadratic sum of the ones in Tables\,\ref{tab:errorcplus} and\,\ref{tab:errorc}.}  
{\scriptsize{
\begin{tabular}{@{}lll   lll  lll  l c@{}}
\hline
\hline
                \bf Nature& \multicolumn{3}{c}{\bf{$\hat f_X$} \bf [keV]} 
                 & \multicolumn{3}{c }{\bf{ $f_X(5.5)$} \bf [keV]}  
                 &  \multicolumn{3}{c}{\bf Mass  [MeV]}                  
                 & Exp.
                 \\
\cline{2-4} \cline{5-7}\cline{8-10}
                 & \multicolumn{1}{l}{LO} 
                 & \multicolumn{1}{l}{NLO} 
                 & \multicolumn{1}{l }{\bf N2LO} 
                      & \multicolumn{1}{l}{LO} 
                 & \multicolumn{1}{l}{NLO} 
                 & \multicolumn{1}{l }{\bf N2LO} 
                   & \multicolumn{1}{l}{LO} 
                 & \multicolumn{1}{l}{NLO} 
                 & \multicolumn{1}{l}{\bf N2LO} 
                  \\
\hline
\bf {$b$}\bf-quark \\
   $S_b(0^{+})$&4.6&5.0&\bf 5.3(1.1)&16&17&\bf 19(4)&10652&10653&\bf 10654(109)&--  \\
    $A_b(1^{+})$&8.7&9.5&\bf 10(2)&16&18&\bf 19(3)&10730&10701&\bf 10680(172)&$Z_b$\\ 
 $\pi_b(0^{-})$&18&23&\bf 27(3)&62&83&\bf 94(11)&13186&12920&\bf 12695(254)&--  \\
 $V_b(1^{-})$&24&33&\bf 40(5)&45&62&\bf 75(9)&12951&12770&\bf 12610(242)&$Y_b$\\

\hline
\hline
\end{tabular}
}}
\label{tab:4q-resultb}
\end{table}
}
\section{Confrontation with the data and some LO results}\label{sec:confront}
\subsection{Axial-vector $(1^{++})$ states}
As mentioned in the introduction, there are several observed states in this channel.
 In addition to the well-established $X_c(3872)$, we have the $X_c(4147,4273)$ and the $Z_c(3900,4025,4050,4430)$.

 For the non-strange states found from their decays into $J/\psi \pi^+\pi^-$, one can conclude, from the results given in Tables\,\ref{tab:resultc} and \,\ref{tab:4q-resultc}, that the $X_c(3872)$ and $Z_c(3900)$
can be well described with an almost pure  $\bar D^*D$ molecule or/and four quark $[cq\bar c\bar q]$  states,  ($q\equiv u,d$) while the one of the $Z_c(4200,4430)$ might be a $\bar D^*_0D_1$ molecule state. Our results for the $X_c(3872)$ confirm our previous LO results in\,\cite{X1A,X1B,X2}. 

 Assuming that the value of $\sqrt{t_c}\approx (6-7)$ GeV, where the optimal values of the masses have been extracted, are approximately the mass of the 1st radial excitation, one can deduce that the higher masses experimental states cannot be such radial excitations. 

 In the bottom sector, experimental checks of our predictions given in Tables\,\ref{tab:resultb} and \,\ref{tab:4q-resultb}are required. 

One can notice that the values of these masses below the corresponding $\bar DD,\bar BB$-like thresholds are much lower than the ones predicted $\simeq 5.12$ (resp $11.32$) GeV for the $1^{++}$ $\bar cgc$ (resp. $\bar bgb$) hybrid mesons\,\cite{KLEIV1b,GOVAERTS,GOVAERTS1,GOVAERTS2,SNB2}. 
\subsection{Scalar $(0^{++})$ states}
Our analysis in Tables\,\ref{tab:resultc} and \ref{tab:4q-resultc} predicts that:

 The $0^{++}$ $\bar DD,\bar D^*D^*$ molecule  and four-quark non-strange states are almost degenerated with the $1^{++}$ ones and have masses around 3900 MeV. This prediction is comparable with the $Z_c(3900)$ quoted by PDG\,\cite{PDG} as a $0^{++}$ state. 

 The predicted mass of the $\bar D^*_{0}D^*_{0}$ molecule is higher [4402(30) MeV] but is still below the $\bar D^*_{0}D^*_{0}$ threshold. 

\subsection{Vector $(1^{-\pm})$ states}
Our predictions in Tables\,\ref{tab:resultc} to \ref{tab:4q-resultb} for molecules  and four-quark vector states in the range of (5646-5961) MeV are too high compared with the observed $Y_c(4140)$ to $Y_c(4660)$ states. Our N2LO results confirm previous LO ones in\,\cite{X3A,X3B,RAFMOLE1} but
do not support the result in \,\cite{MOLE12b} which are too low. 

Our results  indicate that the observed  states might result from a mixing of the molecule / four-quark  with ordinary quarkonia-states (if the description of these states in terms of molecules and/or four-quark states are the correct one). The NP contribution to this  kind of mixing has been estimated to leading order in\,\cite{RAPHAEL1}. The same conclusion holds for the $Y_b(9898,10260,10870)$ where the predicted unmixed molecule / four-quark states are in the range (12326-12829) MeV. 

As these pure molecule states are well above the physical threshold, they might not be bound states and could not be separated from backgrounds.  Our  results go in lines with the ones of\,\cite{ZAHED}. 
\subsection{Pseudoscalar $(0^{-\pm})$ states}
\nin

One expects from Tables\,\ref{tab:resultc} to \ref{tab:4q-resultb} that the $0^{-\pm}$ molecules will populate the region 5656-6020 (resp 12379-12827) MeV for the charm (resp bottom) channels like in the case of the $1^{-\pm}$ vector states. One can notice that these states are much heavier than the predicted
 $0^-$ hybrid $\bar cgc$ (resp. $\bar bgb$) ones 
$\simeq  3.82$ (resp. $\simeq 10.64$) GeV from QSSR\,\cite{KLEIV2b,GOVAERTS,GOVAERTS1,GOVAERTS2,SNB2}.  Like in the case of vector states, these pseudoscalar states are well above the physical threshold. Therefore, these molecule states should be broad and are difficult to separate from backgrounds. 

One can also notice that the $D^*_0D(0^{--})$ and $(0^{-+})$ states are almost degenerate despite the opposite signs of the $\qq$ and $\la \bar qGq\ra$ contributions to  the spectral functions in the two channels (see\,\ref{app:d0d}).
\subsection{Isospin breakings and almost degenerate states}
In our approach, isospin breakings are controlled by the running light quark mass  $\bar m_d-\bar m_u$ and condensate $\la\bar uu-\bar dd\ra$ differences which are tiny quantities. Their effects are hardly noticeable within the accuracy of our approach. Therefore, one expects that the molecules built from the neutral combination of currents which we have taken in Table\,\ref{tab:current} and from the corresponding charged currents will be degenerate in masses because their QCD expressions are the same in the chiral limit. 
\subsection{Radial excitations}
If one considers the value of the continuum threshold $t_c$, at which the optimal value of the ground state is obtained, 
as an approximate value of the mass of the 1st radial excitation, one expects that the radial excitations are in the region of about 0.4 to 1.6 GeV above the ground state mass. A more accurate prediction can be obtained by combining LSR with Finite Energy Sum Rule (FESR)\,\cite{X1A,X1B,X3A,X3B} where the mass-splitting is expected to be around 250-300 MeV at LO. Among these different observed states, the $Z_c(4430)$ and $X_c(4506,4704)$ could eventually be considered as radial excitation candidates. 
\section{Quark Mass Behaviour of the Decay Constants }\label{sec:coupling}
The couplings or decay constants given in Tables\,\ref{tab:resultc} to \ref{tab:4q-resultb} are normalized
in Eq.\,\ref{eq:coupling} in the same way as $f_\pi=130.4(2)$ MeV through its coupling to the pseudoscalar 
current :  $\la 0|(m_u+m_d)\bar u(i\gamma_5)d|\pi\ra=f_\pi m_\pi^2 \phi_\pi(x)$, where $\phi_\pi(x)$  is the pion field.

One can find from Table\,\ref{tab:resultc} that $f_{DD}\simeq 170(15)$ keV which is about $10^{-3}$ of 
$f_\pi$ and of $f_B\simeq f_D\simeq$ 206(7) MeV\,\cite{SNFB12a,SNFB12b,SNREV14,SNREV15,SNFB13,SNLIGHT,SNFB14}. The same observation holds for the other molecule and four-quark states indicating the weak coupling of these states to the associated interpolating currents. 

 Comparing the size of the couplings in the $c$ and $b$ quark channels in Tables\,\ref{tab:resultc} to \ref{tab:4q-resultb}, one can observe that the ratio decreases by a factor about 10 from the $c$  to the $b$ channels for the $0^{++}$ and $1^{++}$ states  which is about the value of the ratio $(\bar m_c/\bar m_b)^{3/2}$, while it decreases but about a factor 4  for the $0^{--}$ and $1^{--}$ states  which is about the value $(\bar m_c/\bar m_b)$.  These behaviours can be compared with the well-known one of $f_B\sim 1/\bar m_b^{1/2}$ from HQET and can motivate further theoretical studies of the molecule and four-quark couplings. 
\section{Summary and Conclusions}
  We have systematically revisited in this paper the LO estimate of the molecule and four-quark state masses and couplings using QCD Laplace sum rule (LSR) at N2LO of PT and including the non-perturbative (NP) contributions of condensates having dimension $d\leq 6$-8. 

  \b The different PT and NP QCD expressions at LO of the spectral functions corresponding to the interpolating currents given in Tables\,\ref{tab:current} and \,\ref{tab:4qcurrent} used in the analysis are given in integrated compact forms in the Appendices.  They are new and more suitable for a phenomenological analysis than the non-integrated forms given in the existing literature.
 
 \b Due to the technical difficulties for evaluating directly the PT $(\alpha_s)^n$ corrections, we have assumed the factorization of the four-quark spectral function into the convolution of two spectral functions built from bilinear currents. 
 We have tested the accuracy of this assumption in Section\,\ref{sec:factor} leading to the conclusions that it can provide an accurate determination of the hadron masses and decay constants. 
 We expect that, within this assumption, one can reproduce with a good accuracy the full radiative corrections. Indeed, it has been shown in\,\cite{BBAR2} that non-factorizable $\alpha_s$ corrections give small contribution of the order of 10\% of the full $\alpha_s$ one, while it is also known\,\cite{SNB1,SNB2} that radiative corrections partially cancel in the ratio of LSR used to extract the mass of the resonance (Eq.\,\ref{eq:ratioLSR}) within the minimal duality ansatz approximation (MDA) for parametrizing the spectral function. 
 
 \b Our results show that radiative corrections are relatively smaller for the masses than for the couplings which can explain the agreement of our results for the masses with the LO ones given in the literature. However, radiative corrections to some couplings are large which may invalidate some results on the hadronic widths from vertex sum rules where LO value of the decay constants have been used.  
  
\b  Our analysis has been done within stability criteria with respect to the LSR 
variable $\tau$, the QCD continuum threshold $t_c$ and the subtraction constant $\mu$ which have provided successful predictions in different hadronic channels (see e.g. \,\cite{SNB1,SNB2}\,\cite{SNFB12a,SNFB12b,SNREV14,SNREV15,SNFB13,SNLIGHT,SNFB14}). The optimal values of the masses and couplings have been extracted at the same value of these parameters where the stability appears as an extremum and/or inflexion points. The analysis is shown in details in different Sections for transparency such that the readers can appreciate and check explicitly the procedure used for extracting the results.  

\b We have also studied the effects of the choice of the value of the quark masses which definitions (running or pole) are ambiguous at LO. The effects are often large for the coupling as one can inspect in the different figures given in previous Sections. The additional error induced by this ad hoc choice is always bypassed by different authors. 

\b We have estimated the error due to higher order PT given in Tables\,\ref{tab:errorc} to \ref{tab:4q-errorb} by an estimate of the N3LO contribution based on a geometric growth of the numerical coefficients of the $a_s^n$ terms following the works in\,\cite{CNZ1,CNZ2,ZAK1,ZAK2,SNZ}. We can see in the estimate given for each truncation of the PT series from LO to N2LO given in Tables\,\ref{tab:resultc} to \ref{tab:4q-resultb} the good convergence of the PT series.

\b The error due to the high dimension condensates  comes from the $\la\bar qq\ra\la\bar qGq\ra $ condensates (part of the full $d=8$ condensate contributions) where we have assumed the same violation of factorization as that of the $\la\bar qq\ra^2$ dimension-six condensates.  One can deduce from Tables\,\ref{tab:errorc} to \ref{tab:4q-errorb} a good convergence of the OPE. Due to the inaccurate control of the size and contributions of high-dimension condensates, we refrain to include these contributions in our estimate but only consider them as a source of the errors. 

\b The results for the XYZ-like spectra are summarized in Tables\,\ref{tab:resultc} to \ref{tab:4q-resultb},
where one can observe that the N2LO predictions for the masses differ only slightly from the LO ones when
the value of the running mass is used for the latter. However, the size of the meson couplings is strongly affected by the radiative corrections  in some channels which consequently may modify the existing estimates of the meson hadronic widths based on vertex functions.
 
\b One can notice that the masses of the $J^P=1^+,0^+$ states are most of them below the corresponding $\bar DD,~\bar BB$-like thresholds and are compatible with some of the observed $XZ$ masses suggesting that these states can be interpreted as  almost pure molecules or/and four-quark states. 

 \b On the contrary, one also notes that the predictions for the $J^P=1^-,0^-$ states are about 1.5 GeV higher than the observed $Y_c$ mesons masses and (1.7-2.6) GeV higher than the observed $Y_b$ ones. Our results do not favour their interpretation as pure molecule or/and four-quark states. These theoretical predictions are far above the corresponding hadronic threshold which suggest that they might not be bound states and are difficult to separate from backgrounds, results in line with the ones of\,\cite{ZAHED}.
 
\b A confrontation of our results with the observed $XYZ$ states are done in details in Section\,\ref{sec:confront}.

\b Finally, we observe that, normalized to $f_\pi=130$ MeV, the $\bar DD,\bar BB$-like molecule and four-quark states couple weakly to the associated interpolating currents than ordinary $D,B$ mesons ($f_{DD}\approx 10^{-3}f_D$). Our numerical results also indicate that the corresponding decay constants  may behave as $1/ \bar m_b^{3/2}$ (resp. $1/ \bar m_b$) for the $1^{+},0^{+}$ (resp. $1^{-}, 0^{-}$)
states compared to the usual $1/ \bar m_b^{1/2}$ behaviour of $f_B$. These results can stimulate further theoretical studies of the molecule and four-quark state decay constants. 

\appendix 
\section{Molecule Spectral Functions in QCD}
They are defined from Eq.\,\ref{2po} as: $\frac{1}{\pi}{\rm Im}\Pi_{mol}^{(1)}(t)$ for spin 1 particles and $\frac{1}{\pi}{\rm Im}\psi_{mol}^{(s,p)}(t)$ from Eq.\,\ref{2po5} for spin 0 ones. 
In the following, we shall use the notations and definitions:
\begin{table}[hbt]
\setlength{\tabcolsep}{0.6pc}
\footnotesize{
{\begin{tabular}{lllll}
$Q$&$\equiv$&$ c,\, b~$,  &$x ={M_Q^{2}}/{t} $~,&$ v =\sqrt{1-4x}\,~, $  \\
$\lv$&=& $\ln{\frac{(1+v)}{(1-v)}}~,$&$\lp=\lid\left(\frac{1+v}{2}\right)-\lid\left(\frac{1-v}{2}\right).$ \\
\end{tabular}
}
}
\end{table}
{\subsection{$(0^{++})$ $\bar D D,~\bar BB$ Molecules }}
{\footnotesize\begin{eqnarray}
 pert&:& \frac{M_Q^8}{5 \cdot 2^{14} \:\pi^6} 
  \bigg[ v \Big( 480 + \frac{1460}{x} - \frac{274}{x^2} - \frac{38}{x^3} + \frac{1}{x^4} \Big) \nn\\
  && +120 {\cal L}_v \Big( 8x - 1 - 6\:\mbox{Log}(x) - \frac{8}{x} + \frac{2}{x^2} \Big)
  - 1440 {\cal L}_+ \bigg] \nn \\
 \qq&:& \frac{M_Q^5 \qq}{2^{7} \:\pi^4}
  \bigg[ v \Big( 6 - \frac{5}{x} - \frac{1}{x^2} \Big) 
  + 6{\cal L}_v \Big( 2x - 2 + \frac{1}{x} \Big) \bigg] \nn \\
\GG &:& -\frac{M_Q^4 \gGG}{3 \cdot 2^{11} \:\pi^6} 
 \bigg[ v \Big( 6 - \frac{5}{x} - \frac{1}{x^2} \Big) 
  + 6{\cal L}_v \Big( 2x - 2 + \frac{1}{x} \Big) \bigg] \nn\\
 \qGq&:& -\frac{3M_Q^3 \qGq}{ 2^{7} \:\pi^4} 
  \bigg[ v \Big( 1 - \frac{3}{x} \Big) + {\cal L}_v \Big( 2x + 1 + \frac{1}{x} \Big) \bigg] \nn\\
\qq^2 &:& \frac{M_Q^2 \:\rho \qq^2 \:v}{2^4 \:\pi^2}  \nn\\
  %
\GGG &:& \frac{M_Q^2 \gGGG}{ 3\cdot 2^{14} \:\pi^6} 
  \bigg[ v \Big( 6 - \frac{25}{x} + \frac{1}{x^2} \Big) + 
  6{\cal L}_v \Big( 2x + 2 + \frac{1}{x} \Big) \bigg] \nn\\
 {\qq\qGq} &:& -\frac{\qq \qGq}{2^5 \:\pi^2} \:v
  \bigg[ 1 - \frac{M_Q^2 \tau}{x} (1 - M_Q^2 \tau ) \bigg] 
\end{eqnarray} 
}
{ \subsection{$(0^{++})$ $\bar D^\ast D^\ast,~\bar B^\ast B^\ast$ Molecules }}
\vspace*{-0.cm}
{\footnotesize
\begin{eqnarray}
{pert} &:& \frac{M_Q^8}{5 \cdot 2^{12} \:\pi^6} 
  \bigg[ v \Big( 480 + \frac{1460}{x} - \frac{274}{x^2} - \frac{38}{x^3} + \frac{1}{x^4} \Big) \nn\\
  && +120 {\cal L}_v \Big( 8x - 1 - 6\:\mbox{Log}(x) - \frac{8}{x} + \frac{2}{x^2} \Big)
  - 1440 {\cal L}_+ \bigg] \nn 
  \eea
  \bea
    \qq &:& \frac{M_Q^5 \qq}{2^{6} \:\pi^4}
  \bigg[ v \Big( 6 - \frac{5}{x} - \frac{1}{x^2} \Big) 
  + 6{\cal L}_v \Big( 2x - 2 + \frac{1}{x} \Big) \bigg] \nn \\
 \GG &:& \frac{M_Q^4 \gGG}{3 \cdot 2^{10} \:\pi^6} 
 \bigg[ v \Big( 6 - \frac{5}{x} - \frac{1}{x^2} \Big) 
  + 6{\cal L}_v \Big( 2x - 2 + \frac{1}{x} \Big) \bigg] \nn\\ 
 {\qGq} &:& \frac{3M_Q^3 \qGq}{ 2^{7} \:\pi^4} 
  \bigg[ \frac{v}{x} - 2 {\cal L}_v \bigg] \nn\\
 {\qq^2} &:& \frac{M_Q^2 \:\rho \qq^2 \:v}{4 \:\pi^2}  \nn\\
  {\GGG} &:& \frac{M_Q^2 \gGGG}{ 3\cdot 2^{12} \:\pi^6} 
  \bigg[ v \Big( 6 - \frac{25}{x} + \frac{1}{x^2} \Big) + 
  6{\cal L}_v \Big( 2x + 2 + \frac{1}{x} \Big) \bigg] \nn\\
 {\qq\qGq}&:& -\frac{\qq \qGq \:v}{8\:\pi^2} \frac{M_Q^4 \tau^2}{x}
\end{eqnarray} 
}
{ \subsection{$(0^{++})$ $\bar D^\ast_0 D^\ast_0,~\bar B^\ast_0 B^\ast_0$ Molecules }}
{\footnotesize
\begin{eqnarray}
{pert}&:& \frac{M_Q^8}{5 \cdot 2^{14} \:\pi^6} 
  \bigg[ v \Big( 480 + \frac{1460}{x} - \frac{274}{x^2} - \frac{38}{x^3} + \frac{1}{x^4} \Big)\nnb\\
 && +120 {\cal L}_v \Big( 8x - 1 - 6\:\mbox{Log}(x) - \frac{8}{x} + \frac{2}{x^2} \Big)
  - 1440 {\cal L}_+ \bigg] \nn\\
 {\qq}&:& -\frac{M_Q^5 \qq}{2^{7} \:\pi^4}
  \bigg[ v \Big( 6 - \frac{5}{x} - \frac{1}{x^2} \Big) 
  + 6{\cal L}_v \Big( 2x - 2 + \frac{1}{x} \Big) \bigg] \nn \\
 {\GG}&:& -\frac{M_Q^4 \gGG}{3 \cdot 2^{11} \:\pi^6} 
 \bigg[ v \Big( 6 - \frac{5}{x} - \frac{1}{x^2} \Big) 
  + 6{\cal L}_v \Big( 2x - 2 + \frac{1}{x} \Big) \bigg]\nnb \\
 {\qGq}&:& \frac{3M_Q^3 \qGq}{ 2^{7} \:\pi^4} 
  \bigg[ v \Big( 1 - \frac{3}{x} \Big) + {\cal L}_v \Big( 2x + 1 + \frac{1}{x} \Big) \bigg]\nnb \\
{\qq^2}&:& \frac{M_Q^2 \:\rho \qq^2 \:v}{2^4 \:\pi^2}  \nn\\
{\GGG}&:& \frac{M_Q^2 \gGGG}{ 3\cdot 2^{14} \:\pi^6} 
  \bigg[ v \Big( 6 - \frac{25}{x} + \frac{1}{x^2} \Big) + 
  6{\cal L}_v \Big( 2x + 2 + \frac{1}{x} \Big) \bigg] \nnb\\
{\qq\qGq}&:& -\frac{\qq \qGq}{2^5 \:\pi^2} \:v
  \bigg[ 1 - \frac{M_Q^2 \tau}{x} (1 - M_Q^2 \tau ) \bigg] ,
\end{eqnarray} 
}
 \subsection{$(1^{+\pm})$ $\bar D^\ast D,~\bar B^\ast B$ Molecules }
 The QCD spectral functions of the $1^{++}$ and $1^{+-}$ states are the same.
{\footnotesize
\beqa
pert &:& \frac{M_Q^8}{5 \cdot 3 \cdot 2^{15} \:\pi^6} 
  \bigg[ v \Big( 840x + 140 + \frac{5248}{x} - \frac{1164}{x^2} - \frac{182}{x^3} + \frac{5}{x^4} \Big) \nn\\ 
  && +120 {\cal L}_v \Big( 14x^2 + 15 - 18\:\mbox{Log}(x) - \frac{32}{x} + \frac{9}{x^2} \Big)
  - 4320 {\cal L}_+ \bigg] \nn \\
\qq &:& \frac{M_Q^5 \qq}{3 \cdot 2^{10} \:\pi^4}
  \bigg[ v \Big( 60x + 82 - \frac{94}{x} - \frac{21}{x^2} \Big) 
  + 24{\cal L}_v \Big( 5x^2 + 6x - 9 + \frac{5}{x} \Big) \bigg] \nn \\
  %
  \GG &:& -\frac{M_Q^4 \gGG}{3^2 \cdot 2^{14} \:\pi^6} 
  \bigg[ v \Big( 60x - 62 + \frac{26}{x} + \frac{3}{x^2} \Big) + 
  24{\cal L}_v \Big( 5x^2 - 6x + 3 - \frac{1}{x} \Big) \bigg] \nn\\
 \qGq &:& -\frac{M_Q^3 \qGq}{ 3\cdot 2^{10} \:\pi^4} 
  \bigg[ v \Big( 66x + 11 - \frac{86}{x} \Big) + 6{\cal L}_v \Big( 22x^2 + 9 + \frac{4}{x} \Big) \bigg] \nn\\
 \qq^2 &:& \frac{M_Q^2 \:\rho \qq^2 v}{16 \pi^2} \nn\\
  \GGG &:& \frac{M_Q^2 \gGGG}{ 3^2\cdot 2^{16} \:\pi^6} 
  \bigg[ v \Big( 132x + 22 - \frac{190}{x} + \frac{9}{x^2} \Big) + 
  24{\cal L}_v \Big( 11x^2 + 3 + \frac{2}{x} \Big) \bigg] \nn\\
{\qq\qGq} &:& -\frac{\qq \qGq}{2^6 \:\pi^2} \:v 
  \bigg[ 1-\frac{M_Q^2 \tau}{x} \left( 1 - 2M_Q^2 \tau \right) \bigg] 
\enqa 
}
%
 \subsection{$(1^{+\pm})$ $\bar D^\ast_0 D_1,~\bar B^\ast_0 B_1$ Molecules }
The QCD spectral functions of the $1^{++}$ and $1^{+-}$ states are the same.
{\footnotesize
\begin{eqnarray}
 {pert} &:& \frac{M_Q^8}{5 \cdot 3 \cdot 2^{15} \:\pi^6} 
  \bigg[ v \Big( 840x + 140 + \frac{5248}{x} - \frac{1164}{x^2} - \frac{182}{x^3} + \frac{5}{x^4} \Big) \nn\\ 
  && +120 {\cal L}_v \Big( 14x^2 + 15 - 18\:\mbox{Log}(x) - \frac{32}{x} + \frac{9}{x^2} \Big)
  - 4320 {\cal L}_+ \bigg] \nn \\
\qq &:& -\frac{M_Q^5 \qq}{3 \cdot 2^{10} \:\pi^4}
  \bigg[ v \Big( 60x + 82 - \frac{94}{x} - \frac{21}{x^2} \Big) 
  + 24{\cal L}_v \Big( 5x^2 + 6x - 9 + \frac{5}{x} \Big) \bigg] \nn \\
  \GG&:& -\frac{M_Q^4 \gGG}{3^2 \cdot 2^{14} \:\pi^6} 
  \bigg[ v \Big( 60x - 62 + \frac{26}{x} + \frac{3}{x^2} \Big) + 
  24{\cal L}_v \Big( 5x^2 - 6x + 3 - \frac{1}{x} \Big) \bigg] \nn\\
  \qGq&:& \frac{M_Q^3 \qGq}{ 3\cdot 2^{10} \:\pi^4} 
  \bigg[ v \Big( 66x + 11 - \frac{86}{x} \Big) + 6{\cal L}_v \Big( 22x^2 + 9 + \frac{4}{x} \Big) \bigg] \nn\\
  %
\qq^2 &:& \frac{M_Q^2 \:\rho \qq^2 v}{16 \pi^2} \nn\\
\GGG&:& \frac{M_Q^2 \gGGG}{ 3^2\cdot 2^{16} \:\pi^6} 
  \bigg[ v \Big( 132x + 22 - \frac{190}{x} + \frac{9}{x^2} \Big) + 
  24{\cal L}_v \Big( 11x^2 + 3 + \frac{2}{x} \Big) \bigg] \nn\\
  \qq\qGq&:& -\frac{\qq \qGq}{2^6 \:\pi^2} \:v
  \bigg[ 1-\frac{M_Q^2 \tau}{x} \left( 1 - 2M_Q^2 \tau \right) \bigg] 
\end{eqnarray}
}
 \subsection{$(0^{-\pm})$ $\bar D^\ast_0 D,~\bar B^\ast_0 B$ Molecules }\label{app:d0d}
The QCD spectral functions of the $0^{--}$ and $0^{-+}$ states are the same.
{\footnotesize
\begin{eqnarray}
{pert} &:& \frac{M_Q^8}{5 \cdot 2^{14} \:\pi^6} 
  \bigg[ v \Big( 480 +  \frac{1460}{x} - \frac{274}{x^2} - \frac{38}{x^3} + \frac{1}{x^4} \Big)\nnb\\
&&  + 120 {\cal L}_v \Big( 8x - 1 - 6\:\mbox{Log}(x) - \frac{8}{x} + \frac{2}{x^2} \Big)
  - 1440 {\cal L}_+ \bigg]  \nnb\\ 
  {\qq} &:& 0  \nnb\\
  %
 {\GG}&:& -\frac{M_Q^4 \gGG}{3 \cdot 2^{11} \:\pi^6} 
  \bigg[ v \Big( 6 - \frac{5}{x} - \frac{1}{x^2} \Big) + 
  6{\cal L}_v \Big( 2x - 2 + \frac{1}{x} \Big) \bigg] \nnb\\
{\qGq} &:& 0  \nnb\\ 
{\qq^2}&:& -\frac{M_Q^2 \:\rho \qq^2 \:v}{16 \:\pi^2}  \nnb\\ 
{\GGG}&:& \frac{M_Q^2 \gGGG}{ 3\cdot 2^{14} \:\pi^6} 
  \bigg[ v \Big( 6 - \frac{25}{x} + \frac{1}{x^2} \Big) + 
  6{\cal L}_v \Big( 2x + 2 + \frac{1}{x} \Big) \bigg]  \nnb\\ 
{\qq\qGq}&:& \frac{\qq \qGq}{2^5 \:\pi^2} \:v
  \bigg[ 1 - \frac{M_Q^2 \tau}{x} (1 - M_Q^2 \tau ) \bigg]~.
\end{eqnarray}
}
 \subsection{$(0^{-\pm})$ $\bar D^\ast D_1,~\bar B^\ast B_1$ Molecules }
The QCD spectral functions of the $0^{--}$ and $0^{-+}$ states are the same.
{\footnotesize\begin{eqnarray}
{pert}&:&\frac{M_Q^{8}}{5\,2^{12}\pi ^6}\bigg[v\left(480+\frac{1460}{x}-\frac{274}{x^2}-\frac{38}{x^3}+\frac{1}{x^4}\right)\nnb\\
&&+120\lv \left(8x-1-6\ln(x)-\frac{8}{x}+\frac{2}{x^2}\right)-1440\lp \bigg],\nnb\\
{\qq}&:&0\, \nnb\\
{\GG} &:&\frac{M_Q^{4}}{3\,2^{10}\pi ^6}\gg \left[v\left(6-\frac{5}{x}-\frac{1}{x^2}\right)+6\lv \left(2x-2+\frac{1}{x}\right)\right], \nnb\\
{\mix}&:& 0\, \nnb \\
{\qq^2}&:&-\frac{M_Q^{2}}{4\pi ^2}\rho\,\qq^2\, v,\nnb \\
{\GGG}&:&\frac{M_Q^{2}}{3\, 2^{12}\pi ^6}  \ggg \left[v\left(6-\frac{25}{x}+\frac{1}{x^2}\right)+6\lv \left(2x+2+\frac{1}{x}\right)\right] ,\nnb \\
{\qq \mix}&:&\frac{1}{4\pi^2}\qq \mix \frac{x}{v}(x+M^2_Q\tau).\nnb\\
\end{eqnarray}
}
\subsection{$(1^{--})$ $\bar D^\ast_0 D^\ast,~\bar B^\ast_0 B^\ast$ Molecules }
The factorized expression corresponds to the value $\epsilon = 0$ and 
the full one to $\epsilon = 1$.
{\footnotesize
\begin{eqnarray}
 pert&:& \frac{M_Q^8}{5 \cdot 3 \cdot 2^{15} \:\pi^6} 
  \bigg[ v \Big( 840x + 140 + \frac{5248}{x} - \frac{1164}{x^2} - \frac{182}{x^3} + \frac{5}{x^4} \Big) \nn\\
  && +120 {\cal L}_v \Big( 14x^2 + 15 - 18\:\mbox{Log}(x) - \frac{32}{x} + \frac{9}{x^2} \Big)
  - 4320 {\cal L}_+ \bigg] \nn \\
  && - \frac{\epsilon \:M_Q^8}{5 \cdot 3 \cdot 2^{14} \:\pi^6} 
  \bigg[ v \Big( 420x - 1730 - \frac{4966}{x} - \frac{477}{x^2} - \frac{6}{x^3} \Big) \nn\\
  && + 60 {\cal L}_v \Big( 14x^2 - 60x - 12 + 12(3+1/x)\:\mbox{Log}(x) + \frac{56}{x} + \frac{3}{x^2} \Big)
  + 1440(3+1/x) {\cal L}_+ \bigg] \nn \\
\qq&:& -\frac{(1-\epsilon) M_Q^5 \qq}{3 \cdot 2^{10} \:\pi^4}
  \bigg[ v \Big( 60x - 62 + \frac{26}{x} + \frac{3}{x^2} \Big) 
  + 24{\cal L}_v \Big( 5x^2 - 6x + 3 - \frac{1}{x} \Big) \bigg] \nn \\
\GG &:& -\frac{(1-\epsilon)M_Q^4 \gGG}{3^2 \cdot 2^{14} \:\pi^6} 
  \bigg[ v \Big( 60x - 62 + \frac{26}{x} + \frac{3}{x^2} \Big) + 
  24{\cal L}_v \Big( 5x^2 - 6x + 3 - \frac{1}{x} \Big) \bigg] \nn\\
\qGq &:& \frac{M_Q^3 \qGq}{ 3\cdot 2^{10} \:\pi^4} 
  \bigg[ v \Big( 66x + 11 - \frac{50}{x} \Big) + 6{\cal L}_v \Big( 22x^2 - 3 + \frac{4}{x} \Big) \bigg] \nn\\
  &&  - \frac{3 \epsilon \:M_Q^3 \qGq}{ 2^{10} \:\pi^4} 
  \bigg[ 3v (2x - 1) + 2{\cal L}_v (6x^2 - 4x + 1) \bigg] \nn\\
 \qq^2 &:& -\frac{M_Q^2 \:\rho \qq^2}{3 \cdot 2^6 \:\pi^2} \:v
  \Big( 12 + \epsilon(4 - 1/x) \Big)  \nn
         \eea
  \bea
 \GGG&:& \frac{M_Q^2 \gGGG}{ 3^2\cdot 2^{16} \:\pi^6} 
  \bigg[ v \Big( 132x + 22 - \frac{190}{x} + \frac{9}{x^2} \Big) + 
  24{\cal L}_v \Big( 11x^2 + 3 + \frac{2}{x} \Big) \bigg] \nn\\
  && - \frac{\epsilon \:M_Q^2 \gGGG}{ 3^2\cdot 2^{16} \:\pi^6} 
  \bigg[ v \Big( 204x - 182 + \frac{2}{x} + \frac{3}{x^2} \Big) + 
  24{\cal L}_v \Big( 17x^2 - 18x + 9 - \frac{1}{x} \Big) \bigg] \nn\\
{\qq\qGq} &:& \frac{\qq \qGq}{2^6 \:\pi^2} \:v
  \bigg[ 1 - \frac{M_Q^2 \tau}{x} \Big( 1 - 2M_Q^2 \tau \Big) - \epsilon (x + M_Q^2 \tau) \bigg] 
\end{eqnarray} \label{eq:app8}
}
\subsection{$(1^{-+})$ $\bar D^\ast_0 D^\ast,~\bar B^\ast_0 B^\ast$ Molecules }
{\footnotesize 
\bea
{pert}&:& \frac{M_Q^8}{5 \cdot 3 \cdot 2^{15} \:\pi^6} 
  \bigg[ v \Big( 1680x - 3320 - \frac{4684}{x} - \frac{2118}{x^2} - \frac{194}{x^3} + \frac{5}{x^4} \Big) \nnb\\
  && + 120 {\cal L}_v \bigg( 28x^2 - 60x + 3 + 6\Big(3+\frac{2}{x}\Big)\:\mbox{Log}(x) + 
  \frac{24}{x} + \frac{12}{x^2} \bigg) + 1440 \Big(3+\frac{2}{x} \Big) {\cal L}_+ \bigg]  \nnb\\
{\qq}&:& -\frac{M_Q^5 \qq}{3 \cdot 2^{9} \:\pi^4}
  \bigg[ v \Big( 60x - 62 + \frac{26}{x} + \frac{3}{x^2} \Big) 
  + 24{\cal L}_v \Big( 5x^2 - 6x + 3 - \frac{1}{x} \Big) \bigg] \nnb \\
{\GG}&:& -\frac{M_Q^4 \gGG}{3^2 \cdot 2^{13} \:\pi^6} 
  \bigg[ v \Big( 60x - 62 + \frac{26}{x} + \frac{3}{x^2} \Big) + 
  24{\cal L}_v \Big( 5x^2 - 6x + 3 - \frac{1}{x} \Big) \bigg] \nnb\\
{\qGq}&:& \frac{M_Q^3 \qGq}{ 3\cdot 2^{9} \:\pi^4} 
  \bigg[ v \Big( 60x - 8 - \frac{25}{x} \Big) + 12{\cal L}_v \Big( 10x^2 - 3x + \frac{1}{x} \Big) \bigg] \nnb
      \eea
  \bea
{\qq^2}&:& -\frac{M_Q^2 \:\rho \qq^2}{3 \cdot 2^6 \:\pi^2} \:v
  \Big( 8 + \frac{1}{x} \Big)  \nnb\\
{\GGG}&:& \frac{M_Q^2 \gGGG}{ 3^2 \cdot 2^{14} \:\pi^6} 
  \bigg[ v \Big( 84x - 40 - \frac{47}{x} + \frac{3}{x^2} \Big) + 
  6{\cal L}_v \Big( 28x^2 - 18x + 12 + \frac{1}{x} \Big) \bigg] \nnb\\
{\qq\qGq}&:& \frac{\qq \qGq}{2^6 \:\pi^2} \:v
  \bigg[ 1 + x - \frac{M_Q^2 \tau}{x} \Big( 1 - x - 2M_Q^2 \tau \Big) \bigg] ~.
  \eea
  }
Note that one also can obtain this  expression 
of the ($1^{-+}$) $D^\ast_0 D^\ast$ molecule by the choice $\epsilon = -1$ in Eq.\, A.8 for ($1^{--}$) $D^\ast_0 D^\ast$ molecule.
\subsection{($1^{--}$) $\bar D D_1,\bar BB_1$ Molecules }\label{app:dd1}
\footnotesize{
\begin{eqnarray}
{pert} &:& \frac{M_Q^8}{3 \cdot 2^{15} \:\pi^6} 
  \bigg[ v \Big( 720 + \frac{3036}{x} - \frac{42}{x^2} - \frac{34}{x^3} + \frac{1}{x^4} \Big) \nn\\
  && +24 {\cal L}_v \bigg( 60x + 27 - 6\Big(9+\frac{2}{x}\Big)\:\mbox{Log}(x) - \frac{88}{x} + \frac{6}{x^2} \bigg)
  - 288 \Big(9+\frac{2}{x} \Big) {\cal L}_+ \bigg] \nn \\
{\qq} &:& 0 \nn \\
 {\GG} &:& 0 \nn\\
  {\qGq} &:& -\frac{M_Q^3 \qGq}{ 3\cdot 2^{9} \:\pi^4} 
  \bigg[ v \Big( 6x + 19 - \frac{25}{x} \Big) + 6{\cal L}_v \Big( 2x^2 + 6x - 3 + \frac{2}{x} \Big) \bigg] \nn\\
 {\qq^2}&:& -\frac{M_Q^2 \:\rho \qq^2}{3 \cdot 2^6 \:\pi^2} \:v
  \Big( 16 - \frac{1}{x} \Big)  \nn\\
 {\GGG}&:& -\frac{M_Q^2 \gGGG}{ 3\cdot 2^{15} \:\pi^6} 
  \bigg[ v \Big( 12x - 34 + \frac{32}{x} - \frac{1}{x^2} \Big) + 
  12{\cal L}_v \Big( 2x^2 - 6x + 2 - \frac{1}{x} \Big) \bigg] \nn\\
{\qq\qGq}&:& \frac{\qq \qGq}{2^6 \:\pi^2} \:v
  \bigg[ 1 - x - M_Q^2 \tau \Big( 1 + \frac{1}{x} - \frac{2M_Q^2 \tau}{x} \Big) \bigg]~. 
  \end{eqnarray} 
}
{\normalsize \subsection{($1^{-+}$) $\bar D D_1,\bar BB_1$ Molecules }\label{app:dd12}}
\footnotesize{
\bea
 {pert}&:& \frac{M_Q^8}{5 \cdot 3 \cdot 2^{15} \:\pi^6} 
  \bigg[ v \Big( 1680x - 3320 - \frac{4684}{x} - \frac{2118}{x^2} - \frac{194}{x^3} + \frac{5}{x^4} \Big) \\
  && + 120 {\cal L}_v \bigg( 28x^2 - 60x + 3 + 6\Big(3+\frac{2}{x}\Big)\:\mbox{Log}(x) + 
  \frac{24}{x} + \frac{12}{x^2} \bigg) + 1440 \Big(3+\frac{2}{x} \Big) {\cal L}_+ \bigg]  \nnb\\
 {\qq}&:& \frac{M_Q^5 \qq}{3 \cdot 2^{9} \:\pi^4}
  \bigg[ v \Big( 60x - 62 + \frac{26}{x} + \frac{3}{x^2} \Big) 
  + 24{\cal L}_v \Big( 5x^2 - 6x + 3 - \frac{1}{x} \Big) \bigg] \nnb \\
 {\GG}&:& -\frac{M_Q^4 \gGG}{3^2 \cdot 2^{13} \:\pi^6} 
  \bigg[ v \Big( 60x - 62 + \frac{26}{x} + \frac{3}{x^2} \Big) + 
  24{\cal L}_v \Big( 5x^2 - 6x + 3 - \frac{1}{x} \Big) \bigg]\nnb \\
{\qGq}&:& -\frac{M_Q^3 \qGq}{ 3\cdot 2^{9} \:\pi^4} 
  \bigg[ v \Big( 60x - 8 - \frac{25}{x} \Big) + 12{\cal L}_v \Big( 10x^2 - 3x + \frac{1}{x} \Big) \bigg]\nnb \\
 {\qq^2}&:& -\frac{M_Q^2 \:\rho \qq^2}{3 \cdot 2^6 \:\pi^2} \:v
  \Big( 8 + \frac{1}{x} \Big) \nnb \\
 {\GGG}&:& \frac{M_Q^2 \gGGG}{ 3^2 \cdot 2^{14} \:\pi^6} 
  \bigg[ v \Big( 84x - 40 - \frac{47}{x} + \frac{3}{x^2} \Big) + 
  6{\cal L}_v \Big( 28x^2 - 18x + 12 + \frac{1}{x} \Big) \bigg]\nnb \\
 {\qq\qGq}&:& \frac{\qq \qGq}{2^6 \:\pi^2} \:v
  \bigg[ 1 + x - \frac{M_Q^2 \tau}{x} \Big( 1 - x - 2M_Q^2 \tau \Big) \bigg] ~.
\eea
}.
\normalsize\section{Four-Quark Spectral Functions }
\label{app:4q}
They are defined from Eq.\,\ref{2po} as: $\frac{1}{\pi}{\rm Im}\Pi_{mol}^{(1)}(t)$ for spin 1 particles and $\frac{1}{\pi}{\rm Im}\psi_{mol}^{(s,p)}(t)$ from Eq.\,\ref{2po5} for spin 0 ones. 
\subsection{$(0^{+})$ Scalar State}
{\footnotesize
\begin{eqnarray}
{pert} &:& \frac{(1+k^2) M_c^8}{5\cdot3\cdot 2^{12} \:\pi^6} \bigg[ 
	v \bigg( 480 + \frac{1460}{x} - \frac{274}{x^2} - \frac{38}{x^3} + \frac{1}{x^4} \bigg)\nnb\\
	&&+ 120{\cal L}_v \bigg( 8x - 1 - 6 \log x - \frac{8}{x} + \frac{2}{x^2} \bigg)  - 1440 \:{\cal L}_+
	\bigg] \nnb\\
	\hspace{1cm}
%
{\qq}&:& \frac{(1-k^2) M_c^5 \qq}{3 \cdot 2^5 \:\pi^4} \bigg[ 
	v \bigg( 6 - \frac{5}{x} - \frac{1}{x^2} \bigg)
	+ 6{\cal L}_v \bigg( 2x - 2 + \frac{1}{x} \bigg)
	\bigg]\nnb\\
%
{\GG} &:& -\frac{(1+k^2) M_c^4 \gGG}{ 3^2 \cdot 2^{11} \:\pi ^6} \bigg[ 
	v \bigg( 6 - \frac{5}{x} - \frac{1}{x^2} \bigg)
	+ 6{\cal L}_v \bigg( 2x - 2 + \frac{1}{x} \bigg)
	\bigg]\nnb\\
%
{\qGq} &:& -\frac{(1-k^2) M_c^3 \qGq}{2^7 \:\pi^4} \bigg[
	v \bigg( 2 - \frac{7}{x} \bigg)
	+ 2{\cal L}_v \bigg( 2x + 2 + \frac{1}{x} \bigg)
	\bigg]\nnb
	\eea
	\bea
%
{\qq^2} &:& \frac{(1+k^2) M_c^2 \rho\qq^2 \:v}{12 \:\pi^2}\nnb\\
%
{\GGG} &:& \frac{(1+k^2) M_c^2 \gGGG}{3^2 \cdot 2^{12} \:\pi^6} \bigg[
	v \bigg( 6 - \frac{25}{x} + \frac{1}{x^2} \bigg)
	+ 6{\cal L}_v \bigg( 2x + 2 + \frac{1}{x} \bigg)
	\bigg] \nnb\\
%
{\qq  \qGq} &:& -\frac{(1+k^2) \qq \qGq}{3 \cdot 2^{4} \:\pi^2} \:v \bigg[
	1 - \frac{M_c^2 \:\tau}{x} \Big( 1 - 2M_c^2 \:\tau \Big)	\bigg]~,
\end{eqnarray}}
where we use the same definitions as in the case of the molecule states; $k$ is the mixing of interpolating currents where $k=0$\,\cite{X3A,X3B} is its optimal
value.
{\subsection{$(1^{+})$ Axial-vector state}
{\footnotesize
\begin{eqnarray}
{pert} &:& \frac{(1+k^2) M_Q^8}{5\cdot3^2\cdot 2^{13} \:\pi^6} \bigg[ 
	v \bigg( 840x + 140 + \frac{5248}{x} - \frac{1164}{x^2} - \frac{182}{x^3} + \frac{5}{x^4} \bigg)\nnb\\
	&&+ 120{\cal L}_v \bigg( 14x^2 + 15 - 18 \log x - \frac{32}{x} + \frac{9}{x^2} \bigg)  - 4320 \:{\cal L}_+
	\bigg] \hspace{1cm}\nnb\\
%
\qq &:& \frac{(1-k^2) M_Q^5 \qq}{3^2 \cdot 2^8 \:\pi^4} \bigg[ 
	v \bigg( 60x + 82 - \frac{94}{x} - \frac{21}{x^2} \bigg)
	+ 24{\cal L}_v \bigg( 5x^2 + 6x - 9 + \frac{5}{x} \bigg)
	\bigg]\nnb\\
%
{\GG} &:& \frac{(1+k^2) M_Q^4 \gGG}{ 3^2 \cdot 2^{11} \:\pi ^6} \bigg[ 
	v \bigg( 6 - \frac{5}{x} - \frac{1}{x^2} \bigg)
	+ 6{\cal L}_v \bigg( 2x - 2 + \frac{1}{x} \bigg)
	\bigg]\nnb\\
%
{\qGq} &:& -\frac{(1-k^2) M_Q^3 \qGq}{3^2 \cdot 2^8 \:\pi^4} \bigg[
	v \bigg( 42x + 7 - \frac{58}{x} \bigg)
	+ 6{\cal L}_v \bigg( 14x^2 + 9 + \frac{2}{x} \bigg)
	\bigg]\nnb\\
%
{\qq^2} &:& \frac{(1+k^2) M_Q^2 \rho\qq^2 \:v}{12 \:\pi^2}\nnb\\
%
{\GGG} &:& \frac{(1+k^2) M_Q^2 \gGGG}{3^3 \: 2^{14} \:\pi^6} \bigg[
	v \bigg( 132x + 22 - \frac{190}{x} + \frac{9}{x^2} \bigg)
	+ 24{\cal L}_v \bigg( 11x^2 + 3 + \frac{2}{x} \bigg)
	\bigg]~\nnb\\
\qq \qGq&:& -\frac{(1+k^2) \qq \qGq}{3 \cdot 2^{5} \:\pi^2} \:v \bigg[
	1 - \frac{M_c^2 \:\tau}{x} \Big( 1 - 4M_c^2 \:\tau \Big)	\bigg]~,
\end{eqnarray}}
%
\subsection{ $(0^{-})$ Pseudoscalar  state}
%
{\footnotesize
\begin{eqnarray}
{pert} &:& \frac{(1+k^2) M_c^8}{5\cdot3\cdot 2^{12} \:\pi^6} \bigg[ 
	v \bigg( 480 + \frac{1460}{x} - \frac{274}{x^2} - \frac{38}{x^3} + \frac{1}{x^4} \bigg)\nnb\\
	&&+ 120{\cal L}_v \bigg( 8x - 1 - 6 \log x - \frac{8}{x} + \frac{2}{x^2} \bigg)  - 1440 \:{\cal L}_+
	\bigg] \hspace{1cm}\nnb\\
%
{\qq} &:& 0\nnb\\
%
{\GG} &:& -\frac{(1+k^2) M_c^4 \gGG}{ 3^2 \cdot 2^{11} \:\pi ^6} \bigg[ 
	v \bigg( 6 - \frac{5}{x} - \frac{1}{x^2} \bigg)
	+ 6{\cal L}_v \bigg( 2x - 2 + \frac{1}{x} \bigg)
	\bigg]\nnb\\
%
{\qGq} &:& 0\nnb\\
{\qq^2}&:& -\frac{(1+k^2) M_c^2 \rho\qq^2 \:v}{12 \:\pi^2}\nnb\\
%
{\GGG} &:& \frac{(1+k^2) M_c^2 \gGGG}{3^2 \cdot 2^{12} \:\pi^6} \bigg[
	v \bigg( 6 - \frac{25}{x} + \frac{1}{x^2} \bigg)
	+ 6{\cal L}_v \bigg( 2x + 2 + \frac{1}{x} \bigg)
	\bigg]\nnb\\
%
{\qq \qGq} &:& \frac{(1+k^2) \qq \qGq}{3 \cdot 2^{4} \:\pi^2} \:v \bigg[
	1 - \frac{M_c^2 \:\tau}{x} \Big( 1 - 2M_c^2 \:\tau \Big)	\bigg]
\end{eqnarray}}
\subsection{$(1^{-})$ Vector State}
%
{\footnotesize
\begin{eqnarray}
pert &:& \frac{(1+k^2) M_Q^8}{5\cdot3^2\cdot 2^{13} \:\pi^6} \bigg[ 
	v \bigg( 840x + 140 + \frac{5248}{x} - \frac{1164}{x^2} - \frac{182}{x^3} + \frac{5}{x^4} \bigg)\nnb\\
	&&+ 120{\cal L}_v \bigg( 14x^2 + 15 - 18 \log x - \frac{32}{x} + \frac{9}{x^2} \bigg)  - 4320 \:{\cal L}_+
	\bigg] \hspace{1cm}\nnb\\
%
\qq&:& \frac{(1-k^2) M_Q^5 \qq}{3^2 \cdot 2^8 \:\pi^4} \bigg[ 
	v \bigg( 60x - 62 + \frac{26}{x} + \frac{3}{x^2} \bigg)
	+ 24{\cal L}_v \bigg( 5x^2 - 6x + 3 - \frac{1}{x} \bigg)
	\bigg] \nnb
	\eea
\bea
%
\GG &:& \frac{(1+k^2) M_Q^4 \gGG}{ 3^2 \cdot 2^{11} \:\pi ^6} \bigg[ 
	v \bigg( 6 - \frac{5}{x} - \frac{1}{x^2} \bigg)
	+ 6{\cal L}_v \bigg( 2x - 2 + \frac{1}{x} \bigg)
	\bigg]\nnb\\
%
\qGq &:& -\frac{(1-k^2) M_Q^3 \qGq}{3^2 \cdot 2^8 \:\pi^4} \bigg[
	v \bigg( 42x + 7 - \frac{22}{x} \bigg)
	+ 6{\cal L}_v \bigg( 14x^2 - 3 + \frac{2}{x} \bigg)
	\bigg]\nnb\\
%
\qq^2&:& -\frac{(1+k^2) M_Q^2 \rho\qq^2 \:v}{12 \:\pi^2}\nnb\\
%
\GGG &:& \frac{(1+k^2) M_Q^2 \gGGG}{3^3 \: 2^{14} \:\pi^6} \bigg[
	v \bigg( 132x + 22 - \frac{190}{x} + \frac{9}{x^2} \bigg)
	+ 24{\cal L}_v \bigg( 11x^2 + 3 + \frac{2}{x} \bigg)
	\bigg]\nnb\\
{\qq  \qGq}&:& \frac{(1+k^2) \qq \qGq}{3 \cdot 2^{5} \:\pi^2} \:v \bigg[
	1 - \frac{M_c^2 \:\tau}{x} \Big( 1 - 4M_c^2 \:\tau \Big)	\bigg]~.
\end{eqnarray}}
%

\end{document}